\documentclass[traditabstract]{aa}
\usepackage{graphicx}
\usepackage{url}
\usepackage{lscape}
\usepackage{longtable}
\usepackage{multirow}
\usepackage{txfonts}
\newcommand{\hi}{H\textsc{i}}   

\begin{document}
   \title{Cold gas properties of the \textit{Herschel} Reference Survey. I. $^{12}$CO(1-0) and H{\sc i} data \thanks{Tables 1, 2, 10, 11, 12 
   are available in electronic form at the CDS via anonymous ftp to cdsarc.u-strasbg.fr(130.79.128.5) or via http://cdsweb.u-strasbg.fr/cgi-bin/qcat?J/A+A/}}
  \author{A. Boselli\inst{1},
	  L. Cortese\inst{2,3},
  	  M. Boquien\inst{1,4}
         }
\authorrunning{Boselli et al.}
\titlerunning{$^{12}$CO(1-0) and H{\sc i} data of the HRS}	 
	 
\institute{Aix Marseille Universit\'e, CNRS, LAM (Laboratoire d'Astrophysique de Marseille) UMR 7326, 13388, Marseille, France \email{Alessandro.Boselli@lam.fr}
           \and Centre for Astrophysics \& Supercomputing, Swinburne University of Technology, Mail H30, PO Box 218, Hawthorn, VIC 3122, Australia \email{lcortese@swin.edu.au}
           \and European Southern Observatory, Karl-Schwarzschild Str. 2, 85748 Garching bei Muenchen, Germany
           \and Institute of Astronomy, University of Cambridge, Madingley Road, Cambridge CB3 0HA, UK \email{mboquien@ast.cam.ac.uk}
}

   \date{}

 
  \abstract
   {We present new $^{12}$CO(1-0) observations of 59 late-type galaxies belonging to the \textit{Herschel} Reference Survey (HRS), a complete K-band-selected, volume-limited 
   (15 $\lesssim$ $D$ $\lesssim$ 25 Mpc) sample of nearby galaxies spanning a wide range in morphological type and luminosity. We studied different recipes to correct 
   single-beam observations of nearby galaxies of different sizes and inclinations for aperture effects. This was done by comparing
   single-beam and multiple-beam observations along the major axis, which were corrected for aperture effects using different empirical or analytical prescriptions, to integrated maps of several 
   nearby galaxies, including edge-on systems observed by different surveys. 
   The resulting recipe is an analytical function determined by assuming that late-type galaxies are 3D exponentially
   declining discs with a characteristic scale length $r_{CO}$ = 0.2 $r_{24.5}$, where $r_{24.5}$ is the optical, $g$- (or B-) band isophotal radius at the 24.5 mag arcsec$^{-2}$
   (25 mag arcsec$^{-2}$), as well as a scale height $z_{CO}$ = 1/100 $r_{24.5}$. 
   Our new CO data are then combined with those available in the literature to produce the most updated catalogue of CO observations for the HRS, 
   now including 225 out of the 322 galaxies of the complete sample.
   The 3D exponential disc integration is applied to all the galaxies of the sample to measure their total CO fluxes, which are later transformed into molecular gas
   masses using a constant and a luminosity-dependent $X_{CO}$ conversion factor. We also collect H{\sc i} data for 315
   HRS galaxies from the literature and present it in a homogenised form. 
     }
   {}
   {}
   {}
   {}
   {}


   \maketitle
%

\section{Introduction}

The \textit{Herschel} Reference Survey (HRS) is a complete sample of nearby galaxies defined to study the physical properties of the interstellar medium (ISM)
in galaxies of different morphological type and luminosity (Boselli et al. 2010a). Composed of 322 galaxies, it has recently been observed by \textit{Herschel}
with the SPIRE (Ciesla et al. 2012) and PACS (Cortese et al. in prep.) instruments. To provide the community with the largest possible set of homogeneous data, our team 
has undertaken several observational campaigns or collected data from the literature to cover the widest possible range in wavelengths. These data include deep UV imaging with GALEX (Cortese et al. 2012a; 
Boselli et al. 2011), mid-infrared imaging with IRAC on \textit{Spitzer} and \textit{WISE} (Ciesla et al. in prep.), MIPS \textit{Spitzer} photometry (Bendo et al. 2012), H$\alpha$ imaging 
(Boselli et al. in prep.), and medium-resolution integrated optical spectroscopy (Boselli et al. 2013a). \\

This paper presents new CO observations of 59 HRS galaxies obtained at the Kitt Peak 12m radiotelescope. These new CO
data, combined with those available for both the other HRS objects and for other nearby galaxies recently mapped in the $^{12}$CO(1-0) line
with various instruments, are used to compare different prescriptions to correct for aperture-effects single-beam observations.
The new CO data, combined with those already available in the literature for the rest of the sample, are homogenised to produce a complete catalogue
of CO fluxes and realistic uncertainties for all the observed HRS. We also collected H{\sc i} data from the literature and homogenised it
for nearly the whole sample. These CO and H{\sc i} data are crucial for any detailed study of the physical properties of the ISM. 

The cold gas is the dominant phase of the ISM in late-type galaxies. It exceeds in mass the dust component by $\sim$ a factor of 100-200 (Sodroski et al. 1994). This cold gas component 
can be observed easily in local galaxies. The atomic hydrogen (H{\sc i})
can be observed directly through the emission of the spin inversion line at 21 cm (1420 Mhz). Because of its symmetric structure, the molecular hydrogen (H$_2$)
has no permanent electric dipole moment. Dipole rotational transitions are thus strongly forbidden, making it very hard to directly observe
the cold phase of this molecule, which in late-type galaxies has generally a temperature of $\sim$ 10 K.    
For this reason, the molecular hydrogen mass is generally determined through observing of the second most aboundant molecule in the
cold ISM, carbon monoxide, under the assumption that CO is a good tracer of H$_2$ (Young \& Scoville 1991). Indeed it has been shown that the 
dynamical mass of the giant molecular clouds is tightly related to the intensity of the CO line. The most commonly used method 
of determining the molecular hydrogen mass is based on observing of the $^{12}$CO(1-0) rotational line at 2.6 mm (115 GHz)
and assuming a given conversion factor $X_{CO}$  between the intensity of the CO line and the H$_2$ column density (see Boselli 2011).\\

In recent years, several works have questioned the existence of a standard conversion factor, but rather indicated that $X_{CO}$ might change 
with the different physical conditions characterising the properties of the ISM, such as the hardness of the ionising and non-ionising stellar radiation fields, 
and the metallicity and the density of the gas (Boselli et al. 2002, Bell et al. 2007, Bolatto et al. 2008, Liszt et al. 2010, Leroy et al. 2011, 
Shetty et al. 2011a,b, Narayanan et al. 2012, Bolatto et al. 2013, Sandstrom et al. 2013). We thus determined molecular gas masses using both a constant
or a luminosity-dependent $X_{CO}$ conversion factor. As constant value we used the one determined in the Milky Way using $\gamma$-ray absorption 
data from COBE ($X_{CO}$ = 2.3 10$^{20}$ cm$^{-2}$/(K km s$^{-1}$), or else, if expressed as $\alpha_{CO}$ = 3.6 M$_{\odot}$/(K km s$^{-1}$
pc$^{-2}$), Strong et al. 1988), which is generally assumed to be representative of environments similar to those encountered in the solar neighbourhood in terms
of metallicity and radiation field. As variable conversion factor, we adopt the calibration proposed by Boselli et al. (2002)
based on the H-band luminosity, log $X_{CO}$ [cm$^{-2}$/(K km s$^{-1}$)] = -0.38 $\times$ log $L_H$ [$L_{H\odot}$] + 24.23. 
We chose this calibration because it can be determined easily for all HRS galaxies, for which an H-band 
magnitude is available from 2MASS. Furthermore, this calibration is a mean calibration compared to those proposed in the literature (Bolatto et al. 2013).\\

The paper is structured as follows. Section 2 describes the HRS sample, while Sects. 3 and 4 present the new CO observations. In Sect.
5 we discuss and apply different aperture correction techniques to the whole set of CO data available in the literature for the entire HRS sample,
and we thus derive total CO fluxes and molecular gas masses. Section 6 gives homogenised H{\sc i} data for the whole sample. A brief conclusion is given in Sect. 7.
Despite the presence of several complete samples of nearby galaxies with CO data (Sage 1993; FCRAO survey, Young et al. 1995; Boselli et al. 1997; Sauty et al. 2003; 
Lisenfeld et al. 2011, Saintonge et al. 2011), the HRS is the only one with the complete set of far-infrared and sub-millimetric \textit{Herschel} data necessary
for a detalied study of the properties of the ISM of normal galaxies. At the same time, it is a statistically significant, complete sample of nearby galaxies spanning a 
wide range in morphological type and stellar mass, making it ideally suited to study the cold gas properties of normal galaxies. It can be used, for instance,
to extend to lower stellar masses (by a factor of $\sim$ 10) the reference work of Saintonge et al. (2011) on the COLD GASS sample, which only includes massive galaxies.
We thus devote three other specific works to the analysis of the gas properties of the HRS using this unique set of data. One, already published, analyses the effects
of the environment on the H{\sc i} scaling relations of the sample (Cortese et al. 2011).
A second one is focussed on determining the mean total and molecular gas scaling relations 
of the HRS (Boselli et al. 2013b), while the last one treats the effects of the environment on the molecular gas content of cluster objects (Boselli et al. 2013c).
The interested reader can find the results of other works based on the combined analysis of the \textit{Herschel} and the other multifrequency data in 
Cortese et al. (2012b; dust scaling relations along the Hubble sequence), Smith et al. (2012; dust properties of early-type galaxies), Boselli et al. (2012;
far infrared colours), Boquien et al. (2012, 2013; dust attenuation properties in resolved galaxies) or in the publication of several papers 
during the science demonstration phase of the instrument (Boselli et al. 2010b; 
Cortese et al. 2010; Eales et al. 2010; Gomez et al. 2010; Pohlen et al. 2010; Sauvage et al. 2010).

\section{The sample}

The \textit{Herschel} Reference Survey (HRS) is an \textit{Herschel}/SPIRE guaranteed time key project aimed at observing a complete, 
K-band-selected (K $\leq$ 8.7 mag for early-types, K $\leq$ 12 for type $\geq$ Sa), volume-limited (15$\lesssim$ $D$ $\lesssim$ 25 Mpc)
sample of nearby galaxies. The HRS survey, as well as the selected sample, are extensively presented in Boselli et al. (2010a). Briefly, the sample
is composed of 322 galaxies out of which 260 are late-type systems\footnote{With respect to the original sample 
given in Boselli et al. (2010a), we removed the galaxy HRS 228 whose new redshift indicates it as a background object.
We also revised the morphological type for 6 galaxies: 
NGC 5701, now classified as Sa; NGC 4438 and NGC 4457, now Sb; NGC 4179, now S0; VCC 1549, now dE; and NGC 4691 now Sa.}. 
Galaxies were selected in the K-band taken as a proxy for galaxy stellar mass (Gavazzi et al. 1996). 
The sample includes objects in environments of 
different densities, from the core of the Virgo cluster to loose groups and fairly isolated systems. As defined, the present
sample is ideal for any statistical study of the mean galaxy population of the nearby universe.\\
The galaxies observed in this work are late-type systems of K-band magnitude brighter than $K$$<$ 10, expressely selected to complete the sample down
to this K-band magnitude limit (142/151 of the late-types). Including our new observations, 225 out of the 322 HRS galaxies (143 detections), 
168/260 for late-type (134 detected), and 57/62 for early-type systems (9 detected) have $^{12}$CO(1-0) data.\\
We also compile and homogenise 21 cm H{\sc i} data from the literature for all galaxies in the HRS. H{\sc i} data are available for 315 out of the 322
galaxies of the sample. The whole HRS sample with its main characteristics is given in Table \ref{TabHRS}, arranged as follows:

\begin{itemize}
\item {Column 1: \textit{Herschel} Reference Sample (HRS) name, from Boselli et al. (2010a).}
\item {Column 2: Zwicky name, from the Catalogue of Galaxies and of Cluster of Galaxies (CGCG; Zwicky et al. 1961-1968).}
\item {Column 3: Virgo Cluster Catalogue (VCC) name, from Binggeli et al. (1985).}
\item {Column 4: Uppsala General Catalog (UGC) name (Nilson 1973).}
\item {Column 5: New General Catalogue (NGC) name (Dreyer 1888).}
\item {Column 6: Index Catalogue (IC) name (Dreyer 1908).}
\item {Columns 7 and 8: J2000 right ascension and declinations, from NED.}
\item {Column 9: Morphological type, from NED, or from our own classification if not available.}
\item {Column 10: Distance, in Mpc. Distances have been determined from the recessional velocity assuming a Hubble constant $H_0$ = 70 km s$^{-1}$ Mpc$^{-1}$
for galaxies outside the Virgo cluster, and assumed to be 23 Mpc for galaxies belonging to the Virgo B cloud, 17 for the other Virgo members (Gavazzi et al. 1999).}
\item {Column 11: Total K-band magnitude ($K_S$$_{tot}$), from 2MASS (Skrutskie et al. 2006).}
\item {Column 12: $g$-band optical isophotal diameter (24.5 mag arcsec$^{-2}$), from Cortese et al. (2012a). For the HRS galaxies without SDSS images, the $g$-band isophotal diameter
was determined from the relation $r_{24.5}(g)$ = 0.871($\pm$0.017)$r_{25}(B)$ + 6.041($\pm$2.101) (Spearman correlation coefficient $\rho$ = 0.92), where $r_{25}(B)$
is the radius given in NED. This relation was determined using the HRS galaxies with both sets of data.}
\item {Column 13: inclination of the galaxy, determined using the prescription based on the morphological type described in Haynes \& Giovanelli (1984)
and the $i$-band ellipticity given in Cortese et al. (2012a).}
\item {Column 14: Heliocentric radial velocity (in km s$^{-1}$), from H{\sc i} data when available, otherwise from NED.}
\item {Column 15: Cluster or cloud membership, from Gavazzi et al. (1999) for Virgo and Tully (1988) or Nolthenius (1993) whenever available, 
or from our own estimate otherwise. }
\item {Column 16: Code to indicate whether H{\sc i} data are available (1) or not (0). }
\item {Column 17: Code to indicate whether CO data are available (1) or not (0).}
\end{itemize}

\section{Observations}

CO observations were carried out during four remote-observing runs from 
the Laboratoire d'Astrophysique de Marseille in spring 2008 and 2009 
using the NRAO Kitt Peak 12 m telescope
\footnote{The Kitt Peak 12-m telescope was operated by the Arizona Radio Observatory}.
One hundred forty-one hours were allocated to this project, out of which $\sim$ 20 hours lost for bad weather conditions or technical problems.
At 115 GHz [$^{12}$CO(1--0)], the telescope's 
half-power beam width (HPBW) is 55", which corresponds to 5.3 kpc 
at the average distance of 20 Mpc of the HRS galaxies. 
Weather conditions were fairly good, with typical 
zenith opacities of 0.15-0.25. 
The pointing accuracy was checked every 
night by broad band continuum observations of nearby planets or the radiogalaxy 3C273, with an 
average error of 5" rms. We used a dual-polarisation SIS mixer, with a 
receiver temperature for each polarisation of about $T_{sys}$=300-400 K 
(in $T^*_R$ scale) at the elevation of the sources. We used a dual beam-switching 
procedure, with two symmetric reference positions offset by 4' in 
azimuth. The backend was a 256 channel filter bank spectrometer with channel 
width of 2 MHz. Each six-minute scan began by a chopper wheel 
calibration on a load at ambient temperature, with an OFF-ON-ON-OFF set of pointings for each scan. Galaxies were observed 
at their nominal coordinates listed in Table 1. Fefteen objects with an optical diameter exceeding three arcminutes
were also mapped along the major axis, with two one-beam off positions. The face-on galaxy HRS 42 (NGC 3596)
was observed along a cross with one beam offset. 
The total integration time was on average 120 minutes ON+OFF (i.e. 60 
minutes on the source), yielding rms noise levels of about 3 mK (in the 
$T^*_R$ scale) after velocity smoothing to 15.7 km s$^{-1}$. The baselines 
were flat owing to the use of beam-switching, thereby requiring only 
linear baselines to be subtracted. 
The antenna temperature $T^*_R$ was corrected for telescope and 
atmospheric losses. In the following analysis we use the $T_R^*$ scale. This scale can be transformed into
the main-beam brightness temperature scale, $T_{mb}$, with $T_{mb}$=$T^*_R$/0.84 
(where the main beam efficiency is $\eta$$_{mb}$=0.54 and the forward 
scattering and spillover efficiency $\eta$$_{fss}$=0.68).
The $T_R^*$ scale can be converted into flux using 39 Jy/K.

\section{Results}

\subsection{Results of our observations}

The $^{12}$CO(1--0) spectra of all the observed galaxies, reduced with the 
CLASS package (Forveille et al. 1990), are shown in Fig. 1. The 
observational results are listed in Table 2. Of the 59 observed galaxies, 13 
were not detected. Table 2 is arranged as follows:

\begin{itemize}

\item {Column 1: HRS name.} 
\item {Columns 2 and 3: R.A. and Dec. pointing offset, in arcsec.}
\item {Column 4: detection flag, 1 for detected, 0 for undetected sources.}
\item {Column 5: observing run: 1,2 and 3 are for the three 2008 runs, 4 for the 2009 run. }
\item {Column 6: rms noise, in mK, on the $T^*_R$ scale, measured after the spectra are smoothed to a velocity resolution $\delta V_{CO}$=15.7 km s$^{-1}$}.
\item {Column 7: number of scans (ON+OFF). Each scan is six minutes long.}
\item {Column 8: Intensity of the $I(CO)$ line ($I(CO)$=$\int$$T^*_R dv$) in K km s$^{-1}$ (area 
definition, which corresponds to the area under the profile of the line measured in between the 
upper and lower limits in velocity within which the line is detected. Galaxies are considered as detected whenever the signal-to-noise, defined as 
$S/N = \frac{I(CO)}{\Delta I(CO)}$
is greater than 2, where $\Delta I(CO)$ is defined in eq. (2). 
For undetected galaxies, the reported value is an upper limit determined as follows: 

\begin{equation}
{I(CO)=5 \sigma (W_{HI} \delta V_{CO})^{1/2}	 {\rm K~km~s^{-1}}}
\end{equation}

where $\sigma$ is the rms noise of the spectrum, $W_{HI}$ the H{\sc i} line width, and 
$\delta V_{CO}$ the spectral resolution (here taken at $\delta V_{CO}$=15.7 km s$^{-1}$).
For galaxies with $W_{HI}$ unavailable, the 
H{\sc i} width has been determined assuming a standard $W_{HI}$ = 300 sin($i$)  
km s$^{-1}$, where $i$ is the galaxy inclination, with a minimum value of 
$W_{HI}$ = 50 km s$^{-1}$ for almost edge-on systems. }
\item {Column 9: Error on the intensity of the CO line, $\Delta$$I(CO)$, 
computed as 

\begin{equation}
{\Delta I(CO)=2 \sigma (W_{CO} \delta V_{CO})^{1/2} 
{\rm {K~km~s ^{-1}}}}
\end{equation}

where $\sigma$ is the rms noise of the spectrum, $W_{CO}$ 
the CO linewidth (given in Col. 11), and $\delta V_{CO}$ the spectral resolution. }
\item {Column 10: Heliocentric velocity determined from the CO line (Gaussian 
fit), in km s$^{-1}$ (optical definition $v$=$cz$=$\Delta\lambda/\lambda_0$). 
The estimated error is 
comparable to the resolution, thus $\sim$ 15 km s$^{-1}$.}
\item {Column 11: Full width at zero level of the CO line, 
in km s$^{-1}$, with an estimated 
error of $\sim$ 20 km s$^{-1}$. For galaxies with a suffix $a$ the width of the CO line is the full width half maximum (FWHM) determined from a Gaussian fit.
The width of the CO line generally corresponds to the
width of the H{\sc i} line, indicated by the red horizontal line in Fig. 1 .}
\item {Column 12: Peak temperature, in $T_R^*$ scale (K).}
\end{itemize}

\subsection{Uncertainty on the $I(CO)$ data}

IN column 9 $\Delta I(CO)$ gives the error on the intensity of the CO line as determined from the typical rms of the observed galaxies.
A different estimate of this uncertainty can be determined by comparing independent measurements of the same galaxy.
All galaxies observed in this work are new CO observations and thus do not have any 
similar data in the literature. The only exception is HRS 48 (NGC 3631), observed at the beginning of each run for checking the tuning of the instrument on 
a CO-bright source and for testing the repeatability of the observations. We thus have three different CO observations of the same galaxy that can be compared. The same galaxy also has two other
independent single-beam observations, one from the FCRAO survey (Young et al. 1995) and a second one from the CO survey of nearby galaxies done with the Onsala radio telescope by Elfhag et al. (1996).
Table \ref{TabHRS48} gives the different $I(CO)$ values obtained within our observing runs and in the literature. Given that the data obtained by the FCRAO (Young et al. 1995) and the Onsala surveys have been
obtained in different beams and on different temperature scales, we also compare the total extrapolated CO fluxes obtained as described in the following sections. 
Table \ref{TabHRS48} shows that by using exactly the same telescope configuration we have differences from one to another observing run in the CO flux estimate of HRS 48 as large as $\simeq$ 30 \%. 
The comparison with other data available in the literature, here done on the extrapolated CO fluxes of HRS 48 to remove any first-order dependence on the aperture correction and
telescope temperature scale, gives differences with the mean value obtained in this work of the order of $\simeq$ 8-13 \%. These differences are significantly smaller than 
those observed within our own independent data for the bright HRS 48. We recall, however, that for this particular object, 
our three independent observations were done with a significantly smaller number of scans ($\sim$ 6 scans each)
than our typical observations of the other HRS galaxies ($\sim$ 20 scans). Given that the uncertainty on the CO intensity $I(CO)$ depends on the rms of the spectrum (see eq. 2), 
we thus expect that for HRS 48, the quite large dispersion in our data is partly due to the low integration time. Table \ref{TabHRS48} shows, however, that the difference between our own
observations of HRS 48 are larger than the expected uncertainties. Given the peaked distribution of the emitting molecular gas in galaxies, it is conceivable that small pointing offstes 
can be the origin of this discrepancy. In other words, the relation given in the previous section to estimate the uncertainty on single-beam observations slightly underestimates $\Delta I(CO)$.
The error on the CO central beam observations done in this work should thus be of the order of 30\%.

\section{CO data from the literature}

\subsection{Unit transformation}

The $^{12}$CO(1-0) data available in the literature come from a wide variety of references and for this reason are often given on different scales, as summarised in Table \ref{Tabtel}. 
Data coming from the FCRAO telescope are generally presented on the $T_A^*$ scale (observed source antenna temperature corrected for atmospheric attenuation, radiative loss, and rearward scattering
and spillover), those from the 12 m Kitt Peak telescope on the $T_R^*$ scale (observed source antenna temperature corrected for atmospheric attenuation, radiative loss, and rearward and forward scattering
and spillover),
while those from the IRAM 30, the SEST, and the Onsala telescopes on the $T_{mb}$ scale (source brightness temperature as measured by the main diffraction beam of the telescope). 
The different temperature scales can be transformed into a common scale
following the prescription of Kutner \& Ulich (1981). The $T_R^*$ temperature can be transformed into the antenna temperature as 

\begin{equation}
{T_R^* = T_A^*/\eta_{fss}},
\end{equation}
\noindent
where $\eta_{fss}$ is the forward spillover and scattering efficiency, while the main beam temperature can be obtained as 

\begin{equation}
{T_{mb} = T_A^*/\eta_{mb}},
\end{equation}
\noindent
where $\eta_{mb}$ is the main beam efficiency.
For the definition of a complete and consistent set of data for the HRS galaxies 
it is thus necessary that all these observations are homogenised on a commun scale. We thus decided to provide, for all galaxies, $^{12}$CO(1-0) fluxes in units of Jy km s$^{-1}$.
For a source with a Gaussian profile observed with a Gaussian beam of similar size, the constant necessary to transform main beam temperatures (in K) into fluxes (in Jy) is
given by the relation (Wilson et al. 2009) :

\begin{equation}
{S(Jy)/T_{mb}(K) = 8.18 \times 10^{-7} \left(\frac{\theta}{\rm{arcsec}}\right)^2 \left(\frac{\nu}{\rm{GHz}}\right)^2},
\end{equation}
\noindent
where $\theta$ is the FWHM of the beam and $\nu$ is the frequency.
For sources extended with respect to the beam, as is the case for all HRS galaxies, the transformation of main beam temperatures into Jy is much more complex and requires the
antenna pattern and the distribution of the emitting source (Wilson et al. 2009). Since a priori the CO distribution of the emitting galaxies within the antenna pattern is unknown, 
astronomers generally transform antenna temperatures into CO fluxes using a constant whose typical value for each telescope, determined with eq. 5, is listed in Table \ref{Tabtel}. 
For consistency with previous works, however, we adopt the same constants as used in the literature and as given in the original references where the 
CO data have been published. We prefer to use the values published in the original references just to make it easy any direct comparison with the already published data.
These conversion constants might differ ($\simeq$ 15\%) slightly from those given in Table \ref{Tabtel}, which are mean values determined for galaxies at redshift $z$=0.  

\subsection{Integrated data of mapped galaxies}

\subsubsection{Total CO fluxes}

Some of the 225 HRS galaxies with CO data have been completely mapped either with a multibeam detector or in interferometric mode 
and have thus high-quality integrated CO fluxes. Of these, 18 objects have been observed by Kuno et al. (2007) 
with the 25 beams BEARS array mounted on the 45 m Nobeyama Radio Observatory telescope, 19 objects by Chung et al. (2009a)
with the 16 beams SEQUOIA array on the 13.7m FCRAO radio telescope, and 2 other galaxies with the IRAM 30
m radio telescope by our team (see below). Ten galaxies have also been observed during the BIMA SONG survey of nearby 
galaxies with the Owens Valley Radio Observatory in interferometric mode by Helfer et al. (2003). To these, we can
add 71 galaxies observed along the major axis during the FCRAO CO survey of nearby galaxies by Young et al. (1995).
Another object, NGC 4565, has also been partly mapped with the IRAM 30 m (Neininger et al. 1996) and 
with the 45 m Nobeyama (Sofue \& Nakai 1994) radiotelescopes.
Because of the complete coverage of the stellar disc, in particular for the single-dish observations done at the Nobeyama, IRAM, and FCRAO radio telescopes, 
we consider that all these data are of much higher quality than 
other single-beam observations that require aperture corrections. We thus decided to use mapped observations whenever possible instead of single beam data.\\

   \begin{figure*}
   \centering
   \includegraphics[width=0.48\textwidth]{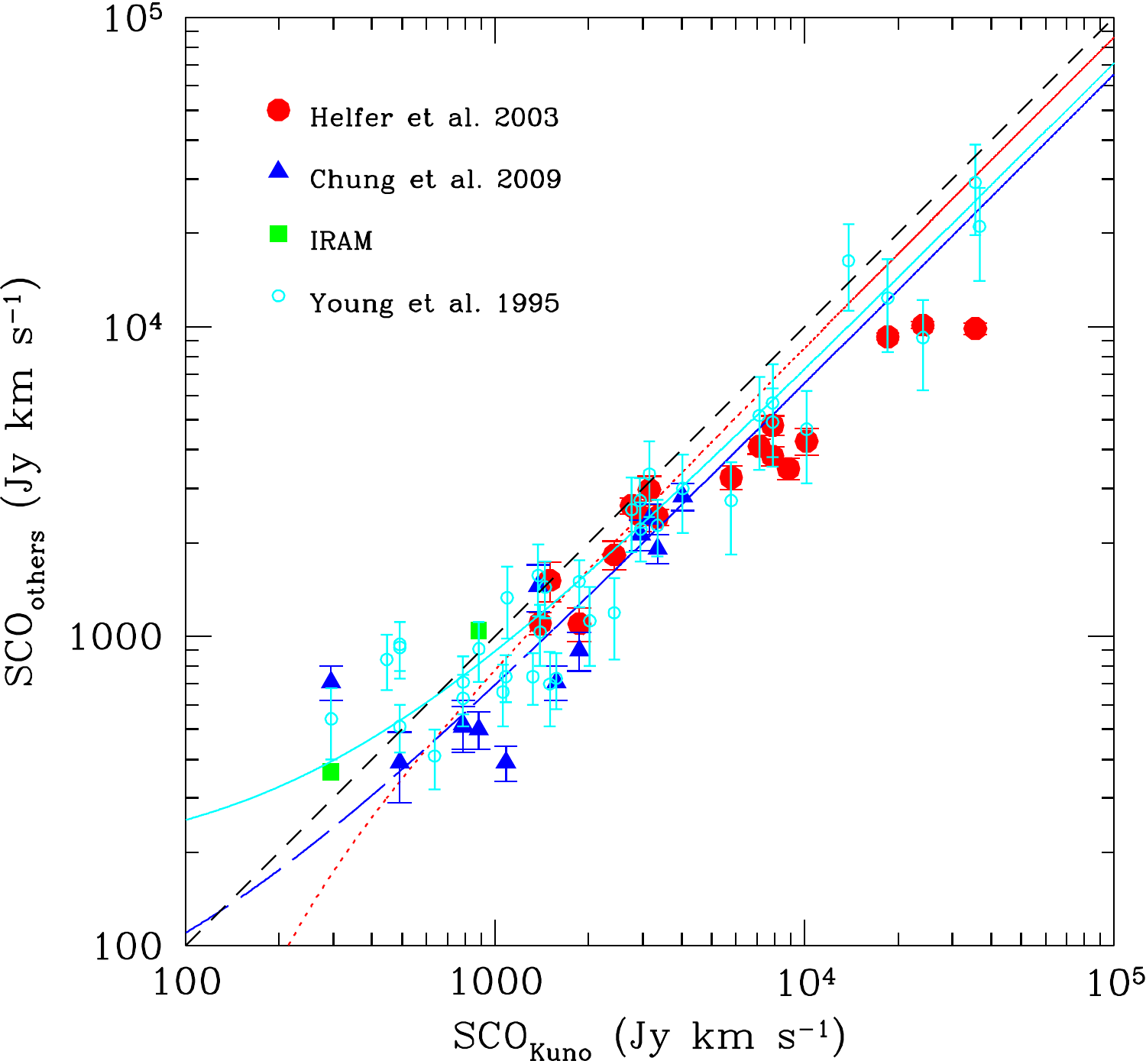}
   \includegraphics[width=0.48\textwidth]{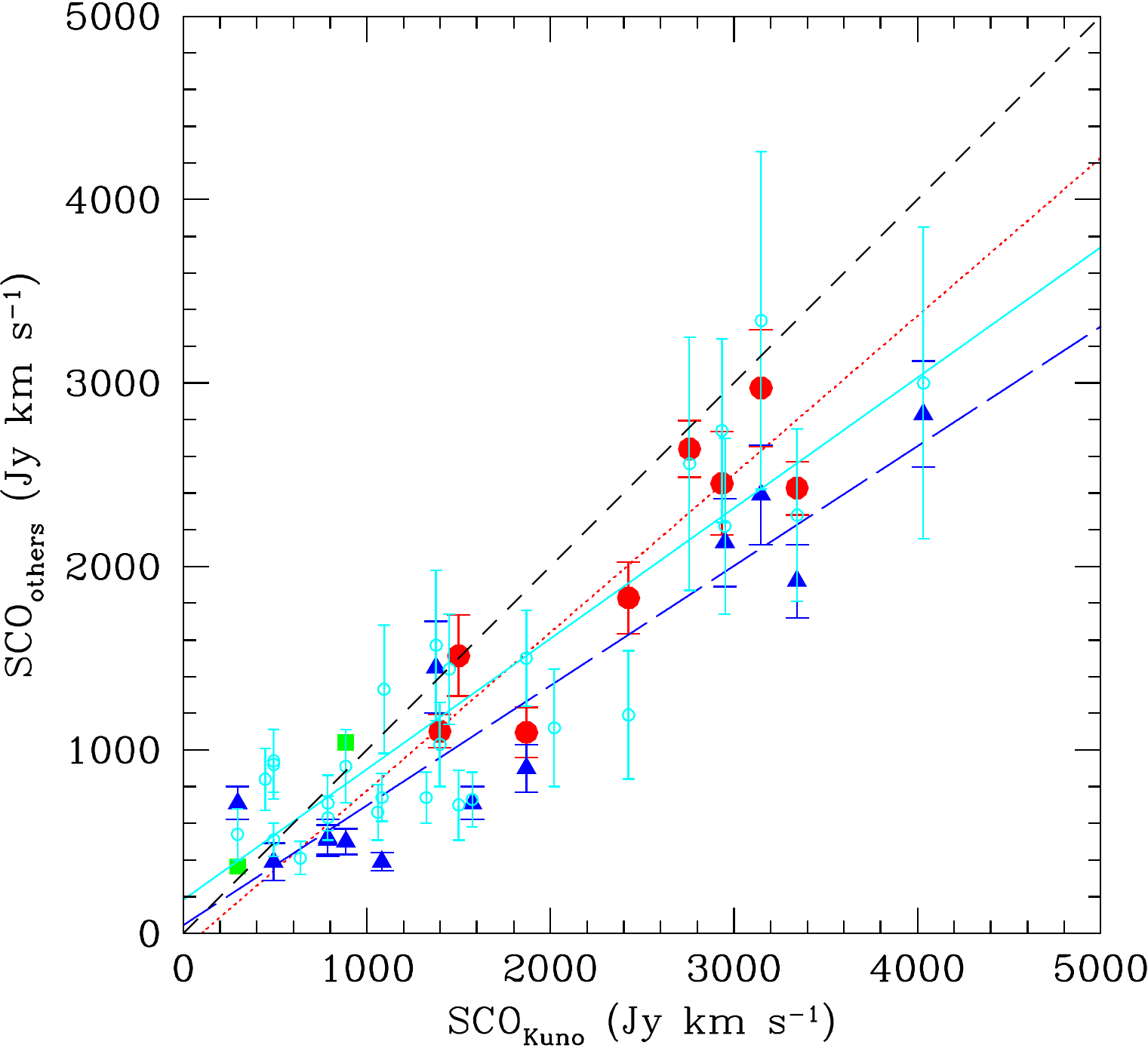}
   \caption{Comparison between the integrated $SCO$ fluxes determined by Kuno et al. (2007) and those determined by Helfer et al. (2003; red filled dots),
   Chung et al. (2009a; blue filled triangles), our own observations of the two galaxies NGC 4548 and NGC 4579 with the IRAM radiotelescope (green filled squares)
   and the extrapolated fluxes of the FCRAO survey (Young et al. 1995; cyan empty circles) on logarithmic scale for all galaxies in common (left) and on linear scale 
   for those objects with $SCO$ in the range 100-5000 Jy km s$^{-1}$ sampled by the HRS (right). The linear best fit to the data, determined assuming the Kuno et al. (2007) data as
   independent variable in the range 100-5000 Jy km s$^{-1}$, are shown by the red dotted line (Helfer et al. 2003), 
   the blue long dashed line (Chung et al. 2009a) and the cyan solid line (Young et al. 1995).
   The short dashed black line shows the 1:1 relationships.}
   \label{integrate}%
   \end{figure*}

A few of these objects have integrated data taken by more than one survey. We compared the different sets of data to isolate those 
that we consider of higher quality. To do that, we cross-matched the different surveys, 
including galaxies outside the HRS sample (22 extra objects) to increase the statistics, and compared the different sets of data in Fig. \ref{integrate}.
All data were first transformed into CO fluxes (in Jy km s$^{-1}$) using the prescriptions indicated in the original papers for a correct comparison.
For the FCRAO data of Young et al. (1995) we used the extrapolated $SCO$ fluxes provided in this reference in the comparison. 
Figure \ref{integrate} shows that the multibeam observations of Kuno et al. (2007) and those obtained by our team using the IRAM radio telescope 
agree within 20\% ($SCO_{Kuno}/SCO_{IRAM}$=0.83$\pm$0.03), while a certain scatter is present when the data of Kuno et al. (2007) are compared with
those obtained by the other teams. All the other datasets give $SCO$ fluxes that are generally smaller than those of Kuno et al. (2007).
The difference between the Kuno et al. (2007) and those obtained by the other surveys does not seem to depend on the total flux of the emitting sources, with the 
exception of the BIMA SONG data of Helfer et al. (2003). Here there is a clear deviation from the one-to-one relation at high $SCO$ fluxes, that could be easily
ascribed to a non complete coverage of the stellar disc in the more extended sources combined with a slightly lower sensitivity to the diffuse emission of interferometric observations 
with respect to the multibeam observations done by Kuno et al. (2007). The difference with Chung et al. (2009a) might also come from 
a higher sensitivity of the Nobeyama radio telescope with respect to the FCRAO. The difference with the $SCO$ fluxes of Young et al. (1995) 
probably comes from the rough extrapolation technique applied to the major axis CO data. The difference between these two sets of data is, however, remarkably
small, indicating that overall the total $SCO$ flux of spiral galaxies can be fairly well deduced using a very simple observing technique combined with an appropriate aperture
correction. \\
This observational evidence suggests that the set of data of Kuno and collaborators be considered as reference, and 
the others be corrected using best fitting relations to minimise the scatter in the comparison with Kuno et al. (2007). To avoid any systematic effect in the correction due to
size, we fit the Helfer et al. (2003) and Young et al. (1995) vs. Kuno et al. (2007) CO flux relations using only CO data in the range
100 $<$ $SCO$ $<$ 5000 Jy km s$^{-1}$, which is the range in the CO flux covered by the HRS galaxies. No restriction in the Chung et al. (2009a) data
is required given that the fluxes of all their galaxies are within this range. 
The best linear fits to the data, shown in Fig. \ref{integrate}, 
are given in Table \ref{Tabfitintegrati}.

For the Young et al. (1995) data, however, the proposed correction does not significantly decrease the mean ratio and the scatter in the $SCO_{Kuno}/SCO_{Young}(corrected)$ ratio. 
Indeed, the mean ratio $SCO_{Kuno}/SCO_{Young}$ determined using the full set of original data published by Young et al. (1995) for all galaxies in common between these two surveys 
is $SCO_{Kuno}/SCO_{Young}$ = 1.10 $\pm$ 0.44, 
while the one determined using the corrected data is $SCO_{Kuno}/SCO_{Young}(corrected)$ = 0.90 $\pm$ 0.34. The majority of the HRS galaxies mapped by Young et al. (1995) 
without any other higher quality data from Kuno et al. (2007), Helfer et al. (2003), or Chung et al. (2009a), however, have only three detected beams along the major axis, and
their total CO flux is $SCO_{Kuno}$ $<$ 5000 Jy km s$^{-1}$. If we restrict the comparison of the Young et al. (1995) and Kuno et al. (2007) data to the subsample of HRS galaxies
in common, the ratio determined with the originally published data, $SCO_{Kuno}/SCO_{Young}$ = 1.29 $\pm$ 0.51, can be compared to that determined using the corrected data, 
$SCO_{Kuno}/SCO_{Young}(corrected)$ = 1.33 $\pm$ 0.55. The proposed correction does not introduce any significant improvement in the data. 
We thus decided not to apply any correction to the FCRAO Young et al. (1995) data of the HRS galaxies.\\

   \begin{figure}
   \centering
   \includegraphics[width=\columnwidth]{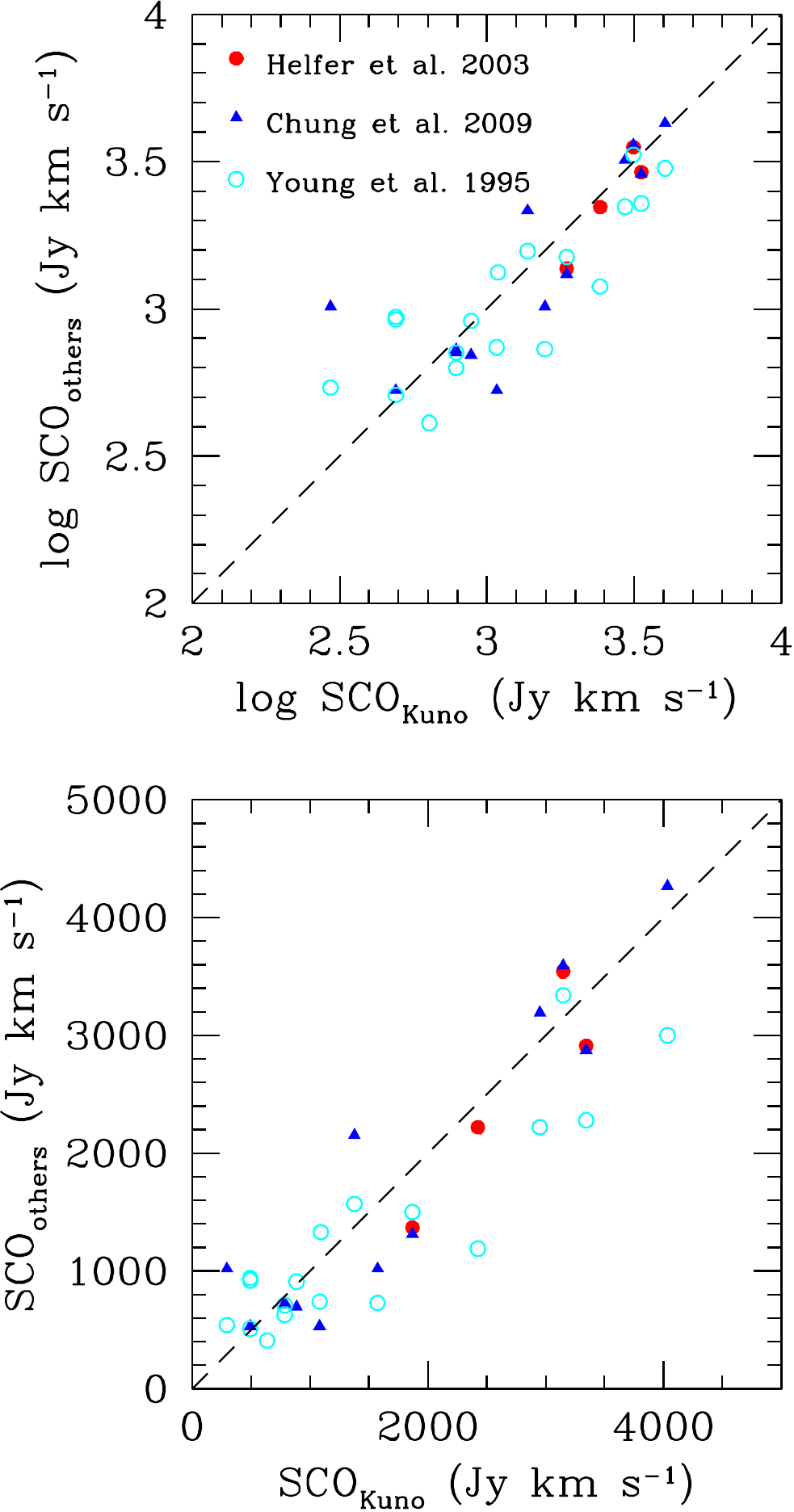}
   \caption{Comparison between the integrated $SCO$ flux determined by Kuno et al. (2007) and those determined by Helfer et al. (2003; red filled dots),
   Chung et al. (2009a; blue filled triangles), corrected as described in the text, and the extrapolated fluxes of the FCRAO survey 
   (Young et al. 1995; cyan empty circles), for all HRS galaxies in common, on logarithmic (upper panel) and on linear scales
   (lower panel). The short dashed black line shows the 1:1 relationships.}
   \label{integratecor}%
   \end{figure}

Figure \ref{integratecor} shows the relationship between the different integrated CO fluxes corrected using the prescriptions given in Table \ref{Tabfitintegrati}
and those of Kuno et al. (2007) for the HRS galaxies with 100 $<$ $SCO$ $<$ 5000 Jy km s$^{-1}$. Given the dispersion in the corrected relations seen in Fig. 
\ref{integratecor} and in Table \ref{Tabfitintegrati}, we decided to take 
the Kuno et al. (2007), Helfer et al. (2003), Chung et al. (2009a), and Young et al. (1995) data in order of priority 
(whenever multiple datasets are available)\footnote{The only exception is the galaxy HRS 91, NGC 4192, 
for which we adopt the flux given in Young et al. (1995) since much closer to three independent estimates from single beam observations. The flux of Kuno et al. (2007)
is indeed underestimated by a factor of $\sim$ 2, as indicated by the position that this galaxy would take in the $M(H_2)$ vs. $M_{star}$ scaling relation
given in Paper III.}. The quality of the FCRAO data of Young et al. (1995), however,
strongly depends on the number of detected beams along the major axis. When only three beam observations are available, the comparison with the Kuno et al. (2007) data 
indicates that the dispersion in the relation is higher than that we obtained using only the central beam, and we extrapolated it using our own prescriptions (see below).
We thus decided to use the extrapolated CO fluxes of Young et al. (1995) only when more than three detections along the major axis are available; otherwise,
we consider the galaxies as observed only in the central beam and measure their total emission using our own prescriptions. \\
We thus end up with 37 HRS galaxies with integrated data, 17 from Kuno et al. (2007), 0 from Helfer et al. 
(2003)\footnote{All HRS galaxies in the BIMA SONG survey have been observed by Kuno et al. (2007).}, 6 from Chung et al. (2009a), and 13 from Young et al. (1995). 
To these we can add a galaxy, NGC 4565 (HRS 213), partly mapped by Neininger et al. (1996) 
with the IRAM radiotelescope and by Sofue \& Nakai (1994) with the Nobeyama radiotelescope. By correcting their own independent sets of data to extend the measure of the CO
flux to the whole galaxy, they end up with a total flux of $SCO$ = 1672.2 Jy km s$^{-1}$ and $SCO$ = 1643 Jy km s$^{-1}$, respectively. There is also an earlier estimate of
the CO flux of this galaxy by Richmond \& Knapp (1986) from observations taken with the Bell Laboratories radiotelescope, which gives $SCO$ = 4487 Jy km s$^{-1}$. Our
own estimate based on the extrapolation of the central single beam as described in sect. 5.3, gives $SCO_{3D}$ = 1550 Jy km s$^{-1}$, a value much closer to the
estimates of Neininger et al. (1996) and Sofue \& Nakai (1994). We thus decided to take for this object $SCO$ = 1672.2 Jy km s$^{-1}$.

\subsubsection{Uncertainties on integrated data}

Table \ref{Tabfitintegrati} and Fig. \ref{integratecor} can also be used to estimate the typical uncertainty on the integrated data. The comparison 
between Kuno et al. (2007) with our own observations of two HRS galaxies done with the IRAM radio telescope indicates that 
$SCO_{Kuno}/SCO_{IRAM}$ = 0.83 $\pm$ 0.03. Despite the lack of statistics, this comparison indicates that the two sets of data are consistent within $\sim$ 17 \%. 
This uncertainty should be shared between the two different sets of data. The uncertainty on the 
Kuno et al. (2007) CO fluxes is thus assumed to be of the order of 12 \%. The uncertainty in the Helfer et al. (2003) data is also comparable. The one in the Chung et al. (2009a)
data is  $\sim$ 40 \%, while in the Young et al. (1995) extrapolated data slightly larger ($\sim$ 45 \%). It is hard to estimate the uncertainty on the CO
flux of NGC 4565 given that the three independent measurements come from the extrapolation of observations of a fraction of this edge-on galaxy.
Two independent measurements are consistent within 2\%, and the third one is different by a factor of $\sim$ 3. We thus arbitrarily assume for 
this source an uncertainty of $\sim$ 30 \%.

\subsection{Single-beam observations of the centres of galaxies}

\subsubsection{Extrapolation of the central beam}

The majority of the 225 HRS galaxies with CO data have only one single-beam observation in their central position. For these objects the beam of the telescope
does not necessarily cover the whole surface of the stellar disc. These CO data must thus be corrected for aperture effects to derive total CO fluxes  
of the observed galaxies. The extrapolation of the central beam observation can be done using either empirical relations calibrated on nearby mapped galaxies or
analytic functions known to reproduce the radial distribution of the CO emission well.
The first method has been proposed by Saintonge et al. (2011), who simulated the observation of the galaxies mapped by Kuno et al. (2007) with Gaussian beams
of different sizes. Using this technique, the total CO flux $SCO_{Saintonge}(tot)$ of a galaxy with a central beam flux $SCO(CB)$ is given by the relation

\begin{equation}
{SCO_{Saintonge}(tot) = \frac{SCO(CB)}{1.094-0.176*N.Beam+0.00968*N.Beam^2}},
\end{equation}
\noindent
where $N.Beam$ is the size of the galaxy in number of beams:

\begin{equation}
{N.beam = \frac{D_{25}(B)}{\theta}},
\end{equation}
\noindent
with $D_{25}(B)$ the 25 mag arcsec$^{-2}$ isophotal $B$-band diameter\footnote{For the sample of galaxies with integrated values in the literature
used for comparing the different aperture corrections, $D_{25}$ is the B-band isophotal diameter 
given in NED.}, and $\theta$ the FWHM of the beam of the radiotelescope (see Table \ref{Tabtel}), both measured in arcseconds. \\
The second approach has recently been proposed by Lisenfeld et al. (2011). Considering that the CO emission of nearby mapped galaxies is represented well by an 
exponential disc of scale length $r_{CO}$: 

\begin{equation}
{SCO(r) = SCO(0)e^{-r/r_{CO}}},
\end{equation}
\noindent
where $SCO(0)$ is the CO emission in the centre of the galaxy. The total CO flux of an exponential disc is given by the relation:

\begin{eqnarray}
SCO_{2D}(tot)&=&\int_{0}^{2\pi}\int_{0}^{\infty} r SCO(r) d r d\theta,\nonumber\\
&=& \int_{0}^{\infty} 2\pi r SCO(0)e^{-r/r_{CO}}d r,\nonumber\\
&=&2\pi r_{CO}^2 SCO(0).
\end{eqnarray}

As discussed in Lisenfeld et al. (2011), the scale length of the CO emitting disc, $r_{CO}$, is correlated well with the optical scale length of the stellar disc 
and with $r_{25}$, the optical 25 mag arcsec$^{-2}$ isophotal radius\footnote{For the sample of galaxies with integrated values in the literature
used for comparing the different aperture corrections, $r_{25}$ is the isophotal radius derived from the optical diameter 
given in NED, which generally corresponds to the B-band isophotal diameter at 25 mag arcsec$^{-2}$.
For the HRS galaxies we use the $g$-band isophotal radius $r_{24.5}(g)$ given in Table \ref{TabHRS}. }. The observations of the THINGS sample of nearby objects done by Leroy et al. (2008)
indicates that, on average, $r_{CO}/r_{25}$ = 0.2. Lisenfeld et al. (2011) derive a consistent value using different sets of data, from
the BIMA SONG survey of Regan et al. (2001), to the sample of Nishiyama \& Nakai (2001) observed with the Nobeyama radiotelescope, to the FCRAO galaxies
mapped by Young et al. (1995) along the major axis. The CO emission of the observed galaxies within the central beam can be determined by 
convolving the CO intensity profile with a Gaussian beam:

\begin{eqnarray}
SCO_{2D}(CB)&=&4 SCO(0) \int_{0}^{\infty} \int_{0}^{\infty}  \exp\left(-ln(2)\left[\left(\frac{2x}{\theta}\right)^2 + \left(\frac{2y\cos(i)}{\theta}\right)^2\right]\right)\nonumber\\
&&\exp\left(-\frac{\sqrt{x^2+y^2}}{r_{CO}}\right) dxdy,
\end{eqnarray}
\noindent
where $i$ is the inclination of the disc. The integral given in eq. (10) can be solved numerically, and $SCO(0)$ can be determined by comparing eq. (10) with
the CO flux observed in the central beam. With eq. (9), it can be used to derive the total $SCO_{2D}(tot)$ of the observed galaxies.

Our sample includes a large number of edge-on galaxies such as NGC 4565. For these objects the integral given in eq. (10) does not consider that
the molecular gas disc also has a given thickness in the z direction orthogonal to the plane of the disk. We thus modify the prescription of Lisenfeld et al. (2011) to
take into account that the molecular gas disc has a 3D-distribution. Consistently with what is generally assumed for the dust distribution 
in edge-on galaxies (Xilouris et al. 1999; De Looze et al. 2012), we assume an exponential distribution even in the z direction:

\begin{equation}
{SCO(r,z) = SCO(0)e^{-r/r_{CO}} e^{-|z|/z_{CO}}},
\end{equation}
\noindent
where $z_{CO}$ is the scale height of the disc. The total CO flux is then

\begin{eqnarray}
SCO_{3D}(tot)&=&\int_{-\infty}^{\infty} \int_{0}^{\infty} \int_{0}^{2\pi} r SCO(r,z) d \theta dr dz,\nonumber\\
&=& \int_{-\infty}^{\infty} \int_{0}^{\infty} 2\pi r SCO(0)e^{-r/r_{CO}} e^{-|z|/z_{CO}} d r d z,\nonumber\\
&=&4\pi r_{CO}^2 z_{CO} SCO(0).
\end{eqnarray}

\begin{figure}
   \centering
   \includegraphics[width=1\columnwidth]{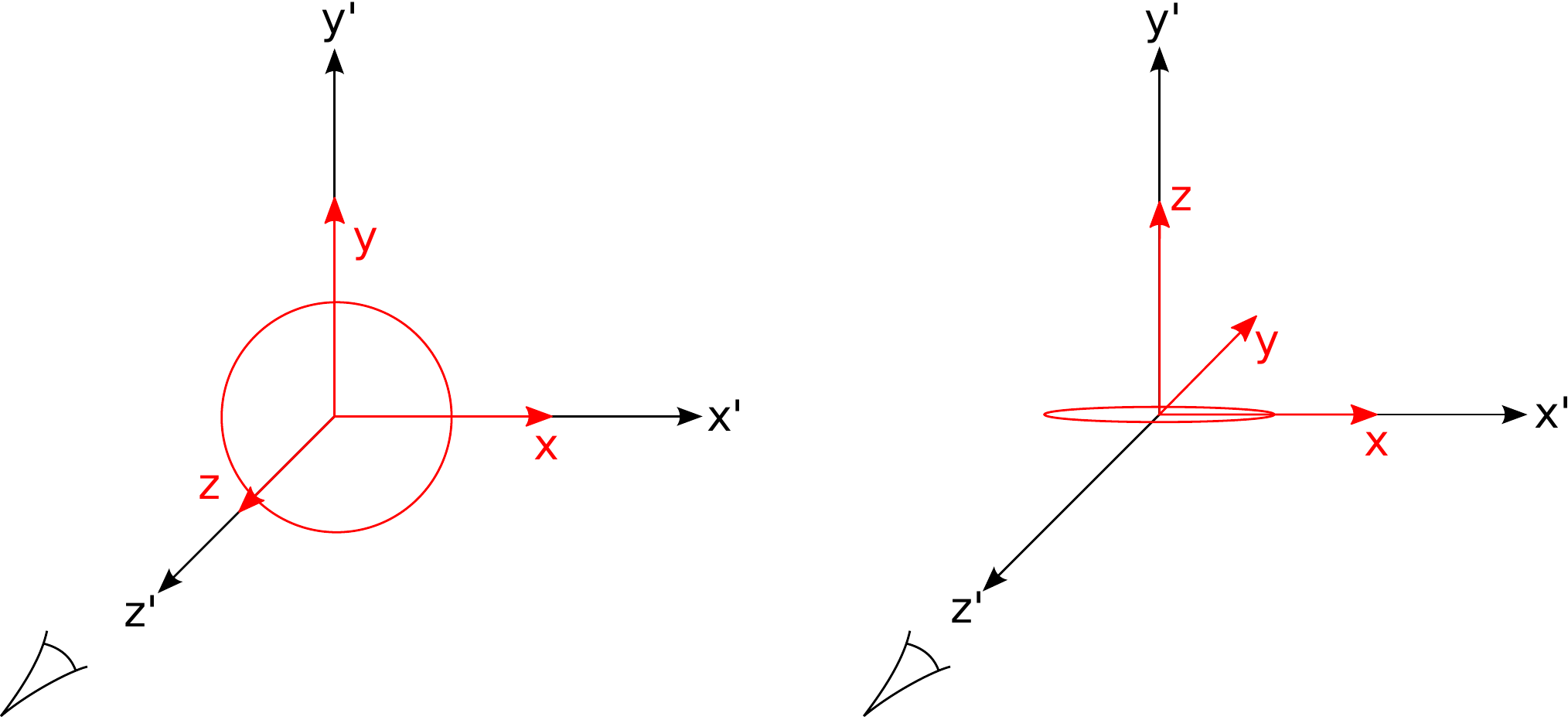}
   \caption{Definition of the referential of the galaxy (black) and that of the observer (red) in the case of a face--on (left) and an edge--on (left) galaxy. 
   The referential of the observer corresponds to the referential of the galaxy rotated by angle $i$ around the $x$ axis.}
   \label{fig:axes}%
   \end{figure}

To compute the total CO flux of the galaxy, we first compute the flux detected in the central beam. We do so in the referential of the observer 
defined by the orthogonal axes $x'$, $y'$, and $z'$, where the last is the coordinate along the line of sight (Fig.~\ref{fig:axes}).
The referential of the observer corresponds to the referential of the galaxy rotated by angle $i$. We obtain the following equation in cartesian coordinates:

\begin{eqnarray}
{SCO_{3D}(CB) = 4 SCO(0) \int_{-\infty}^{\infty} \int_{0}^{\infty} \int_{0}^{\infty} \exp\left[-4ln(2)\frac{x'^2+y'^2}{\theta^2}\right]}\nonumber\\
{\exp\left[-\frac{\sqrt{x'^2+\left(y'\cos(i)-z'\sin(i)\right)^2}}{r_{CO}}-\frac{|y'\sin(i)+z'\cos(i)|}{z_{CO}}\right] d x' d y' d z'}.
\end{eqnarray}

Mapped CO observations of edge-on galaxies are needed to estimate the typical scale height of the CO disc in spirals.
NGC 891 is the only edge-on galaxy fully mapped in the $^{12}$CO(1-0) line (Scoville et al. 1993; Yim et al. 2011). The most recent BIMA-SONG observations of this object have
revealed that the scale height of the CO disc is $z_{CO}$ $\simeq$ 0.185 kpc (Yim et al. 2011\footnote{The analytic function adopted in this work to reproduce the z-scale distribution of
the molecular gas is Gaussian, not exponential, thus not directly comparable to the one adopted in this work.}) 
or, equivalently, $z_{CO}$/$r_{25}$ $\sim$ 1/101. Previous observations of the same galaxy
have given a z-scale ranging from 0.160 kpc in the nucleus to $z_{CO}$ = 0.276 kpc at the edge (Scoville et al. 1993).
To check whether these values of $z_{CO}$ are representative of normal, spiral galaxies, 
we compared them to the typical scale height of the dust component of other edge-on galaxies. Indeed, given the tight correlation between dust and
molecular gas, we can reasonably assume that the tickness of the dusty disc is comparable to that of the molecular gas. The scale height of the dust disc can be measured using either
far-infrared (dust in emission) or optical (dust in absorption) images. From the analysis of the \textit{Herschel}/SPIRE images of the same galaxy, Bianchi \& Xilouris (2011) have
determined that $z_{dust}$ = 0.200 kpc, consistently with $z_{CO}$ $\simeq$ 0.185 kpc determined by Yim et al. (2011). Using energy transfer models adapted to reproduce the far-infrared emission 
observed by \textit{Herschel} of the edge-on galaxy NGC 4565, De Looze et al. (2012) derive $z_{dust}$ = 2.5 arcsec, corresponding to $z_{dust}$/$r_{25}$ $\sim$ 1/194. 
The analysis of the optical images of seven nearby edge-on galaxies done by Xilouris et al. (1997, 1999) gives values of $z_{dust}$/$r_{25}$ ranging from 1/50 to 1/184, as summarised in Table
\ref{Tabedgeon}. The mean value determined from all these data is $z_{dust}$/$r_{25}$ $\simeq$ 1/99, consistent with $z_{CO}$/$r_{25}$ $\sim$ 1/101
determined by Yim et al. (2011) in NGC 891. A similar value ($z_{dust}$/$r_{25}$ = 1/108) has been also obtained from the direct analysis of seven edge-on galaxies 
observed by \textit{Herschel} (Verstappen et al. 2013). Since $r_{24.5}(g)$ $\simeq$ $r_{25}(B)$, we assume $z_{CO}$/$r_{25}(B)$ = $z_{CO}$/$r_{24.5}(g)$ = 1/100.

To understand which among these three different recipes is considered the most likely to estimate the total CO emission of galaxies, we applied them to all the objects observed by Kuno et al.
(2007) for which single-beam observations are available and compared the extrapolated fluxes to those measured using the CO maps.

   \begin{figure*}
   \centering
   \includegraphics[width=1\textwidth]{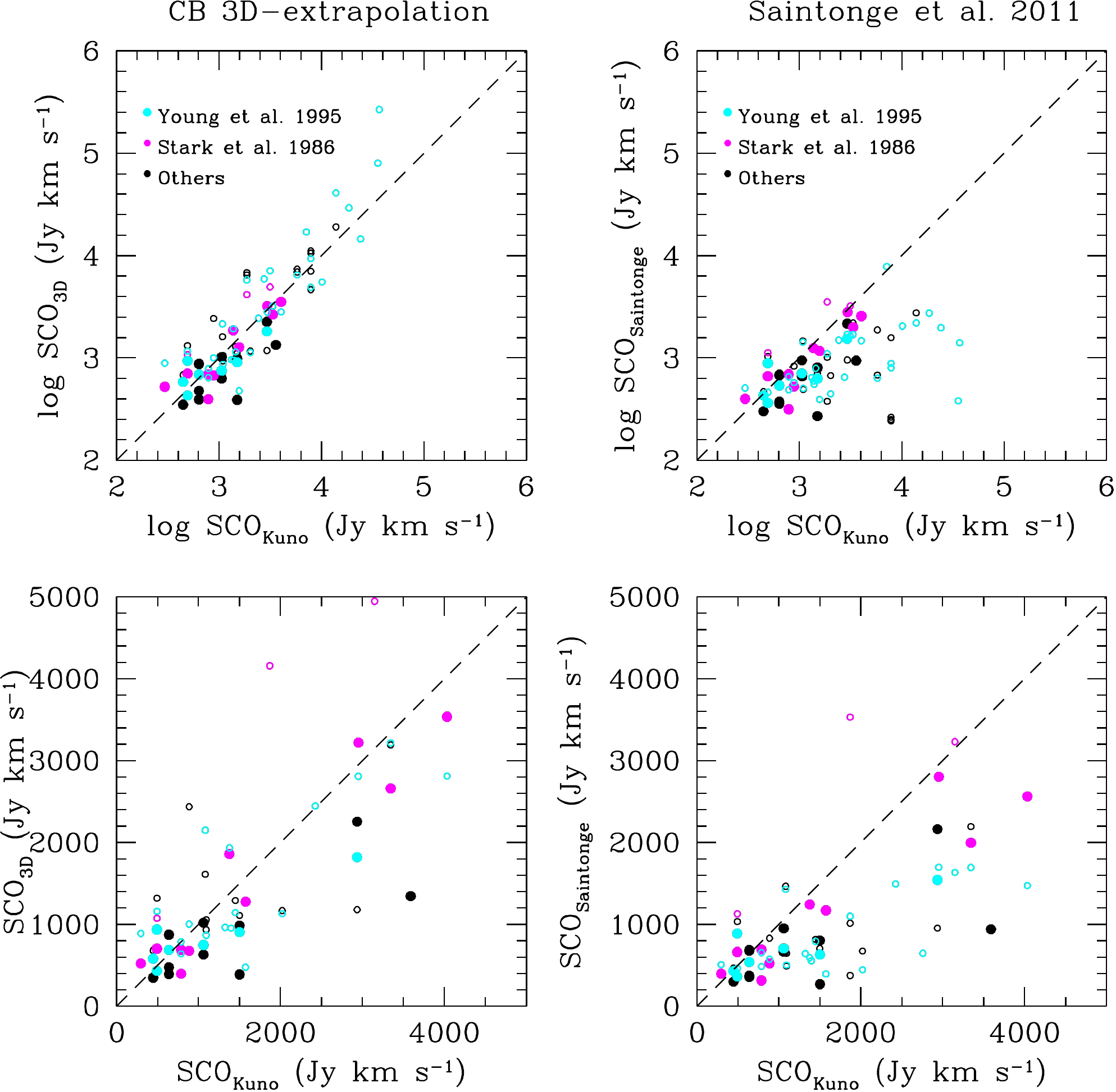}
   \caption{Comparison of single-beam observations corrected for aperture effects using the prescription based on a 3D-distribution of the molecular gas
   (eq. 11; left panels) and that of Saintoinge et al. (2011) (eq. 6; right panels)
   on logarithmic (upper panels) and linear (in the range 0-5000 Jy km s$^{-1}$ sampled by the HRS galaxies) scales (lower panels) 
   for galaxies with integrated data from Kuno et al. (2007).
   Different symbols are used for galaxies observed in the FCRAO survey of Young et al. (1995; cyan), Stark et al. (1986; magenta), and other references (black). 
   Filled dots are for galaxies 
   with $D_{25}(B)/\theta$ $\leq$ 5, empty symbols for those objects with $D_{25}(B)$/$\theta$ $>$ 5.
   The short dashed black line shows the 1:1 relationships.}
   \label{singlecorrection}%
   \end{figure*}

\begin{figure}
   \centering
   \includegraphics[width=\columnwidth]{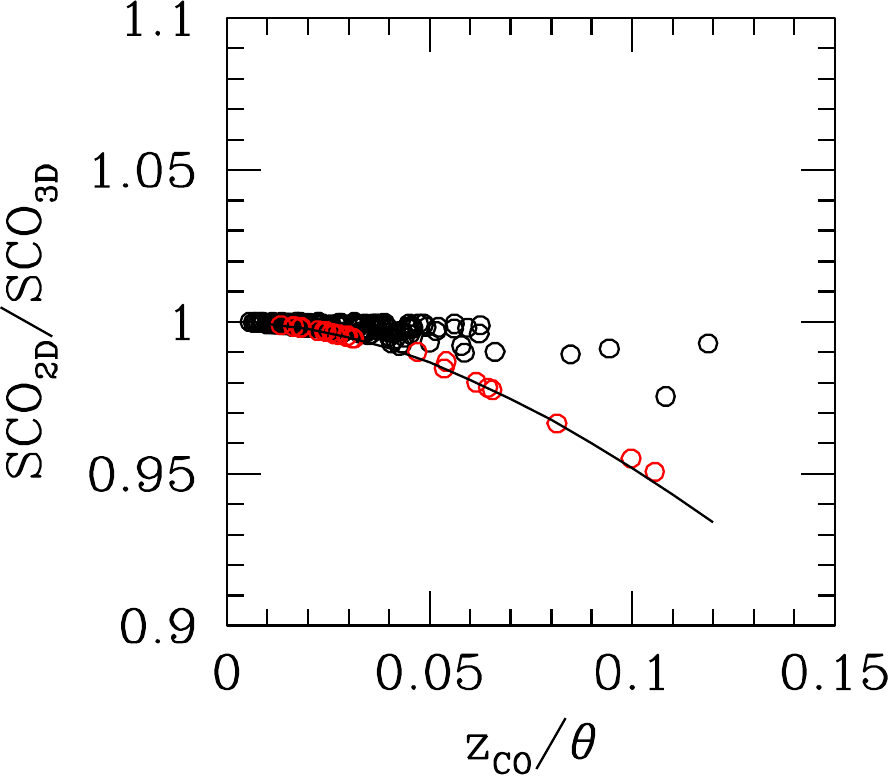}
   \caption{Relationship between the ratio of the 2D (eq. 9) vs. 3D (eq. 12) aperture corrections $SCO_{2D}/SCO_{3D}$ and the ratio of the 
   molecular gas scale height of the disc to the beam size $z_{CO}/\theta$ for the HRS galaxies of the sample. Black open circles are for galaxies with
   an inclination $\leq$ 80 deg, red symbols for edge-on systems ($i$ $>$ 80 deg). The black solid line gives the relation determined for 
   model edge--on galaxies.}
   \label{2D3D}
   \end{figure}

Figure \ref{singlecorrection} shows the relationship between the CO fluxes extrapolated from central-beam observations as described in this text and the total CO fluxes 
given in Kuno et al. (2007) for those galaxies with single-beam observations available in the literature. The comparison uses the extrapolation prescription 
proposed by Saintonge et al. (2011) (eq. 6, right panels) and the one proposed in this work based on a 3D-distribution of the molecular gas (eq. 11). The extrapolation 
recipe proposed by Lisenfeld et al. (2011) gives basically the same results as the one proposed in this work, with the exception of edge-on galaxies where it underestimates the total CO
flux by a few percent (see Figure \ref{2D3D}).

Figure \ref{singlecorrection} and Table \ref{Tabfitintegrati3D} show that both the 2D- (eq. 8) and 
the 3D- (eq. 11) analytic prescriptions given above are more appropriate than the empirical relation of Saintonge et al. (2011). 
Indeed this last recipe systematically underestimates the total CO flux 
whenever the beam size of the telescope is smaller than one fifth of the optical diameter of the target (empty symbols). 
We have also tested whether the use of a slightly differenty exponential disc scale lengths in eqs. (8) and (11) gives better agreement between the CO fluxes extrapolated
from the central-beam observations and the integrated values of Kuno et al. (2007). Using a higher $r_{CO}/r_{24.5}$ ratio than the canonical value of 0.2 
indicated by Lisenfeld et al. (2011) leads to ratios $SCO_{3D}/SCO_{Kuno}$ $\simeq$ 1 in those galaxies with $SCO_{Kuno}$ $<$ 5000 Jy km s$^{-1}$, 
but at the same time systematically overestimates $SCO_{3D}(tot)$ in the brightest galaxies.
We thus decided to adopt the correction given in eq. (11), keeping $r_{CO}/r_{24.5}$ = 0.2 for all galaxies with only one single-beam observation. 

We can also check whether the assumption of a 3D molecular gas disc with $z_{dust}$/$r_{24.5}$ $\simeq$ 1/100 is appropriate for extrapolating 
single-beam observations of edge-on galaxies. To do that, we extrapolate all single-beam observations of nearby edge-on galaxies with data available in the literature 
for which accurate estimates of the total CO emission are available from complete or partial CO mapping (see Table \ref{TabedgeonCOdata}). Table \ref{TabedgeonCOdata}
indicates that, despite the large difference either in the CO fluxes of the central beam observations or in those determined from interferometry, partially  mapped
or major axis mapped edge-on galaxies, the total CO fluxes determined using the prescriptions given in this work in eq. (8; 2D) or eq. (11; 3D) are close to the broad range
of observational data. Equation (8) gives, as expected, slightly lower values (a few per cent) than eq. (11) just because it assumes an infinitely thin disc in the z
direction (see Fig. \ref{2D3D}). 
The comparison between aperture-corrected and total CO fluxes given in Table \ref{TabedgeonCOdata} suggests 
that the 2D extrapolations underestimate the total CO emission more than the 3D
one does (but only by a few \%). We thus decided to prefer and adopt the 3D recipe given in eq. (11) in this work.

\subsubsection{Uncertainties on single beam extrapolated data}

\begin{figure}
   \centering
   \includegraphics[width=1\columnwidth]{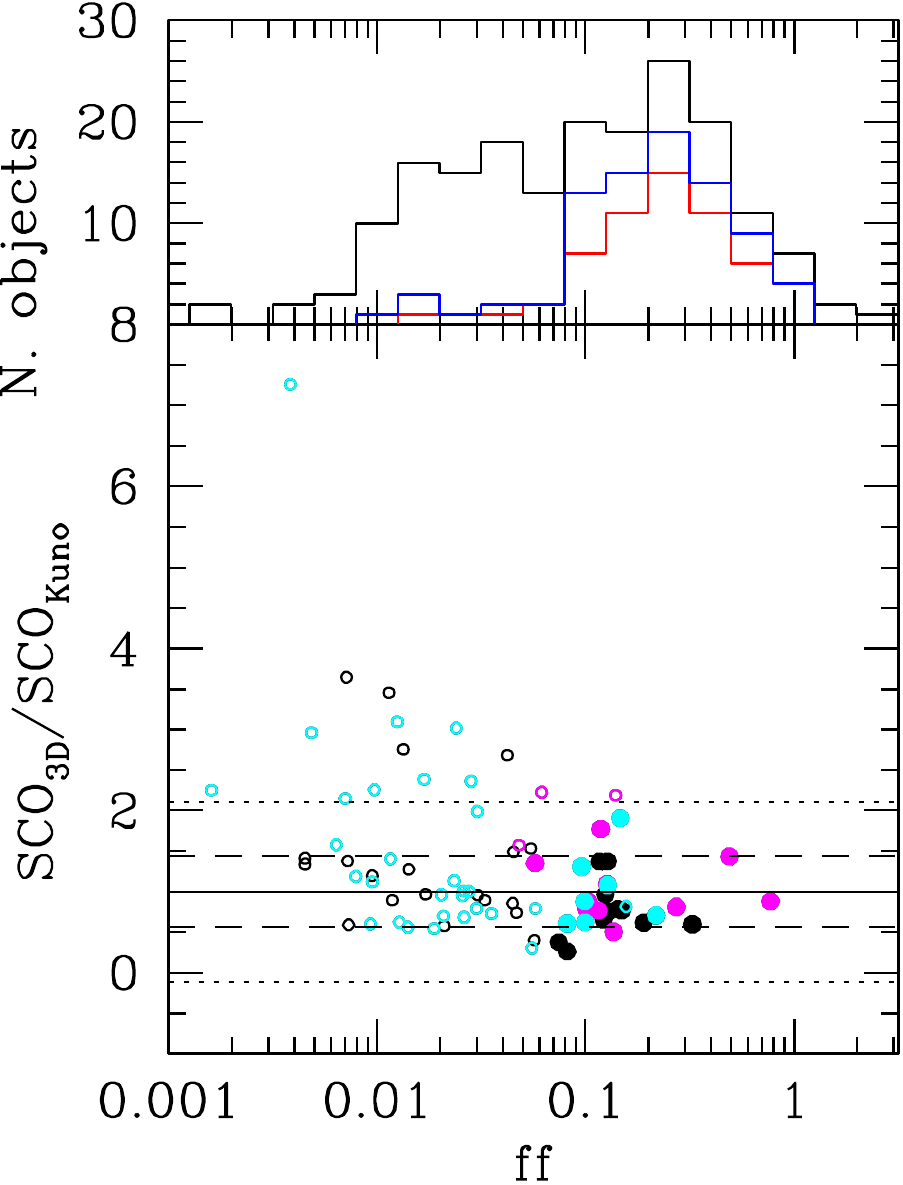}
   \caption{Relationship between the ratio of the total CO flux as determined from the extrapolation of the central beam observation to the total value determined from the maps
   for galaxies with integrated data from Kuno et al. (2007) as a function of the filling factor \textit{ff}, defined as the fraction of the galaxy covered by the central beam (lower panel).
   Different symbols are used for the galaxies observed in the FCRAO survey of Young et al. (1995; cyan), Stark et al. (1986; magenta) and other references (black). 
   Filled dots are for galaxies 
   with $D_{25}(B)/\theta$ $\leq$ 5, empty symbols for those objects with $D_{25}(B)$/$\theta$ $>$ 5.
   The solid line shows the one ratio, the long dashed lines the typical 1 $\sigma$ uncertainty (44 \%) for galaxies where the beam size covers more than 10 \% ~ of the total optical
   surface of the galaxy, the dotted line 1 $\sigma$ uncertainty (114 \%) for galaxies where the beam size covers less than or equal to 10 \% ~ of their surface. The upper panel shows the distribution in 
   the area covered by the central beam for all HRS galaxies with available CO data (black histogram). The blue histogram gives the distribution of all galaxies 
   observed by our team, including previous data (Boselli et al. 1995; 2002, Sauty et al. 2003), while the red one only those observed in this work. }
   \label{errore}%
   \end{figure}

By comparing the flux extrapolated from single-beam observations to the integrated value for galaxies in the Kuno et al. (2007) sample, we
can quantify the typical uncertainty in the total CO emission due to a combined effect of the uncertainty on the $I(CO)$ measurement and of
the aperture correction. Figure \ref{errore} shows the relationship between the ratio of the total CO flux as determined from the 3D-extrapolation 
of the central-beam observation to the total value determined from the maps for galaxies in the Kuno et al. (2007) sample as a function of the surface of the galaxy covered by the telescope
beam. Figure \ref{errore} shows that whenever the beam covers more than 10 \% ~ of the surface of the disc, the mean uncertainty in the total, extrapolated CO flux is 
of the order of 44 \% ($SCO_{3D}/SCO_{Kuno}$ = 1.01 $\pm$ 0.44). In the remaining galaxies, where the beam covers less than 10 \% ~ of the surface of the galaxy, 
the mean uncertainty in the total CO flux estimate is as large as 123 \% ($SCO_{3D}/SCO_{Kuno}$ = 1.46 $\pm$ 1.14).

\subsection{Multiple beam observations}

\begin{figure*}
   \centering
   \includegraphics[width=1\textwidth]{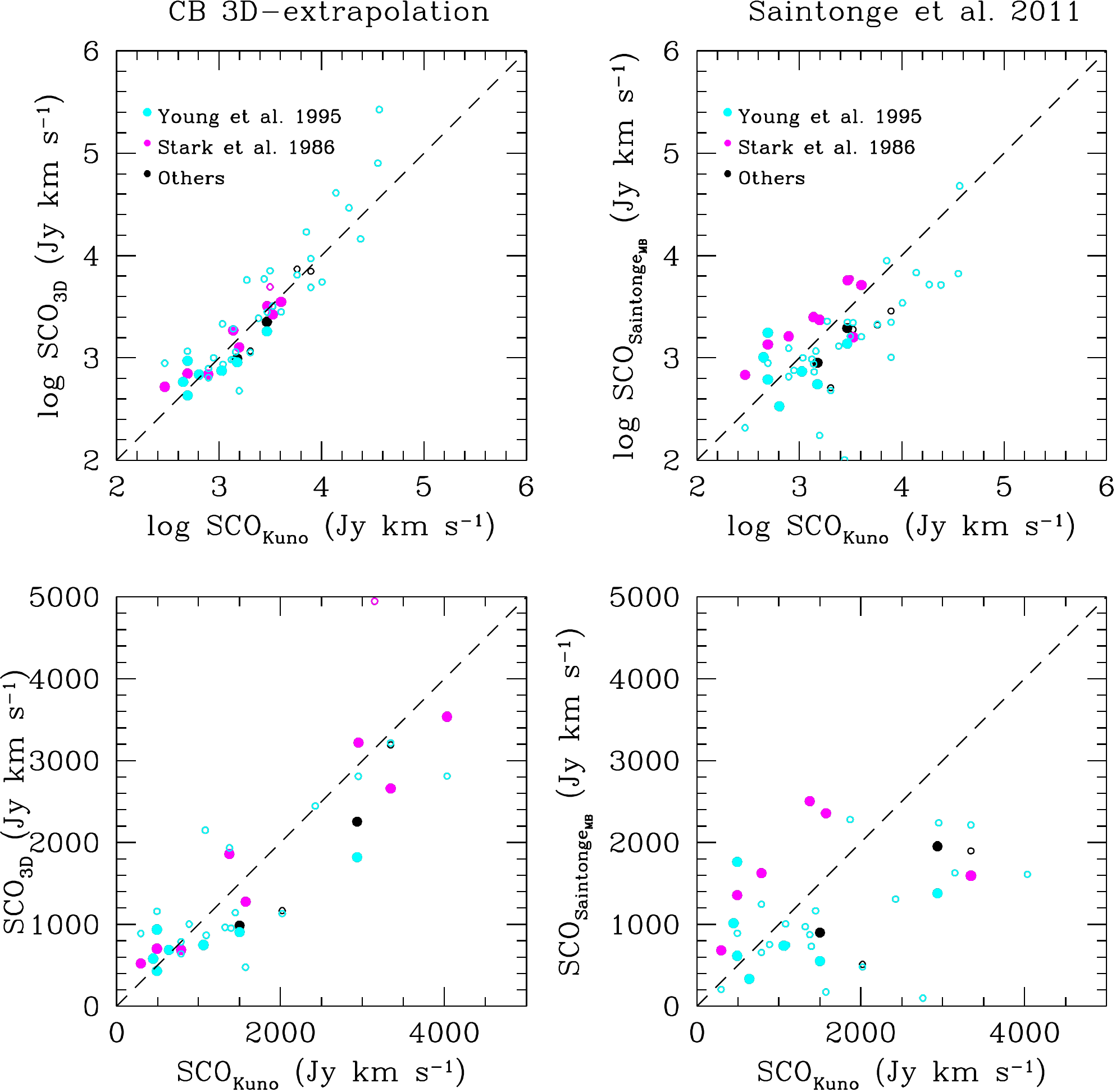}
   \caption{Comparison of multiple beam observations along the major axis corrected for aperture effects using the prescription based on a 3D-distribution of the molecular gas
   and assuming only the central beam (eq. 11; left panels) and that of Saintoinge et al. (2011) (eq. 13; right panels)
   on logarithmic (upper panels) and linear (in the range 100-5000 Jy km s$^{-1}$ sampled by the HRS galaxies) scales (lower panels) 
   for galaxies with integrated data from Kuno et al. (2007).
   Different symbols are used for galaxies observed in the FCRAO survey of Young et al. (1995; cyan), Stark et al. (1986; magenta) and other references (black). 
   Filled dots are for galaxies 
   with $D_{25}/\theta$ $\leq$ 5, empty symbols for those objects with $D_{25}$/$\theta$ $>$ 5.
   The short dashed black line shows the 1:1 relationships. }
   \label{multibeam}%
   \end{figure*}

Some HRS galaxies have multiple observations along the major axis. These data can be combined and extrapolated to determine total CO luminosities 
as done with different techniques found in the literature. Typical examples are the method of Solomon \& Sage (1988) or that of the FCRAO team (Young et al. 1995).
While the former use different polynomials to combine the emission in the different beams, the latter use the radial variation in the CO emission to deduce a CO profile 
(exponential, Gaussian, etc.) that is later used to fit and extrapolate the observations. More recently, Saintonge et al. (2011) have proposed a very simple prescription 
previously determined by simulating the observation through two adjacent beams (one central, the second one beam off-centre) of the Kuno et al. (2007) sample of mapped galaxies. 
The simulations have been done using beams of different sizes. 
Multiple beam observations of the HRS galaxies come mainly from the FCRAO survey of Young et al. (1995), with a few from our own observations.
As previously discussed, the FCRAO gives quite accurate extrapolated fluxes for these objects, values that we adopt here. There are, however, a few objects with 
only two or three detected beams along the major axis, where the extrapolation of the FCRAO is not very accurate. This is probably because, with such a small
sampling, it is hard to accurately deduce a radial profile of the CO emission. Eighteen HRS galaxies have two or three detections along the major axis (from the FCRAO survey, 4, and from our own
observations presented in this work, 14).
It is thus worth understanding whether this radial information can be used to constrain the distribution of the molecular gas along the disc better and thus allow more accurate
determination of the total CO emission. To do this we applied an updated version of the technique used by the COLD GASS survey (Saintonge et al. 2011)
for galaxies observed in the central position ($SCO(CB)$) and in an adjacent beam ($SCO(\textit{adj})$) along the major axis. 
Following this technique, the total CO flux of the galaxy is given by the relation

\begin{equation}
{SCO_{Saintonge_{MB}}(tot) = \frac{SCO(CB)}{1.166-3.557 \times f_{OFF}+3.360 \times f_{OFF}^2}},
\end{equation}
\noindent
where $f_{OFF}$ is the ratio of the flux measured in the two beams:

\begin{equation}
{f_{OFF} = \frac{SCO(CB)}{SCO(\textit{adj})}},
\end{equation} 
\noindent
where $SCO_{Saintonge_{MB}}(tot)$ stands for the total, extrapolated CO flux  determined from multiple beam observations.
The variable $f_{OFF}$ gives an empirical estimate of the slope of the radial variation of the CO emission. We then compared the total CO fluxes 
determined using this relation with the integrated fluxes of Kuno et al. (2007) in Fig. \ref{multibeam}. 
Different and independent sets of data are available from the FCRAO survey of Young et al. (1995), from Stark et al. (1996), from this work, and from several other references in the literature.
Table \ref{Tabmultiple} gives for comparison the mean values of the ratio $SCO_{3D}/SCO_{Kuno}$ and $SCO_{Saintoinge_{MB}}/SCO_{Kuno}$ for those galaxies 
in the Kuno et al. (2007) sample with multiple beam observations along the major axis. Figure \ref{multibeam} and Table \ref{Tabmultiple} indicate that
the CO fluxes extrapolated using eq. (11), hence not taking the CO observations along the major axis into account, are on average more accurate than those extrapolated 
using the multi-beam prescription of Saintonge et al. (2011). These last, indeed, generally underestimate the total CO emission of the observed galaxies, while the 3D extrapolation
is quite accurate if the size of the galaxy does not exceed about five times the FWHM of the telescope beam. 
Most of the galaxies of the HRS, and in particular those observed in this work, match this condition. We thus adopted a 3D-extrapolation correction also for these
objects with two- or three-beam observations along the major axis.

\subsection{The homogenised CO data catalogue}

Single-beam observations of HRS galaxies are available from different references in the literature or from our own observations, including those presented in this work.
Several objects have multiple, independent observations. Overall there are 344 independent pointings on 225 HRS
galaxies. We have collected all these data and listed them in Table \ref{TabCOHRSCB}, arranged as follows

\begin{itemize}

\item {Column 1: HRS name.} 
\item {Column 2: Telescope coded as in Table \ref{Tabtel}.}
\item {Column 3: Number of independent beams for this reference.}
\item {Column 4: Beam size $\theta$, in arcseconds.}
\item {Column 5: Sign for the observation: 1 for detected sources, 0 for undetected objects.}
\item {Column 6: Intensity of the CO line in the central beam, in K km s$^{-1}$, on the temperature scale (listed in column 9) as indicated in the original paper where the data are presented. 
No intensities are given for undetected galaxies.}
\item {Column 7: rms the observations, in mK, expressed on the temperature scale as indicated in the original paper where the data are presented (column 9).}
\item {Column 8: Velocity resolution of the smoothed CO spectra, in km s$^{-1}$.}
\item {Column 9: Temperature scale.}
\item {Column 10: Reference to the data. }
\item {Column 11: Jy/K calibration constant used to transform temperature brightnesses into CO fluxes. These might differ from those reported in Table \ref{Tabtel}
if the CO intensities are expressed on other scales than those generally used for the telescope, or if the authors expressely indicate different calibration factors in their original
papers.}
\item {Column 12: Extrapolated CO flux $SCO_{3D}$, or upper limit, in Jy km s$^{-1}$. CO intensities measured in the central beam, and upper limits are extrapolated to total
values using eq. (11). Upper limits in the central beam are determined as described in sect. (3; eq. 1).}
\item {Column 13: Filling factor of the central beam, defined as the ratio of the surface covered by the beam of the telescope to the optical area of the galaxy.}
\item {Column 14: A flag indicates those data used for measuring the total CO emission of HRS galaxies (flag=1) or those not used (flag=0). }
\item {Column 15: Notes to single-beam observations. 1: assumed in $T_R^*$. 2: transformed from the original table to this scale. 3: partly mapped galaxy, with an 
equivalent beamsize assumed to be $\theta$ = 36 arcsec}.

\end{itemize}

\begin{figure}
   \centering
   \includegraphics[width=1\columnwidth]{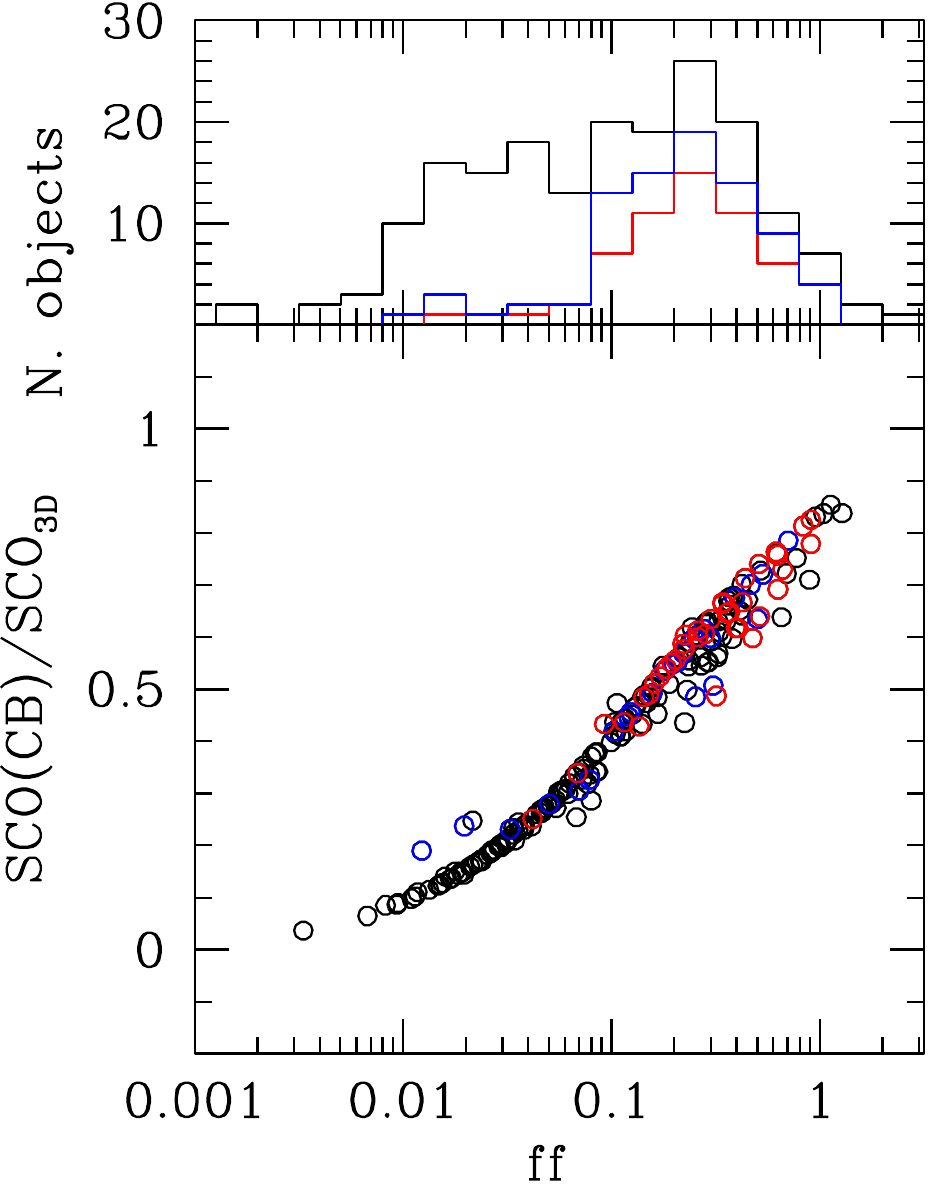}
   \caption{Relationship between the ratio of the central beam to total extrapolated CO flux ratio (aperture correction) 
   as a function of the filling factor \textit{ff}, defined as the fraction of the galaxy 
   covered by the central beam for all the HRS galaxies with available single-beam observations (lower panel).
   Red symbols are for those galaxies observed in this work, blue symbols for other objects observed by our team 
   in previous surveys, while the remaining black circles those with data available in the literature.
   The upper panel shows the distribution in the filling factor for all the HRS galaxis observed in CO. The blue histogram gives the distribution of all galaxies 
   observed by our team, including previous data (Boselli et al. 1995; 2002, Sauty et al. 2003), while the red one only those observed in this work. }
   \label{errorHRS}%
   \end{figure}

This new set of extrapolated data are then cross-matched with those with available complete mapping (see sect. 5.2) to select the best available data. 
Total CO emission from mapped galaxies are taken in the order described in sect. 5.2. If no integrated CO data are available, the total $SCO$
emission is taken from extrapolated single-beam observations. When, for the same galaxy, different sets of data are available,
we give the mean value determined by combining detections. If detections and undetections are available for the same object, we check that the detections are consistent with the upper
limits, and give the detection as final result.\\
Multiple central-beam observations are useful for inferring a realistic uncertainty on the extrapolated CO emission of the HRS galaxies. For those objects with multiple observations,
we quantify the uncertainty by measuring the standard deviation from the mean value determined using only detections. When the uncertainty drops to values lower than 20 \%, we assume 
a conservative error of 20 \%. For galaxies with available integrated CO maps 
from the literature, we assume the standard uncertainties determined as described in sect. 5.2 for the different references. 
For those with only one central beam detection available, we assume a mean uncertainty
determined as described in sect. 5.3 according to the correction factor, namely 44 \% ~ if the surface of the beam covers more than 10\% ~ of the 
total surface of the galaxy, 114 \% if the sampled area is smaller. Figure \ref{errorHRS} shows the ratio of the central beam to the extrapolated CO fluxes, 
which corresponds to the aperture correction, as a function of the surface of the galaxy covered by the telescope beam (filling factor \textit{ff} in Table). Figure \ref{errorHRS}
shows that most of the galaxies observed in this work or by our team in previous observing runs (Boselli et al. 1995; 2002) 
have been covered by the beam of the telescope by more than 10 \% of their 
surface, thus have relatively low errors in the extrapolated CO flux.\\

The final results are given in Table \ref{TabHRSH2}, arranged as follows:

\begin{itemize}
\item {Column 1: HRS name.}
\item {Column 2: Sign for the observation: 1 for detected sources, 0 for undetected objects.}
\item {Column 3: $SCO$ flux, in Jy km s$^{-1}$.}
\item {Column 4: Error on the $SCO$ flux, in Jy km s$^{-1}$.}
\item {Column 5: Signal-to-noise $S/N$, defined as the ratio of the total CO flux to the uncertainty in the CO line, this last defined as in Column 9 of Table 2. 
The $S/N$ is given only for galaxies with single-beam observations.}
\item {Column 6: Filling factor \textit{ff}, defined as the ratio of the surface covered by the beam of the telescope to the optical area of the galaxy.\textit{ff} = $Int$ indicates galaxies with CO integrated
maps, \textit{ff} = $MA$, those objects mapped along the major axis with more than three detected beams (fron the FCRAO survey).} 
\item {Column 7: Logarithm of the molecular hydrogen mass, $M(H_2)_c$, determined assuming the standard Galactic conversion factor 
$X_{CO}$ = 2.3 10$^{20}$ cm$^{-2}$/(K km s$^{-1}$) (Strong et al. 1988). The molecular hydrogen mass of galaxies (in solar units) is determined from the relation:

\begin{equation}
{M(H_2) = 3.9 \times 10^{-17} \times X_{CO} \times SCO \times D^2},
\end{equation}
\noindent
where $D$ is the diatance (in Mpc) given in Table 1.}
\item {Column 8: Logarithm of the molecular hydrogen mass, $M(H_2)_v$, determined by assuming the H-band luminosity dependent conversion factor log $X_{CO}$ = -0.38 $\times$ log $L_H$ + 24.23,
where $X_{CO}$ is in cm$^{-2}$/(K km s$^{-1}$) and $L_H$ in solar units (Boselli et al. 2002).}
\item {Column 8: References to the CO data.} 

\end{itemize}

 \section{H{\sc i} data}

The H{\sc i} integrated data are available for 315 out of the 322 HRS galaxies. We collected them from a wide variety of references in the literature.
Among these, we havefavored those coming from the H{\sc i} survey ALFALFA (Giovanelli et al. 2005) and recently published in Haynes et al. (2011).
An important fraction of the HRS galaxies is not observable from Arecibo. For these galaxies, we preferred the homogenised data
of Springob et al. (2005), then those available in the literature by chosing those with the highest $S/N$, lower rms, or with available spectra 
in the published papers. 
The H{\sc i} data for the HRS are given in Table \ref{TabHI}, arranged as follows
\begin{itemize}

\item {Column 1: HRS name.}
\item {Column 2: Sign for the observation: 1 for detected sources, 0 for undetected objects.}
\item {Column 3: rms, in mJy. Not all references in the literature give this entity, thus the rms is available only for a (large) portion of the HRS galaxies.}
\item {Column 4: H{\sc i} flux $SHI$, in Jy km s$^{-1}$. For undetected galaxies, the $SHI$ is an upper limit to the H{\sc i} flux, defined as

\begin{equation}
{SHI = 5 \times rms \times WHI_{exp} },
\end{equation}
\noindent
where $WHI_{exp}$ = 300 km s$^{-1}$ is the expected H{\sc i} line width of the observed galaxy. 


}

\item {Column 5: Logarithm of the H{\sc i} mass, in M$_{\odot}$, determined from the relation: 

\begin{equation}
{M(HI) = 2.356 \times 10^5 \times D^2 \times SHI},
\end{equation} 
\noindent
where the distance $D$, expressed in Mpc, is taken from Table \ref{TabHRS}. The typical uncertainty on the H{\sc i} mass is $\simeq$ 15 \%.}

\item {Column 6: H{\sc i} line width $WHI$ measured at 50 \% of the peak flux, in km s$^{-1}$, corrected 
for instrumental broadening and cosmological redshift as in Catinella et al. (2012). Observed $WHI_o$ are corrected following the relation:

\begin{equation}
{WHI = \frac{WHI_o - \Delta s}{1+z}},
\end{equation}
\noindent
where the broadening correction $\Delta s$ is given by

\begin{equation}
{\Delta s = 2 \times \delta V_{HI} \times 0.25},
\end{equation}
\noindent
where $\delta V_{HI}$ is the velocity resolution. The ALFALFA data have been corrected using a different prescription. To have homogenised data, we removed 
the ALFALFA correction and applied consistently the correction given in eq. 18-19. 
For some galaxies, the data presented in the original papers cannot 
be homogeneously corrected using this prescription (line width measured not at 50 \% of the intensity peak, lack of any information on the velocity resolution in the data). 
For these galaxies, we report the value given in the literature, but
not corrected for broadening and cosmological redshift. These galaxies can be recognised by the flag given in column 8.}
\item {Column 7: H{\sc i}-deficiency parameter ($\mathrm{\hi}-def$), defined as the difference in logarithmic scale between the expected and the observed
H{\sc i} mass of a galaxy of given angular size and morphological type (Haynes \& Giovanelli 1984). The H{\sc i}-deficiency for all the HRS galaxies 
is determined using the recent calibration of Boselli \& Gavazzi (2009).}
\item {Column 8: Flag for the $WHI$ line width measurement: "1" corrected for broadening and cosmological redshift; "2" uncorrected values.}
\item {Column 9: Reference to the H{\sc i} data.}

\end{itemize}

\section{Conclusion}

We have presented new $^{12}$CO(1-0) observations of 59 late-type galaxies of the \textit{Herschel} Reference Survey done with the 12-metre Kitt Peak radiotelescope.
By comparing the total CO emission of fully mapped nearby galaxies, including edge-on systems, to CO fluxes corrected for aperture effects using different prescriptions
developed in this work or proposed in the literature, we identified the most accurate recipe for extrapolating central single-beam observations of extended galaxies. 
The central-beam observation was corrected for the Gaussian shape of the beam and extrapolated assuming a 3D exponential distribution of the molecular gas.
The typical scale length of the CO emitting disc is $r_{CO}$ = 0.2 $r_{24.5}$, where $r_{24.5}$ is the optical, $g$- (or $B$-band) isophotal radius at the 24.5 mag arcsec$^{-2}$
(25 mag arcsec$^{-2}$), while
its scale height is $z_{CO}$ = 1/100 $r_{24.5}$. This scale height was calibrated on well known nearby edge-on systems and is comparable to the typical scale height of the
dusty disc observed either in absorption using optical images or in emission using far-infrared data.\\
We applied this new recipe to our new CO data and to all the CO data available in the literature to build the most complet set of molecular gas data for the HRS, which is now available
for 225 out of the 322 objects composing the sample. We also collected  and homogenise H{\sc i} data from the literature for 315 HRS galaxies. Both H{\sc i} and CO data are used to estimate the total
H{\sc i} and H$_2$ molecular gas content of the targets. All these data are available on an HRS-dedicated web page on the HeDaM database: http://hedam.lam.fr/. These data are
used to study the molecular and total gas scaling relations of the HRS galaxies (Paper II) and the effects of the environment on the molecular gas phase 
by comparing the molecular gas content of isolated and Virgo cluster galaxies of the sample (Paper III).

\begin{acknowledgements}

We are extremely grateful to L. Ziurys for the generous time allocation at the 12-metre  
Kitt Peak radiotelescope, and to the telescope operators for their assistance during the observations. We want to thank
M. Fossati for helping us to write some procedures used to reduce the CO data.
This research has made use of data from the HRS project. HRS is a Herschel Key Programme utilising 
guaranteed time from the SPIRE instrument team, ESAC scientists and a mission scientist. 
The HRS data was accessed through the Herschel Database in Marseille (HeDaM - http://hedam.lam.fr) 
operated by CeSAM and hosted by the Laboratoire d'Astrophysique de Marseille.
We are grateful to the referee, U. Lisenfeld, for the accurate and detailed reviewing and for the constructive comments that
significantly helped increase the quality of the manuscript.
A.B thanks the ESO visiting programme committee for inviting him
at the Garching headquarters for a two-month stay.
\end{acknowledgements}

\onecolumn
\begin{landscape}
\begin{center} \tiny 
\begin{longtable}{c c c c c c c c c c c c c c c c c}
\caption{Herschel Reference Survey.\label{TabHRS}}\\
\hline \hline 
\multicolumn{1}{c}{HRS} &
\multicolumn{1}{c}{CGCG}&      
\multicolumn{1}{c}{VCC} &    
\multicolumn{1}{c}{UGC} &     
\multicolumn{1}{c}{NGC} &    
\multicolumn{1}{c}{IC}  & 
\multicolumn{1}{c}{RA(2000)} &   
\multicolumn{1}{c}{dec}      & 
\multicolumn{1}{c}{type}     &
\multicolumn{1}{c}{$D$}    & 
\multicolumn{1}{c}{$K_S$ $_{tot}$}& 
\multicolumn{1}{c}{$D_{24.5}(g)$}       &
\multicolumn{1}{c}{incl}  &    
\multicolumn{1}{c}{vel}           &   
\multicolumn{1}{c}{memb}  & 
\multicolumn{1}{c}{HI}    & 
\multicolumn{1}{c}{CO}    \\
\multicolumn{1}{c}{} &
\multicolumn{1}{c}{} &      
\multicolumn{1}{c}{} &    
\multicolumn{1}{c}{} &     
\multicolumn{1}{c}{} &    
\multicolumn{1}{c}{}  & 
\multicolumn{1}{c}{h m s} &   
\multicolumn{1}{c}{$^{\circ}$ ' "}      & 
\multicolumn{1}{c}{}       &
\multicolumn{1}{c}{Mpc}    & 
\multicolumn{1}{c}{mag}    & 
\multicolumn{1}{c}{'}       &  
\multicolumn{1}{c}{deg}	&  
\multicolumn{1}{c}{km s$^{-1}$}           &   
\multicolumn{1}{c}{}    & 
\multicolumn{1}{c}{}    & 
\multicolumn{1}{c}{}    \\
\multicolumn{1}{c}{(1)} &
\multicolumn{1}{c}{(2)}&      
\multicolumn{1}{c}{(3)} &    
\multicolumn{1}{c}{(4)} &     
\multicolumn{1}{c}{(5)} &    
\multicolumn{1}{c}{(6)}  & 
\multicolumn{1}{c}{(7)} &   
\multicolumn{1}{c}{(8)}      & 
\multicolumn{1}{c}{(9)}       &
\multicolumn{1}{c}{(10)}    & 
\multicolumn{1}{c}{(11)}    & 
\multicolumn{1}{c}{(12)}       &    
\multicolumn{1}{c}{(13)}           &   
\multicolumn{1}{c}{(14)}    & 
\multicolumn{1}{c}{(15)}    & 
\multicolumn{1}{c}{(16)}    &
\multicolumn{1}{c}{(17)}    \\ 
\hline
  \\
\endhead 
 \\ 
 \hline
  \multicolumn{16}{c}{{Continued on next page\ldots}} \\
\endfoot

  \\ 
\hline \hline
\endlastfoot
       1&   123035&        - &       -&        -&        -&101739.66& 224835.9&Pec  		   &   16.79 &11.59&	 0.92&   67.09 & 1175&Leo Cl.		       &    0 &  0  \\
       2&   124004&        - &    5588&        -&        -&102057.13& 252153.4&S?  		   &   18.44 &11.03&	 0.73&   44.82 & 1291&Leo Cl.		       &    1 &  0  \\
       3&    94026&        - &    5617&     3226&        -&102327.01& 195354.7&E2:pec;LINER;Sy3	   &   16.70 & 8.57&	 3.15&   34.97 & 1169&Leo Cl.		       &    1 &  1  \\
       4&    94028&        - &    5620&     3227&        -&102330.58& 195154.2&SAB(s)pec;Sy1.5     &   16.40 & 7.64&	 4.80&   61.37 & 1148&Leo Cl.		       &    1 &  1  \\
       5&    94052&        - &       -&        -&      610&102628.37& 201341.5&Sc   		   &   16.71 & 9.94&	 1.96&   77.41 & 1170&Leo Cl.		       &    1 &  0  \\
       6&   154016&        - &    5662&    3245A&        -&102701.16& 283821.9&SB(s)b  		   &   18.89 &11.83&	 2.21&   90.00 & 1322&Leo Cl.		       &    1 &  0  \\
       7&   154017&        - &    5663&     3245&        -&102718.39& 283026.6&SA(r)0:?;HII;LINER  &   18.77 & 7.86&	 3.13&   55.29 & 1314&Leo Cl.		       &    1 &  1  \\
       8&   154020&        - &    5685&     3254&        -&102919.92& 292929.2&SA(s)bc;Sy2   	   &   19.37 & 8.80&	 3.92&   70.43 & 1356&Leo Cl.		       &    1 &  1  \\
\end{longtable}
\end{center}
\end{landscape}

\twocolumn

\clearpage

\onecolumn
\begin{center} \tiny \begin{longtable}{l c c c c c c l c c c c}
\caption[Results of the CO observations.]{Results of the CO observations of the HRS galaxies.\label{tab:hrscoobs}}\\
  \hline \hline \\[-2ex]
\multicolumn{1}{c}{HRS} &
\multicolumn{1}{c}{$\Delta(x)$} &
\multicolumn{1}{c}{$\Delta(y)$} &
\multicolumn{1}{c}{sgn} &
\multicolumn{1}{c}{Run} &
\multicolumn{1}{c}{rms} &
\multicolumn{1}{c}{N. scan} &
\multicolumn{1}{c}{$I(CO)$} &
\multicolumn{1}{c}{$\Delta I(CO)$} &
\multicolumn{1}{c}{Vel} &
\multicolumn{1}{c}{$W_{CO}$} &
\multicolumn{1}{c}{Peak $T$} \\
\multicolumn{1}{c}{} &
\multicolumn{1}{c}{arcsec} &
\multicolumn{1}{c}{arcsec} &
\multicolumn{1}{c}{} &
\multicolumn{1}{c}{} &
\multicolumn{1}{c}{mK} &
\multicolumn{1}{c}{} &
\multicolumn{1}{c}{K km s$^{-1}$} &
\multicolumn{1}{c}{K km s$^{-1}$} &
\multicolumn{1}{c}{km s$^{-1}$} &
\multicolumn{1}{c}{km s$^{-1}$} &
\multicolumn{1}{c}{K} \\
\multicolumn{1}{c}{(1)} &
\multicolumn{1}{c}{(2)} &
\multicolumn{1}{c}{(3)} &
\multicolumn{1}{c}{(4)} &
\multicolumn{1}{c}{(5)} &
\multicolumn{1}{c}{(6)} &
\multicolumn{1}{c}{(7)} &
\multicolumn{1}{c}{(8)} &
\multicolumn{1}{c}{(9)} &
\multicolumn{1}{c}{(10)} &
\multicolumn{1}{c}{(11)} &
\multicolumn{1}{c}{(12)} \\ [0.5ex] \hline
  \\[-1.8ex]
  \endfirsthead
\multicolumn{12}{c}{{\tablename} \thetable{} -- Continued} \\[0.5ex]
  \hline \hline \\[-2ex]
\multicolumn{1}{c}{HRS} &
\multicolumn{1}{c}{$\Delta(x)$} &
\multicolumn{1}{c}{$\Delta(y)$} &
\multicolumn{1}{c}{sgn} &
\multicolumn{1}{c}{Run} &
\multicolumn{1}{c}{rms} &
\multicolumn{1}{c}{N. scan} &
\multicolumn{1}{c}{$I(CO)$} &
\multicolumn{1}{c}{$\Delta I(CO)$} &
\multicolumn{1}{c}{Vel} &
\multicolumn{1}{c}{$W_{CO}$} &
\multicolumn{1}{c}{Peak $T$} \\
\multicolumn{1}{c}{} &
\multicolumn{1}{c}{arcsec} &
\multicolumn{1}{c}{arcsec} &
\multicolumn{1}{c}{} &
\multicolumn{1}{c}{} &
\multicolumn{1}{c}{mK} &
\multicolumn{1}{c}{} &
\multicolumn{1}{c}{K km s$^{-1}$} &
\multicolumn{1}{c}{K km s$^{-1}$} &
\multicolumn{1}{c}{km s$^{-1}$} &
\multicolumn{1}{c}{km s$^{-1}$} &
\multicolumn{1}{c}{K} \\
\multicolumn{1}{c}{(1)} &
\multicolumn{1}{c}{(2)} &
\multicolumn{1}{c}{(3)} &
\multicolumn{1}{c}{(4)} &
\multicolumn{1}{c}{(5)} &
\multicolumn{1}{c}{(6)} &
\multicolumn{1}{c}{(7)} &
\multicolumn{1}{c}{(8)} &
\multicolumn{1}{c}{(9)} &
\multicolumn{1}{c}{(10)} &
\multicolumn{1}{c}{(11)} &
\multicolumn{1}{c}{(12)} \\ [0.5ex] \hline
  \\[-1.8ex]
\endhead 
 \\[-1.8ex] \hline
  \multicolumn{12}{c}{{Continued on next page\ldots}} \\
\endfoot
  \\[-1.8ex] \hline \hline
\endlastfoot
     8  &  0	&      0   & 0  &   1  &  3.00 &      20  &    1.18   &     -   & -	&  -	&   - \\
     9  &  0	&      0   & 1  &   1  &  3.00 &      20  &    1.79   &    0.31 & 1391  &  167  &   7.14E-3  \\
    11  &  0	&      0   & 0  &   4  &  5.00 &      20  &    1.16   &     -   & -	&  -	&   - \\
    16  &  0	&      0   & 1  &   1  &  4.00 &      30  &    1.85   &    0.29 & 1267  &  83	&   1.66E-2  \\
    17  &  0	&      0   & 1  &   2  &  4.00 &      20  &    4.17   &    0.41 & 1293  &  164  &   1.87E-2  \\
    17N &  -35  &    +42   & 1  &   2  &  4.00 &      20  &    1.03   &    0.26 & 1403  &  69$^a$&   1.10E-2  \\
    17S &  +35  &    -42   & 1  &   2  &  3.00 &      20  &    0.62   &    0.16 & 1166  &  44$^a$&   1.33E-2  \\
    31  &  0	&      0   & 0  &   1  &  3.00 &      20  &    1.17   &     -   & -	&  -	&   - \\
\end{longtable}
\end{center}										                  

\twocolumn

\clearpage

\begin{table*}
\caption{Independent observations of HRS 48 (NGC 3631)}
\label{TabHRS48}
{
\[
\begin{tabular}{cccccc}
\hline
\hline
\noalign{\smallskip}
\hline
Run	&	Telescope	& $\theta$	& Units		& $I(CO)$& $SCO_{3D}$	\\
        &                       &   "           &               & K km s$^{-1}$ & Jy  km s$^{-1}$ \\
\hline
2008,1	&	Kitt Peak	&	55	& $T_R^*$	& 6.77$\pm$0.30	& 1003$\pm$44	\\
2008,2	&			&		&		& 7.33$\pm$0.24	& 1086$\pm$36	\\
2009    &			&		&		& 4.90$\pm$0.39	& 726$\pm$58	\\
Mean$^a$&			&		&		& 6.22$\pm$0.25	& 922$\pm$37	\\
Elfhag et al. (1998)	&Onsala	&	34	& $T_{mb}$	& 12.10$\pm$1.41& 1060$\pm$124	\\
Young et al. (1995)	&FCRAO	&	45	& $T_A^*$	& 4.21$\pm$0.49	& 869$\pm$101	\\
\noalign{\smallskip}
\hline
\end{tabular}
\]
Notes: a: determined using the combined spectrum.}
\end{table*}

\clearpage

\begin{table}
\caption{Main parameters of the different radio telescopes}
\label{Tabtel}
{
\[
\begin{tabular}{cccc}
\hline
\hline
\noalign{\smallskip}
\hline
Telescope	& $\theta$(")	& Units		& Jy/K$^2$	\\
\hline
Kitt Peak$^1$	&	55	& $T_R^*$	& 39	\\
IRAM		& 	21	& $T_{mb}$	& 4.8	\\
FCRAO		&	45	& $T_A^*$	& 42	\\
SEST		&	43	& $T_{mb}$	& 19	\\
Onsala		&	34	& $T_{mb}$	& 12	\\
BELL		&	100	& $T_{mb}$	& 108	\\	
\noalign{\smallskip}
\hline
\end{tabular}
\]
Notes: 1: in some references CO fluxes are expressed in $T_{mb}$ scale.\\
2: conversion between K and Jy (see eq. 8).}
\end{table}

\begin{table*}
\caption{Best fit to the relations between integrated CO fluxes: $SCO_{y}$ = a $\times$ $SCO_{Kuno}$ + b}
\label{Tabfitintegrati}
{
\[
\begin{tabular}{ccccc}
\hline
\hline
\noalign{\smallskip}
\hline
$SCO_y$		& a			& b			& $\rho^1$	&$SCO_{Kuno}/SCO_y(corrected)^2$\\
\hline
$SCO_{Helfer}$	& 0.86($\pm$0.17)	& -85.28($\pm$428.62)	& 0.90		& 1.02 $\pm$ 0.18	\\
$SCO_{Chung}$	& 0.65($\pm$0.08)	& 44.84($\pm$157.85)	& 0.93		& 1.10 $\pm$ 0.43	\\
$SCO_{Young}^3$	& 0.71($\pm$0.07)	& 183.61($\pm$140.59)	& 0.88		& 1.12 $\pm$ 0.49	\\
\noalign{\smallskip}
\hline
\end{tabular}
\]
Notes: 1: Spearman correlation coefficient.\\
2: Ratio of the Kuno et al. (2007) to the Helfer et al. (2003), Chung et al. (2009a), and Young et al. (1995) $SCO$ fluxes once
these last have been corrected using the relations given in this table. For comparison, $SCO_{Kuno}/SCO_{IRAM}$ = 0.83 $\pm$ 0.03 for the two HRS
galaxies we observed with the IRAM radio telescope. \\
3: This relation is not used to correct the Young et al. (1995) data (see text).
}
\end{table*}

\begin{table*}
\caption{Disc scale height for edge-on galaxies}
\label{Tabedgeon}
{
\[
\begin{tabular}{ccccccccc}
\hline
\hline
\noalign{\smallskip}
\hline
Name		& $D$	      &  $r_{25}^a$	& $incl$     & $z_{CO}$      & $r_{dust}^b$   & $z_{dust}^b$  &      Reference \\
		& Mpc	      &  kpc		& $^o$ 	    & kpc	    & kpc	     & kpc	    &		    \\
\hline
\noalign{\smallskip}
NGC 891		& 9.5	      &  16.79		& 88.3 	    & 0.160-0.276   & 7.54	     & 0.200,0.240  & 1,2,3,4,5 \\
NGC 4013	& 11.6	      &  7.90		& 89.6 	    & - 	    & 2.60	     & 0.110	    & 4     \\
NGC 4565	& 16.9	      &  35.17		& 87.2 	    & - 	    & 40.36	     & 0.181	    & 7     \\
NGC 5529	& 29.6	      &  24.60		& 87.2 	    & - 	    & 7.70	     & 0.370	    & 4     \\
NGC 5907	& 11.0	      &  18.39		& 87.2 	    & - 	    & 5.30	     & 0.100	    & 4     \\
IC 2531		& 22.0	      &  21.60		& 89.6 	    & - 	    & 8.43	     & 0.210	    & 4     \\
UGC 1082	& 37.0	      &  14.38		& 89.7 	    & - 	    & 5.47	     & 0.290	    & 4     \\
UGC 2048	& 63.0	      &  33.23		& 89.6 	    & - 	    & 16.50	     & 0.500	    & 6     \\
\hline
\end{tabular}
\]
Notes: $a$: from NED; $b$: measured from optical images in the $i$-band. References: 1: Scoville et al. (1993); 
2: Yim et al. (2011); 3: Bianchi \& Xilouris (2011); 4: Xilouris e  al. (1999); 5: Xilouris et al. (1998); 6: Xilouris et al. (1997); 7: De Looze et al. (2012).}
\end{table*}

\begin{table*}
\caption{Comparison between extrapolated single-beam observations and integrated fluxes from Kuno et al. (2007)}
\label{Tabfitintegrati3D}
{
\[
\begin{tabular}{cccc}
\hline
\hline
\noalign{\smallskip}
\hline
Sample											& $SCO_{3D}/SCO_{Kuno}$		& $SCO_{2D}/SCO_{Kuno}$		& $SCO_{Saintonge}/SCO_{Kuno}$\\
\hline
All galaxies										& 1.33$\pm$1.01			& 1.32$\pm$0.98			& 0.64$\pm$0.48			\\
All galaxies with $SCO_{Kuno}$ $<$ 5000 Jy km s$^{-1}$ \& N.Beam$\leq$5			& 0.92$\pm$0.39			& 0.92$\pm$0.39			& 0.77$\pm$0.34			\\
All galaxies with N.Beam$>$5								& 1.55$\pm$1.17			& 1.54$\pm$1.13			& 0.57$\pm$0.53			\\
All galaxies with $SCO_{Kuno}$ $<$ 5000 Jy km s$^{-1}$ \& N.Beam$\leq$5	\& ref FCRAO	& 1.02$\pm$0.47			& 1.01$\pm$0.47			& 0.85$\pm$0.46			\\
\noalign{\smallskip}
\hline
\end{tabular}
\]
}
\end{table*}

\begin{table*}
\caption{Total CO emission of edge-on galaxies}
\label{TabedgeonCOdata}
{
\[
\begin{tabular}{ccccccccc}
\hline
\hline
\noalign{\smallskip}
\hline
Name		& incl 	& $SCO(CB)$	& Ref	& $\theta$	& $SCO_{2D}$	& $SCO_{3D}$	& $SCO_{Int}$			& Ref	\\
		& $^o$ 	& Jy km s$^{-1}$& 	& "     	& Jy km s$^{-1}$& Jy km s$^{-1}$& Jy km s$^{-1}$		&	\\
\hline
\noalign{\smallskip}
NGC 891		& 88.3 	& 606.5		& 1	& 45		& 3440		& 3579		& 5398-7197; 1970; 4570; 5226	& 4$^a$; 5$^b$; 1$^c$; 2$^c$ 	\\
		&	& 1509.3	& 2	& 55		& 7090		& 7288		& 				& 				\\
NGC 4565	& 87.2	& 177.6		& 3	& 34		& 1452		& 1550		& 1672; 1643			& 6$^b$; 7$^b$			\\
NGC 5907	& 87.2	& 173.0		& 1	& 45		& 952		& 985		& 1750; 725; 1230		& 8$^b$; 9$^b$; 1$^c$		\\
		&	& 150.0		& 3	& 34		& 1097		& 1160		& 				& 				\\
\hline
\end{tabular}
\]
Notes: $a$: from interferometric observations; $b$: from partial mapping along the major and minor axes; $c$: from mapping along the major axis; 
References: 1: Young et al. (1995; FCRAO); 2: Sage (1993); 3: Elfhag et al. (1996); 4: Scoville et al. (1993); the flux given in this reference is 5398 
Jy km s$^{-1}$, and is estimated by the authors to represent between 75 and 100\% ~ of the total flux of the galaxy; 
5: Sakamoto et al. (1997); 6: Neininger et al. (1996); 7: Sofue \& Nakai (1994); 8: Dumke et al. (1997); 9:
Sofue \& Nakai (1993). }
\end{table*}

\begin{table*}
\caption{Comparison between total CO fluxes extrapolated using the 3D aperture correction using only the central beam proposed in this work (eq. 11) 
and the Saintonge et al. (2011) multiple-beam prescriptions (eq. 13) for galaxies in the Kuno et al. (2007) sample}
\label{Tabmultiple}
{
\[
\begin{tabular}{ccc}
\hline
\hline
\noalign{\smallskip}
\hline
Sample											& $SCO_{3D}/SCO_{Kuno}$		& $SCO_{Saintonge_{MB}}/SCO_{Kuno}$\\
\hline
All galaxies										& 1.35$\pm$1.08			& 0.54$\pm$0.77			\\
All galaxies with $SCO_{Kuno}$ $<$ 5000 Jy km s$^{-1}$ \& N.Beam$\leq$5			& 1.03$\pm$0.40			& 1.45$\pm$0.95			\\
All galaxies with N.Beam$>$5								& 1.50$\pm$1.27			& 0.65$\pm$0.47			\\
\noalign{\smallskip}
\hline
\end{tabular}
\]
}
\end{table*}

\clearpage

\onecolumn
\begin{center} \tiny \begin{longtable}{l c c c c c c l c c c c c c c}
\caption[Central beam observations of the HRS]{Central beam observations of the HRS.\label{TabCOHRSCB}}\\
\hline \hline \\[-2ex]
\multicolumn{1}{c}{HRS} &
\multicolumn{1}{c}{Tel} &
\multicolumn{1}{c}{N. beam} &
\multicolumn{1}{c}{$\theta$} &
\multicolumn{1}{c}{sign} &
\multicolumn{1}{c}{$I(CO)$} &
\multicolumn{1}{c}{rms} &
\multicolumn{1}{c}{$\delta V_{CO}$} &
\multicolumn{1}{c}{T scale} &
\multicolumn{1}{c}{Ref} &
\multicolumn{1}{c}{Jy/K} &
\multicolumn{1}{c}{$SCO_{3D}$} &
\multicolumn{1}{c}{$ff$} &
\multicolumn{1}{c}{flag} &
\multicolumn{1}{c}{Notes} \\
\multicolumn{1}{c}{(1)} &
\multicolumn{1}{c}{(2)} &
\multicolumn{1}{c}{(3)} &
\multicolumn{1}{c}{(4)} &
\multicolumn{1}{c}{(5)} &
\multicolumn{1}{c}{(6)} &
\multicolumn{1}{c}{(7)} &
\multicolumn{1}{c}{(8)} &
\multicolumn{1}{c}{(9)} &
\multicolumn{1}{c}{(10)} &
\multicolumn{1}{c}{(11)} &
\multicolumn{1}{c}{(12)} &
\multicolumn{1}{c}{(13)} &
\multicolumn{1}{c}{(14)} &
\multicolumn{1}{c}{(15)} \\
\multicolumn{1}{c}{} &
\multicolumn{1}{c}{} &
\multicolumn{1}{c}{} &
\multicolumn{1}{c}{arcsec} &
\multicolumn{1}{c}{} &
\multicolumn{1}{c}{K km s$^{-1}$} &
\multicolumn{1}{c}{mK} &
\multicolumn{1}{c}{km s$^{-1}$} &
\multicolumn{1}{c}{} &
\multicolumn{1}{c}{} &
\multicolumn{1}{c}{Jy K$^{-1}$} &
\multicolumn{1}{c}{Jy km s$^{-1}$} &
\multicolumn{1}{c}{} &
\multicolumn{1}{c}{} &
\multicolumn{1}{c}{} \\ [0.5ex] \hline
   \\[-1.8ex]
\endfirsthead
\multicolumn{15}{c}{{\tablename} \thetable{} -- Continued} \\[0.5ex]
  \hline \hline \\[-2ex]
\multicolumn{1}{c}{HRS} &
\multicolumn{1}{c}{Tel} &
\multicolumn{1}{c}{N. beam} &
\multicolumn{1}{c}{$\theta$} &
\multicolumn{1}{c}{sign} &
\multicolumn{1}{c}{$I(CO)$} &
\multicolumn{1}{c}{rms} &
\multicolumn{1}{c}{$\delta V_{CO}$} &
\multicolumn{1}{c}{T scale} &
\multicolumn{1}{c}{Ref} &
\multicolumn{1}{c}{Jy/K} &
\multicolumn{1}{c}{$SCO_{3D}$} &
\multicolumn{1}{c}{$ff$} &
\multicolumn{1}{c}{flag} &
\multicolumn{1}{c}{Notes} \\
\multicolumn{1}{c}{(1)} &
\multicolumn{1}{c}{(2)} &
\multicolumn{1}{c}{(3)} &
\multicolumn{1}{c}{(4)} &
\multicolumn{1}{c}{(5)} &
\multicolumn{1}{c}{(6)} &
\multicolumn{1}{c}{(7)} &
\multicolumn{1}{c}{(8)} &
\multicolumn{1}{c}{(9)} &
\multicolumn{1}{c}{(10)} &
\multicolumn{1}{c}{(11)} &
\multicolumn{1}{c}{(12)} &
\multicolumn{1}{c}{(13)} &
\multicolumn{1}{c}{(14)} &
\multicolumn{1}{c}{(15)}\\
\multicolumn{1}{c}{} &
\multicolumn{1}{c}{} &
\multicolumn{1}{c}{} &
\multicolumn{1}{c}{arcsec} &
\multicolumn{1}{c}{} &
\multicolumn{1}{c}{K km s$^{-1}$} &
\multicolumn{1}{c}{mK} &
\multicolumn{1}{c}{km s$^{-1}$} &
\multicolumn{1}{c}{} &
\multicolumn{1}{c}{} &
\multicolumn{1}{c}{Jy K$^{-1}$} &
\multicolumn{1}{c}{Jy km s$^{-1}$} &
\multicolumn{1}{c}{} &
\multicolumn{1}{c}{} &
\multicolumn{1}{c}{}  \\ [0.5ex] \hline
  \\[-1.8ex]
\endhead 
 \\[-1.8ex] \hline
  \multicolumn{15}{c}{{Continued on next page\ldots}} \\
\endfoot
  \\[-1.8ex] \hline \hline
\endlastfoot
           3    &  FCRAO  & 	  1   &        45    &    0  &        -     &     3.00  &      15.0    &  $T^*_A$  & 	   3  &        42    &      126.29  &   0.087  & 	 1 &   \\
           3    &  FCRAO  & 	  1   &        45    &    0  &        -     &     8.00  &      15.0    &  $T^*_A$  & 	   1  &        42    &      336.78  &   0.087  & 	 1 &   \\
           3    &  IRAM   & 	  1   &        22    &    0  &        -     &     3.52  &      31.0    &  $T_{mb}$ & 	  23  &       4.7    &       57.82  &   0.021  & 	 1 &   \\
           4    &  IRAM   & 	  1   &        36    &    1  &      54.00   &    10.00  & 	5.2    &  $T_{mb}$ & 	  22  &       4.8    &     1085.37  &   0.038  & 	 1 &   \\
           7    &  IRAM   & 	  1   &        22    &    0  &       0.74   &     3.67  &      31.0    &  $T_{mb}$ & 	  23  &       4.7    &       37.54  &   0.029  & 	 1 &   \\
           7    &  KP12m  & 	  1   &        55    &    0  &        -     &     3.00  &      30.0    &  $T^*_R$  & 	   4  &        39    &       94.70  &   0.182  & 	 1 &   \\
           8    &  KP12m  & 	  1   &        55    &    0  &        -     &     3.00  &      15.7    &  $T^*_R$  & 	   5  &        39    &       91.43  &   0.194  & 	 1 &   \\
           9    &  KP12m  & 	  1   &        55    &    1  &       1.79   &     3.00  &      15.7    &  $T^*_R$  & 	   5  &        39    &      119.08  &   0.220  & 	 1 &   \\
\end{longtable}
References to the CO data:\\
1:  Young et al. (1995),	
2:  Stark et al. (1986),	
3:  Thronson et al. (1989),	
4:  Sage \& Wrobel (1989),	
5:  This work,			
6:  Elfhag et al. (1996),	
7: Albrecht et al. (2007),	
8: Sauty et al. (2003),		
9: Gondhalekar et al. (1998),	
10: Boselli et al. (1995),	
11: Welch \& Sage (2003),	
12: Smith \& Madden (1997),	
13: Sage (1993),		
14: Boselli et al. (2002),	
15: Leroy et al. (2005),	
16: Bregman \& Hogg (1988),	
17: B\"oker et al. (2003),	
18: Jaffe (1987),		
19: Sage et al. (2007),		
20: Kenney, private communication, 
21: Jackson et al. (1989),	
22: Braine et al. (1993),	
23: Young et al. (2011),	
24: Combes et al. (2007),	
25: Wiklind et al. (1995)\\	
Notes: 1: assumed in $T_R^*$. 
2: transformed from the original table to this scale.
3: partly mapped galaxy, with an equivalent beamsize assumed to be $\theta$=36
arcsec.\\		
\end{center}		     							       

\twocolumn

\onecolumn
\begin{center} \tiny 
\begin{longtable}{l c c c c c c c c}
\caption[Integrated CO fluxes and H$_2$ masses of the HRS]{Integrated CO fluxes and molecular hydrogen masses of the HRS of the HRS.\label{TabHRSH2}}\\
\hline \hline \\[-2ex]
\multicolumn{1}{c}{HRS} &
\multicolumn{1}{c}{sgn} &
\multicolumn{1}{c}{$SCO$} &
\multicolumn{1}{c}{err} &
\multicolumn{1}{c}{$S/N$} &
\multicolumn{1}{c}{$ff$} &
\multicolumn{1}{c}{log $M(H_2$)$_c$} &
\multicolumn{1}{c}{log $M(H_2$)$_v$} &
\multicolumn{1}{c}{Ref} \\
\multicolumn{1}{c}{(1)} &
\multicolumn{1}{c}{(2)} &
\multicolumn{1}{c}{(3)} &
\multicolumn{1}{c}{(4)} &
\multicolumn{1}{c}{(5)} &
\multicolumn{1}{c}{(6)} &
\multicolumn{1}{c}{(7)} &
\multicolumn{1}{c}{(8)} &
\multicolumn{1}{c}{(9)} \\
\multicolumn{1}{c}{} &
\multicolumn{1}{c}{} &
\multicolumn{1}{c}{Jy km s$^{-1}$} &
\multicolumn{1}{c}{Jy km s$^{-1}$} &
\multicolumn{1}{c}{} &
\multicolumn{1}{c}{} &
\multicolumn{1}{c}{M$\odot$} &
\multicolumn{1}{c}{M$\odot$} &
\multicolumn{1}{c}{} \\ [0.5ex] \hline
   \\[-1.8ex]
\endfirsthead
\multicolumn{9}{c}{{\tablename} \thetable{} -- Continued} \\[0.5ex]
  \hline \hline \\[-2ex]
\multicolumn{1}{c}{HRS} &
\multicolumn{1}{c}{sgn} &
\multicolumn{1}{c}{$SCO$} &
\multicolumn{1}{c}{err} &
\multicolumn{1}{c}{$S/N$} &
\multicolumn{1}{c}{$ff$} &
\multicolumn{1}{c}{log $M(H_2$)$_c$} &
\multicolumn{1}{c}{log $M(H_2$)$_v$} &
\multicolumn{1}{c}{Ref} \\
\multicolumn{1}{c}{(1)} &
\multicolumn{1}{c}{(2)} &
\multicolumn{1}{c}{(3)} &
\multicolumn{1}{c}{(4)} &
\multicolumn{1}{c}{(5)} &
\multicolumn{1}{c}{(6)} &
\multicolumn{1}{c}{(7)} &
\multicolumn{1}{c}{(8)} &
\multicolumn{1}{c}{(9)} \\
\multicolumn{1}{c}{} &
\multicolumn{1}{c}{} &
\multicolumn{1}{c}{Jy km s$^{-1}$} &
\multicolumn{1}{c}{Jy km s$^{-1}$} &
\multicolumn{1}{c}{} &
\multicolumn{1}{c}{} &
\multicolumn{1}{c}{M$\odot$} &
\multicolumn{1}{c}{M$\odot$} &
\multicolumn{1}{c}{} \\ [0.5ex] \hline
  \\[-1.8ex]
\endhead 
 \\[-1.8ex] \hline
  \multicolumn{9}{c}{{Continued on next page\ldots}} \\
\endfoot
  \\[-1.8ex] \hline \hline
\endlastfoot
           1       &    -  &         -   & 	  -    &      -   & 	  -     &      -     &     -    & 		 -  \\ 
           2       &    -  &         -   & 	  -    &      -   & 	  -     &      -     &     -    & 		 -  \\ 
           3       &    0  &       57.81 & 	  -    &      -   & 	 0.021  &     8.16   &    8.05  & 		23  \\ 
           4       &    1  &     1085.00 &    1230.39  &    21.72 & 	 0.038  &     9.42   &    9.17  & 		22  \\ 
           5       &    -  &         -   & 	  -    &      -   & 	  -     &      -     &     -    & 		 -  \\ 
           6       &    -  &         -   & 	  -    &      -   & 	  -     &      -     &     -    & 		 -  \\ 
           7       &    0  &       37.54 & 	  -    &      -   & 	 0.029  &     8.07   &    7.86  & 		23  \\ 
           8       &    0  &       91.42 & 	  -    &      -   & 	 0.194  &     8.49   &    8.39  & 		 5  \\
\end{longtable}
References to the CO data:\\
1:  Young et al. (1995),	
2:  Stark et al. (1986),	
3:  Thronson et al. (1989),	
4:  Sage \& Wrobel (1989),	
5:  This work,			
6:  Elfhag et al. (1996),	
7: Albrecht et al. (2007),	
8: Sauty et al. (2003),		
9: Gondhalekar et al. (1998),	
10: Boselli et al. (1995),	
11: Welch \& Sage (2003),	
12: Smith \& Madden (1997),	
13: Sage (1993),		
14: Boselli et al. (2002),	
15: Leroy et al. (2005),	
16: Bregman \& Hogg (1988),	
17: B\"oker et al. (2003),	
18: Jaffe (1987),		
19: Sage et al. (2007),		
20: Kenney, private communication, 
21: Jackson et al. (1989),	
22: Braine et al. (1993),	
23: Young et al. (2011),	
24: Combes et al. (2007),	
25: Wiklind et al. (1995),	
26: Kuno et al. (2007),		
27: Chung et al. (2009a),	
28: Neininger et al. (1996).\\	
\end{center}		     							       

\twocolumn

\onecolumn
\begin{center} \tiny 
\begin{longtable}{c c c c c c c c c}
\caption{HI data.\label{TabHI}}\\
\hline \hline \\[-2ex]
\multicolumn{1}{c}{HRS} &
\multicolumn{1}{c}{sgn}&      
\multicolumn{1}{c}{rms} &    
\multicolumn{1}{c}{$SHI$} &     
\multicolumn{1}{c}{}log $M(HI)$ &    
\multicolumn{1}{c}{$WHI$}  & 
\multicolumn{1}{c}{$HI-def$}  &
\multicolumn{1}{c}{flag} &   
\multicolumn{1}{c}{Ref}      \\
\multicolumn{1}{c}{} &
\multicolumn{1}{c}{}&      
\multicolumn{1}{c}{mJy} &    
\multicolumn{1}{c}{Jy km s$^{-1}$} &     
\multicolumn{1}{c}{M$\odot$} &    
\multicolumn{1}{c}{km s$^{-1}$}  & 
\multicolumn{1}{c}{} &   
\multicolumn{1}{c}{} &   
\multicolumn{1}{c}{}      \\
\multicolumn{1}{c}{(1)} &
\multicolumn{1}{c}{(2)}&      
\multicolumn{1}{c}{(3)} &    
\multicolumn{1}{c}{(4)} &     
\multicolumn{1}{c}{(5)} &    
\multicolumn{1}{c}{(6)}  & 
\multicolumn{1}{c}{(7)} &   
\multicolumn{1}{c}{(8)} &   
\multicolumn{1}{c}{(9)}      \\[0.5ex] \hline
  \\[-1.8ex]
\endhead 
 \\[-1.8ex] \hline
  \multicolumn{8}{c}{{Continued on next page\ldots}} \\
\endfoot

  \\[-1.8ex] \hline \hline
\endlastfoot
           1   &       - &       -    &  	 -     &       -    &	   -	&    -     &   -  &	  -  \\ 
           2   &       1 &      2.55  &  	3.32   &      8.42  &	 185	&  0.07    &   1  &	  1  \\ 
           3   &       1 &       -    &  	6.79   &      8.65  &	 388	&  0.38    &   1  &	  3  \\ 
           4   &       1 &       -    &        15.00   &      8.98  &	 475	&  0.57    &   2  &	  4  \\ 
           5   &       1 &      1.14  &  	2.41   &      8.20  &	 286	&  0.66    &   1  &	  2  \\ 
           6   &       1 &      9.42  &  	9.29   &      8.89  &	 177	&  0.62    &   1  &	  2  \\ 
           7   &       0 &      1.05  &  	1.57   &      8.12  &	   -	&  1.02    &   -  &	  2  \\ 
           8   &       1 &     13.74  &        51.37   &      9.66  &	 394	&  0.07    &   1  &	  2  \\
\end{longtable}
References to the HI data:\\
 1: Haynes et al. (2011),
 2: Springob et al. (2005),
 3: Chamaraux et al. (1987),
 4: Davis \& Seaquist (1983),
 5: Bicay \& Giovanelli (1987),
 6: Peterson (1979),
 7: Helou et al. (1982),
 8: Bicay \& Giovanelli (1986),
 9: Lake \& Schommer (1984),
10: Staveley-Smith \& Davies (1988),
11: Krumm \& Salpeter (1979),
12: Helou et al. (1981),
13: O'Neil (2004),
14: Schneider et al. (1992),
15 Noordermeer et al. (2005),
16: Courtois et al. (2009),
17: Richter \& Huchtmeier (1987),
18: Huchtmeier \& Seiradakis (1985),
19: Huchtmeier et al. (2005),
20: Helou et al. (1984),
21: Lu et al. (1993),
22: Bottinelli et al. (1982),
23: Giovanardi et al. (1983a),
24: Wardle \& Knapp (1986),
25: van Driel et al. (2000),
26: Hoffman et al. (1989),
27: Taylor et al. (2012),
28: Fisher \& Tully (1981),
29: Bottinelli et al. (1990),
30: Magri (1994),
31: Hoffman et al. (1987),
32: Haynes \& Giovanelli (1986),
33: Wong et al. (2006),
34: Chung et al. (2009b),
35: Koribalski et al. (2004),
36: Giovanardi et al. (1983b),
37: Haynes et al. (2000),
38: Gavazzi et al. (2005),
39: Theureau et al. (1998),
40: Haynes \& Gioanelli (1991),
41: de Vaucouleurs et al. (1991),
42: Lewis (1987),
43: Huchtmeier \& Richter (1989),
44: Smoker et al. (2000).
\end{center}

\twocolumn

 
   \begin{figure*}
   \centering
   \includegraphics[width=0.22\textwidth]{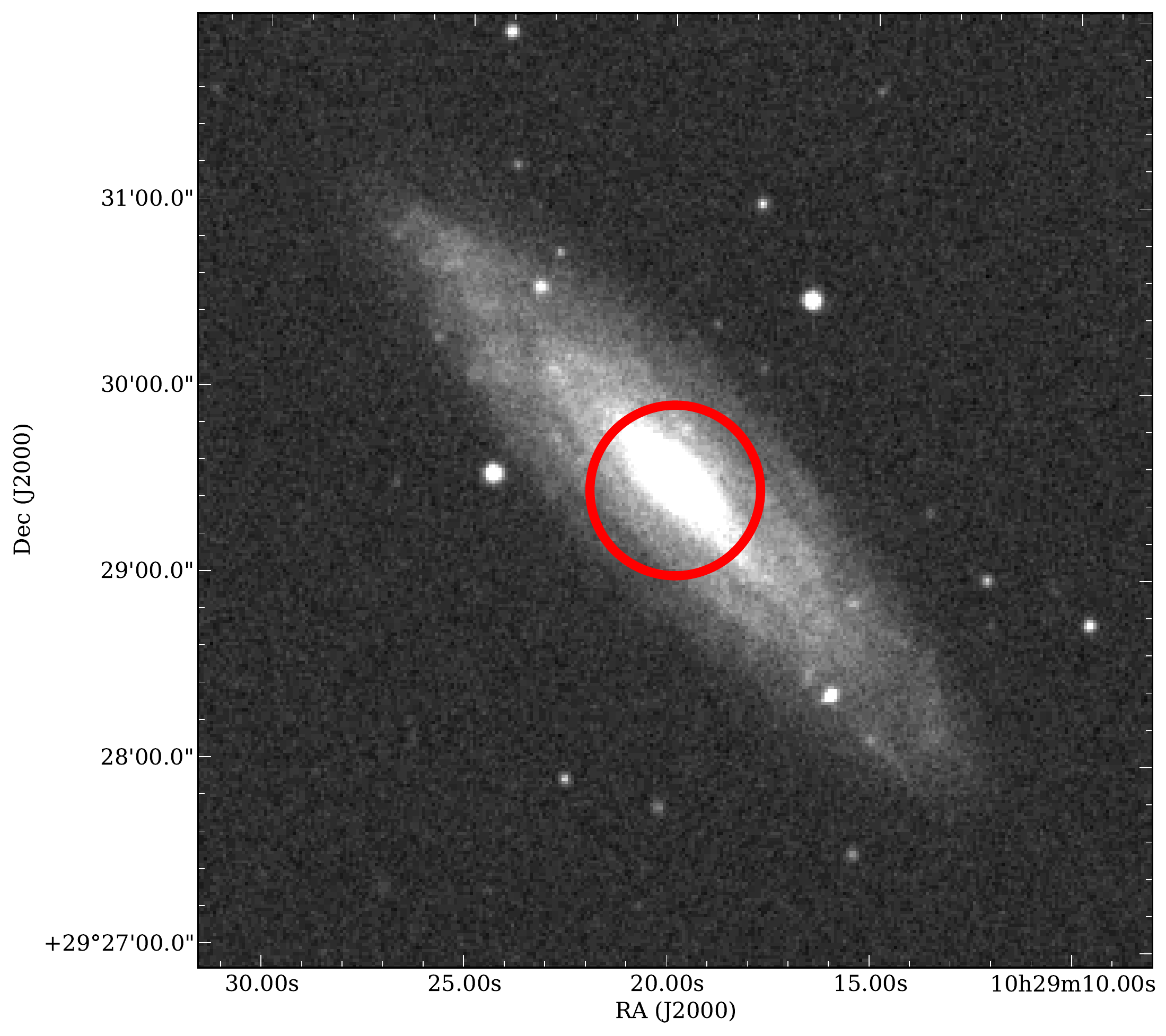}
   \includegraphics[width=0.22\textwidth]{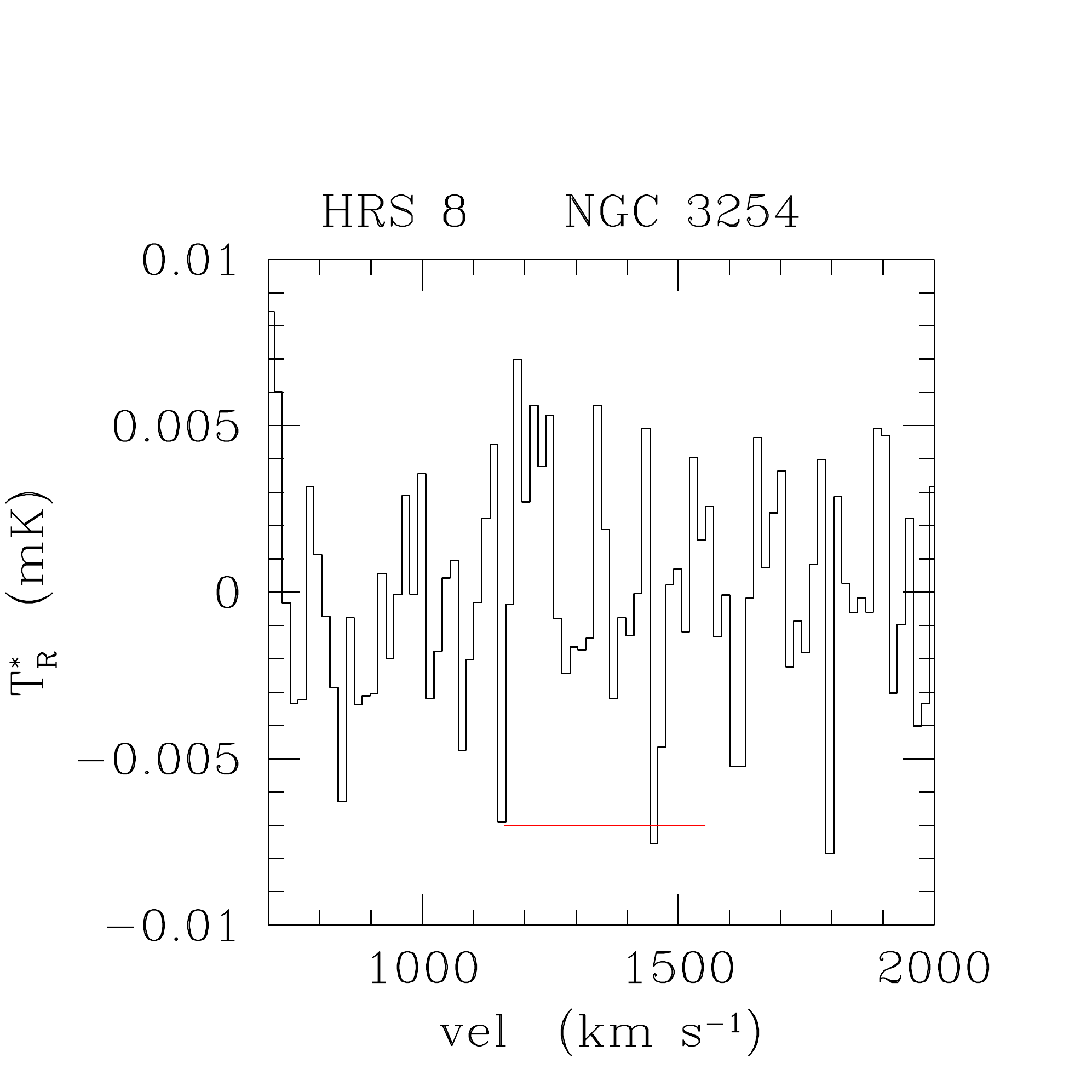}\\
   \includegraphics[width=0.22\textwidth]{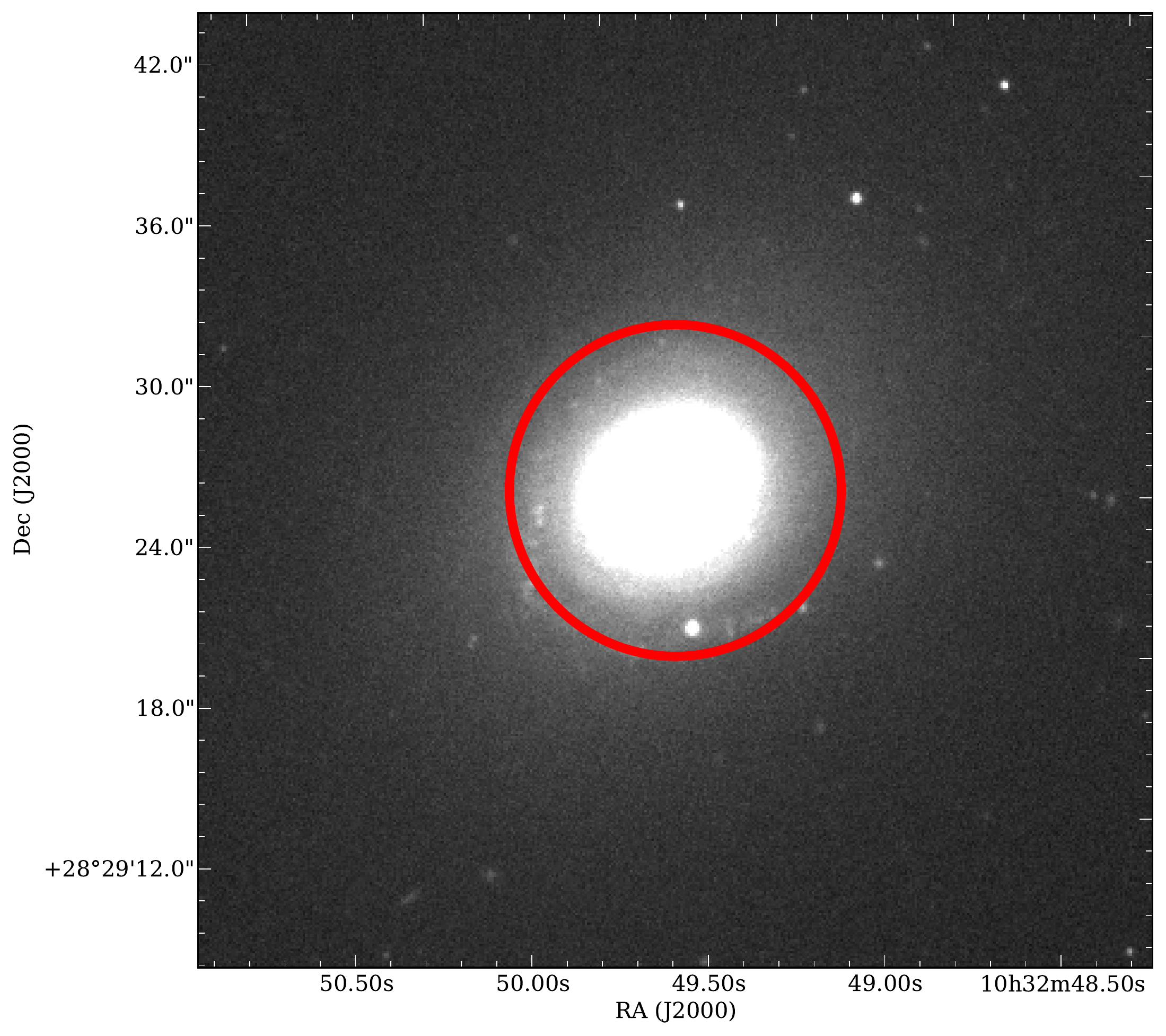}
   \includegraphics[width=0.22\textwidth]{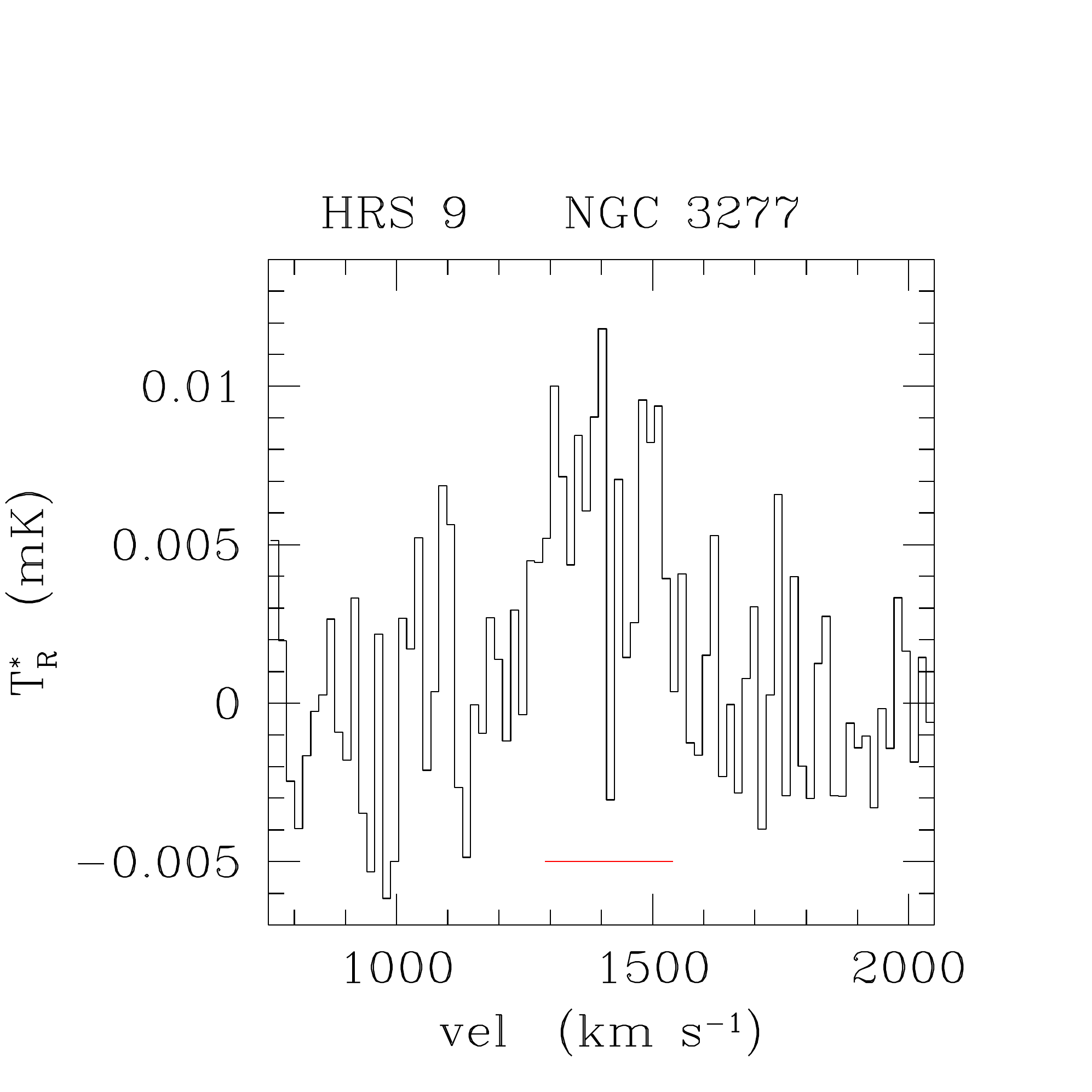}\\
   \includegraphics[width=0.22\textwidth]{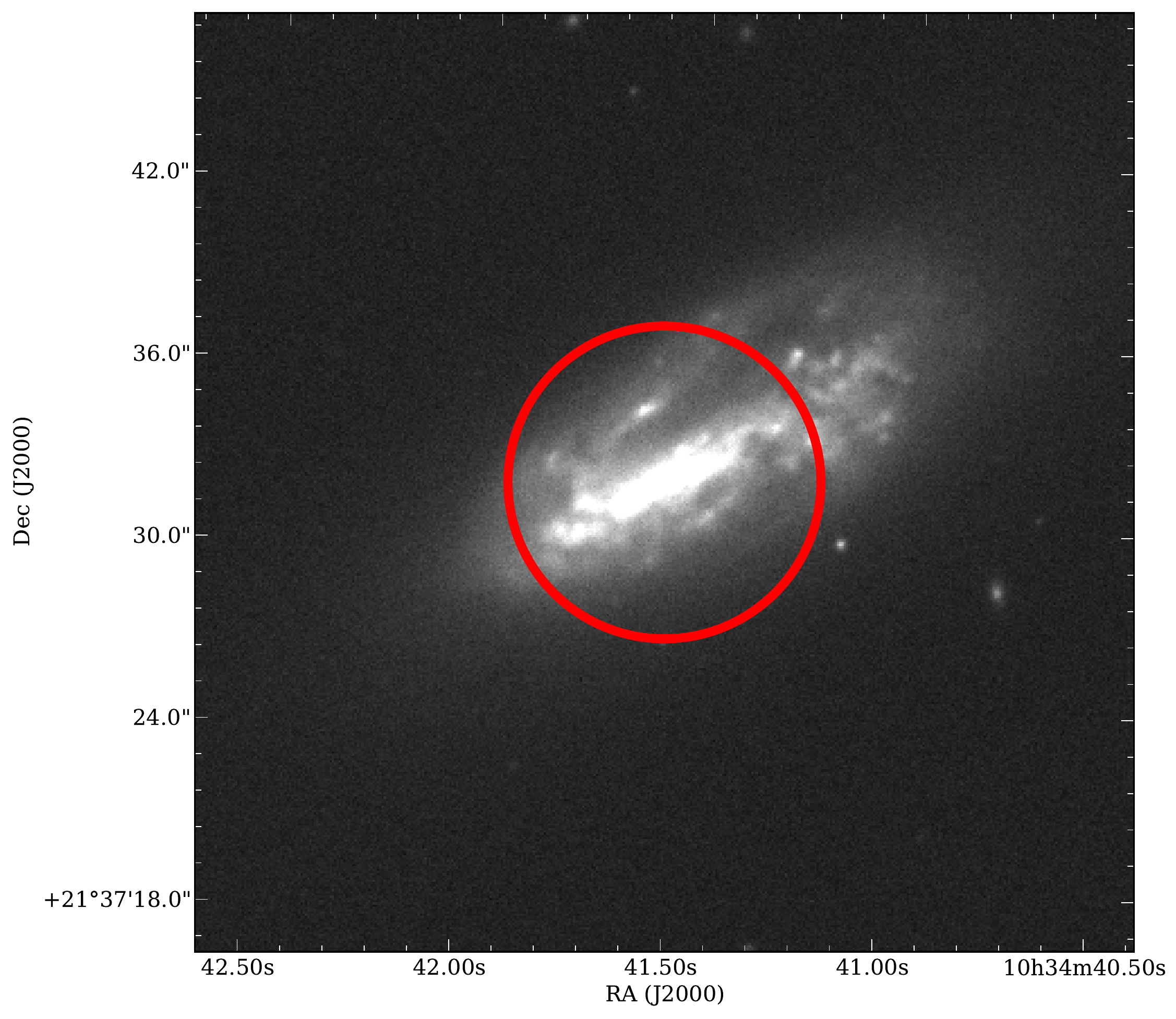}
   \includegraphics[width=0.22\textwidth]{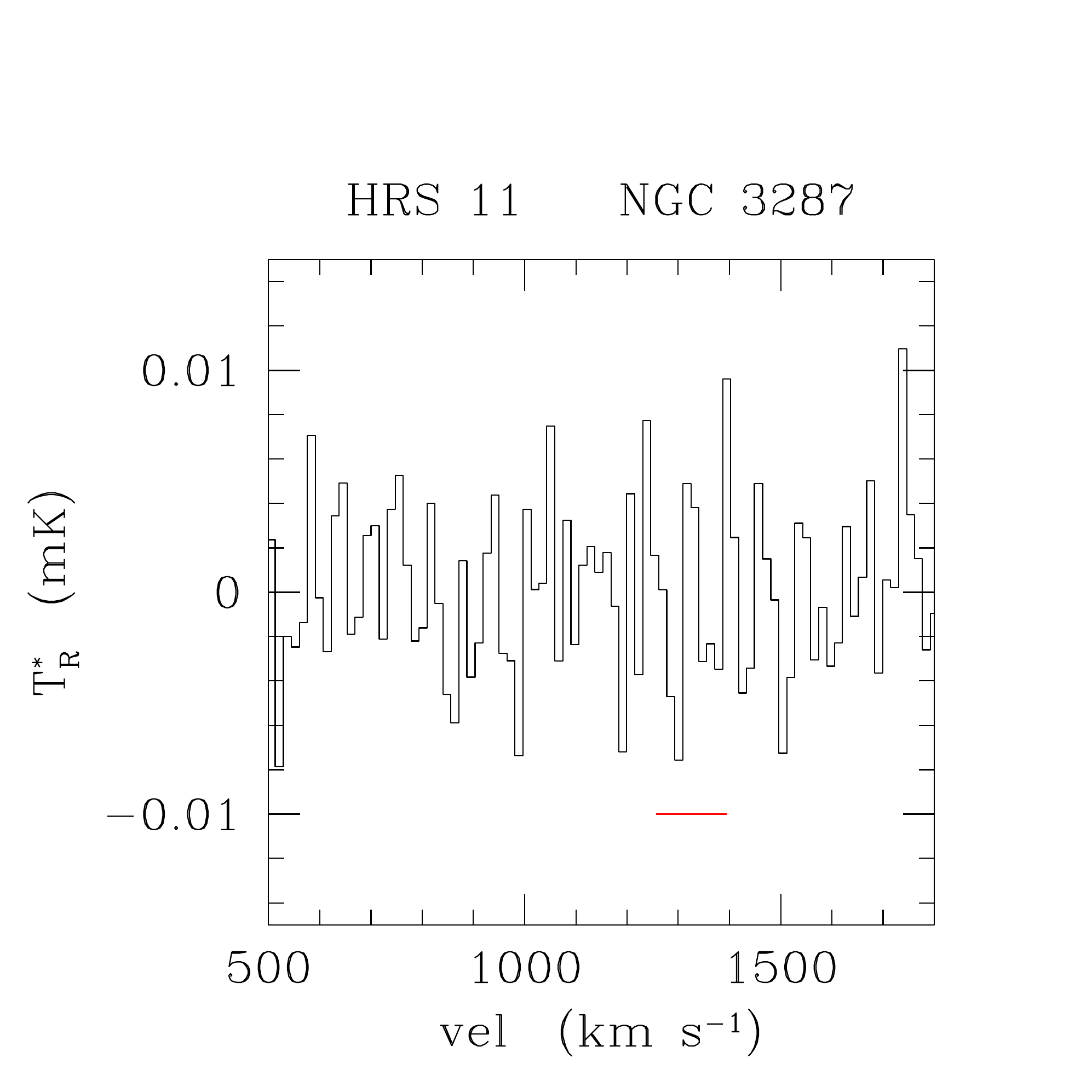}\\
   \includegraphics[width=0.22\textwidth]{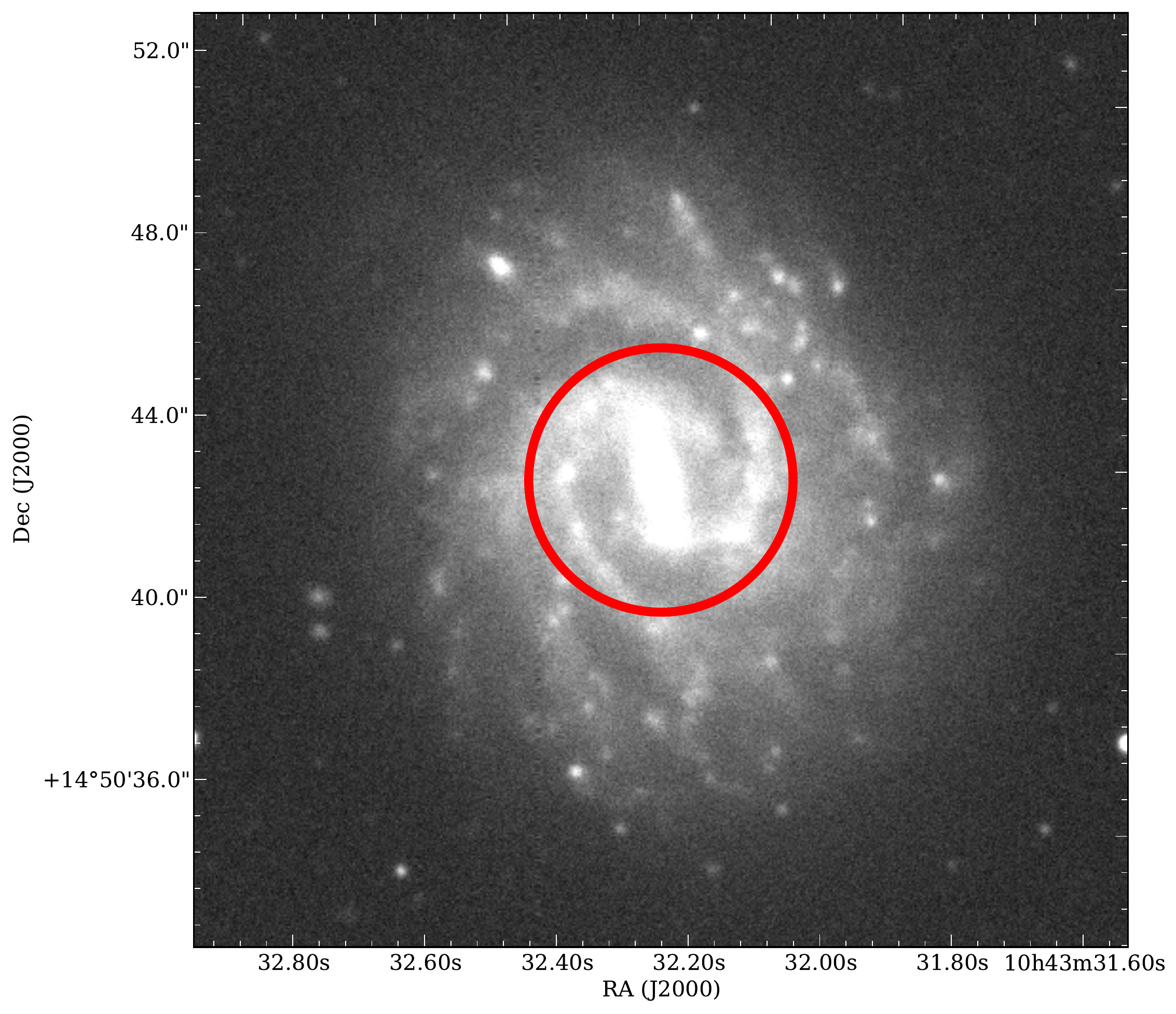}
   \includegraphics[width=0.22\textwidth]{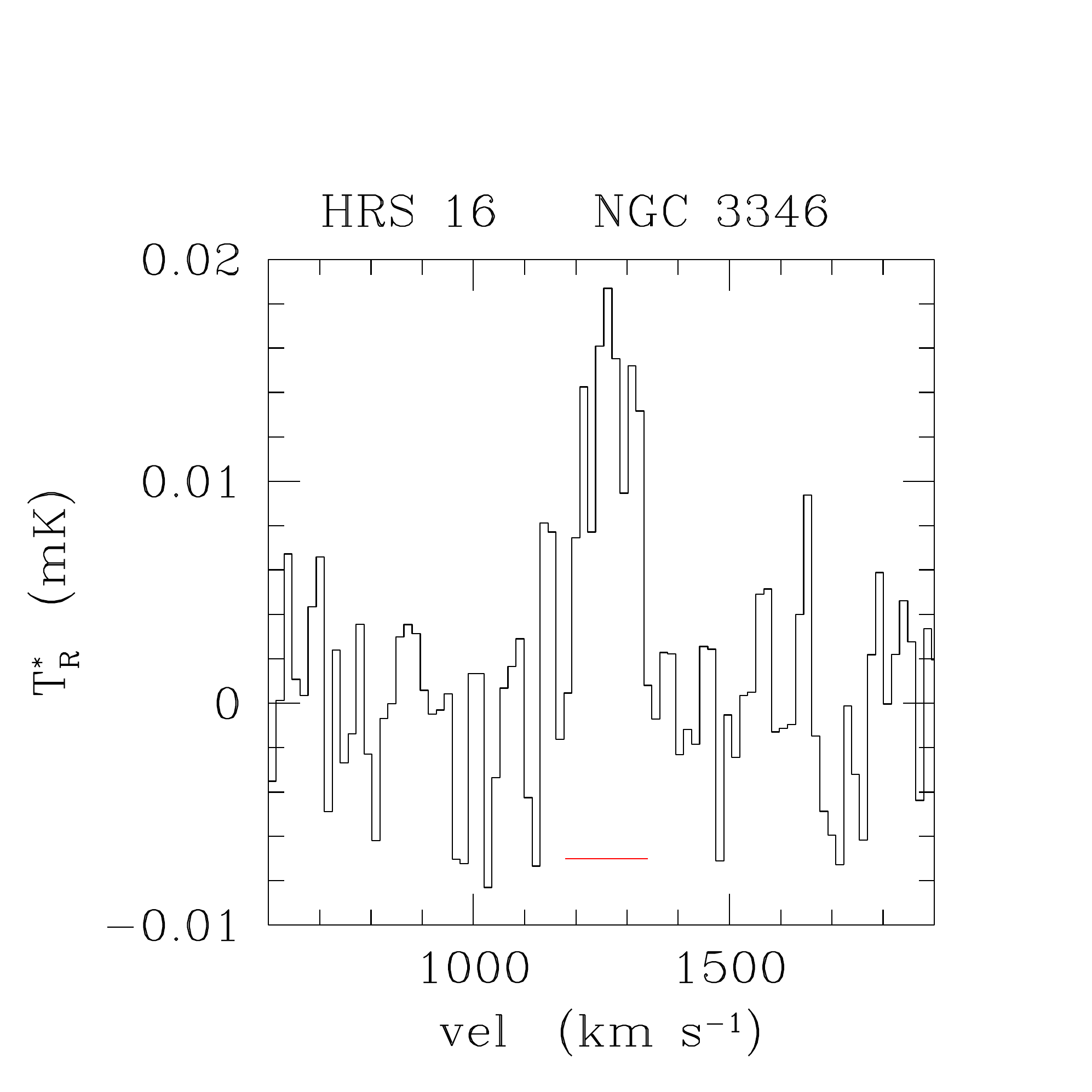}\\
   \includegraphics[width=0.22\textwidth]{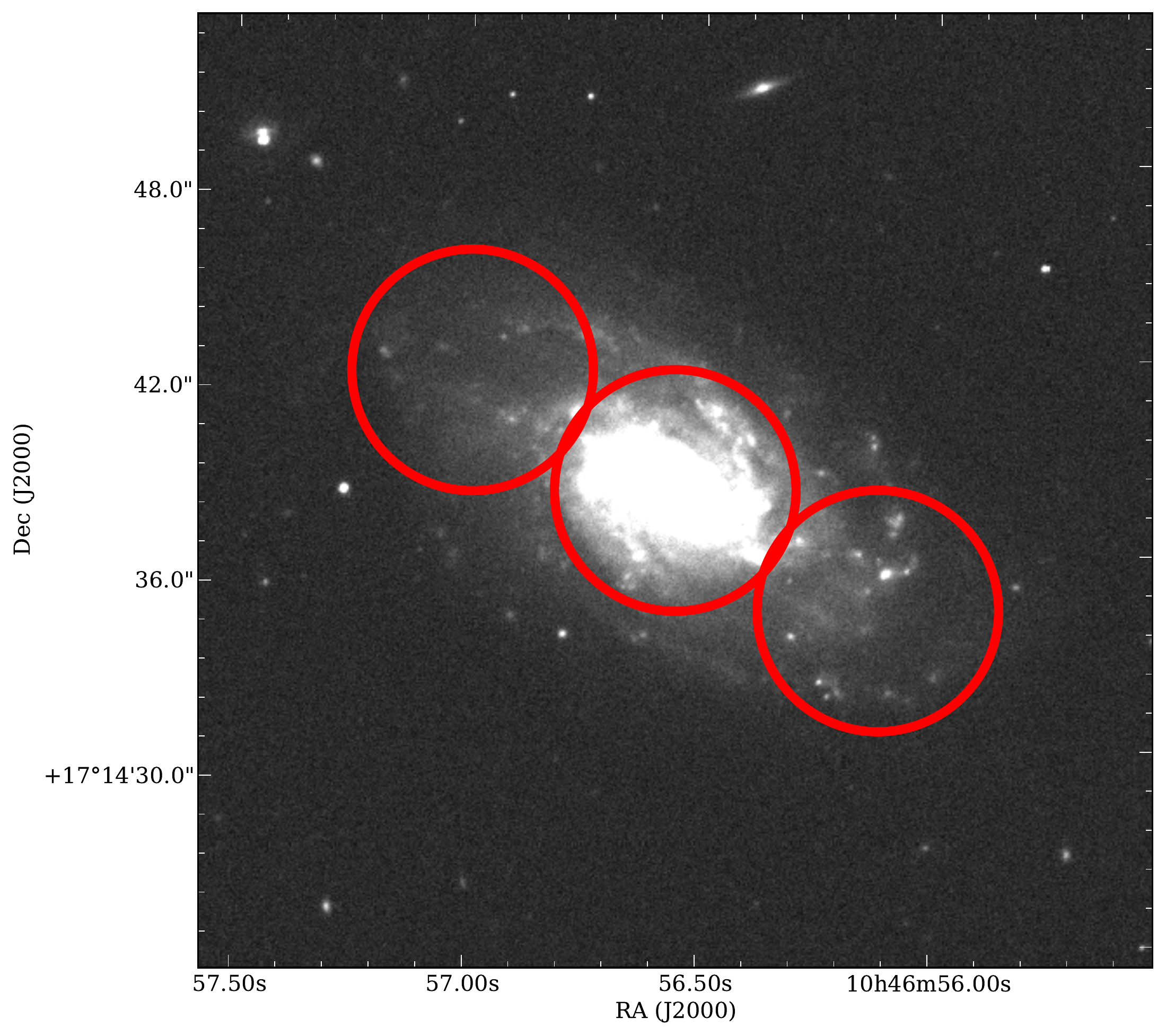}
   \includegraphics[width=0.22\textwidth]{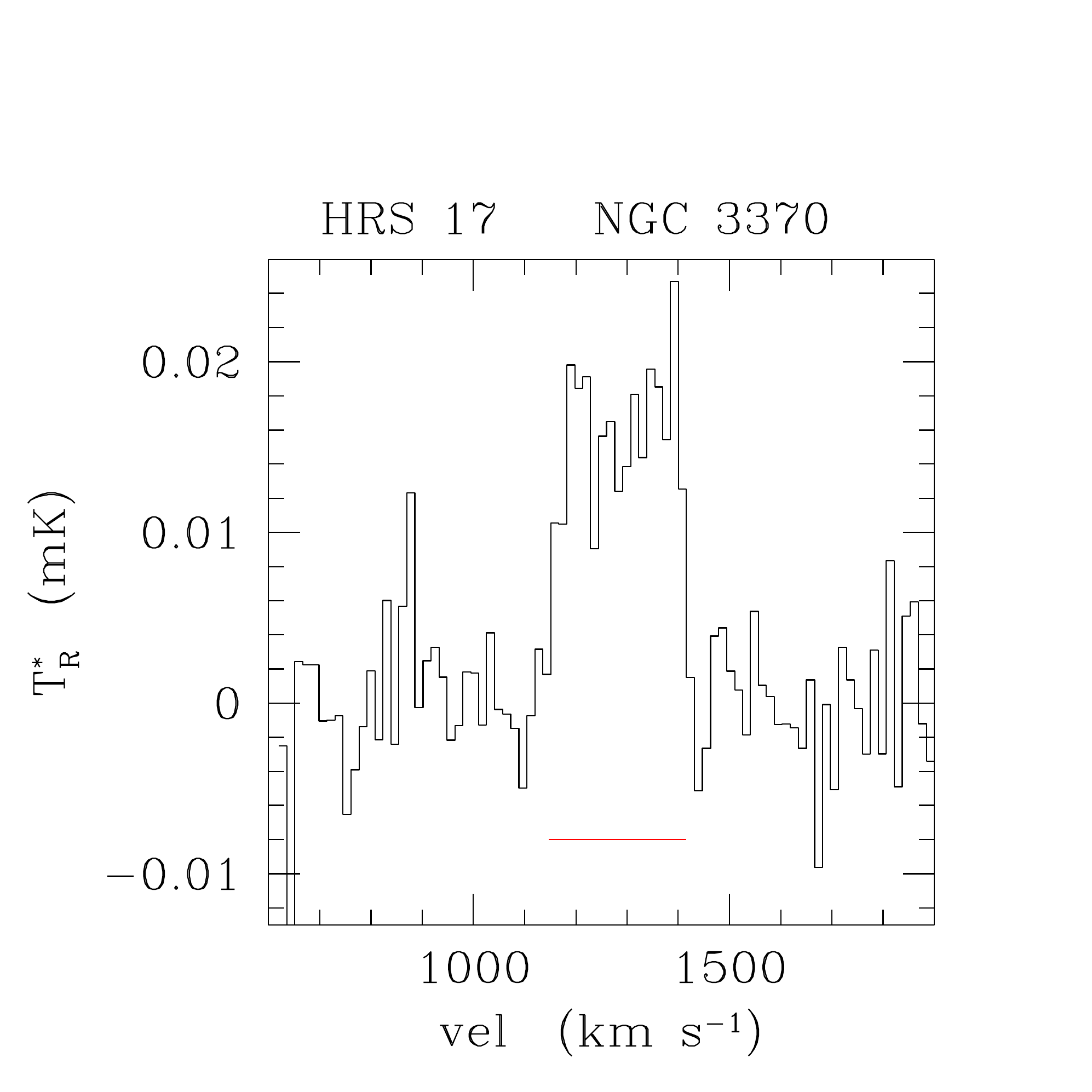}
   \includegraphics[width=0.22\textwidth]{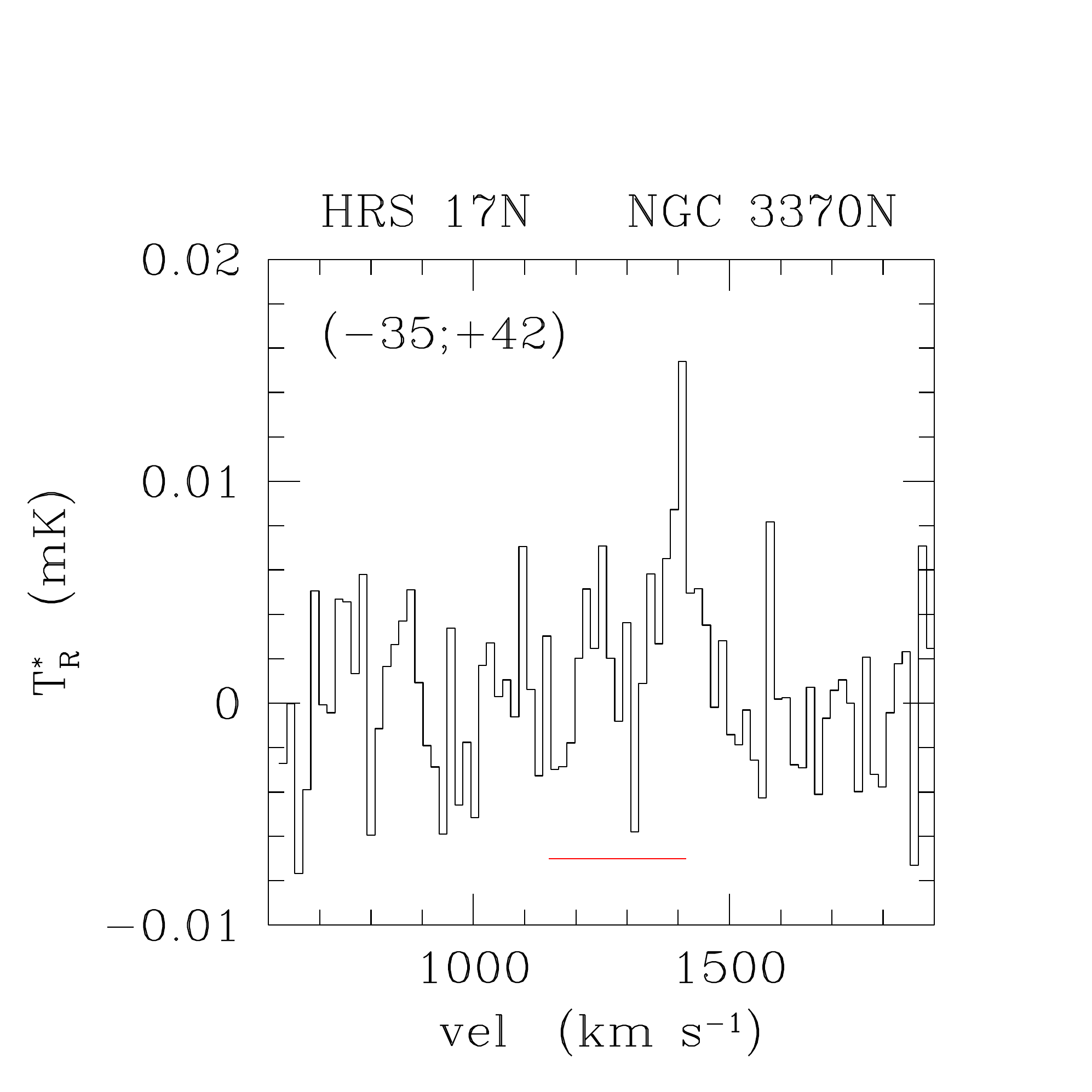}
   \includegraphics[width=0.22\textwidth]{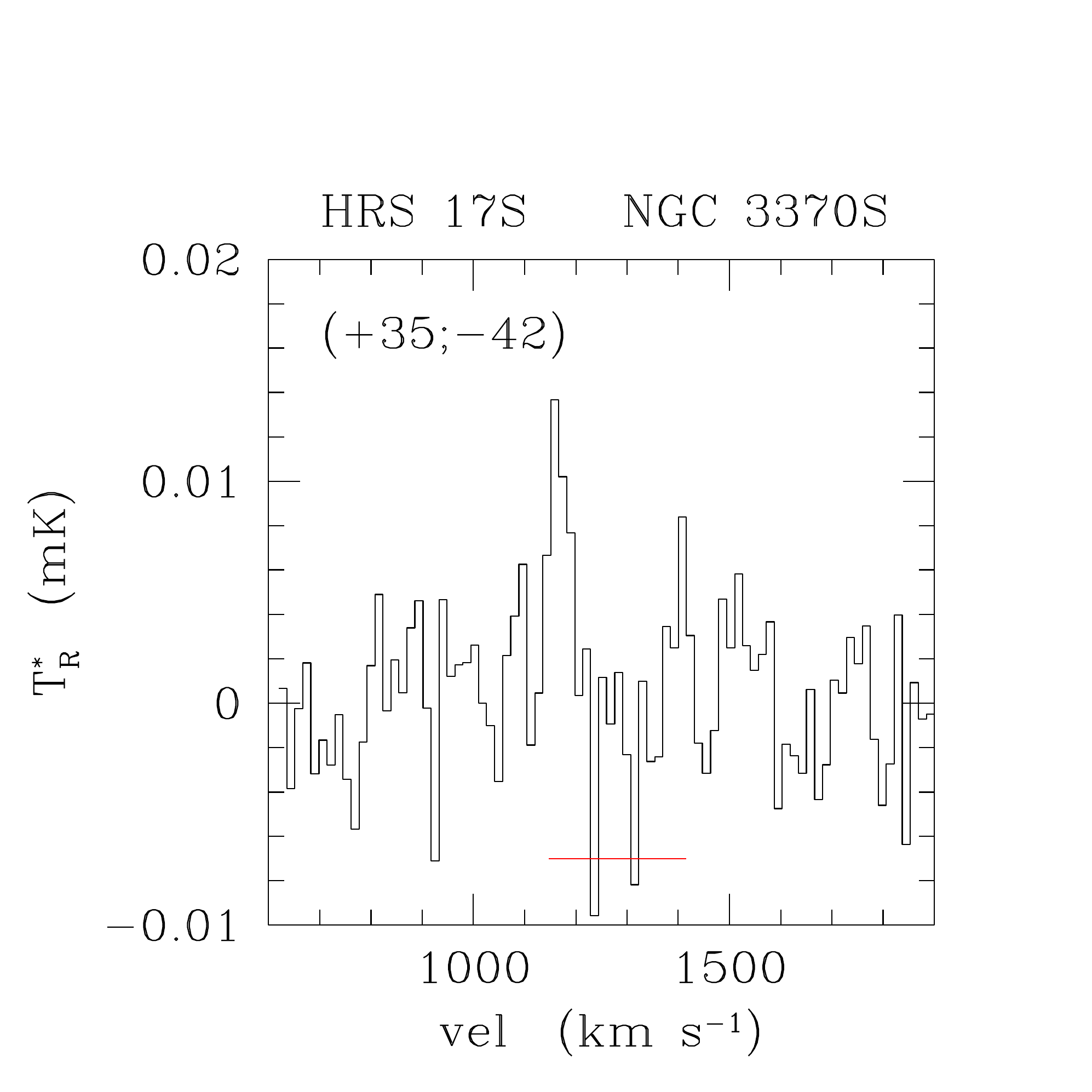}\\
   \caption{CO data of the HRS galaxies observed in this work, in order of increasing HRS name. Left: the $r$-band SDSS image of the observed galaxies with superimposed the 
position covered by the beam of the telescope (red solid line). Right: observed $^{12}$CO(1-0) spectra smoothed to a velocity resolution of 15.7 km s$^{-1}$. Intensities are in T$_R^*$ scale.}
   \label{spettri}%
   \end{figure*}
   \clearpage

   \addtocounter{figure}{-1}
   \begin{figure*}
   \centering
   \includegraphics[width=0.22\textwidth]{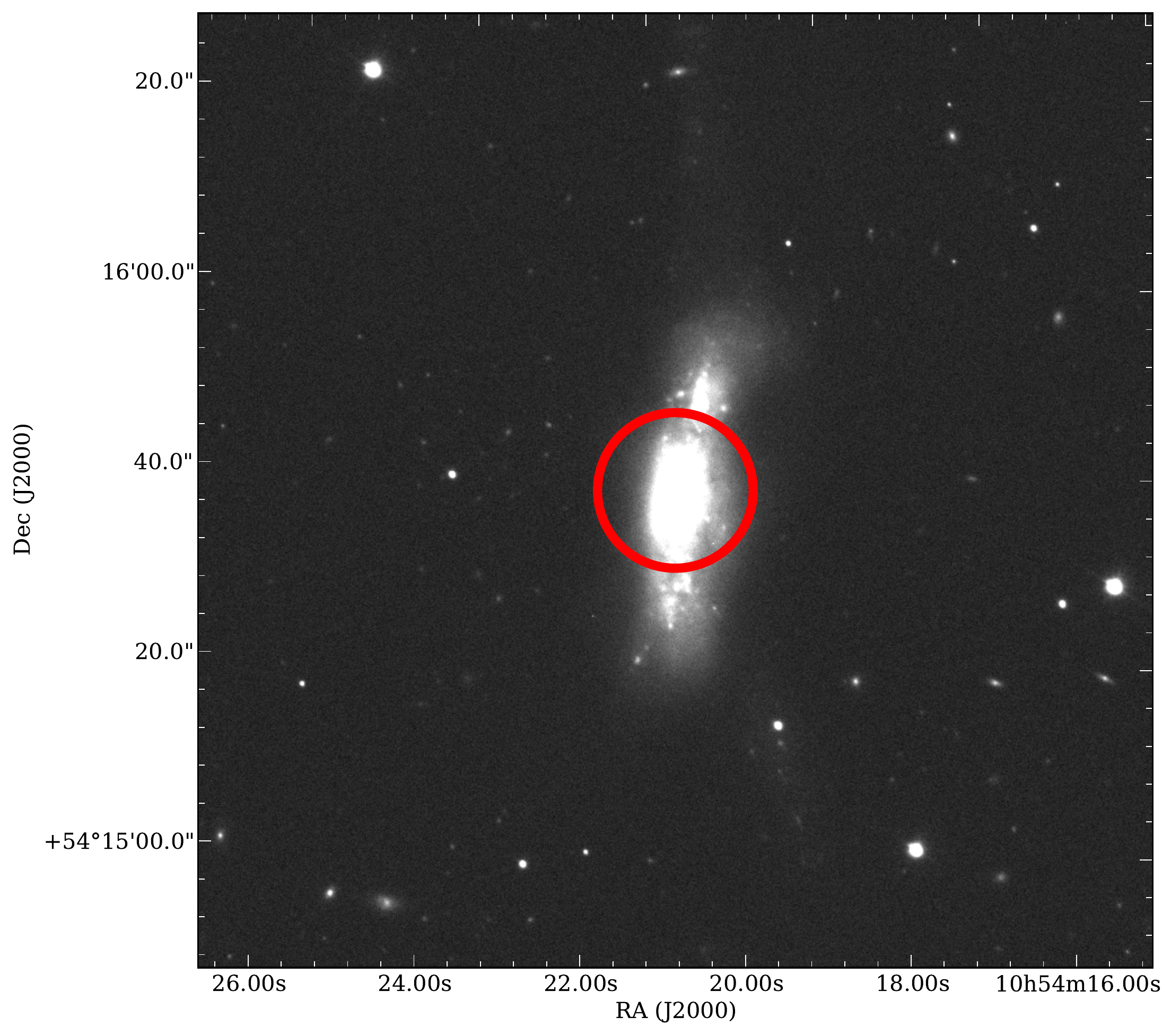}
   \includegraphics[width=0.22\textwidth]{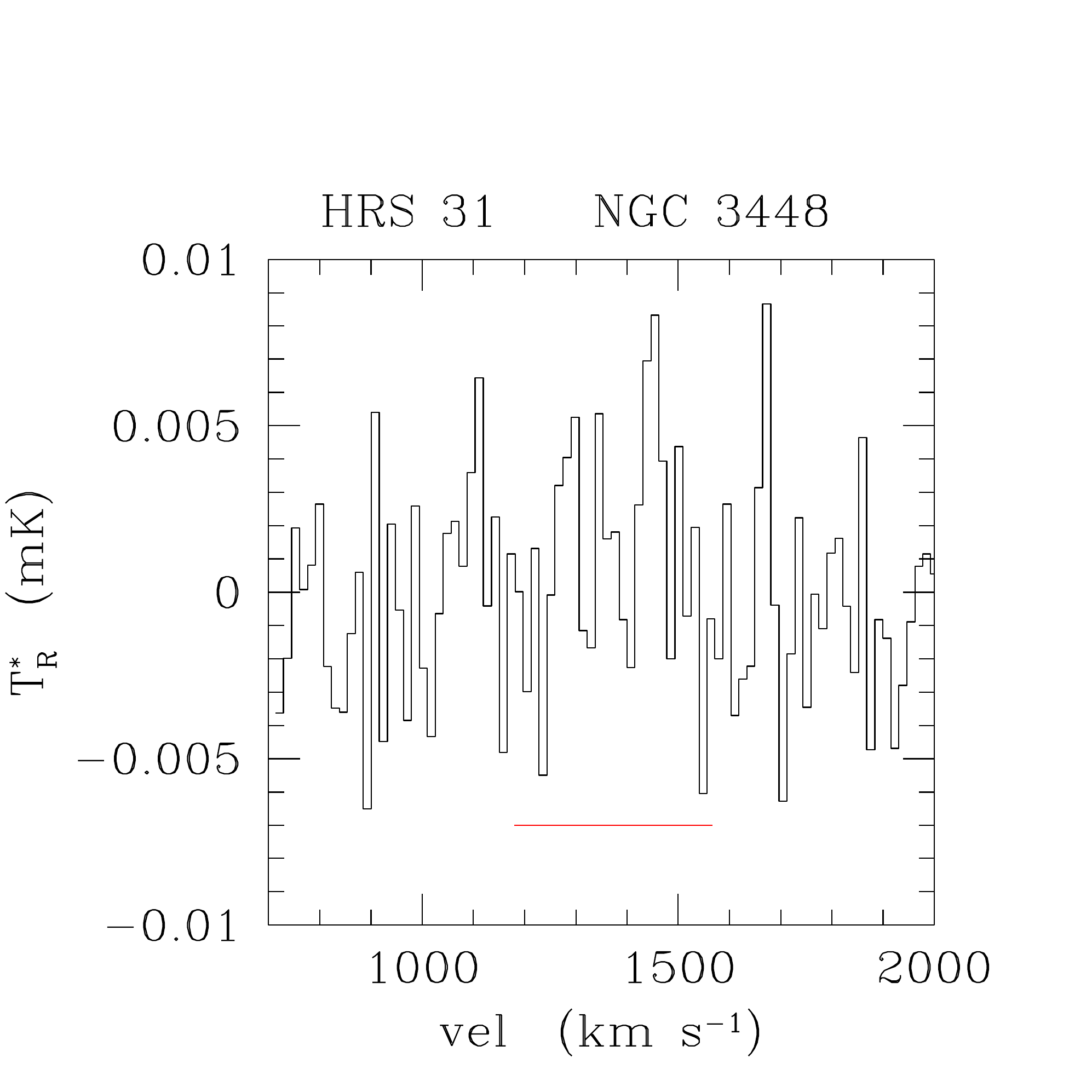}\\
   \includegraphics[width=0.22\textwidth]{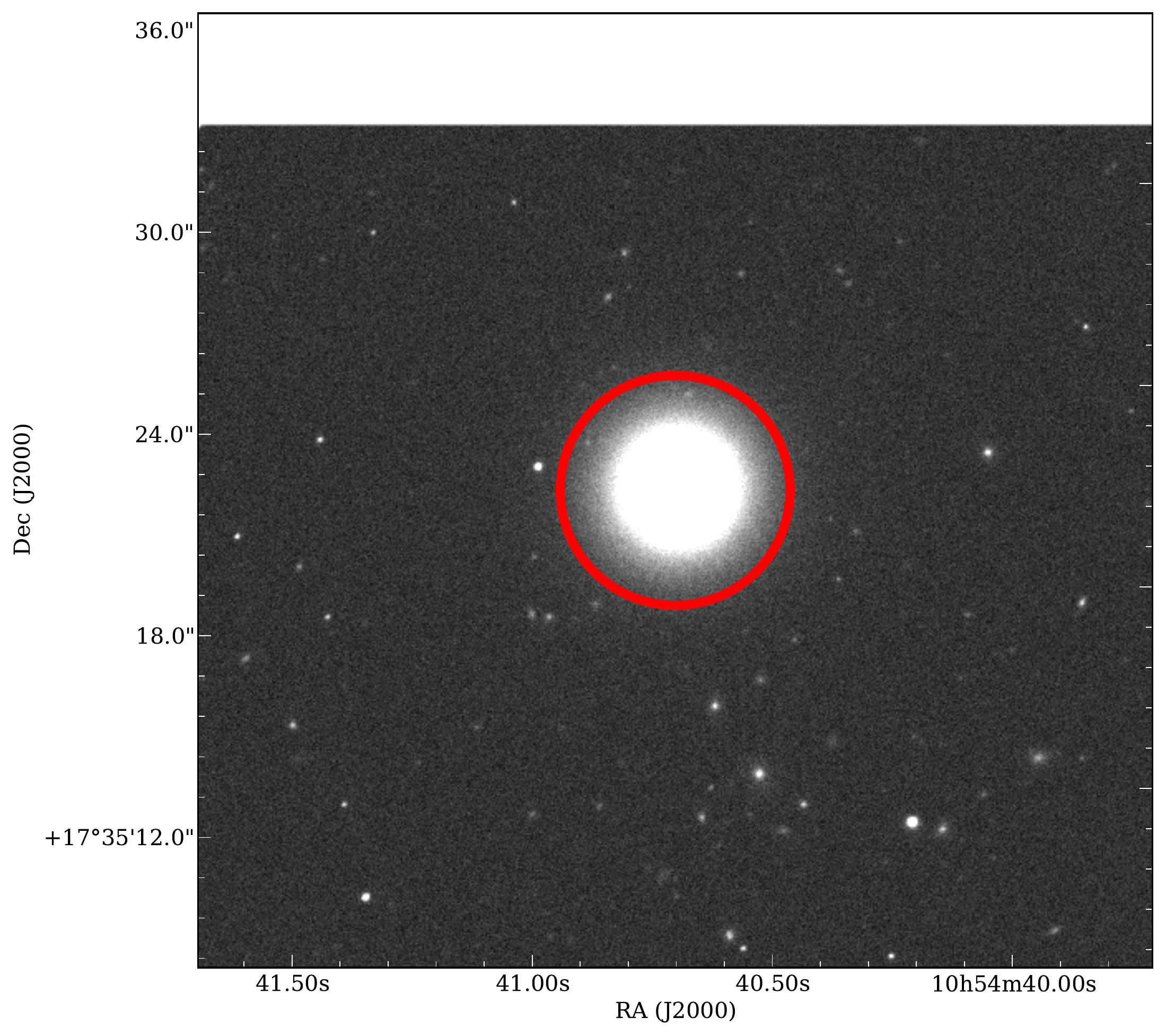}
   \includegraphics[width=0.22\textwidth]{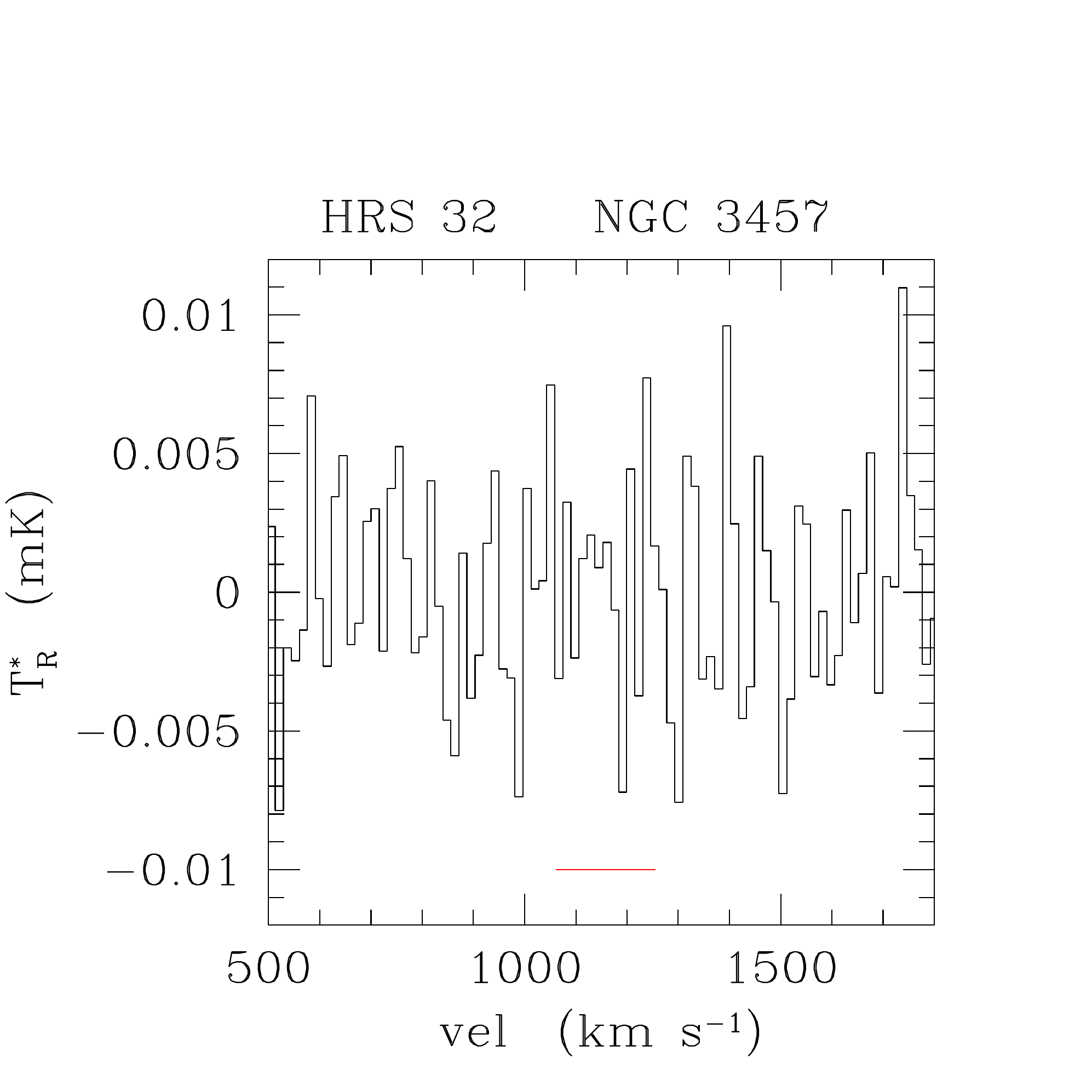}\\
   \includegraphics[width=0.22\textwidth]{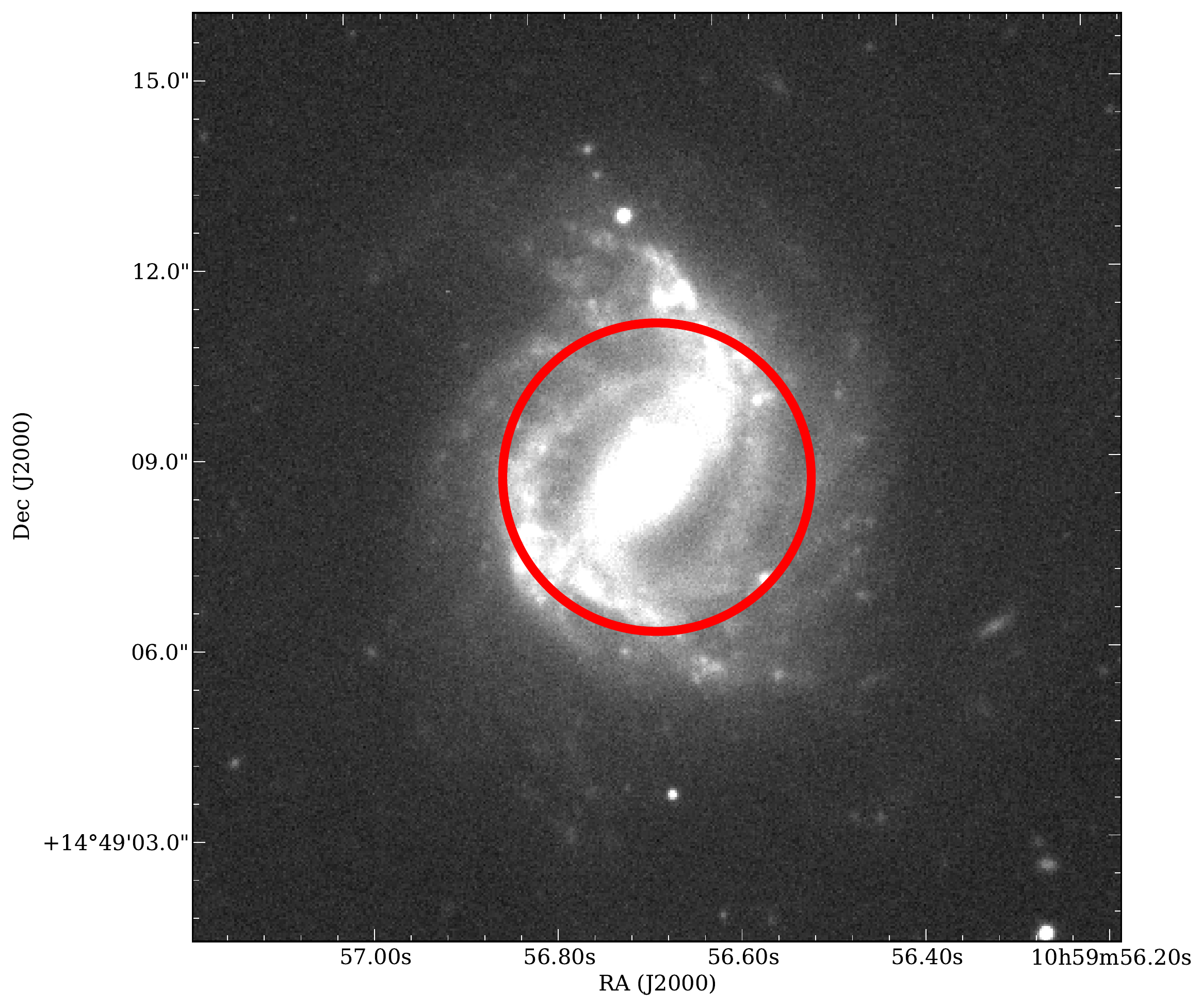}
   \includegraphics[width=0.22\textwidth]{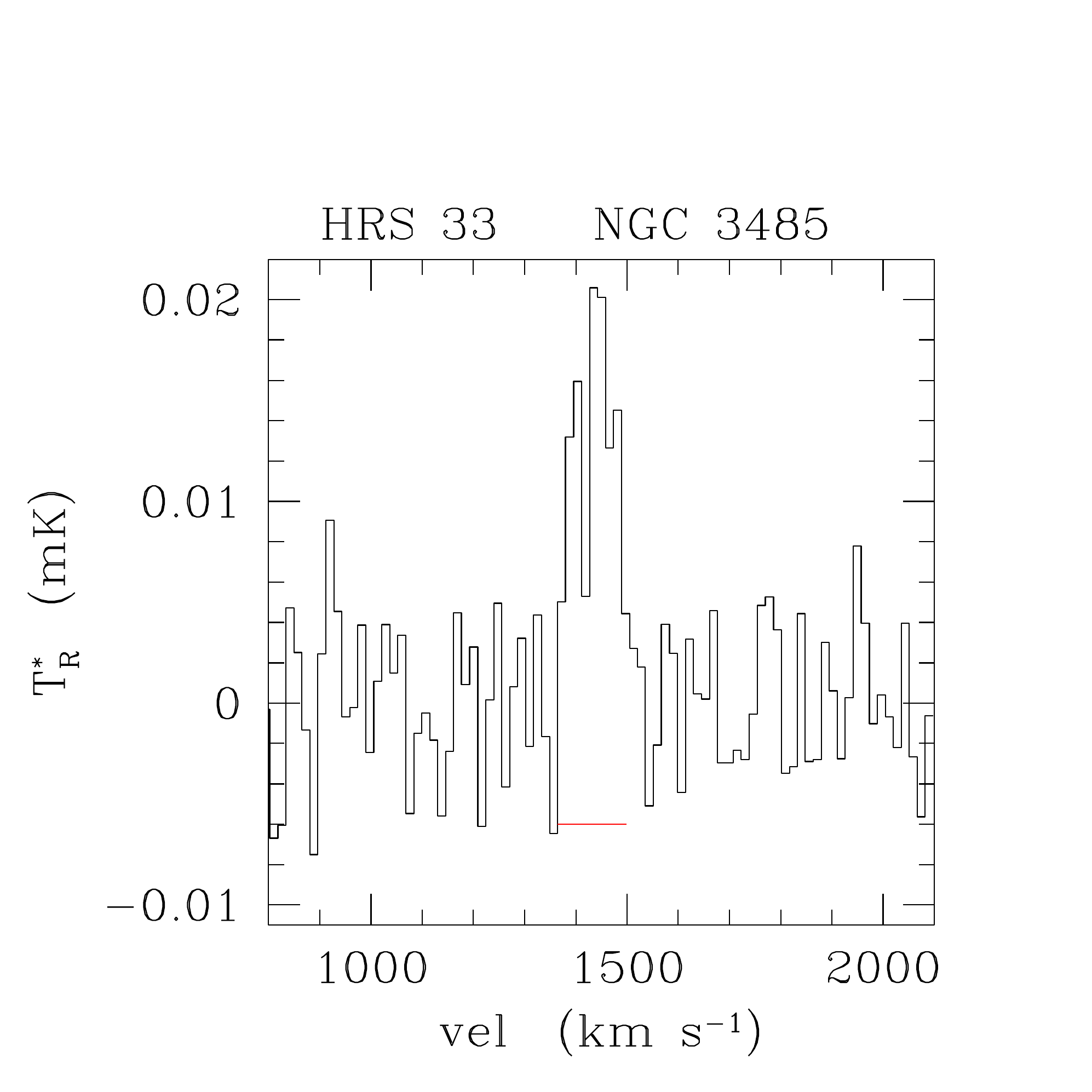}\\
   \includegraphics[width=0.22\textwidth]{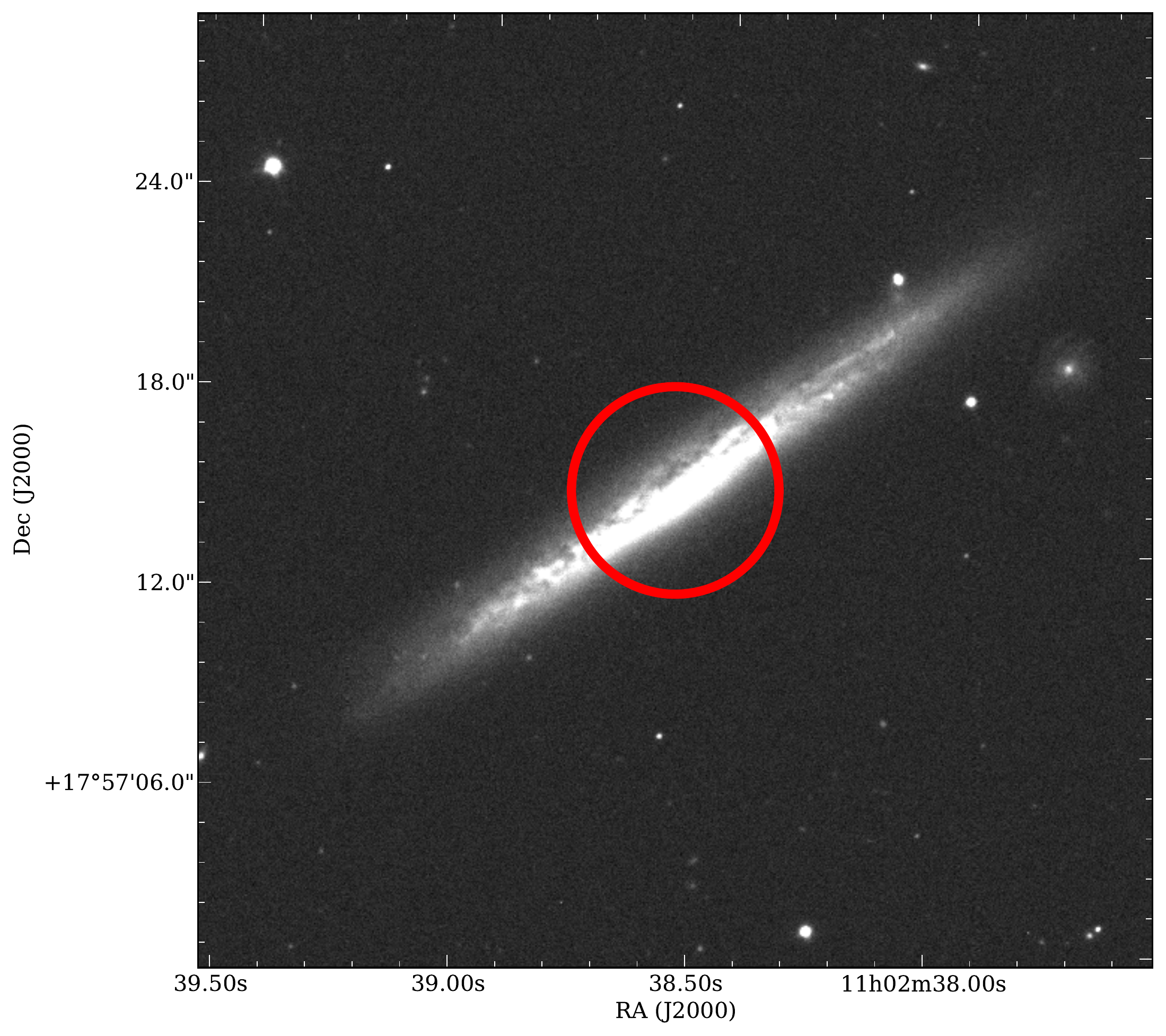}
   \includegraphics[width=0.22\textwidth]{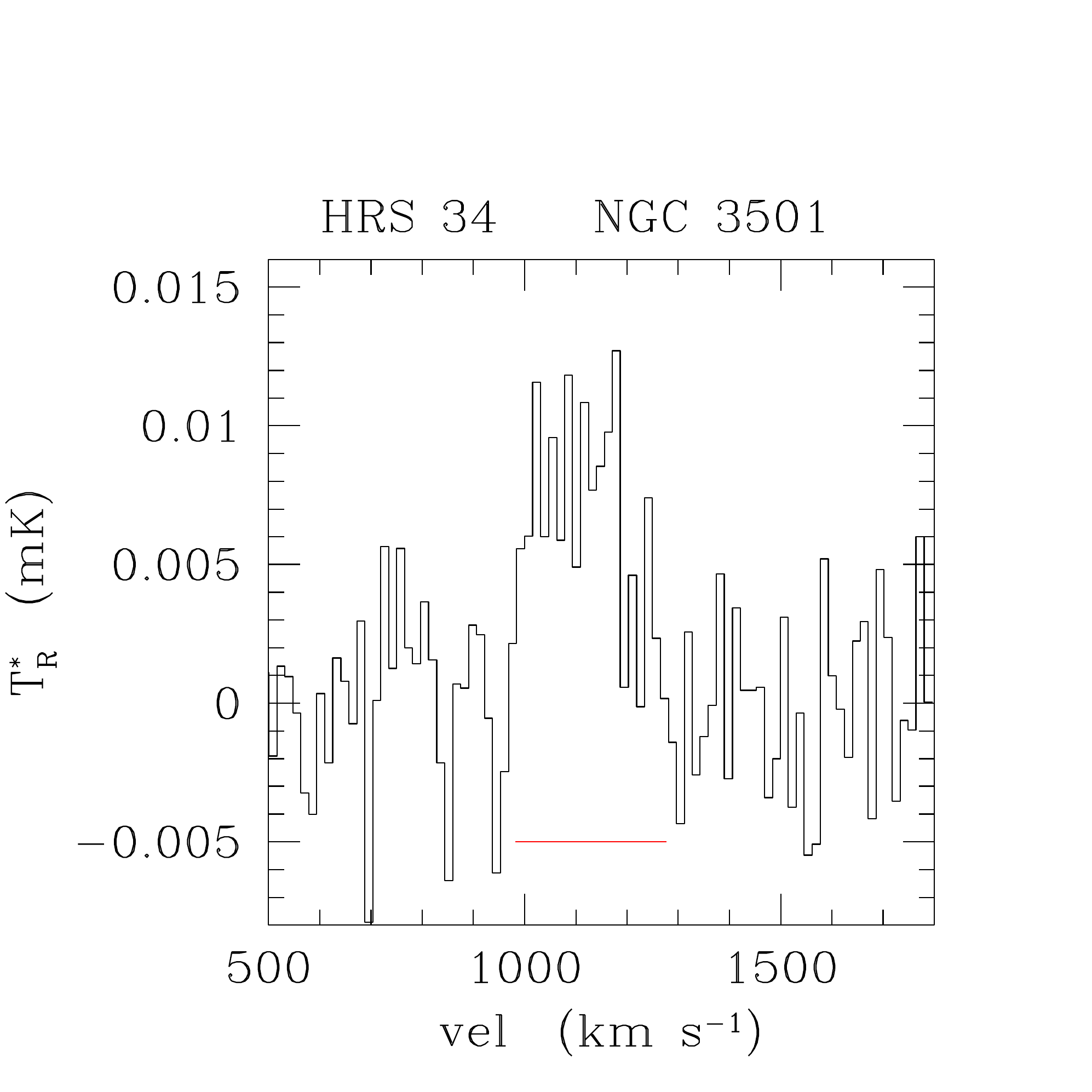}\\
   \includegraphics[width=0.22\textwidth]{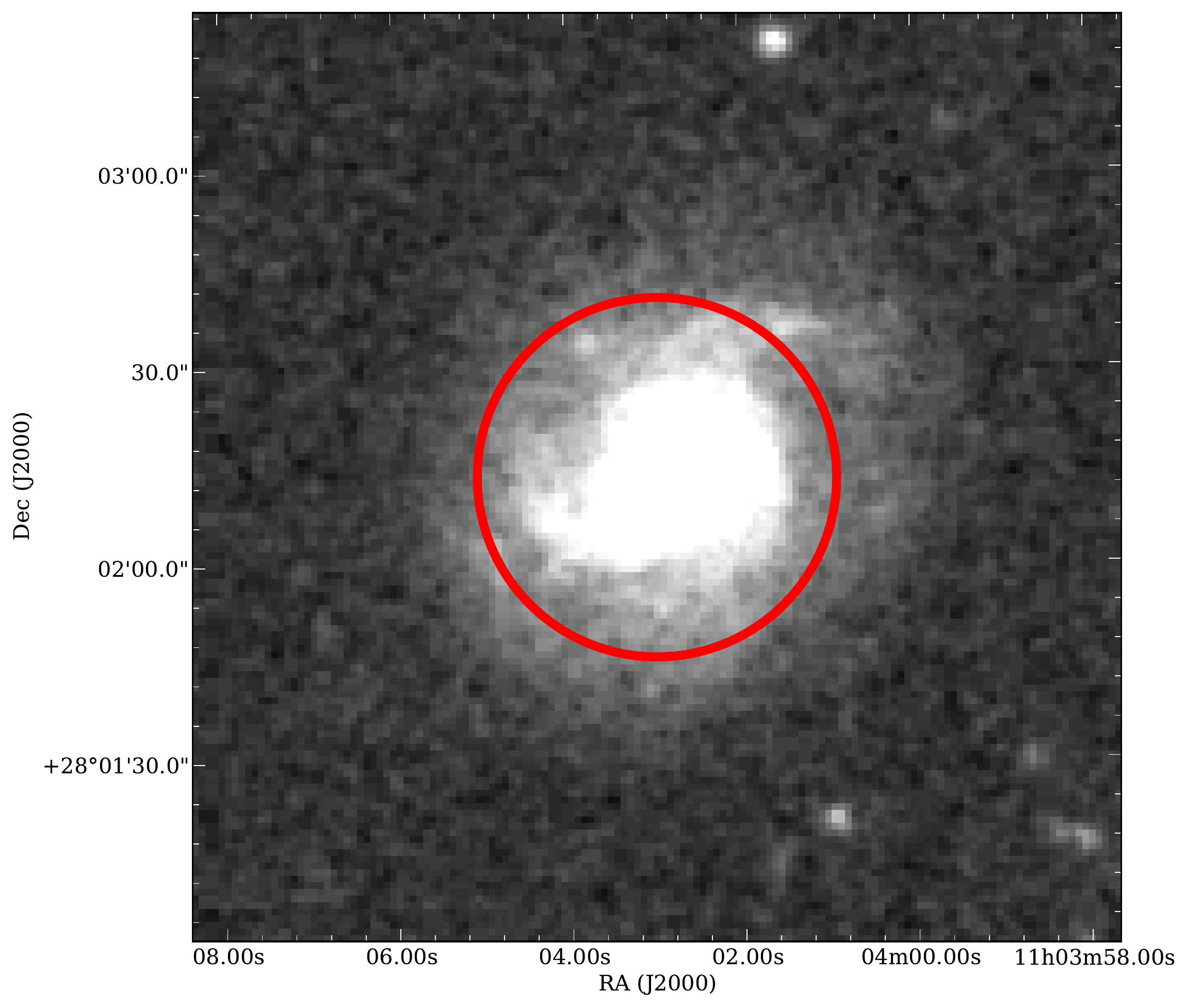}
   \includegraphics[width=0.22\textwidth]{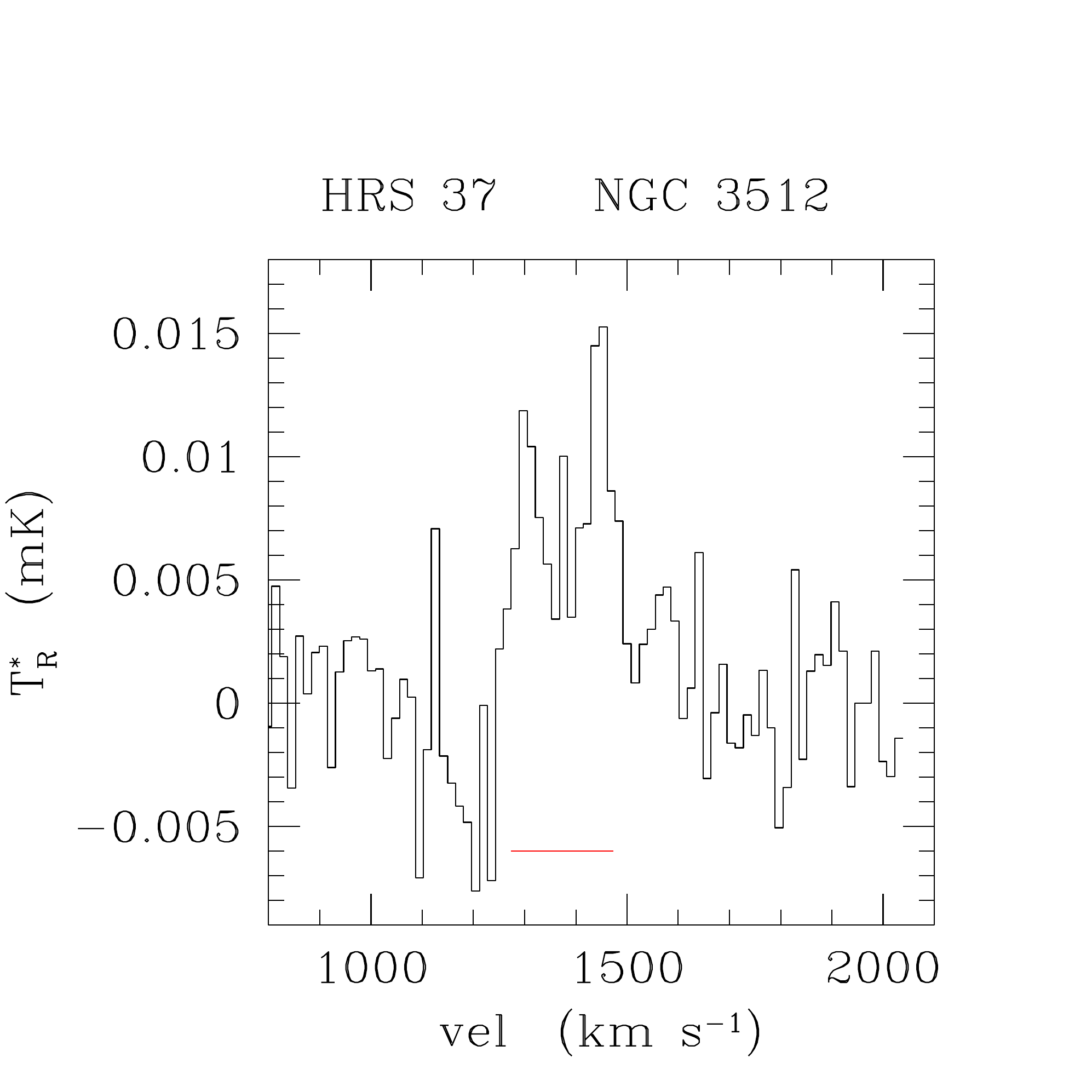}\\
   \caption{Continued.}
   \label{spettri}%
   \end{figure*}
   \clearpage
   
   \addtocounter{figure}{-1}
   \begin{figure*}
   \centering
   \includegraphics[width=0.22\textwidth]{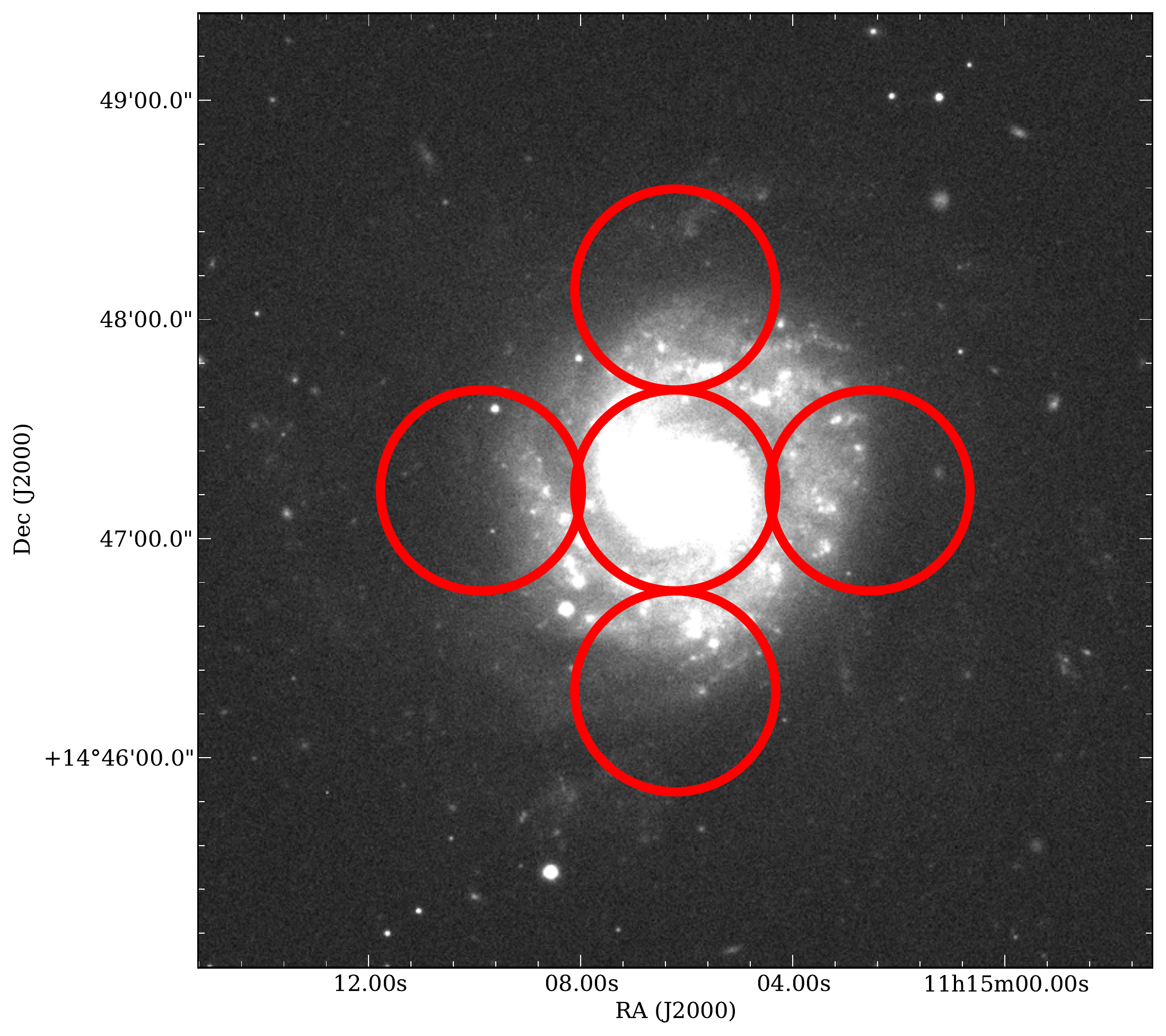}
   \includegraphics[width=0.22\textwidth]{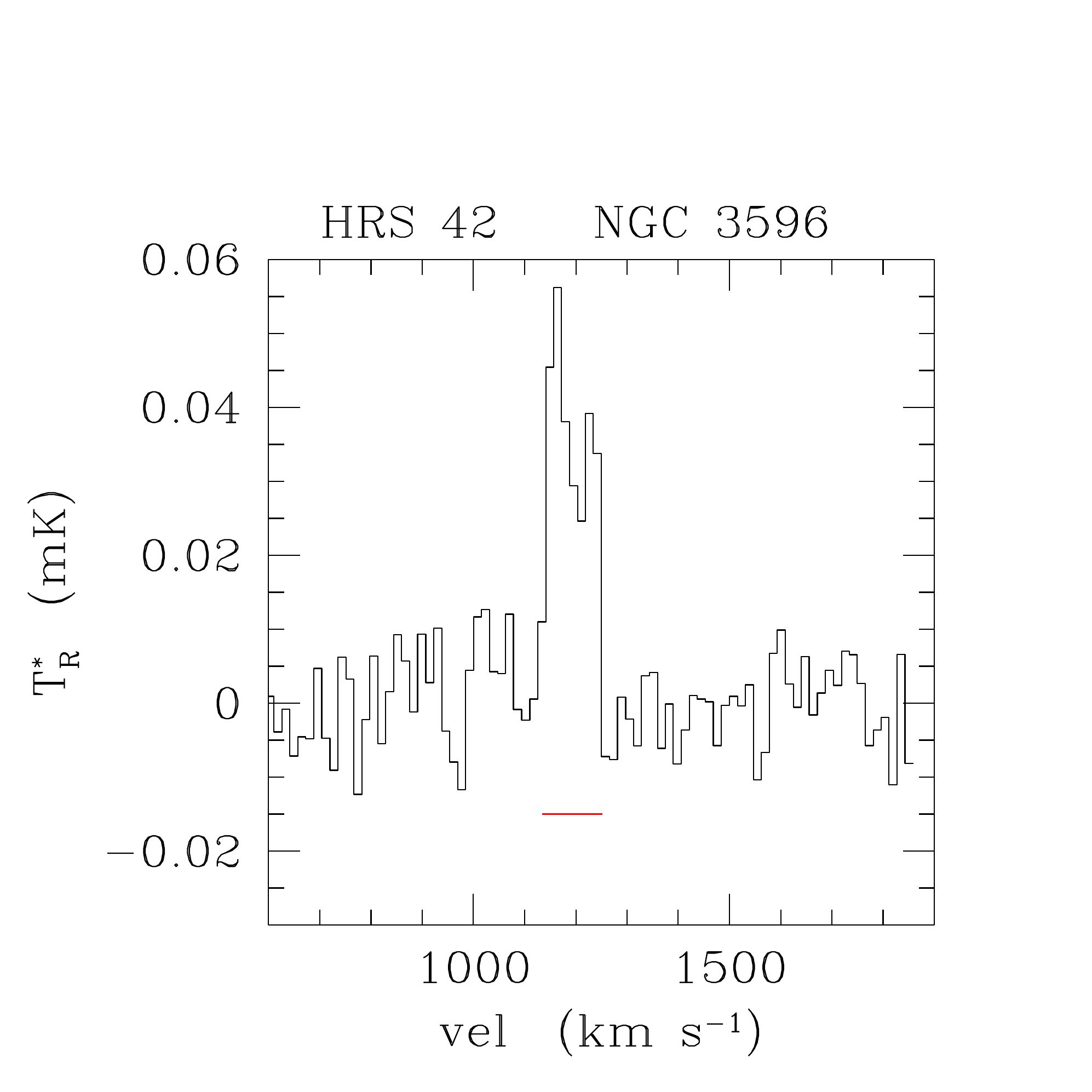}\\
   \includegraphics[width=0.22\textwidth]{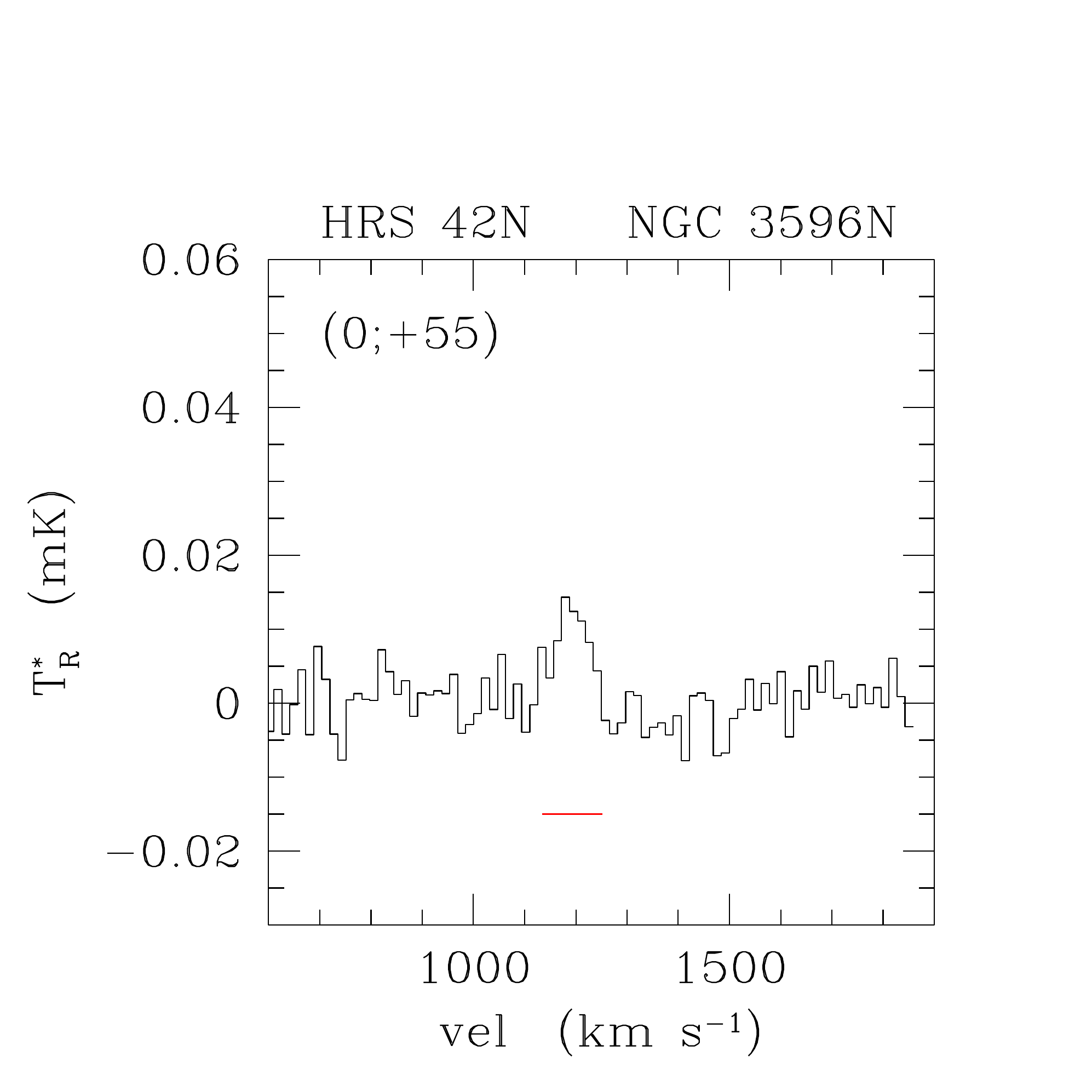}
   \includegraphics[width=0.22\textwidth]{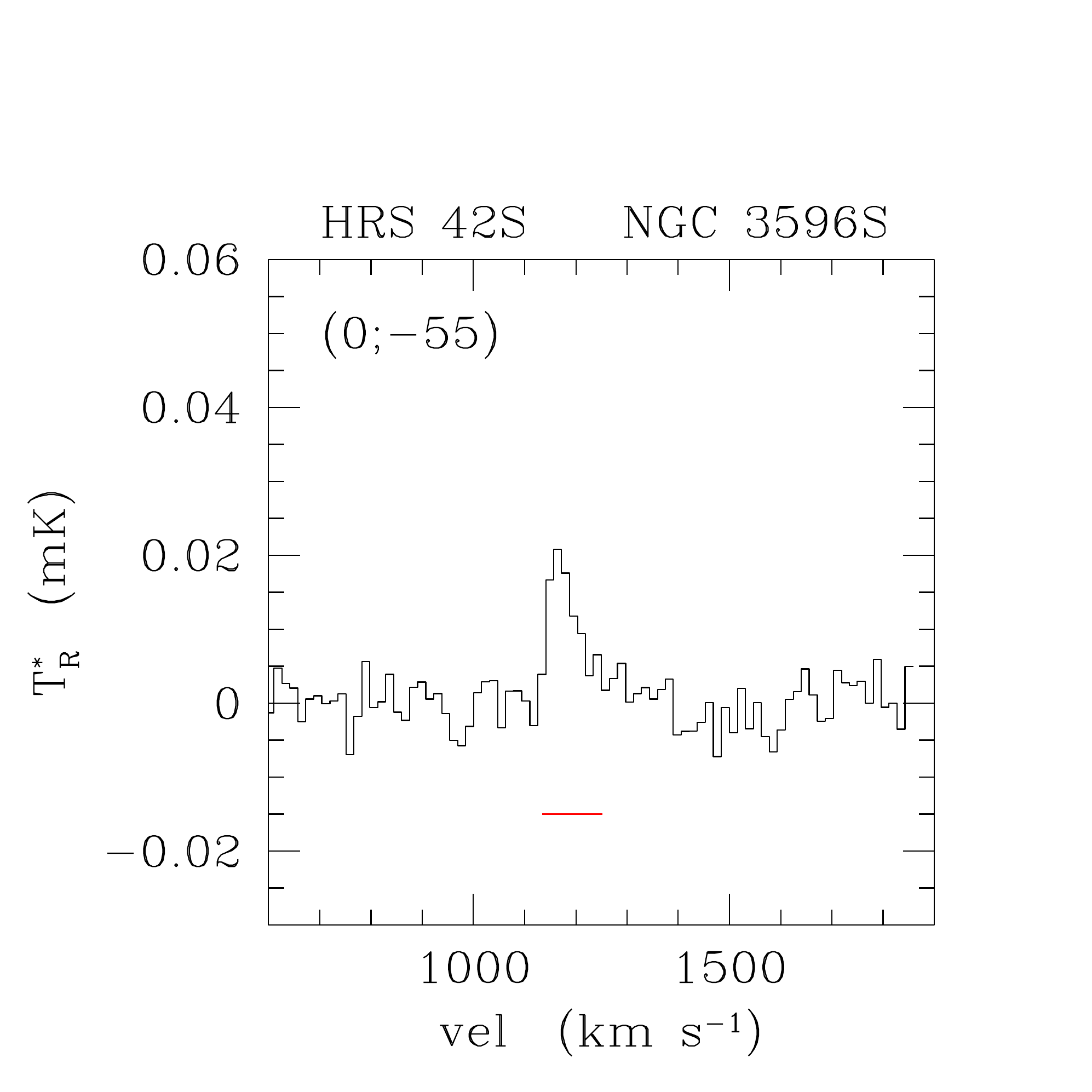}
   \includegraphics[width=0.22\textwidth]{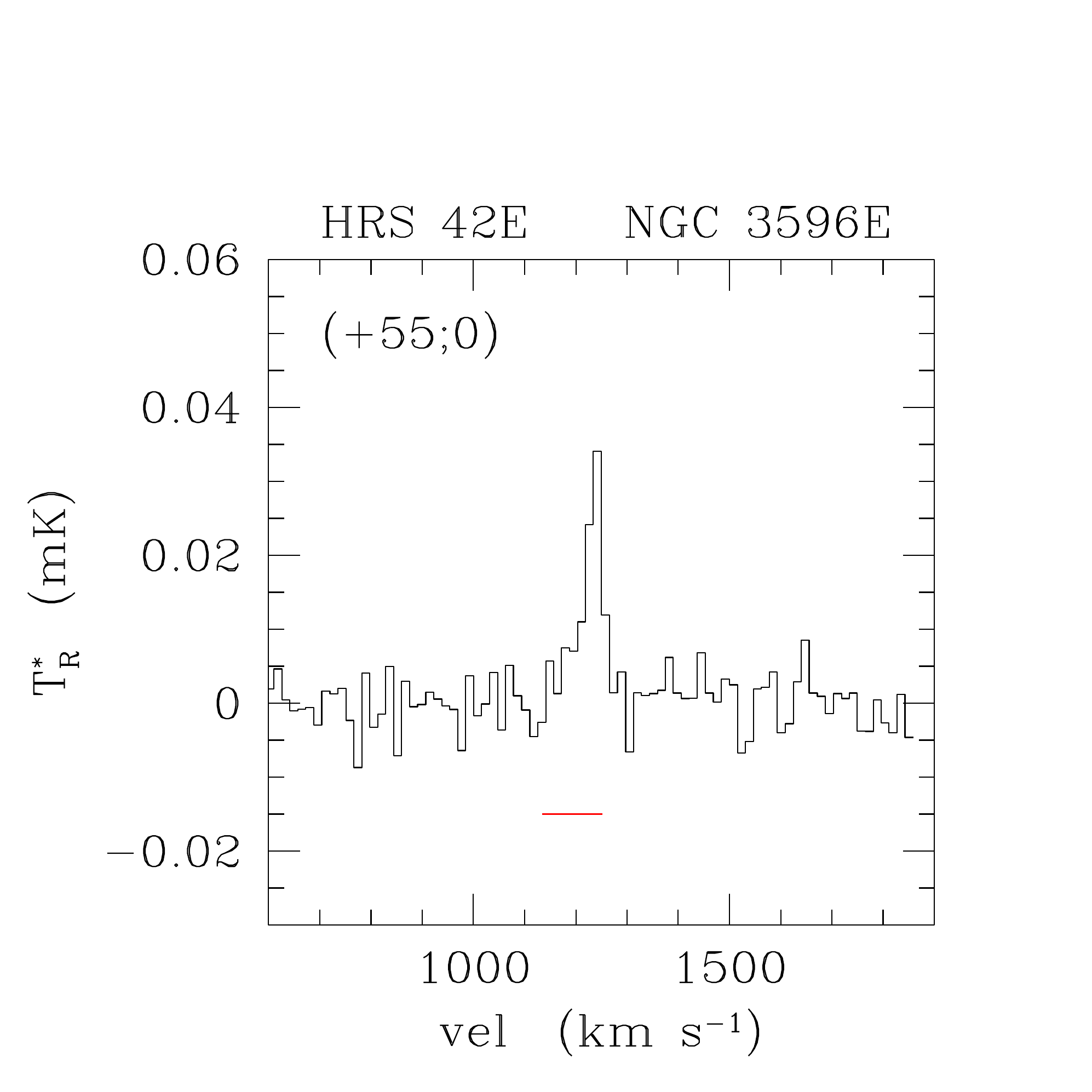}
   \includegraphics[width=0.22\textwidth]{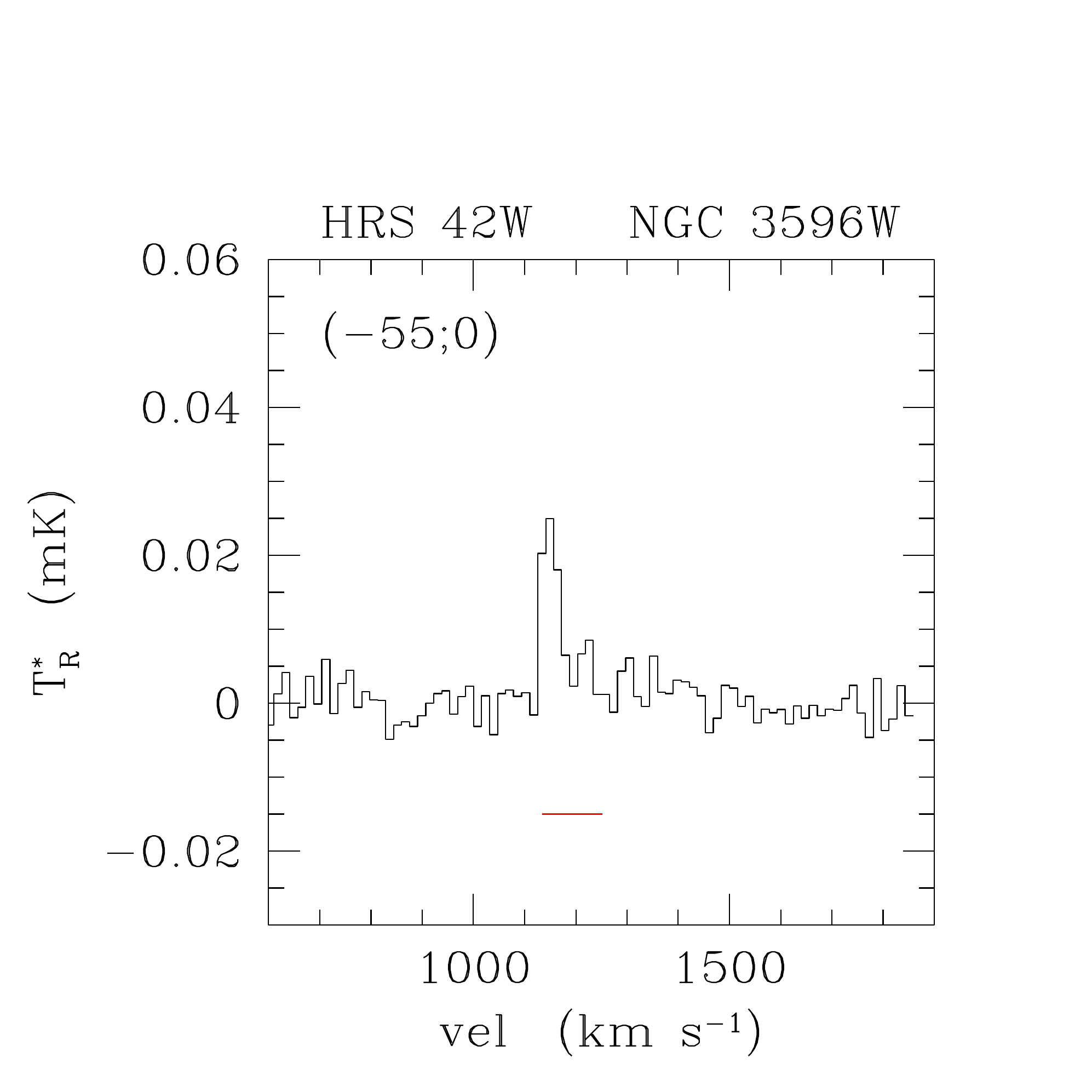}\\
   \includegraphics[width=0.22\textwidth]{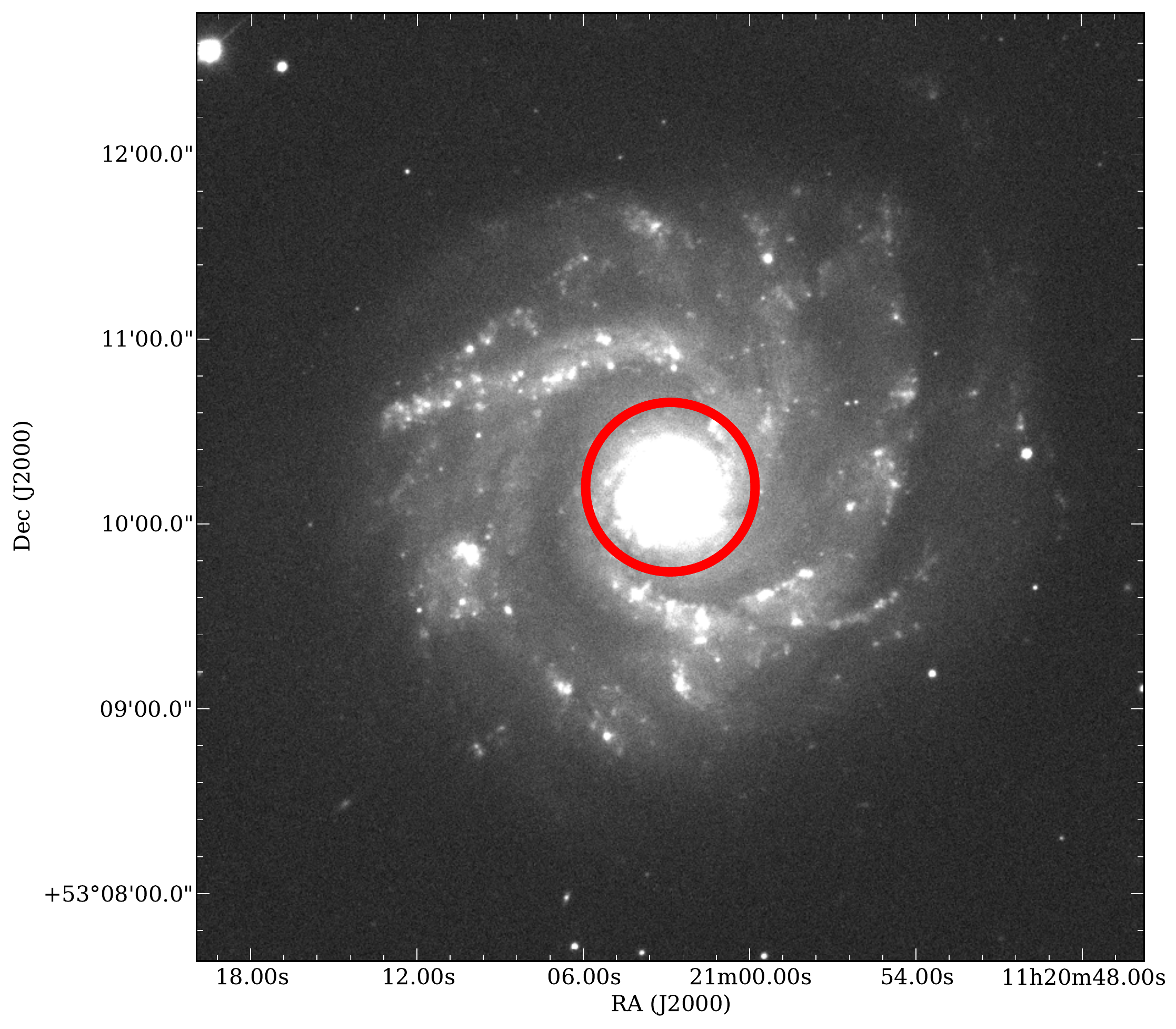}
   \includegraphics[width=0.22\textwidth]{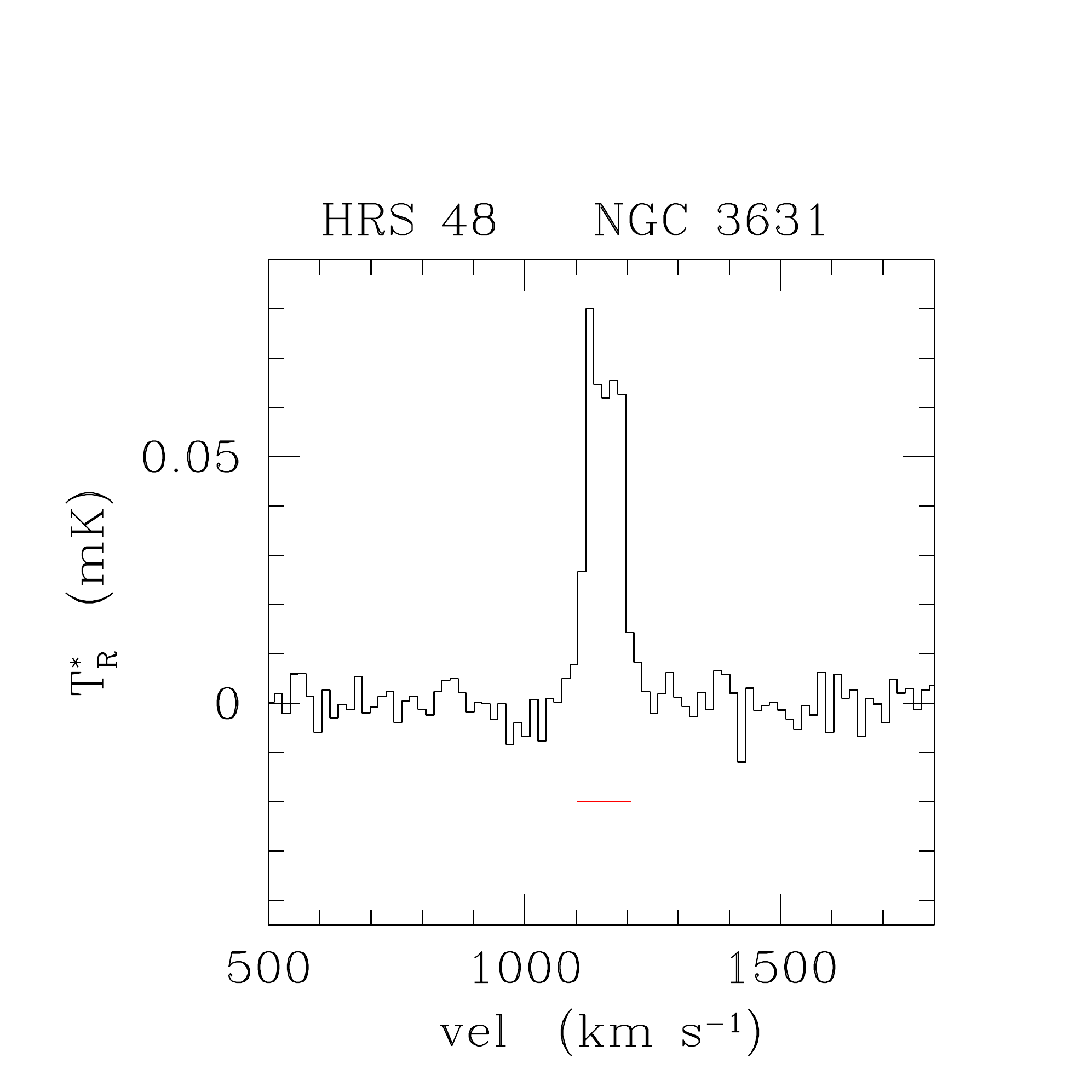}\\
   \includegraphics[width=0.22\textwidth]{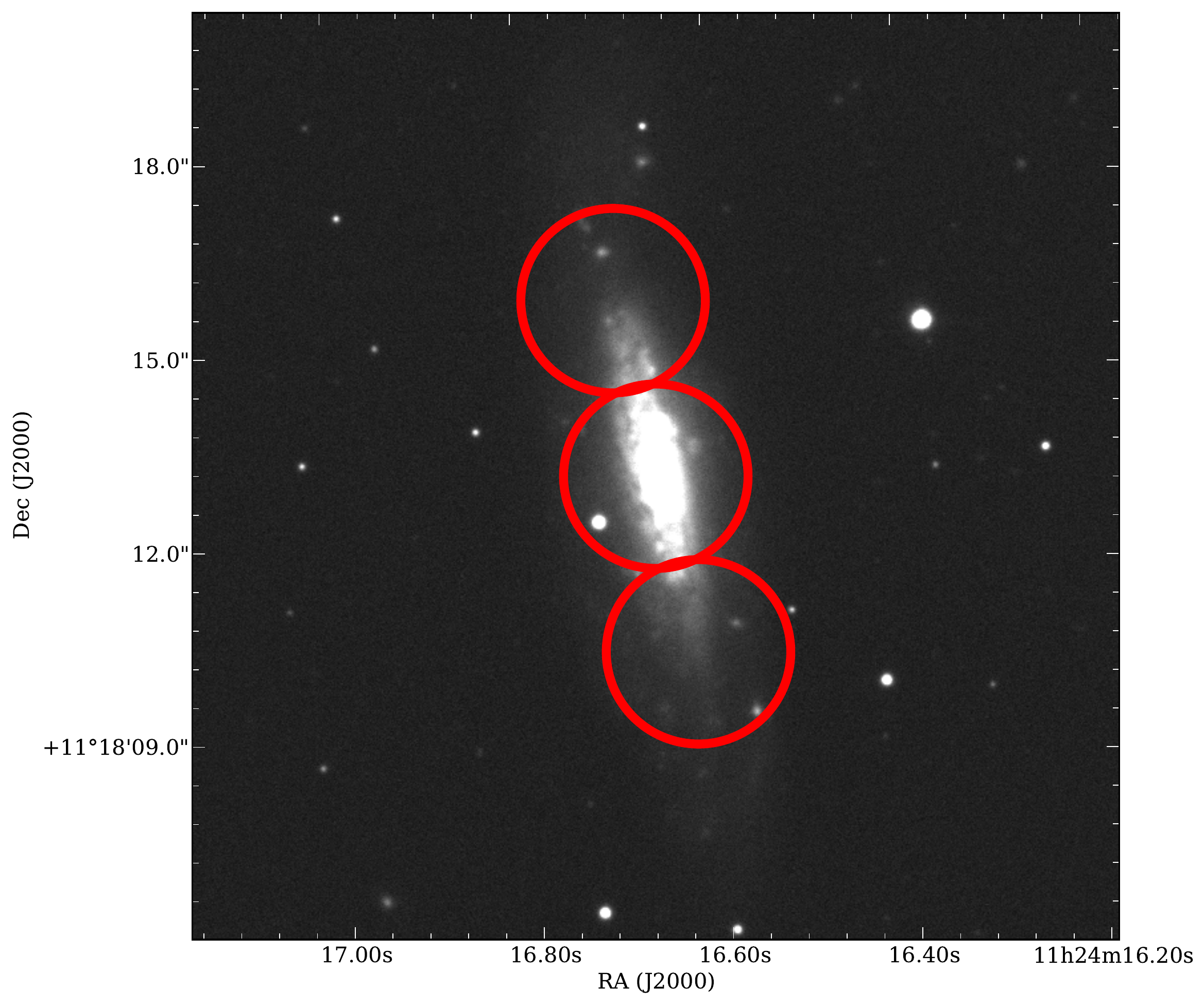}
   \includegraphics[width=0.22\textwidth]{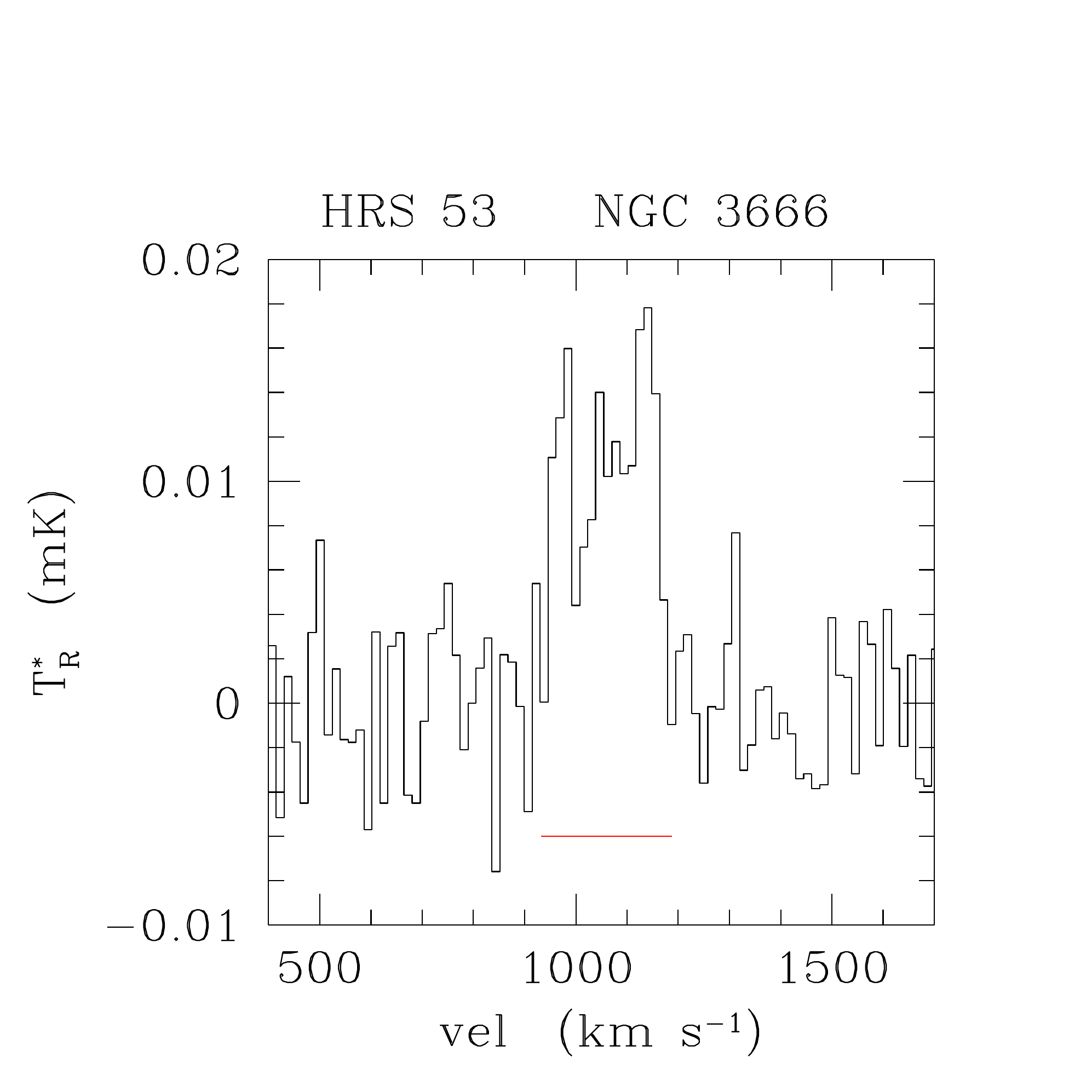}
   \includegraphics[width=0.22\textwidth]{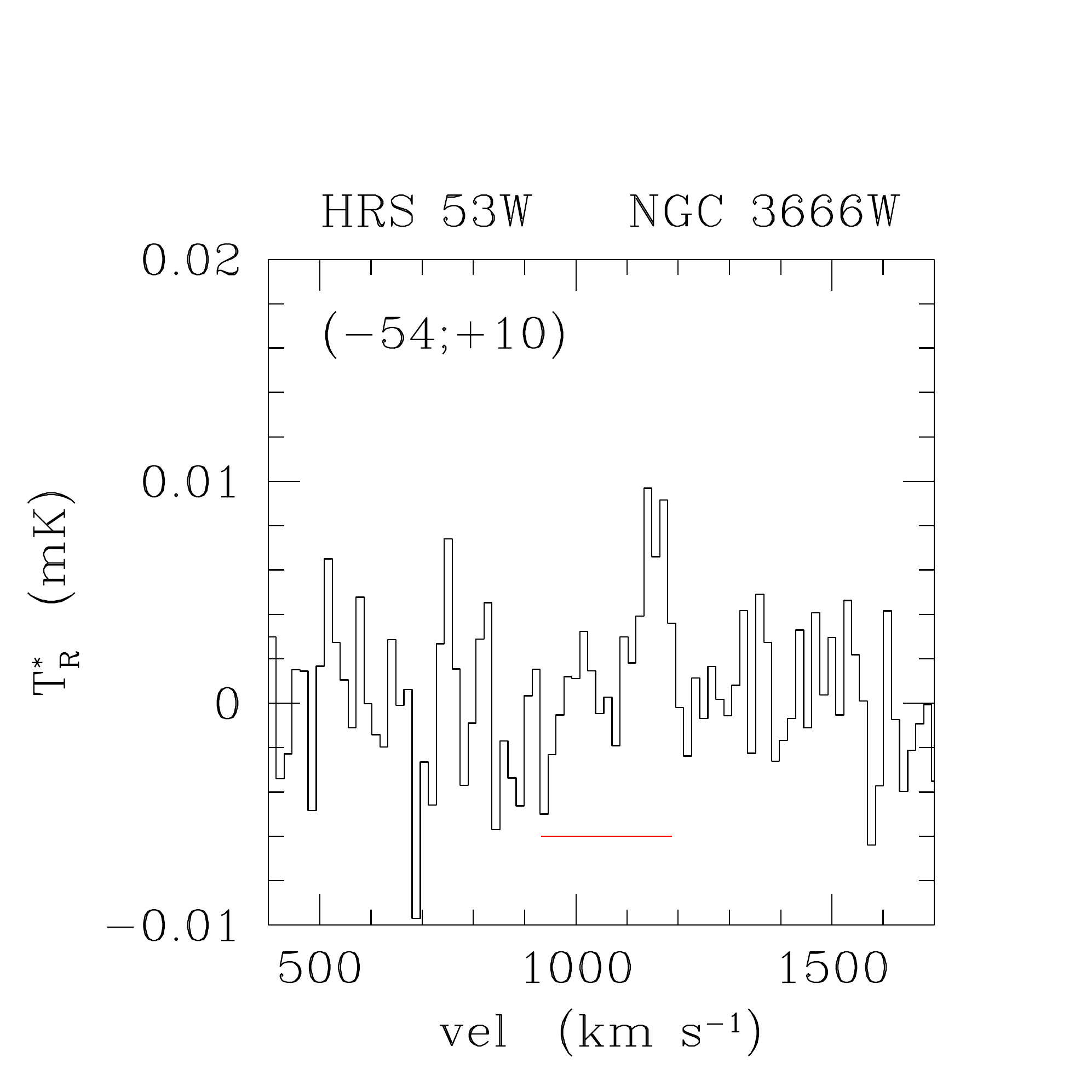}
   \includegraphics[width=0.22\textwidth]{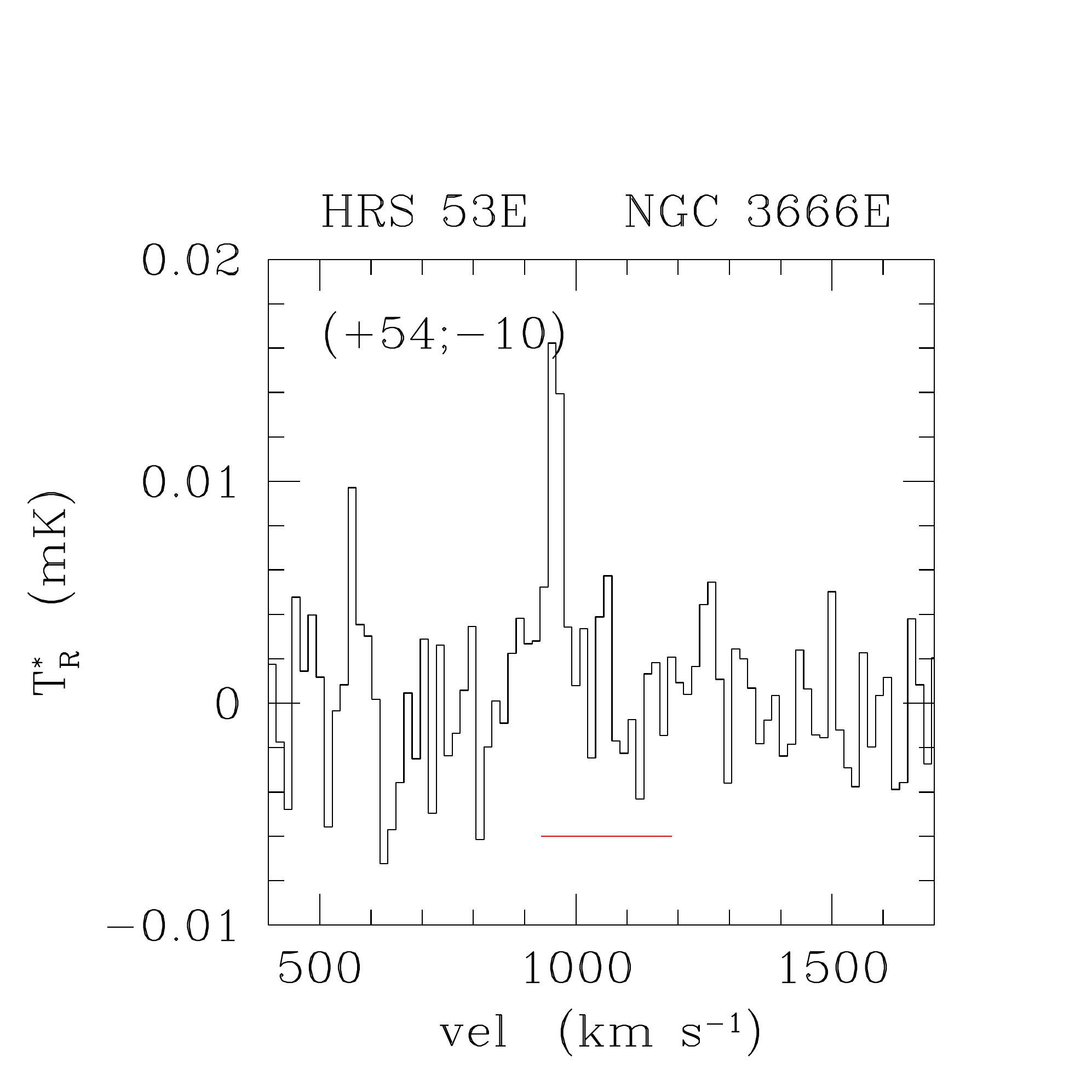}\\
   \includegraphics[width=0.22\textwidth]{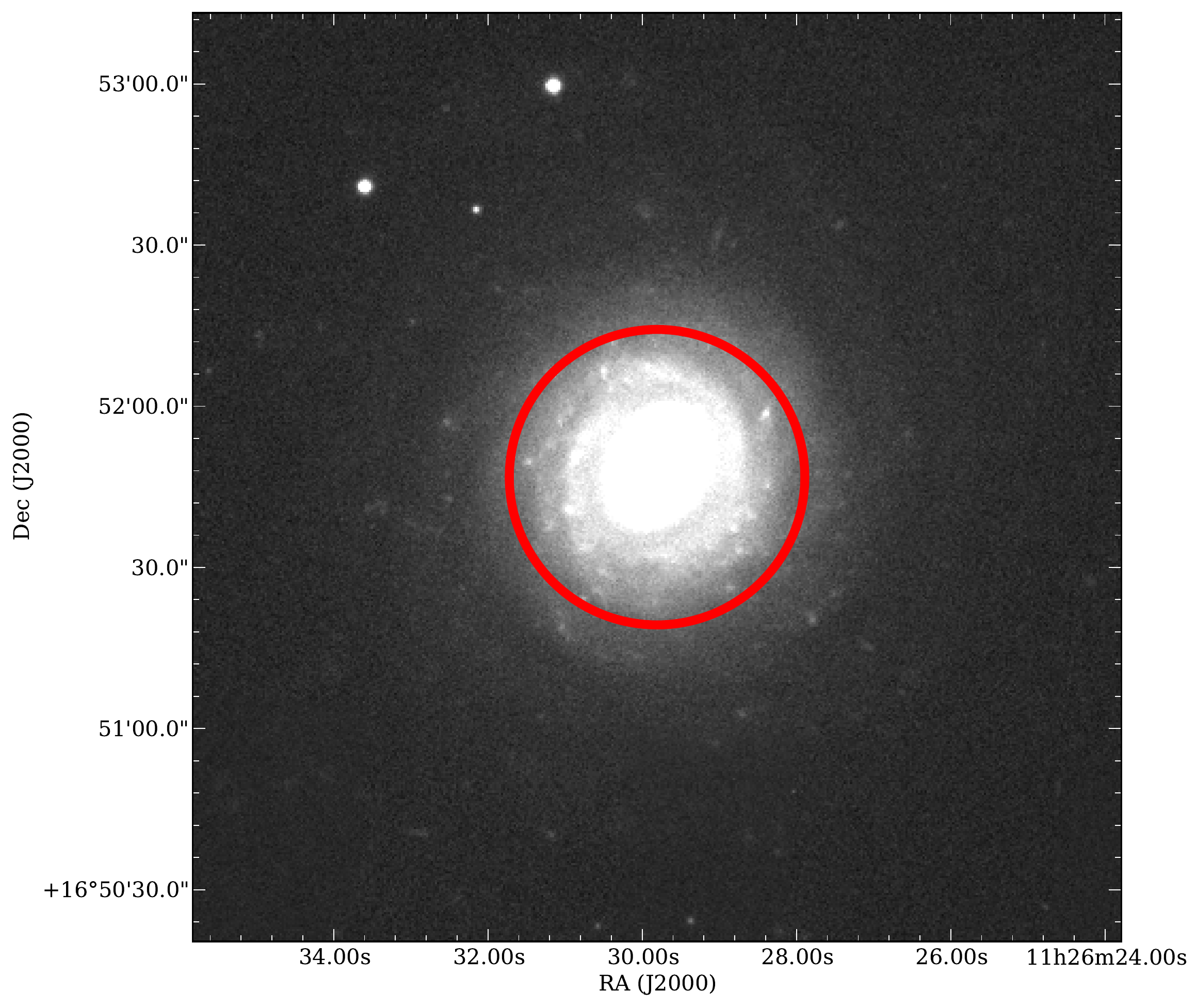}
   \includegraphics[width=0.22\textwidth]{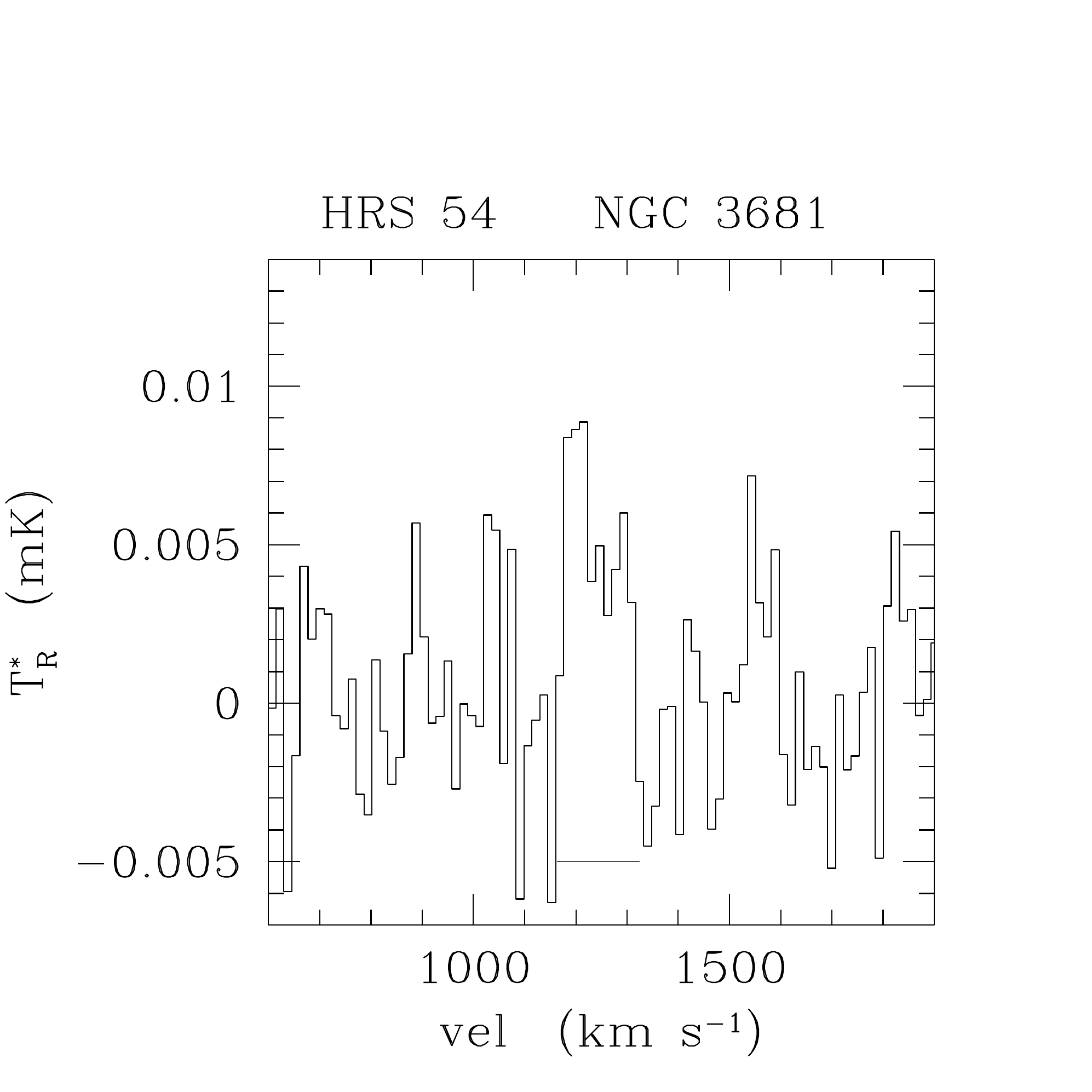}\\
   \caption{Continued.}
   \label{spettri}%
   \end{figure*}
   \clearpage
 
   \addtocounter{figure}{-1}
   \begin{figure*}
   \centering
   \includegraphics[width=0.22\textwidth]{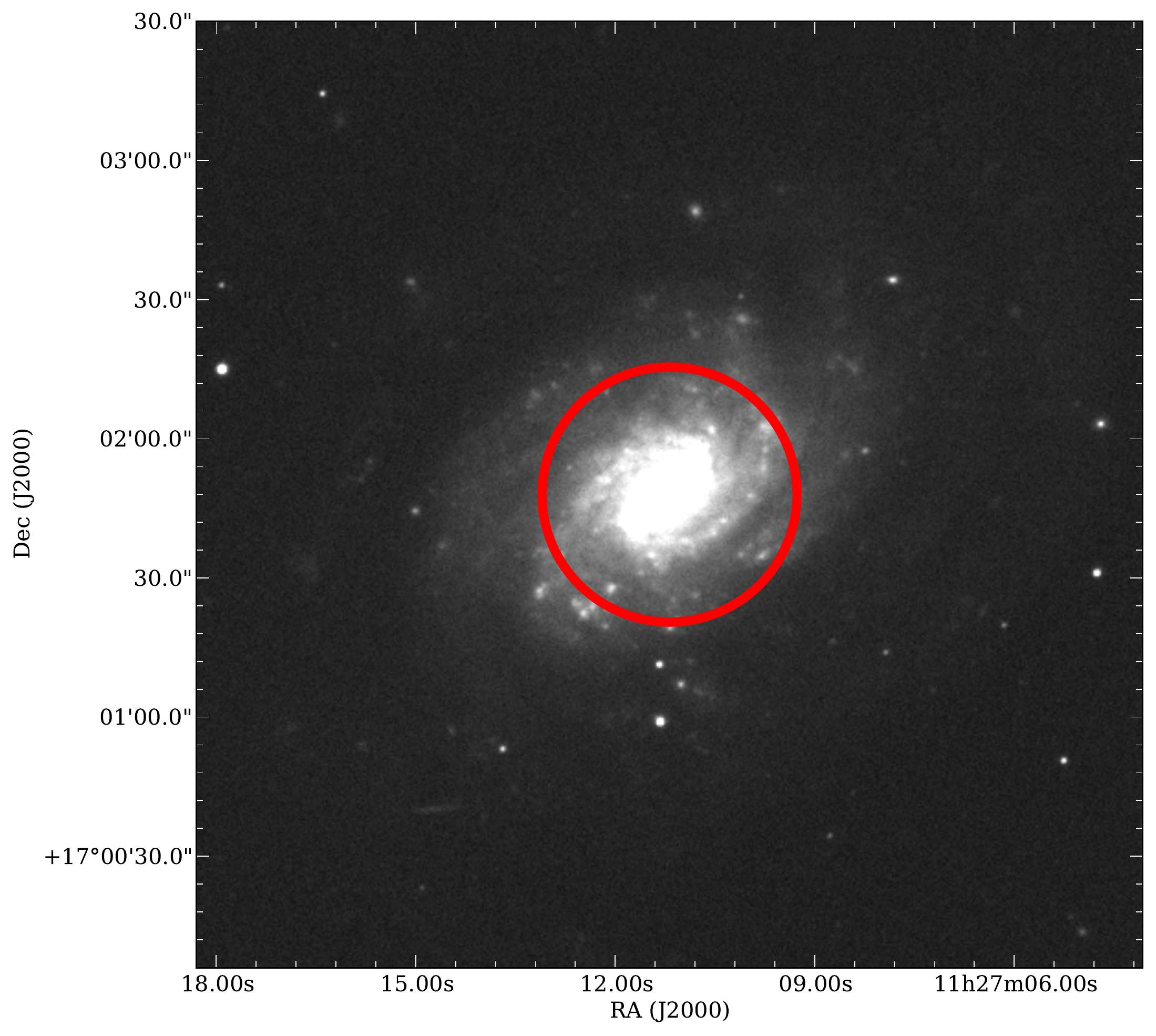}
   \includegraphics[width=0.22\textwidth]{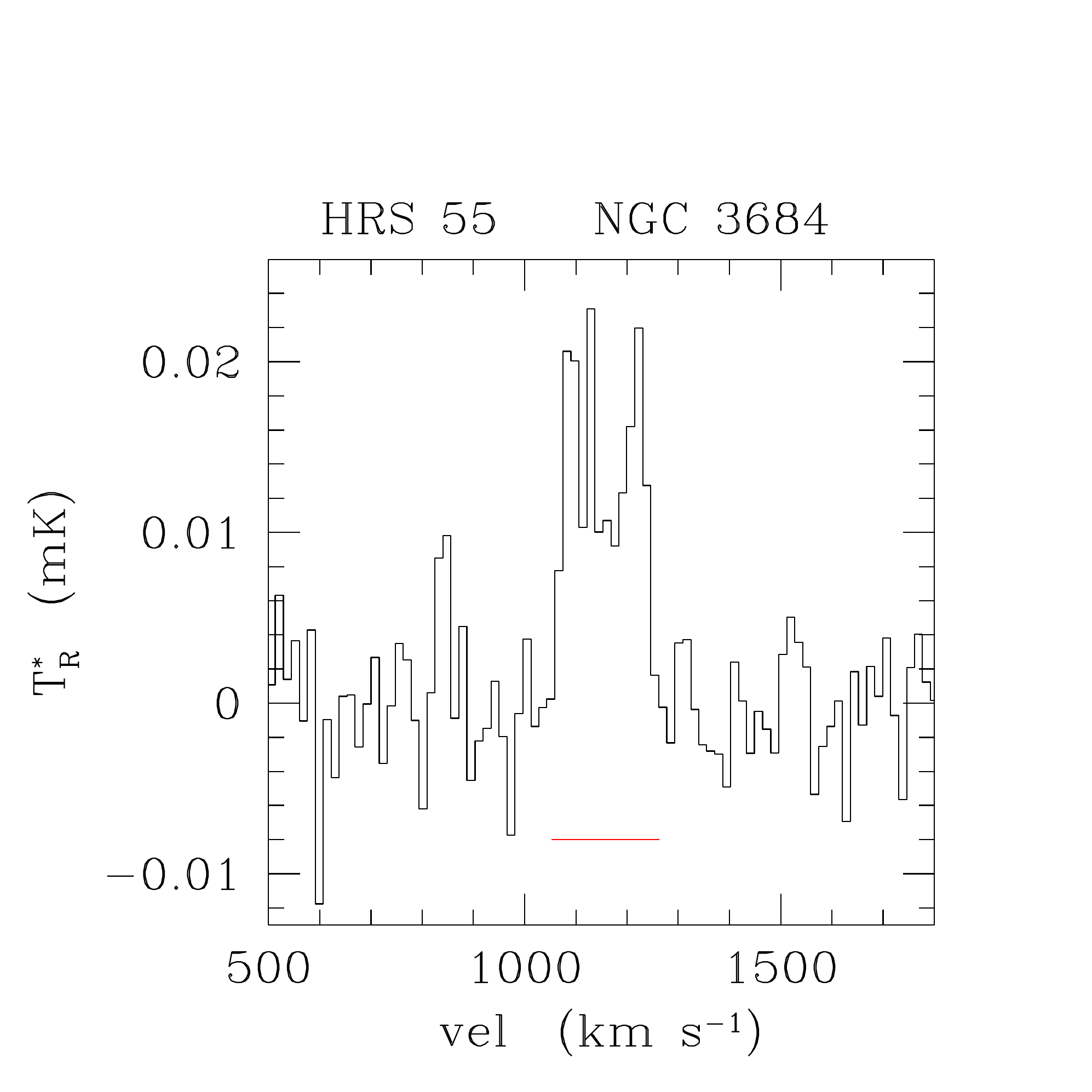}\\
   \includegraphics[width=0.22\textwidth]{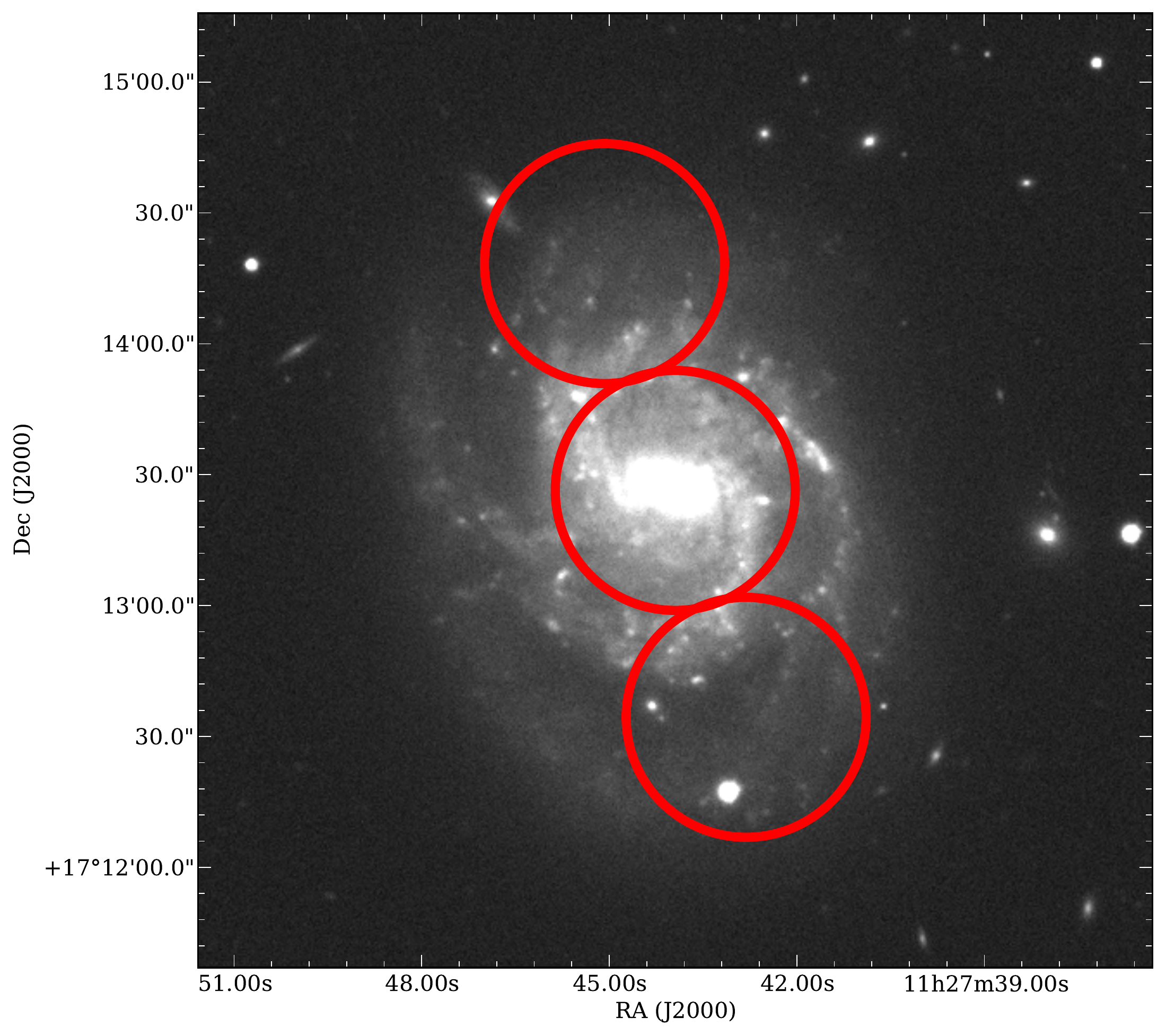}
   \includegraphics[width=0.22\textwidth]{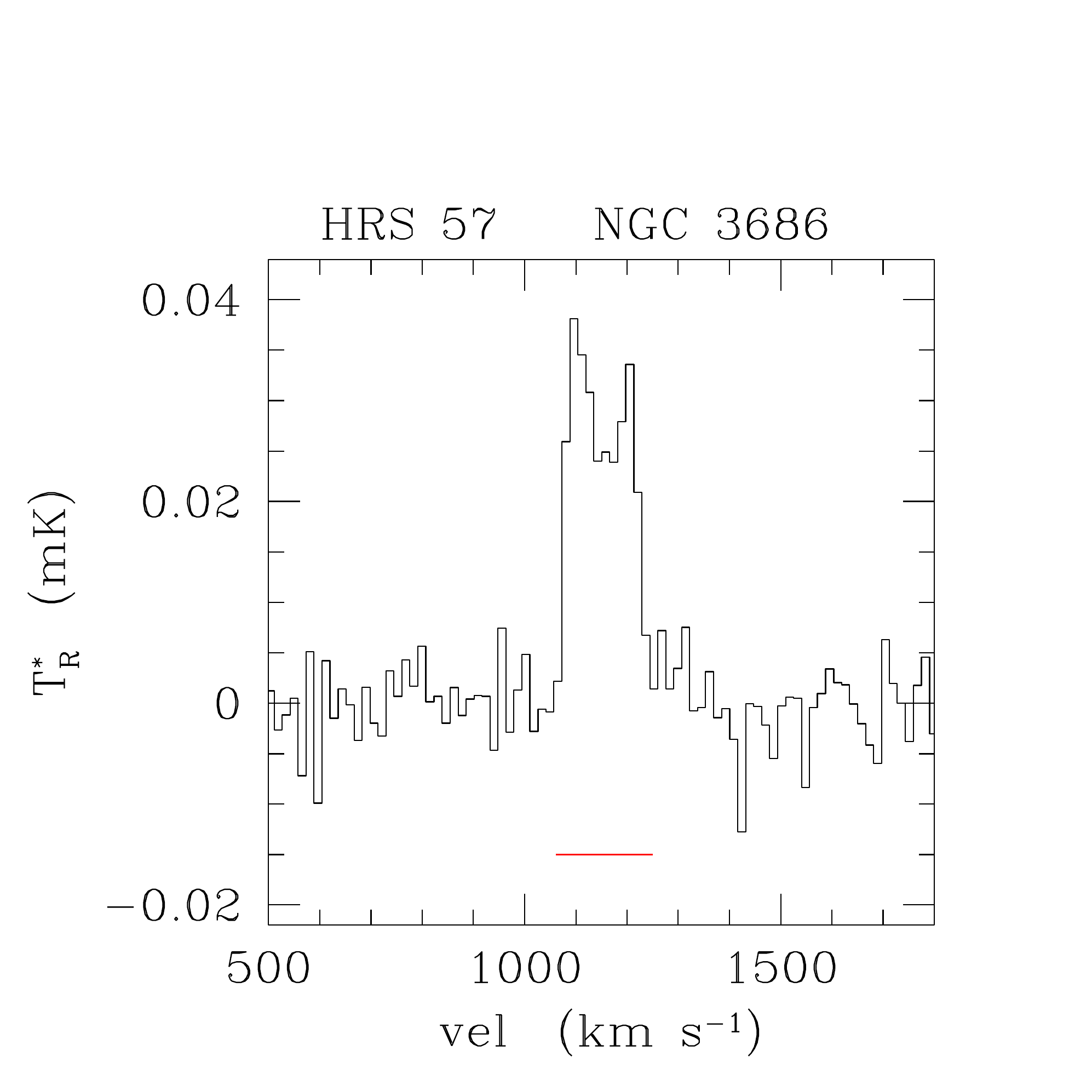}
   \includegraphics[width=0.22\textwidth]{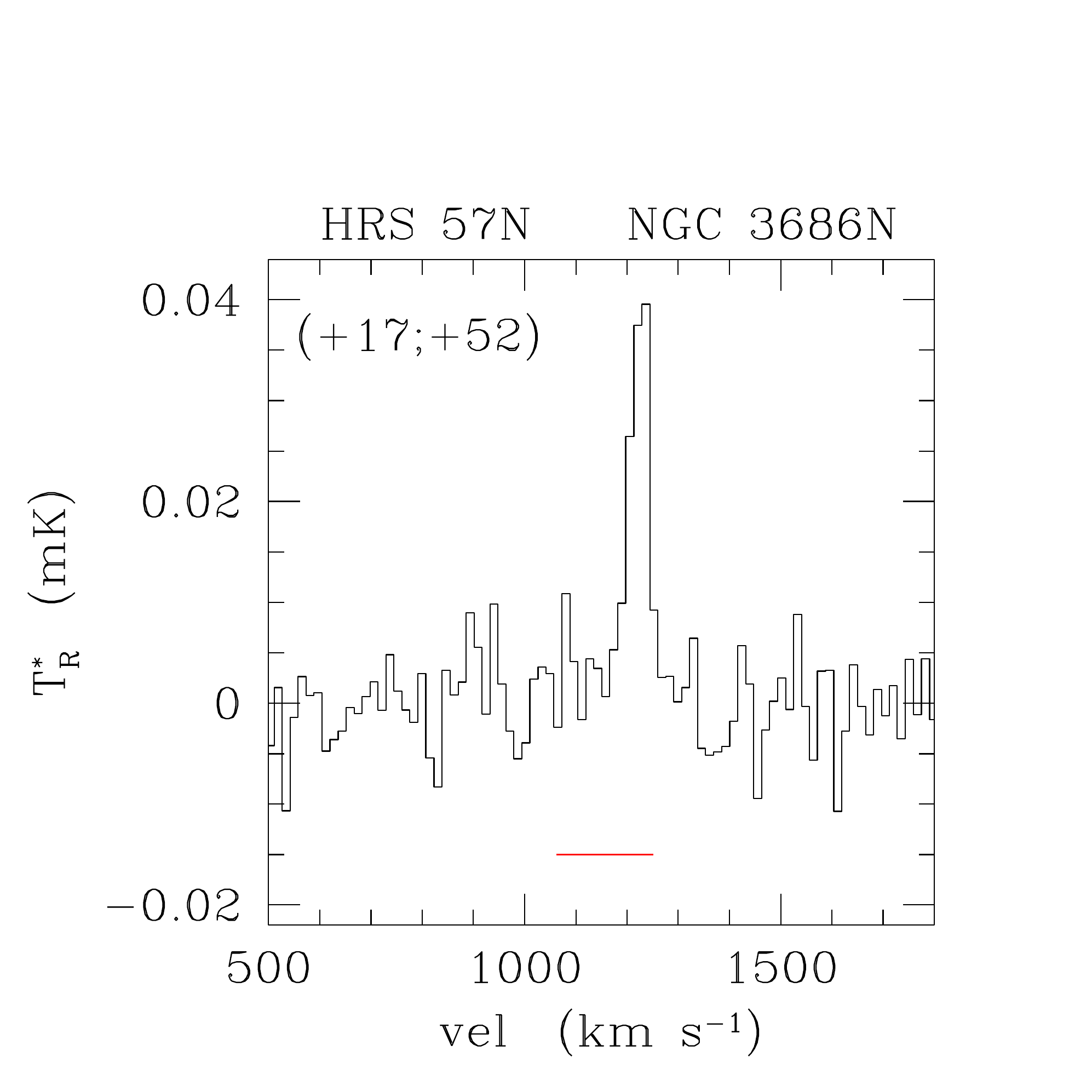}
   \includegraphics[width=0.22\textwidth]{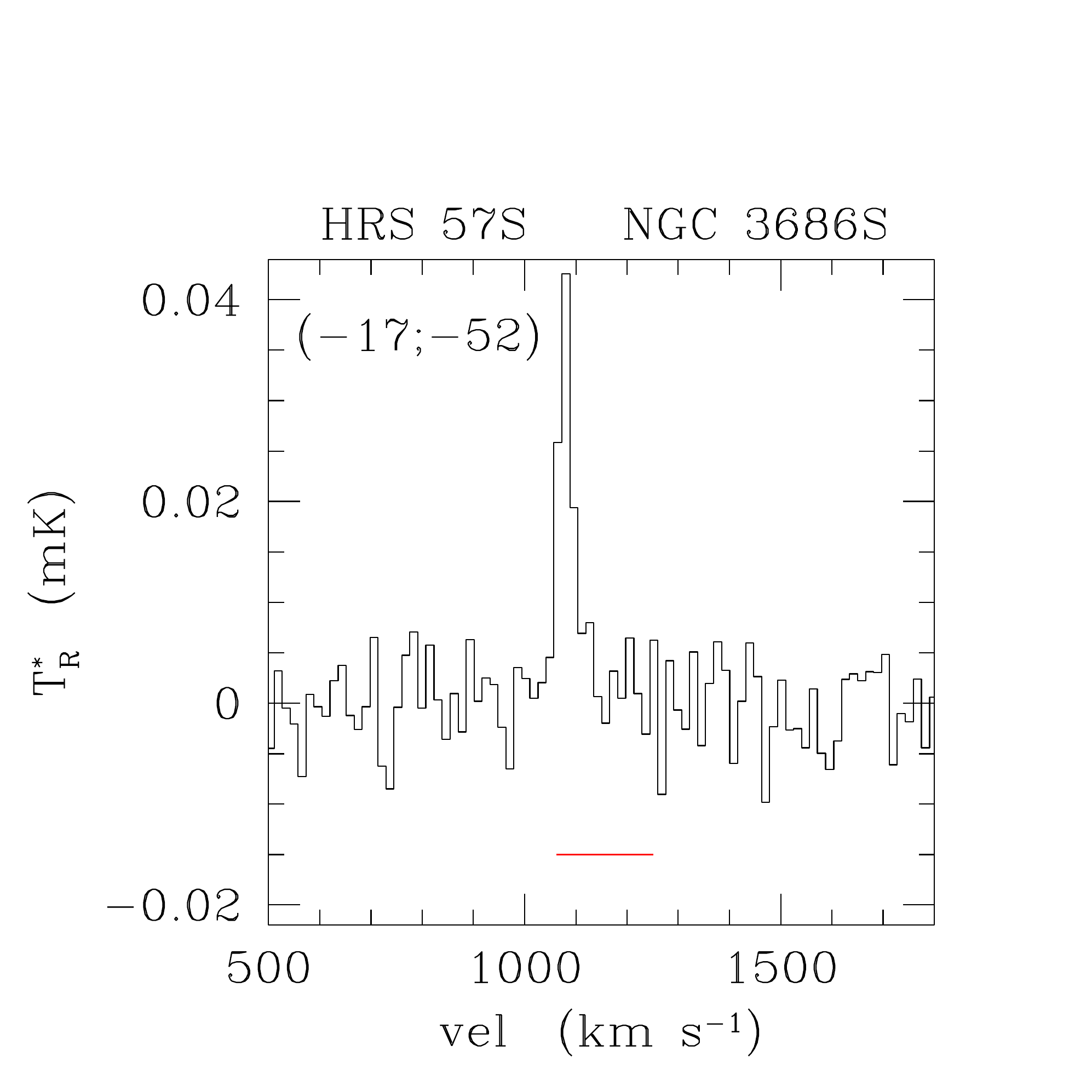}\\
   \includegraphics[width=0.22\textwidth]{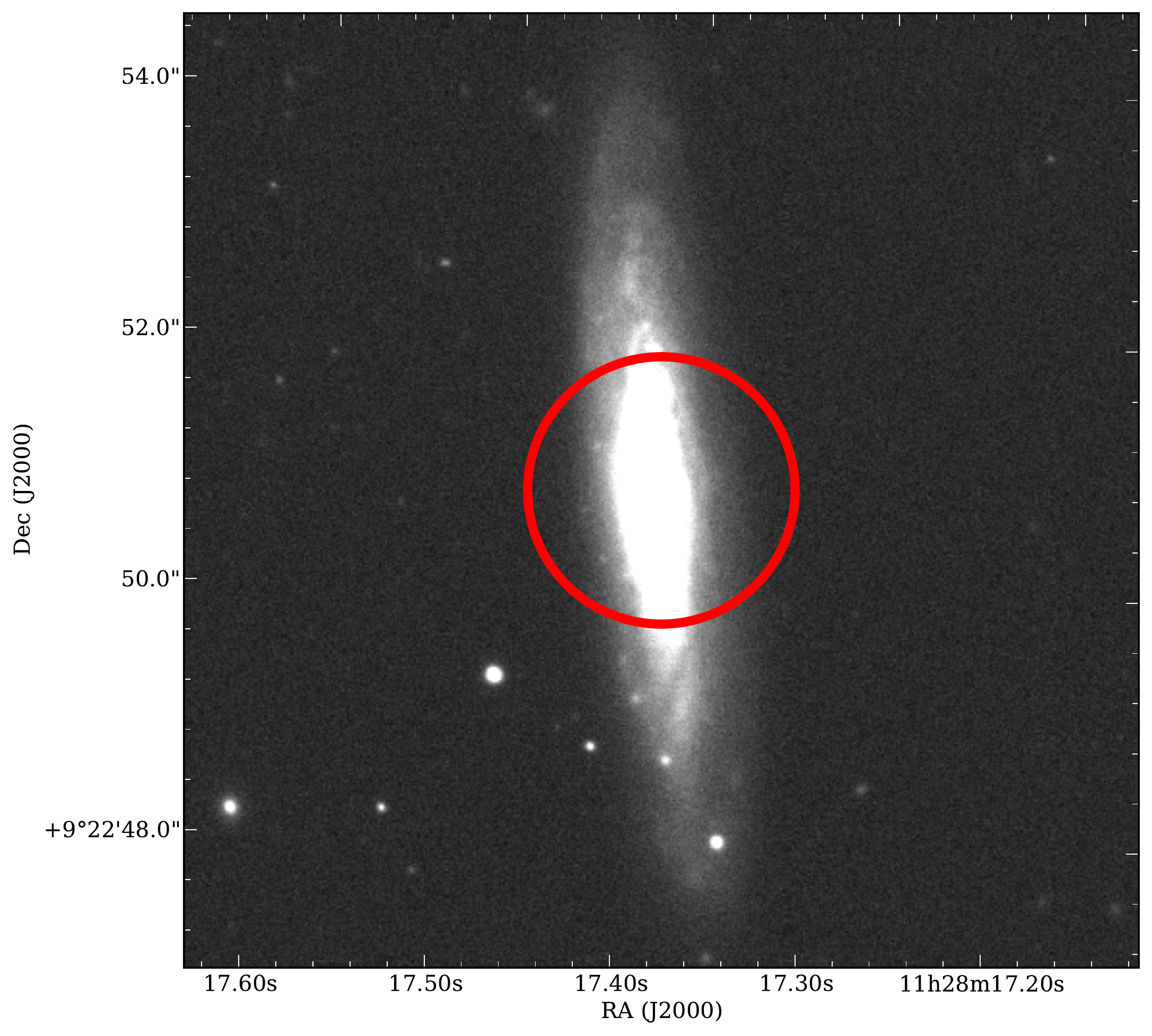}
   \includegraphics[width=0.22\textwidth]{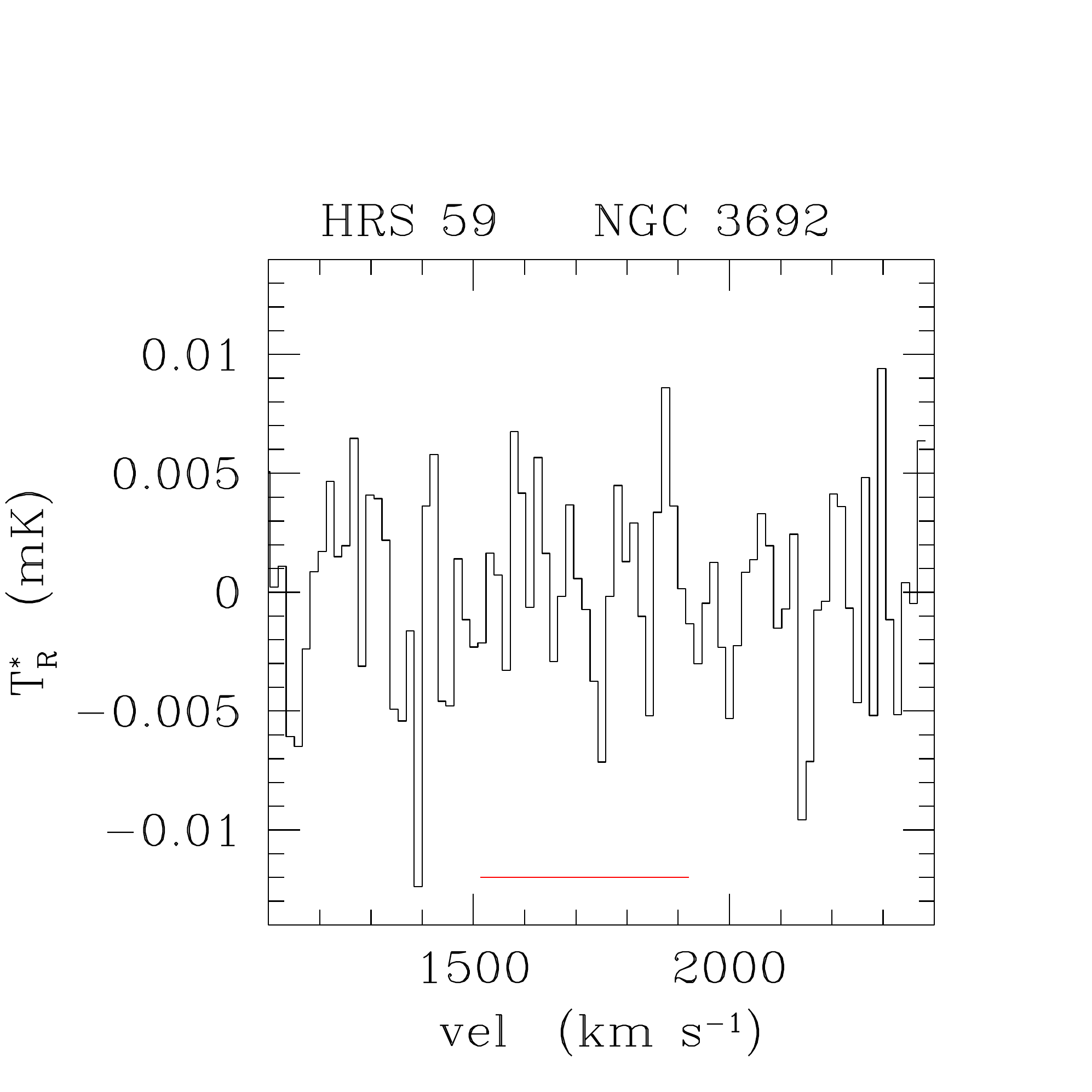}\\
   \includegraphics[width=0.22\textwidth]{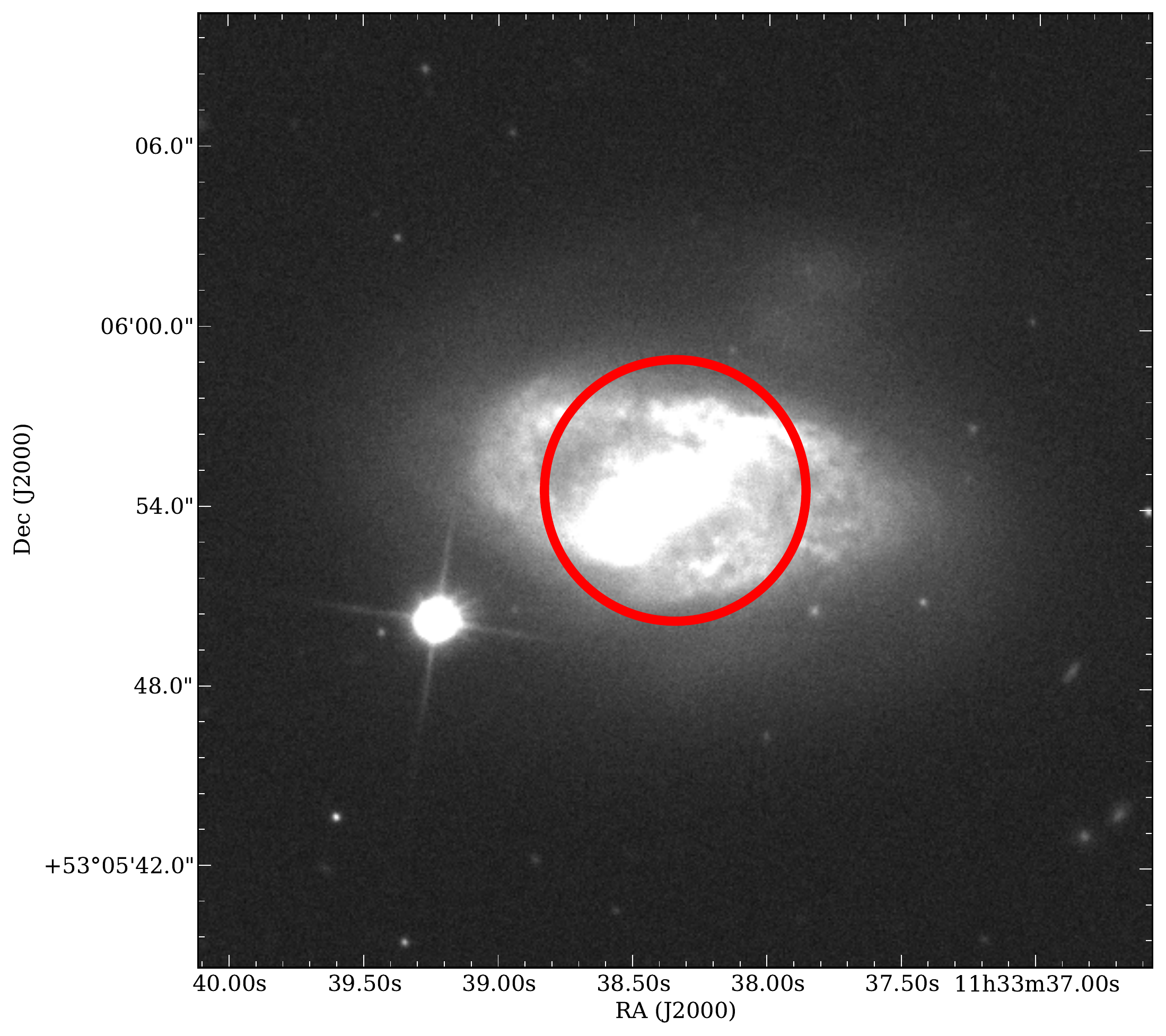}
   \includegraphics[width=0.22\textwidth]{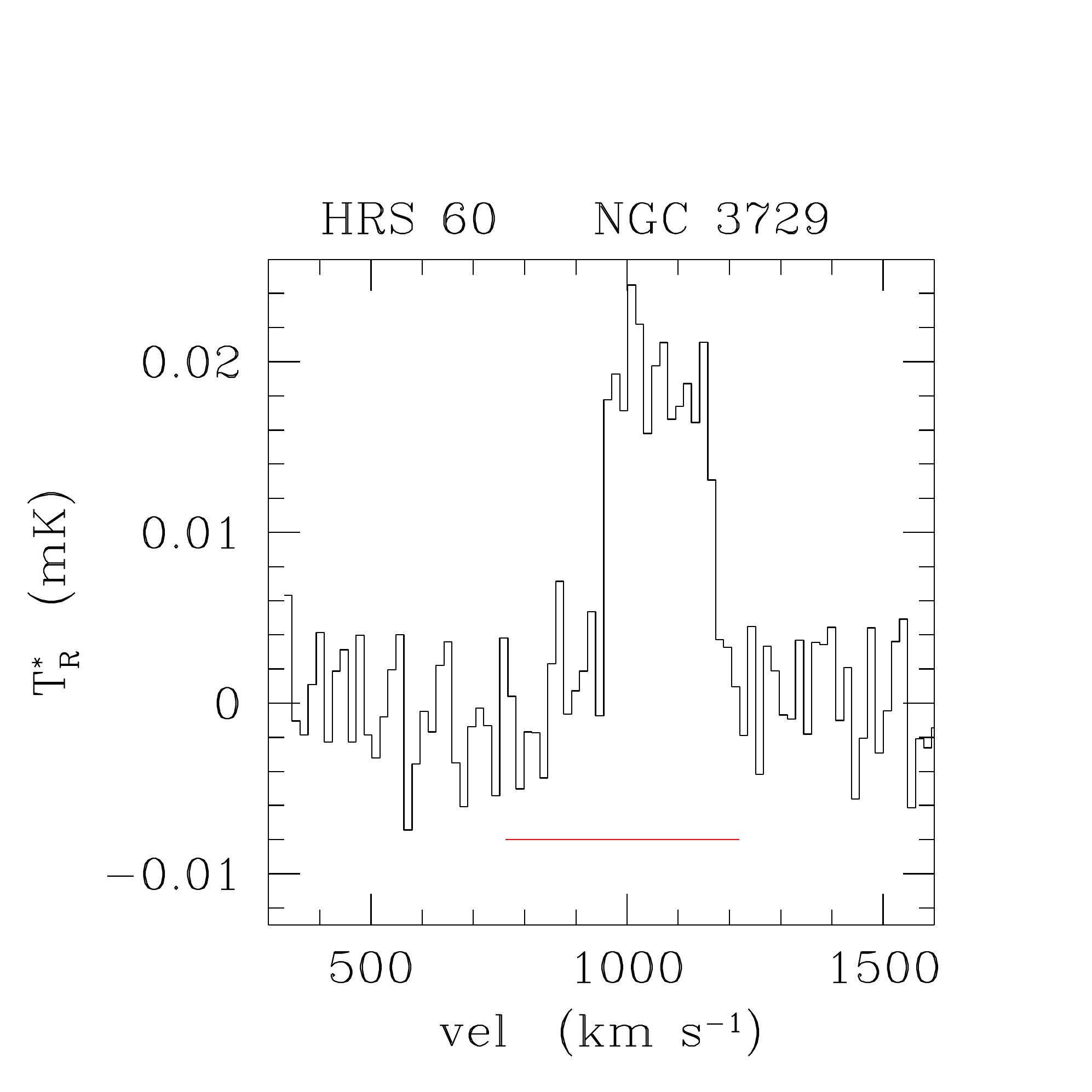}\\
   \includegraphics[width=0.22\textwidth]{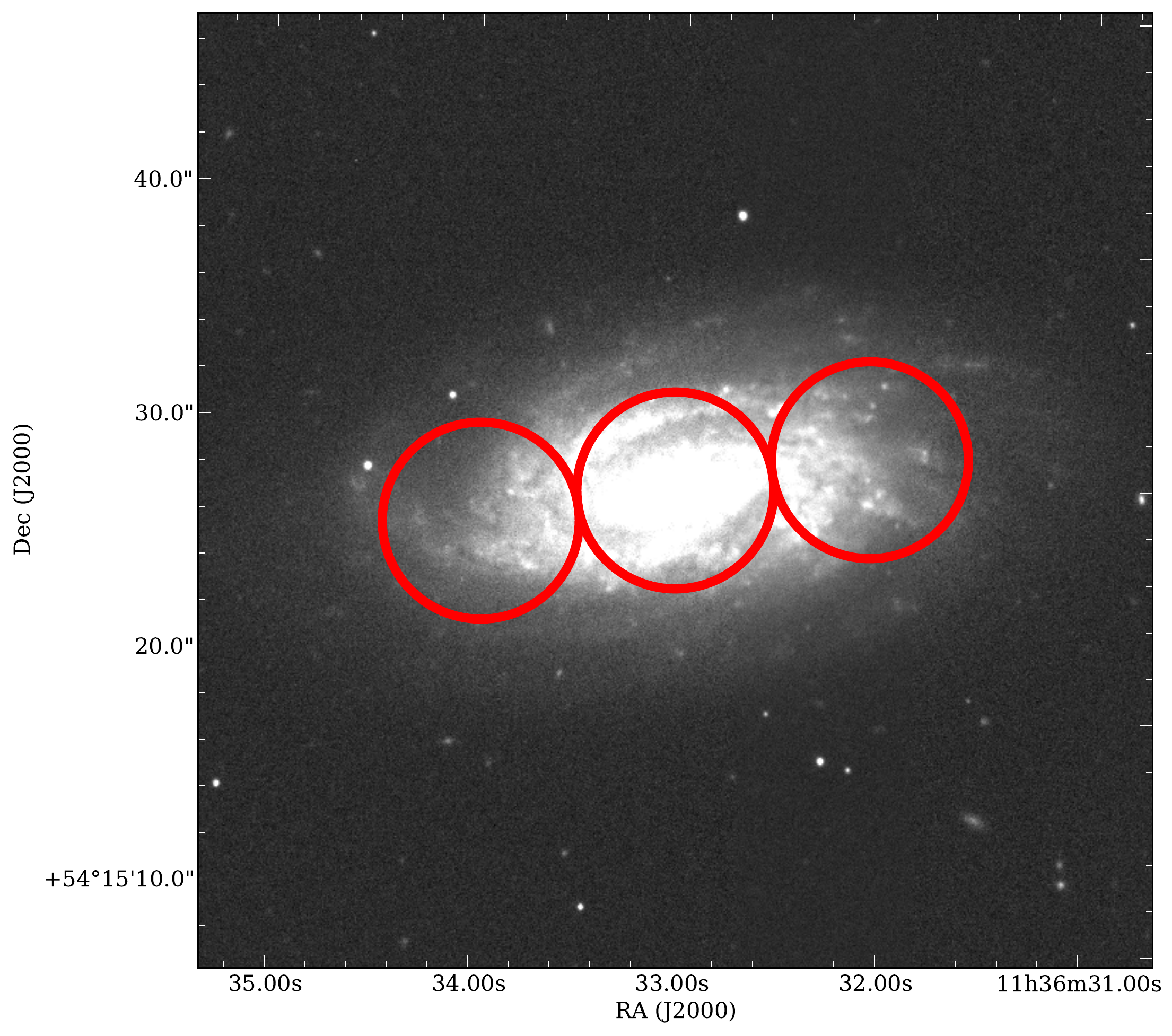}
   \includegraphics[width=0.22\textwidth]{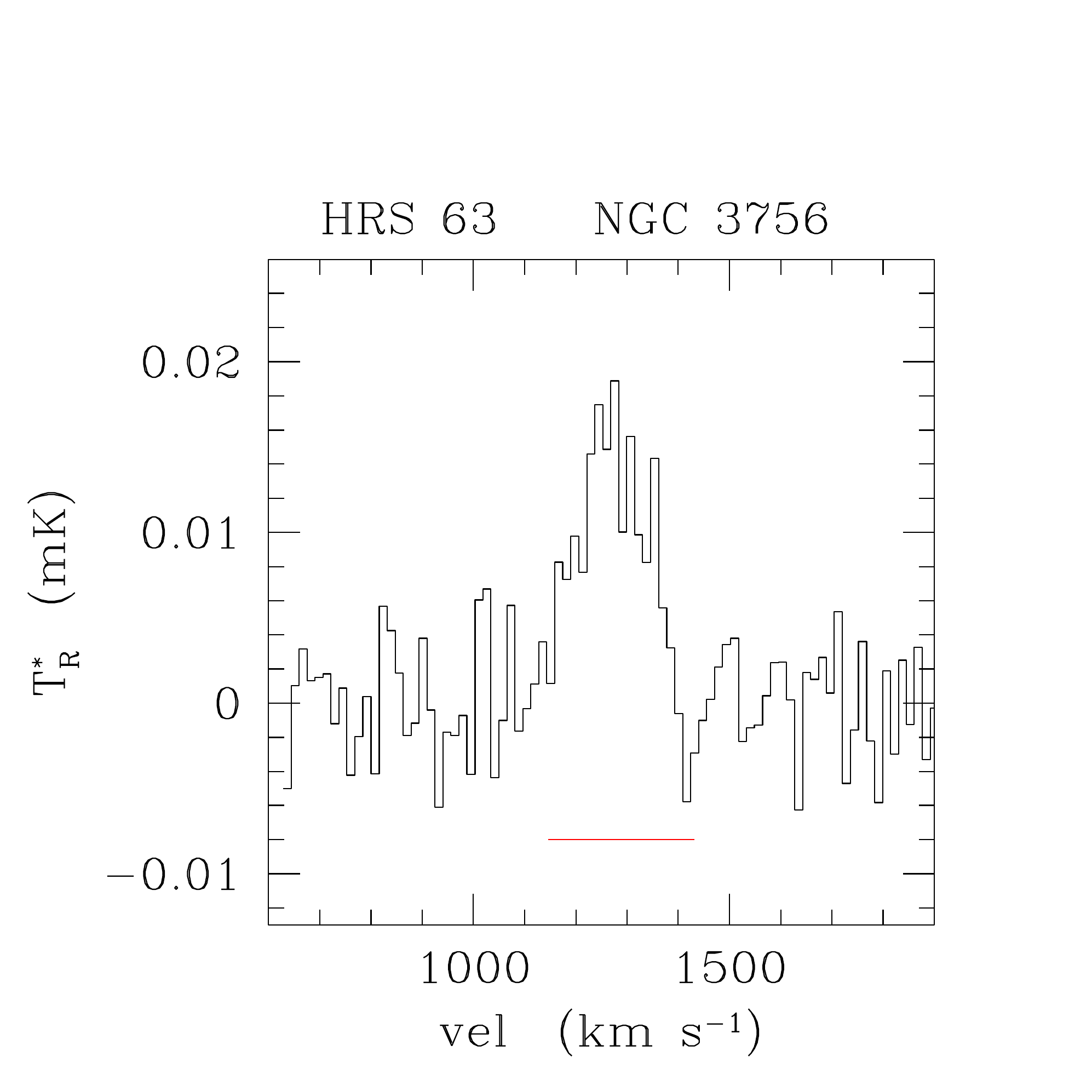}
   \includegraphics[width=0.22\textwidth]{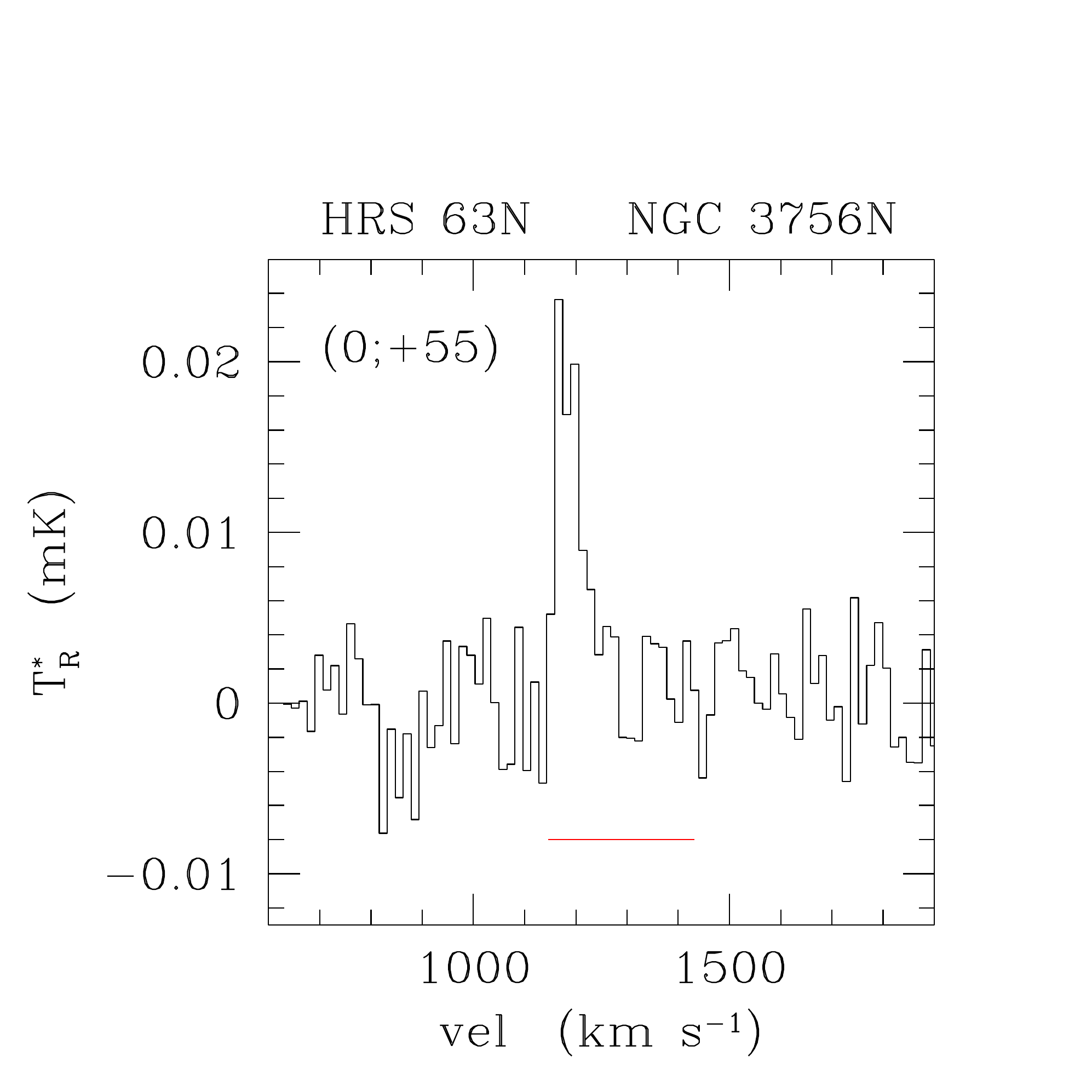}
   \includegraphics[width=0.22\textwidth]{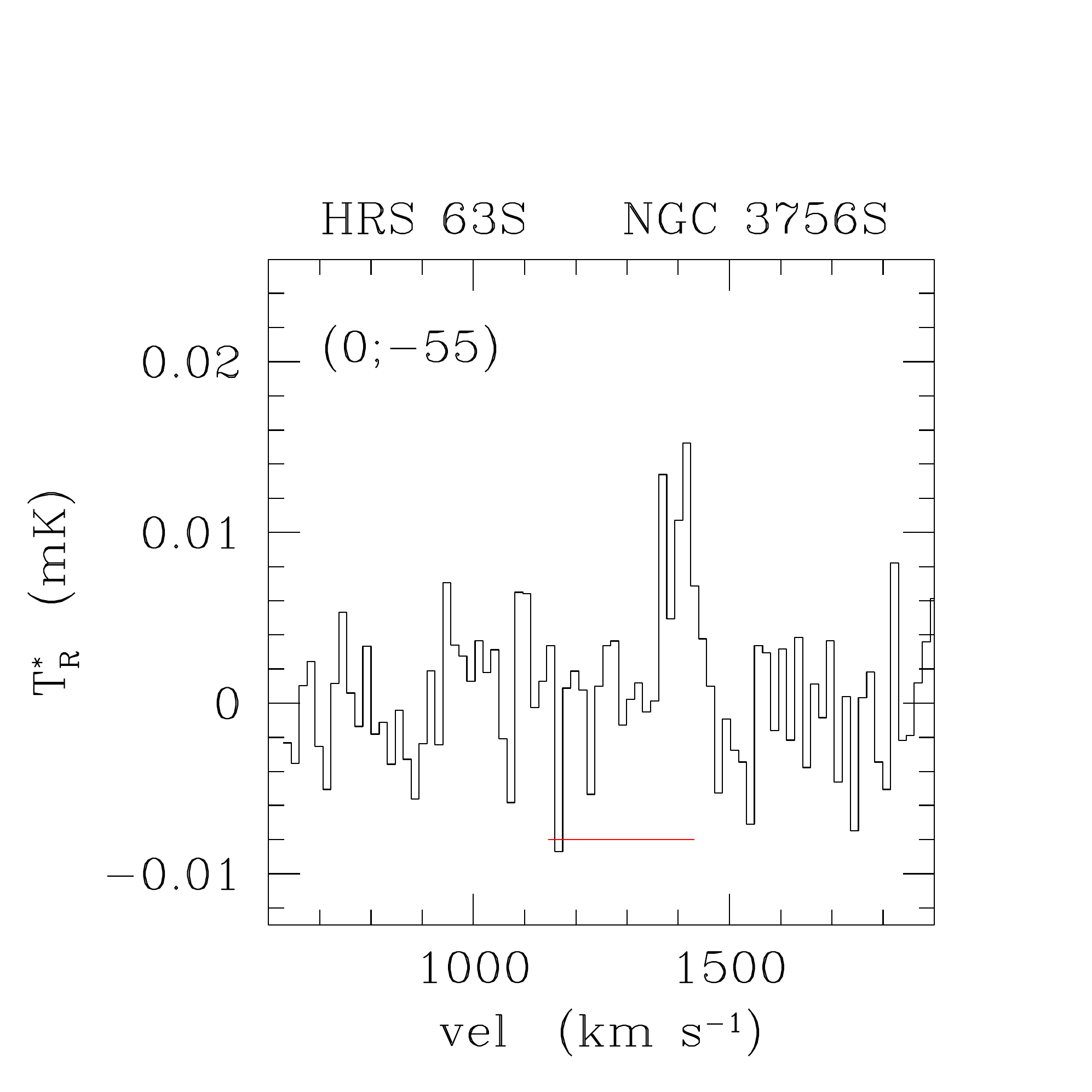}\\
   \caption{Continued.}
   \label{spettri}%
   \end{figure*}
   \clearpage

   \addtocounter{figure}{-1}
   \begin{figure*}
   \centering
   \includegraphics[width=0.22\textwidth]{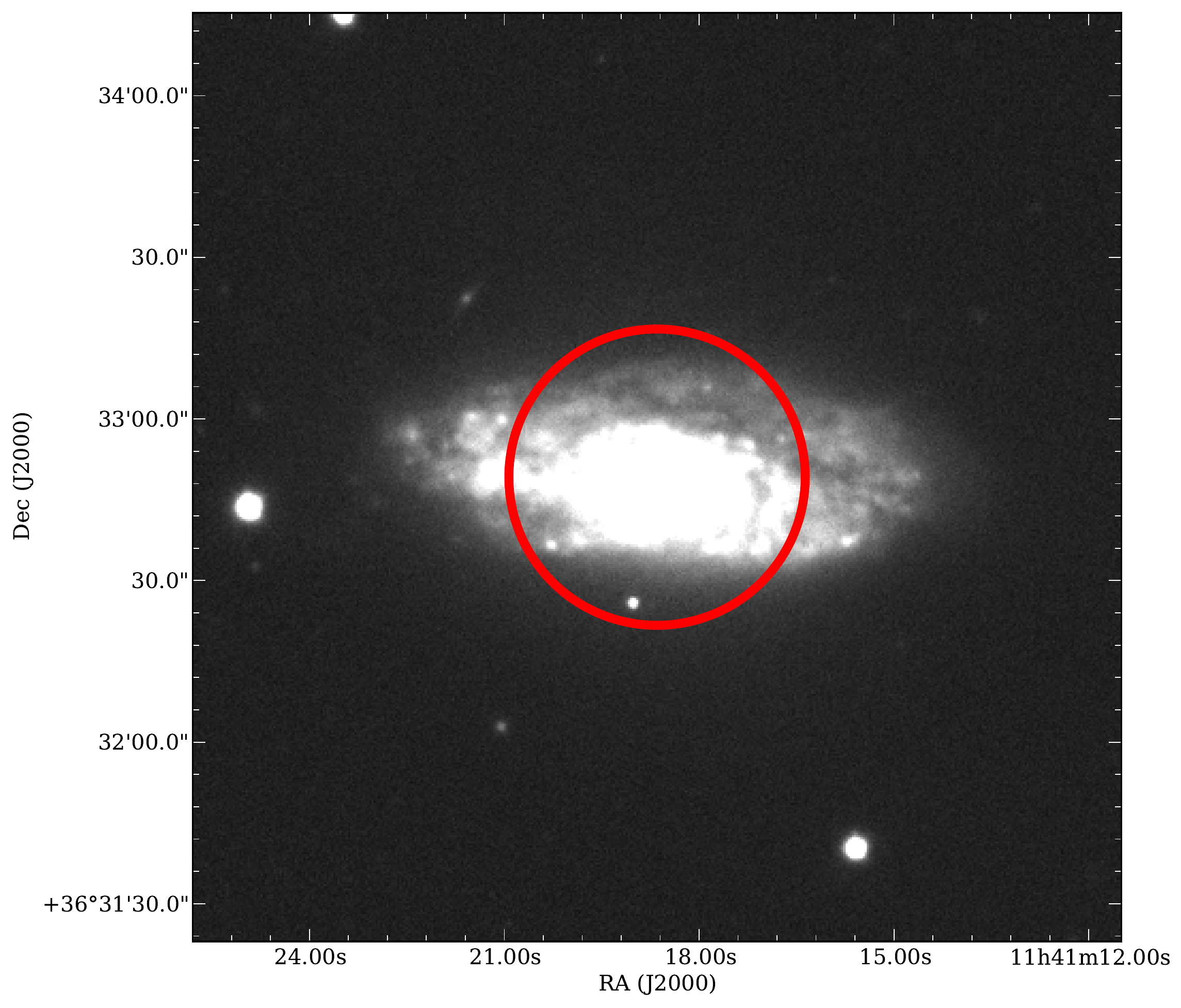}
   \includegraphics[width=0.22\textwidth]{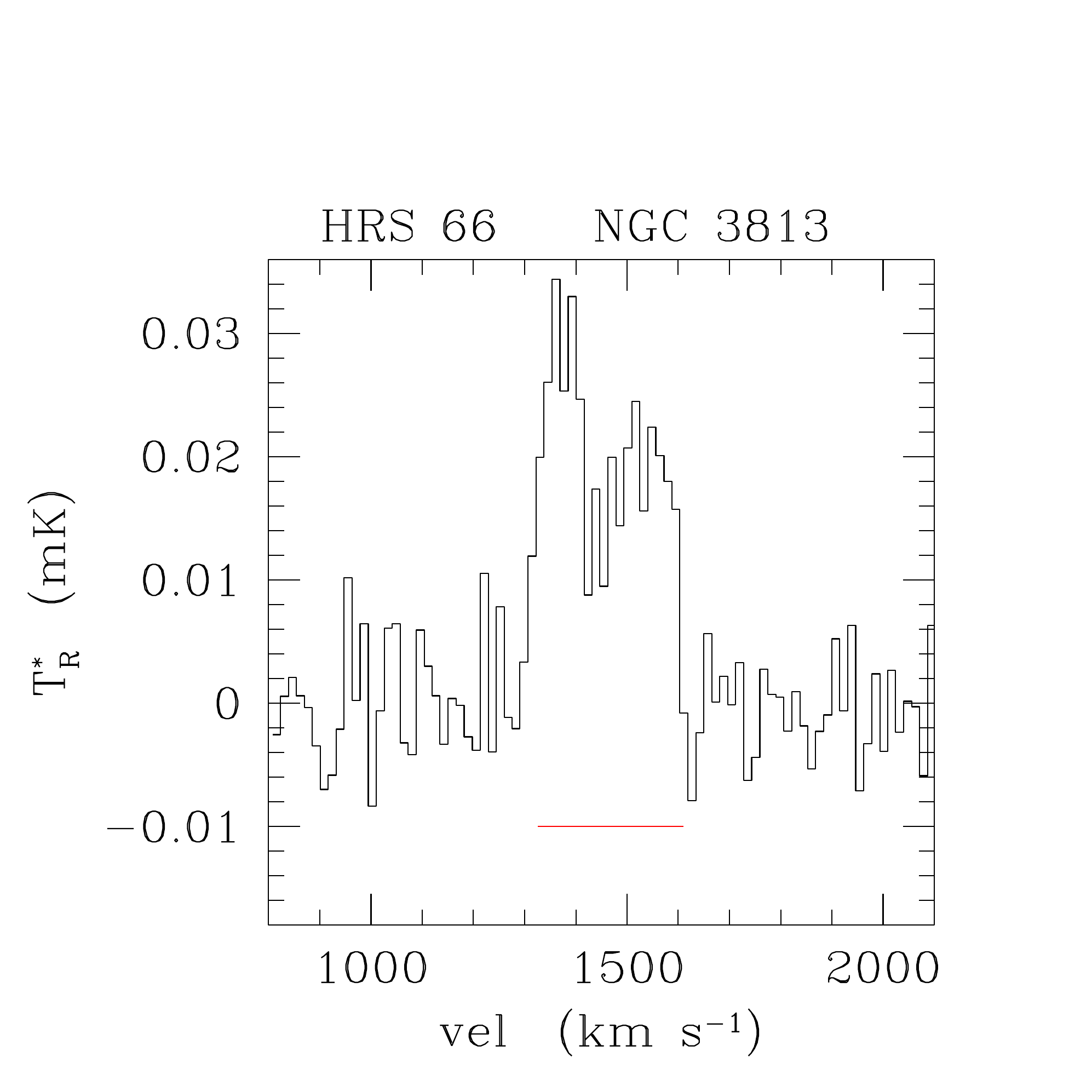}\\
   \includegraphics[width=0.22\textwidth]{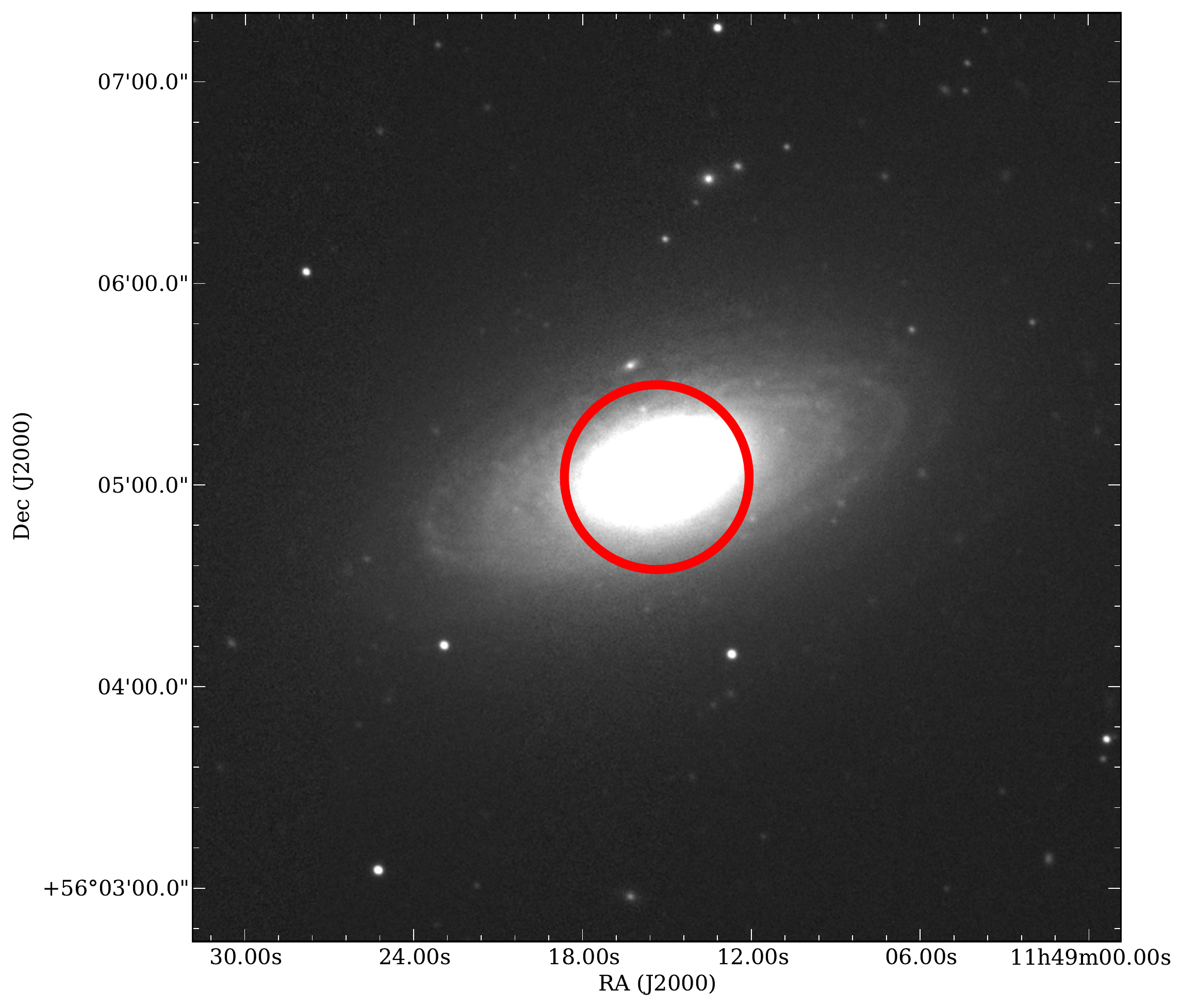}
   \includegraphics[width=0.22\textwidth]{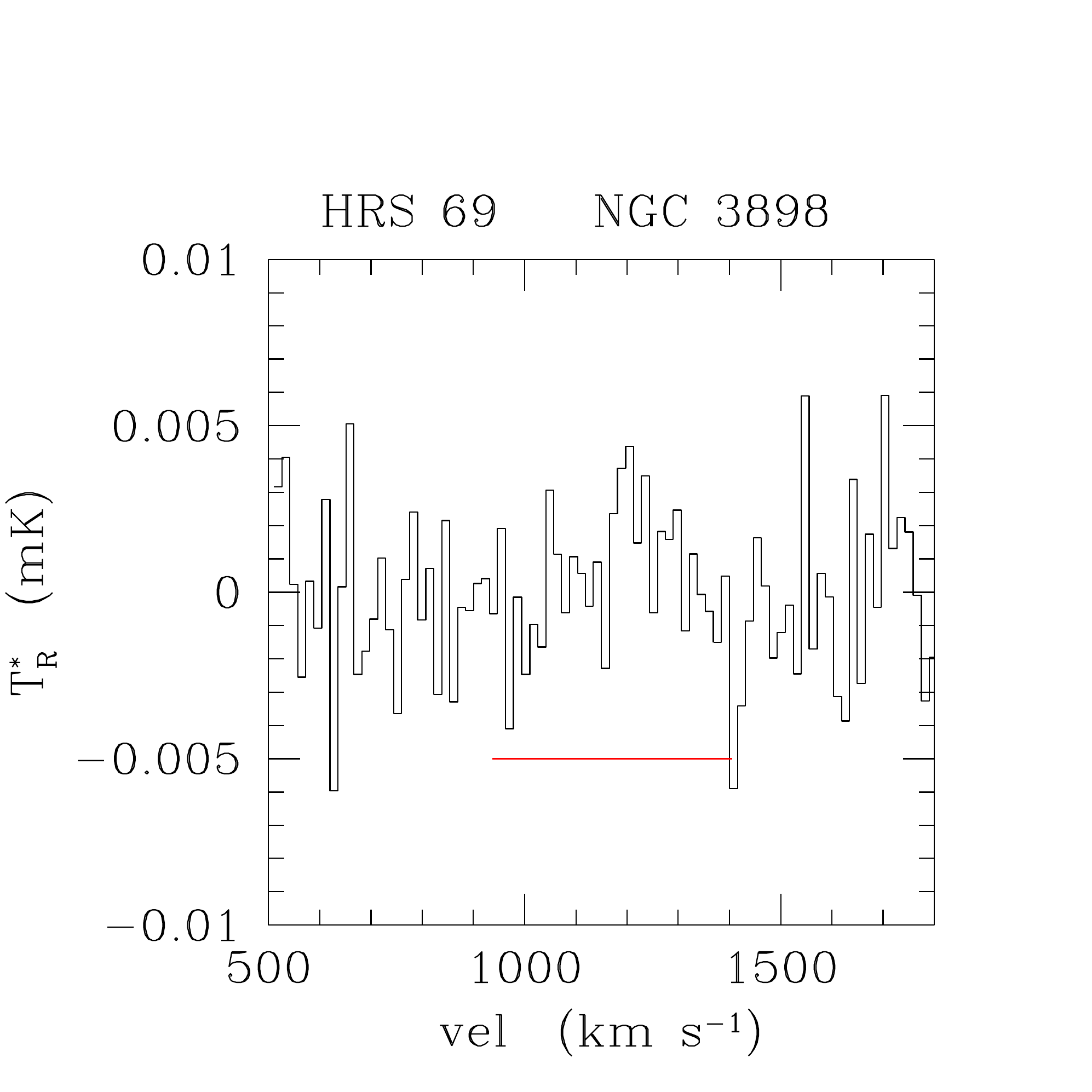}\\
   \includegraphics[width=0.22\textwidth]{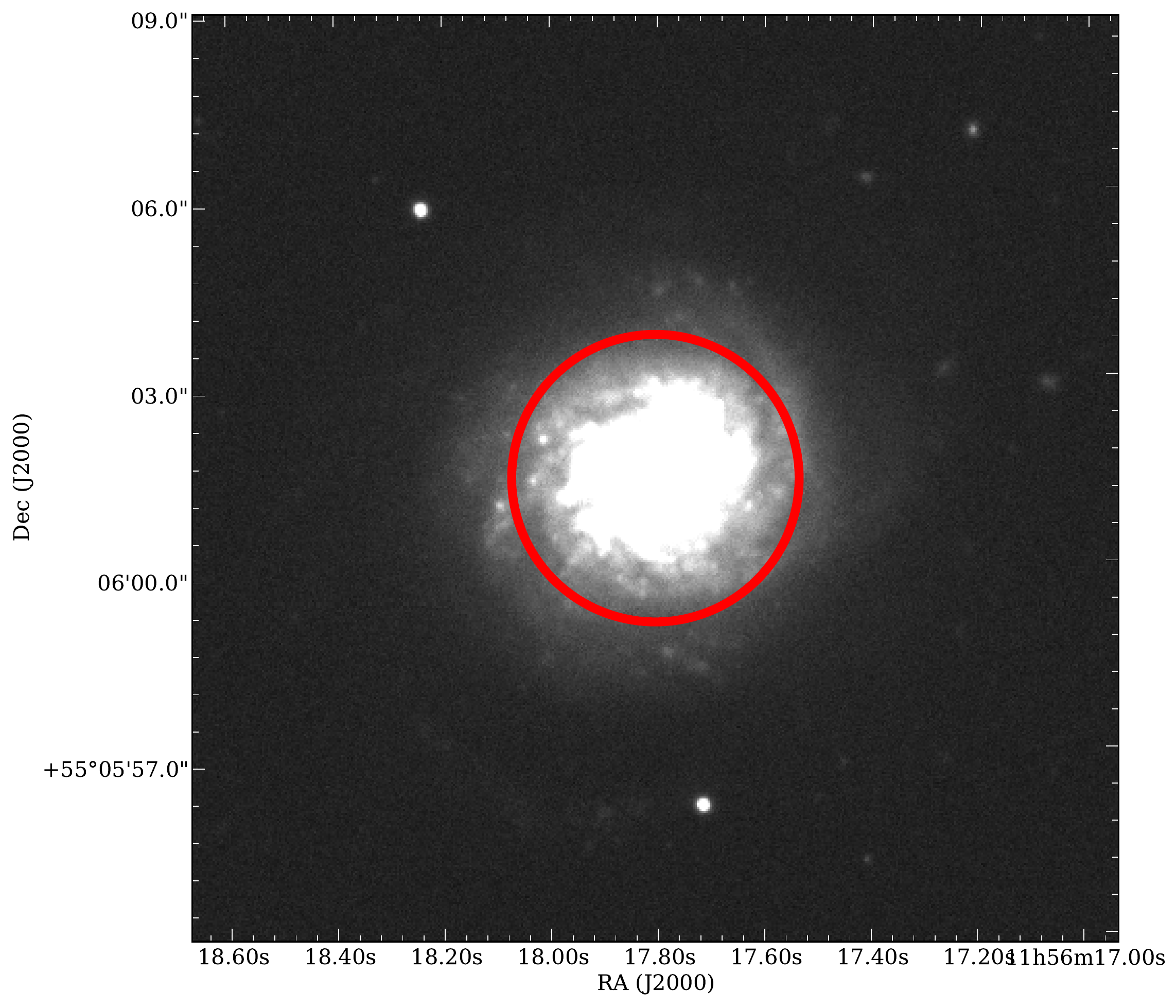}
   \includegraphics[width=0.22\textwidth]{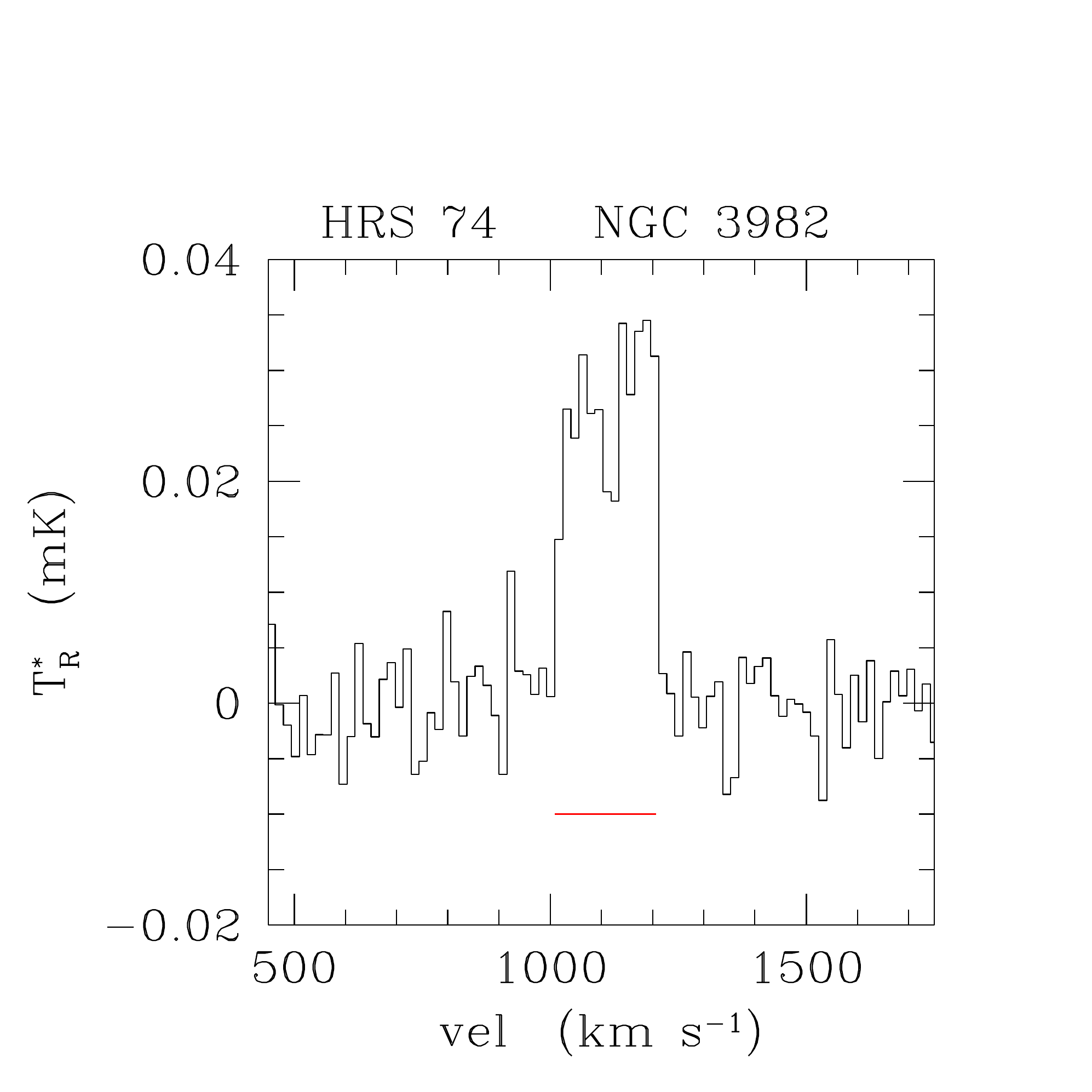}\\
   \includegraphics[width=0.22\textwidth]{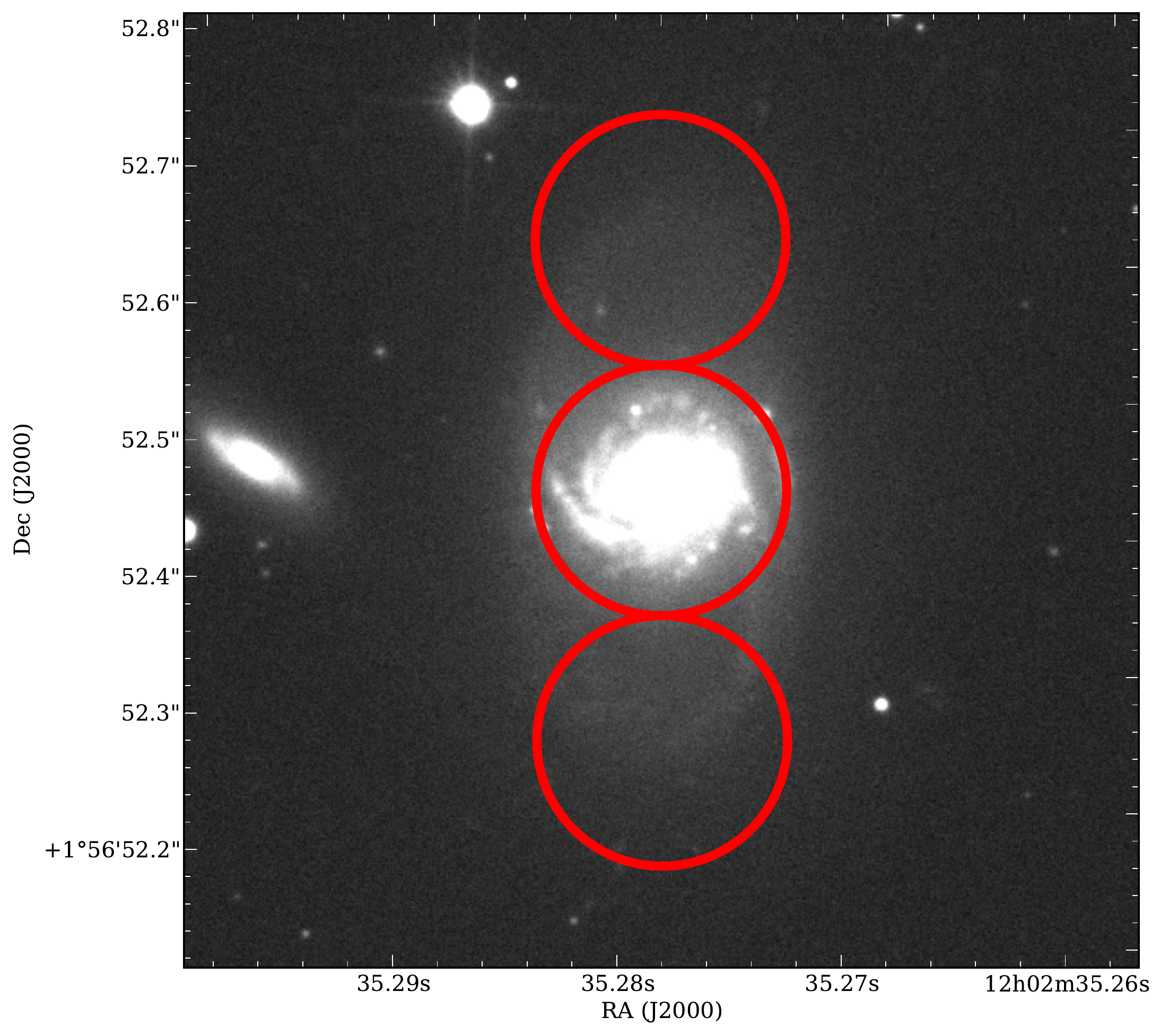}
   \includegraphics[width=0.22\textwidth]{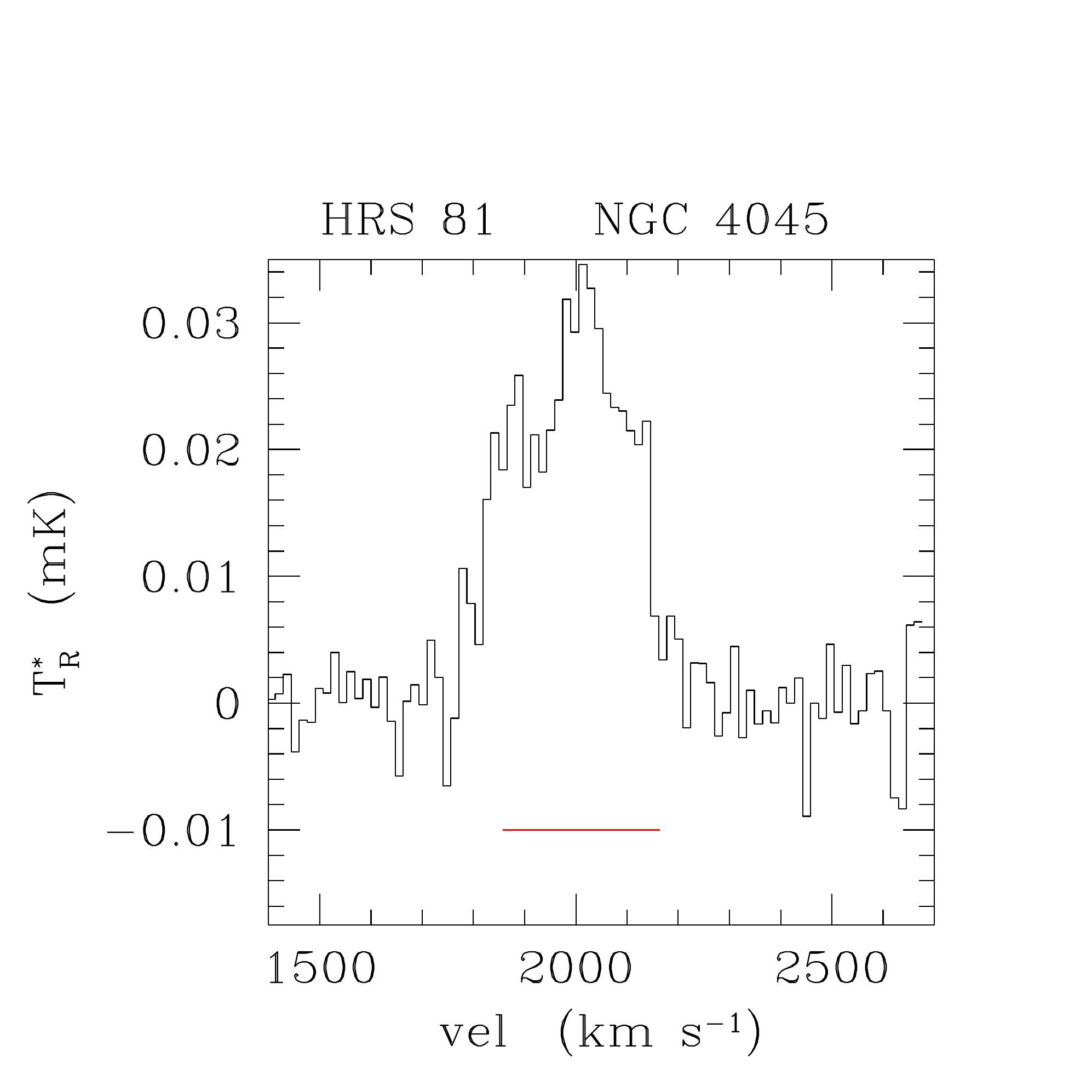}
   \includegraphics[width=0.22\textwidth]{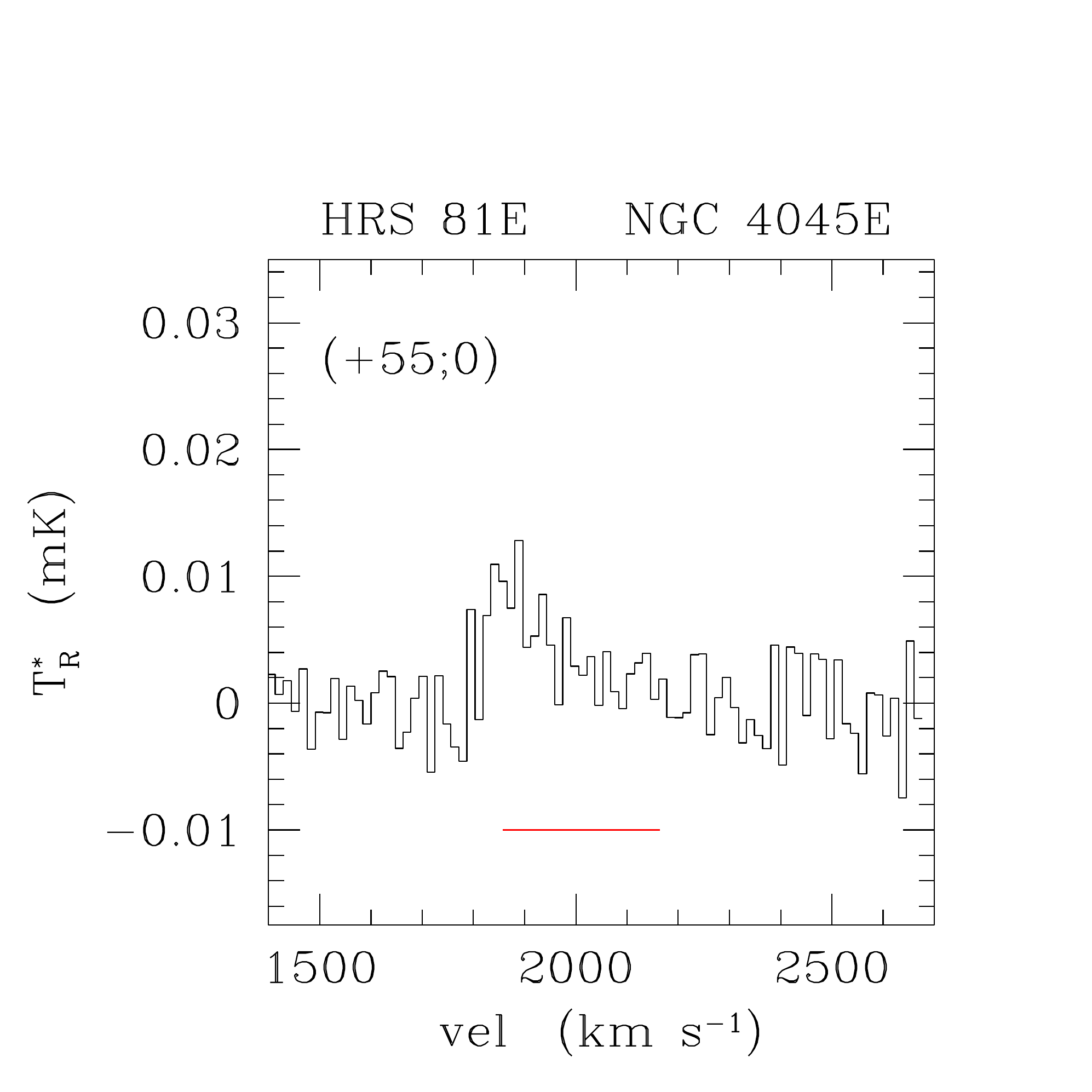}
   \includegraphics[width=0.22\textwidth]{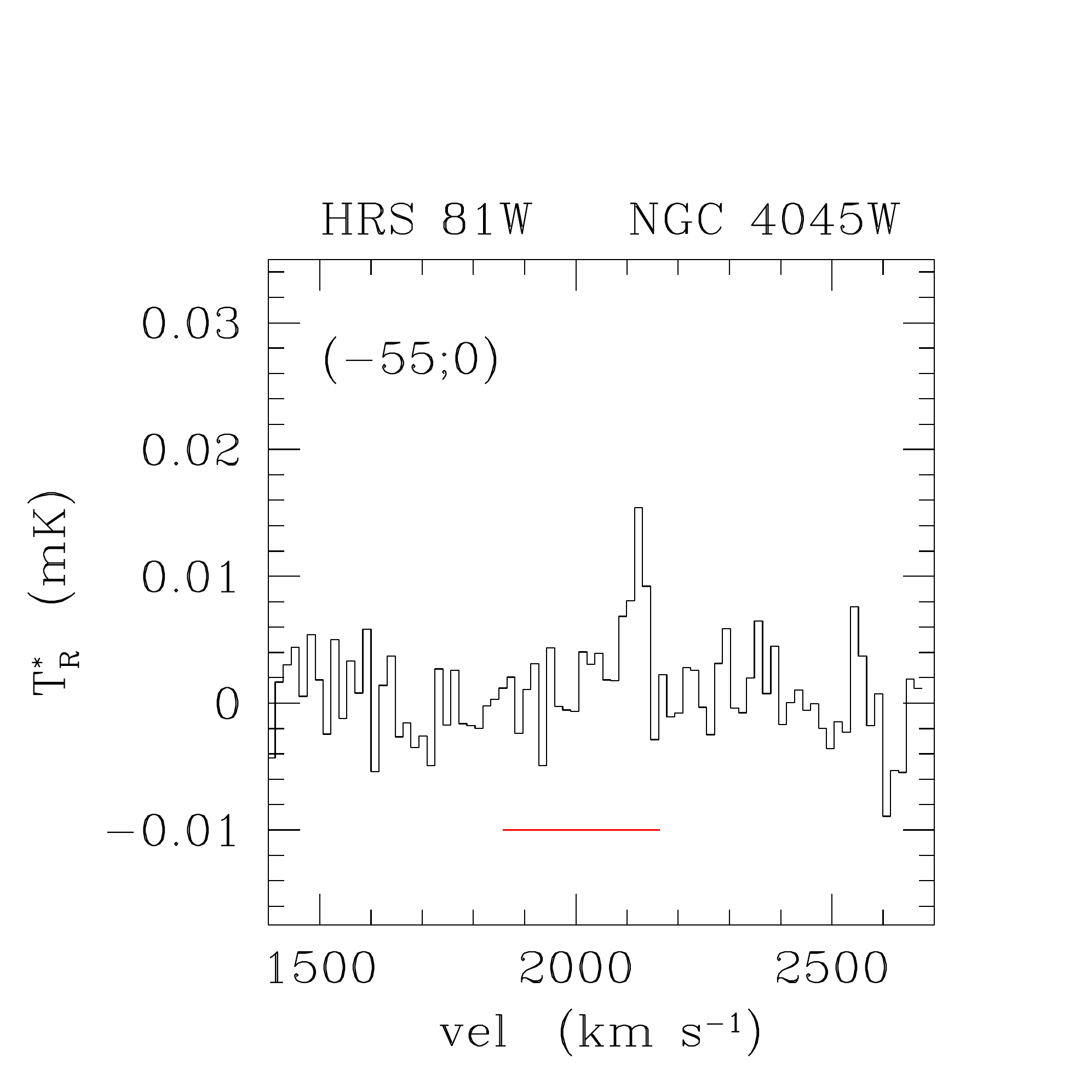}\\
   \includegraphics[width=0.22\textwidth]{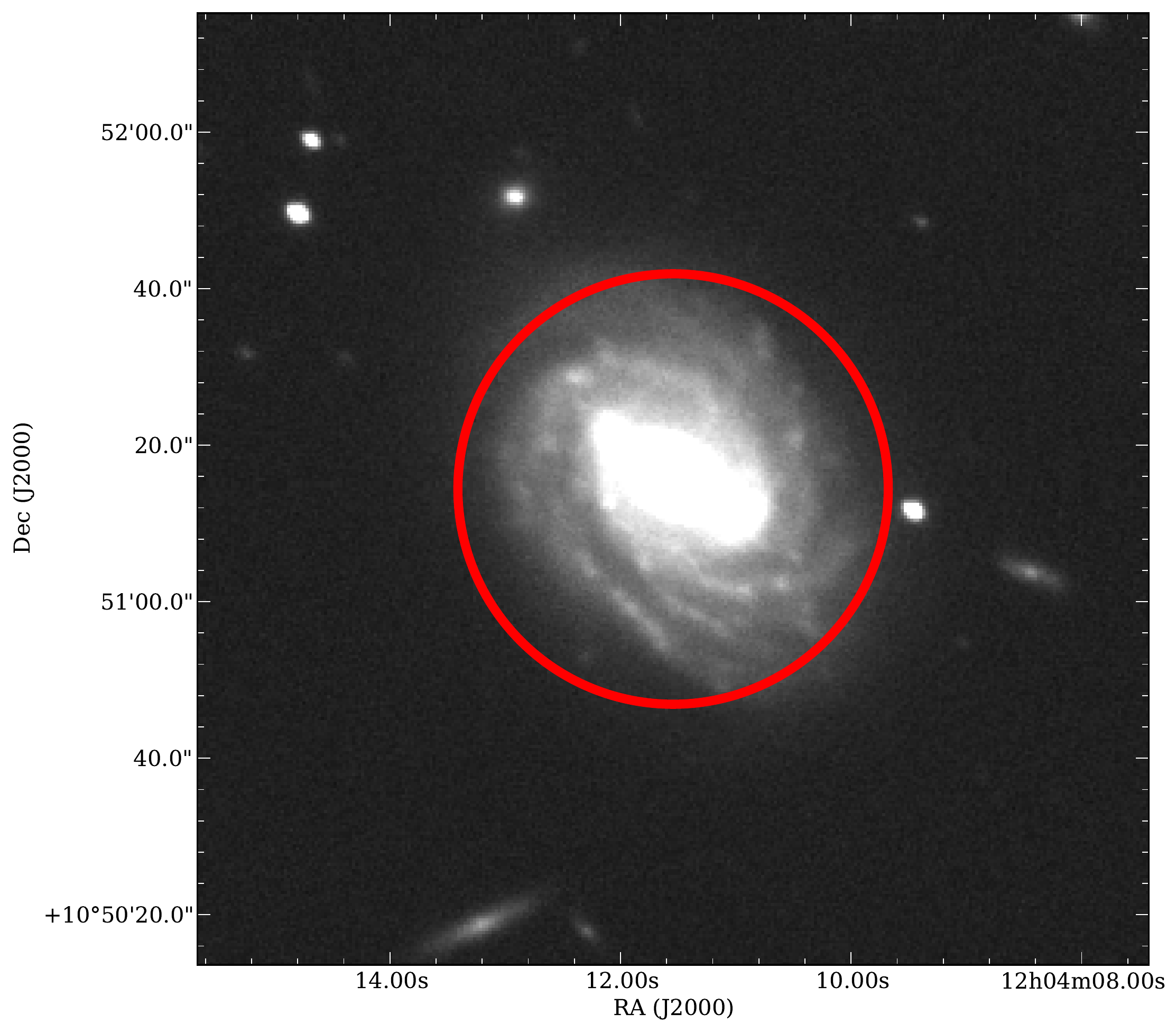}
   \includegraphics[width=0.22\textwidth]{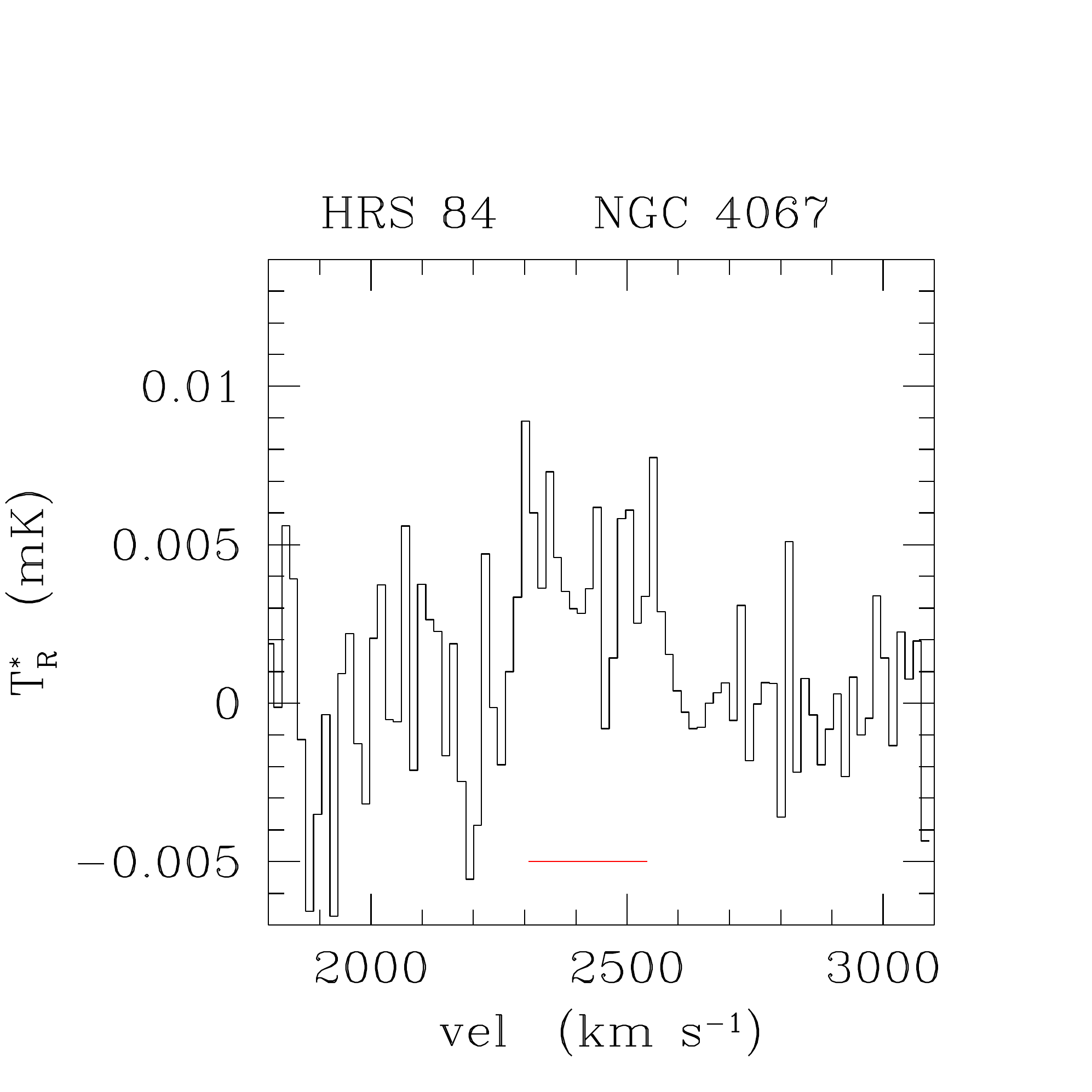}\\
   \caption{Continued.}
   \label{spettri}%
   \end{figure*}
   \clearpage

   \addtocounter{figure}{-1}
   \begin{figure*}
   \centering
   \includegraphics[width=0.22\textwidth]{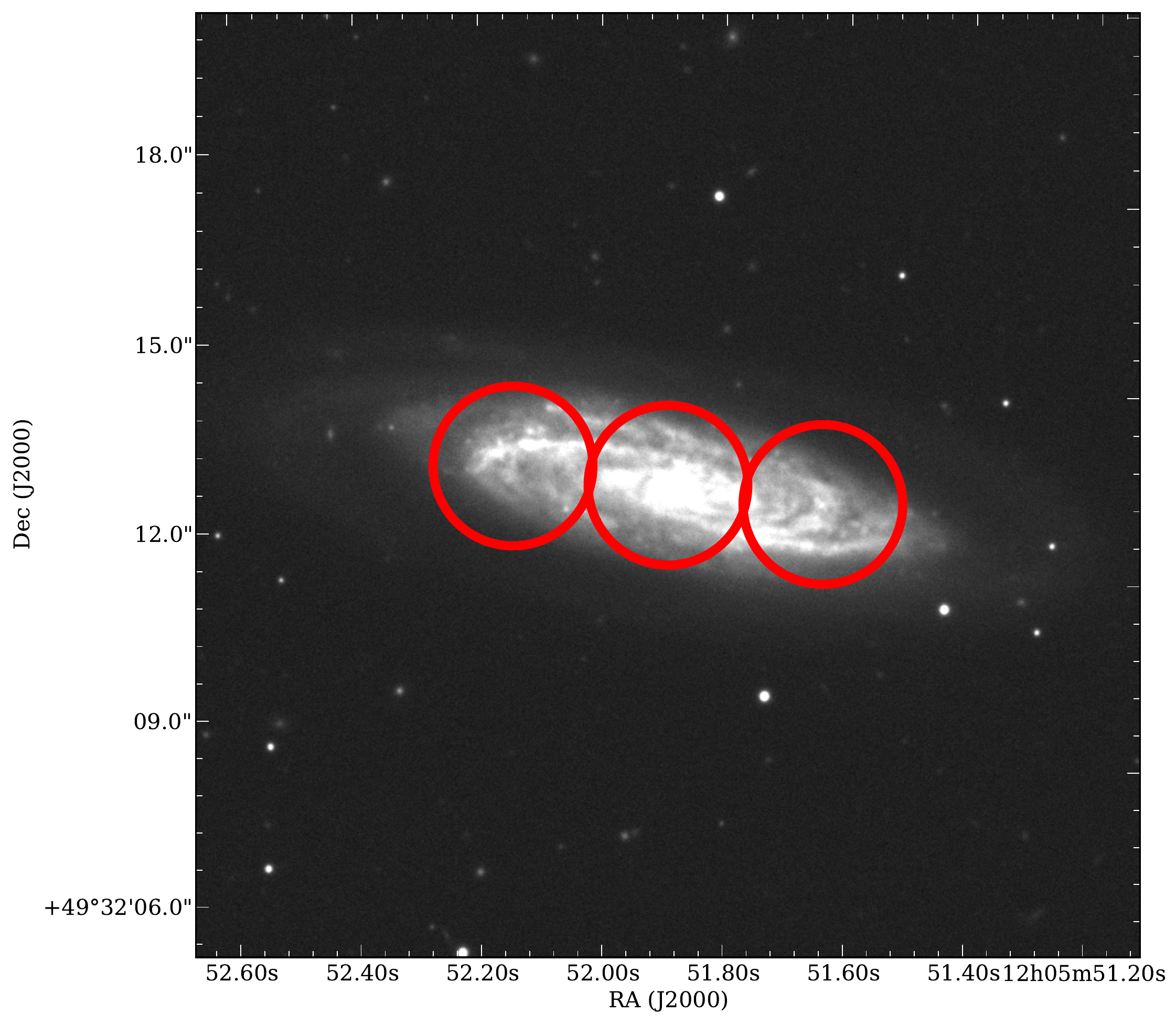}
   \includegraphics[width=0.22\textwidth]{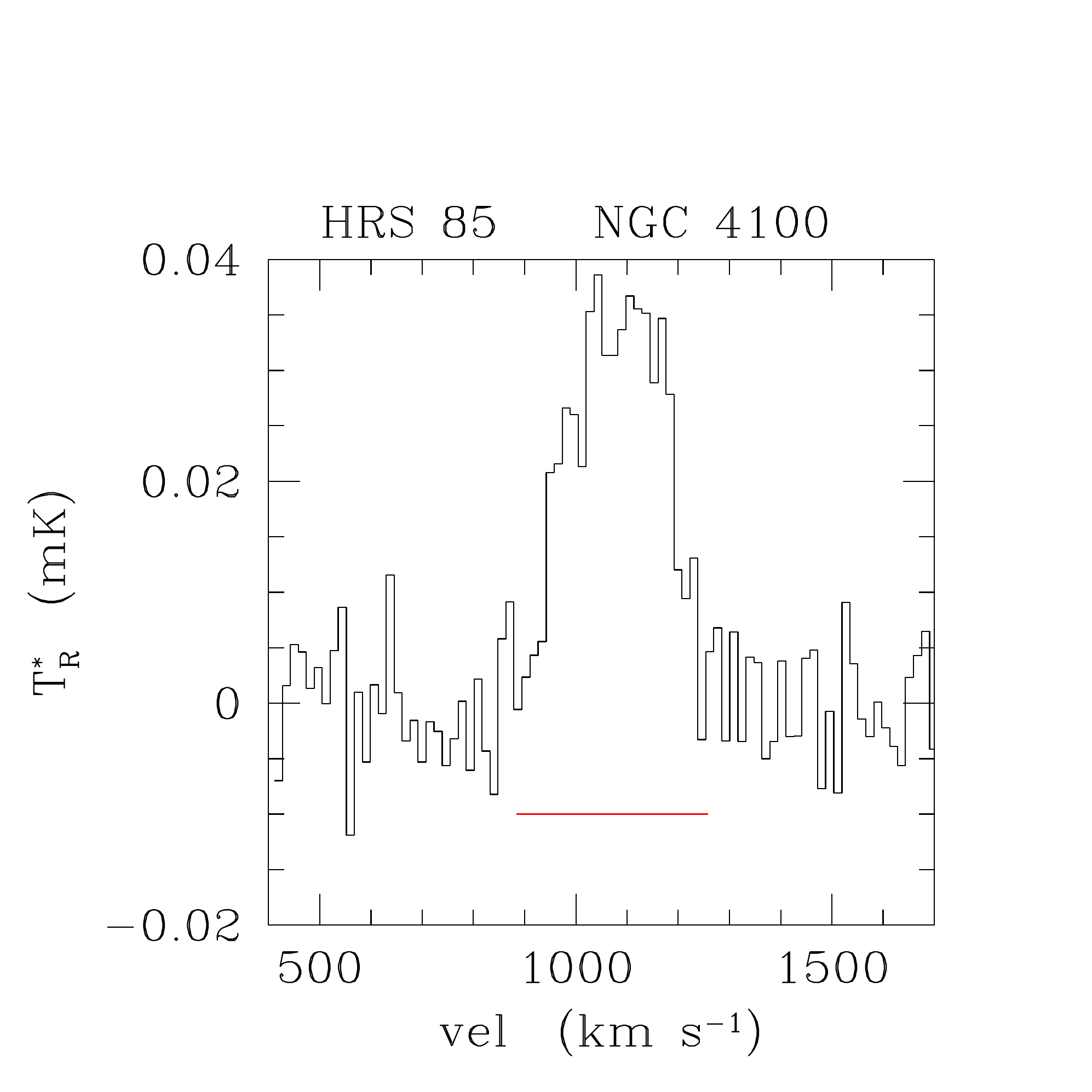}
   \includegraphics[width=0.22\textwidth]{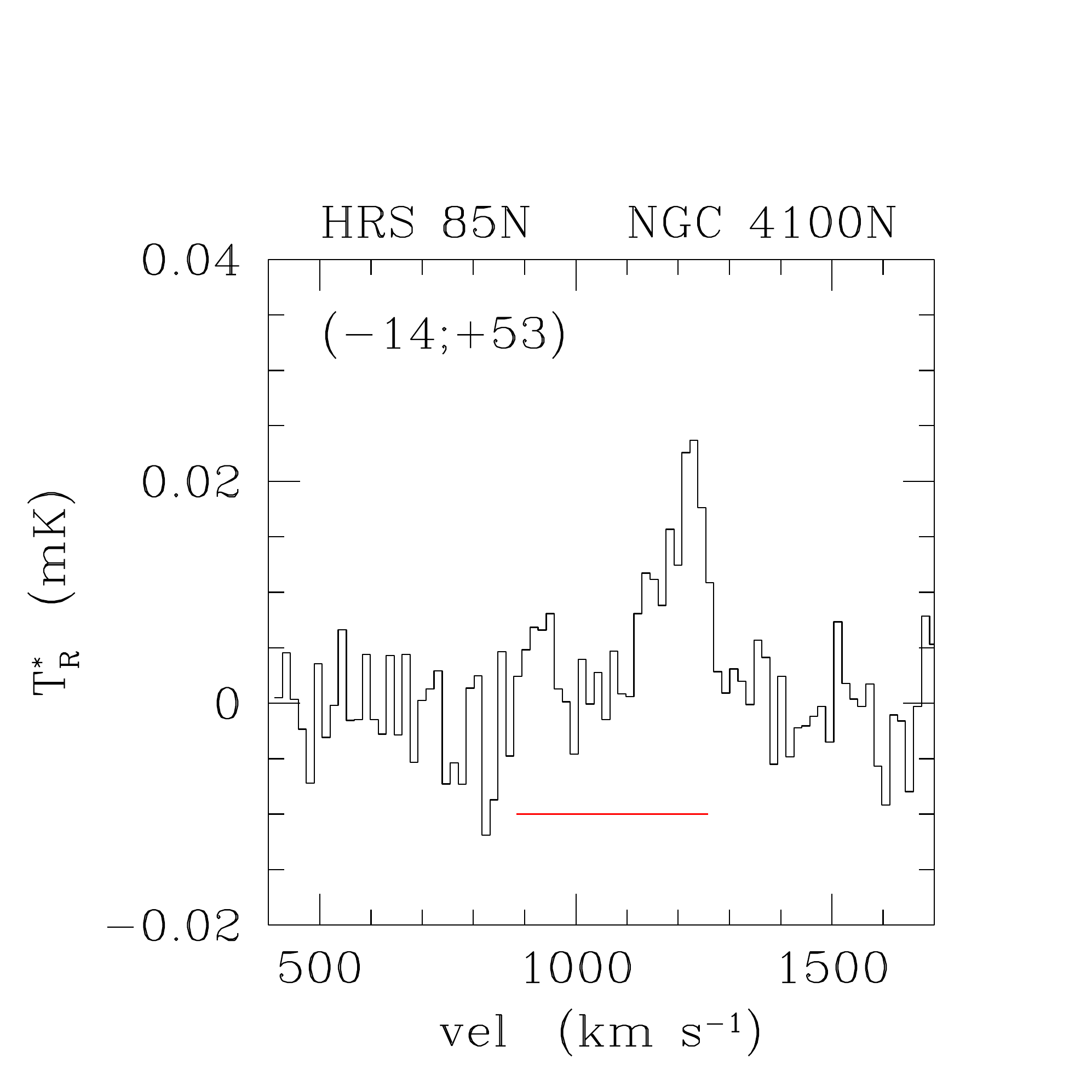}
   \includegraphics[width=0.22\textwidth]{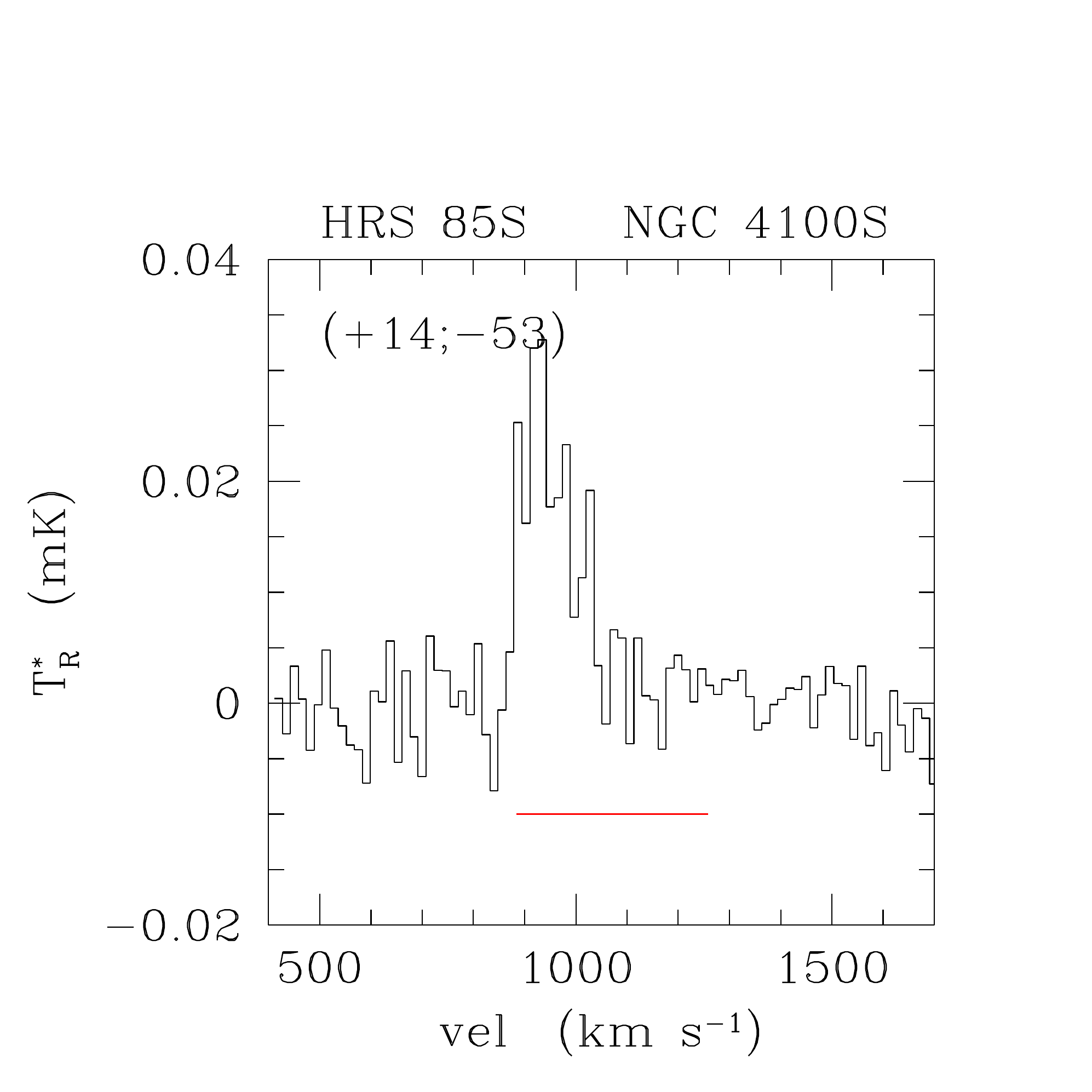}\\
   \includegraphics[width=0.22\textwidth]{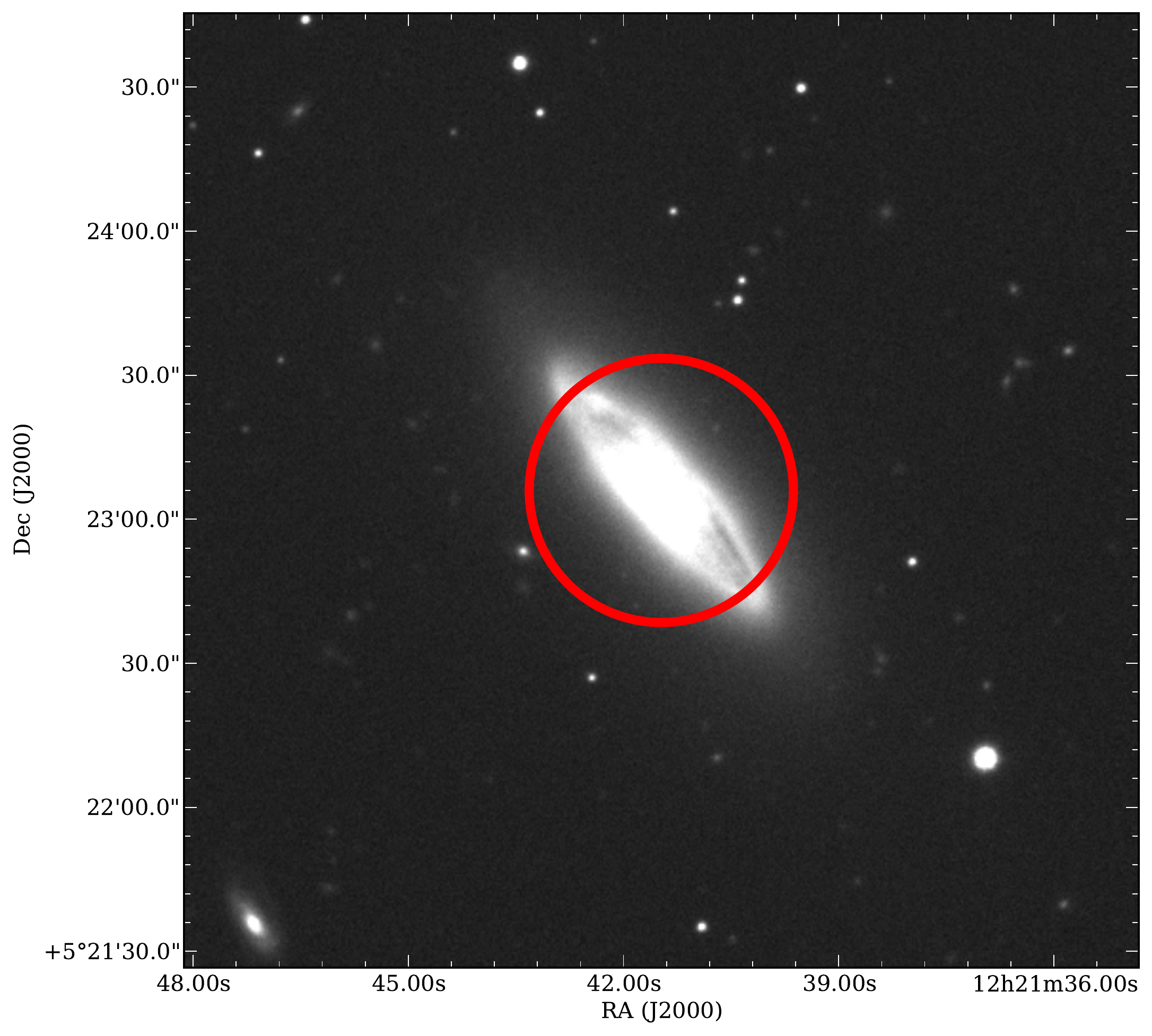}
   \includegraphics[width=0.22\textwidth]{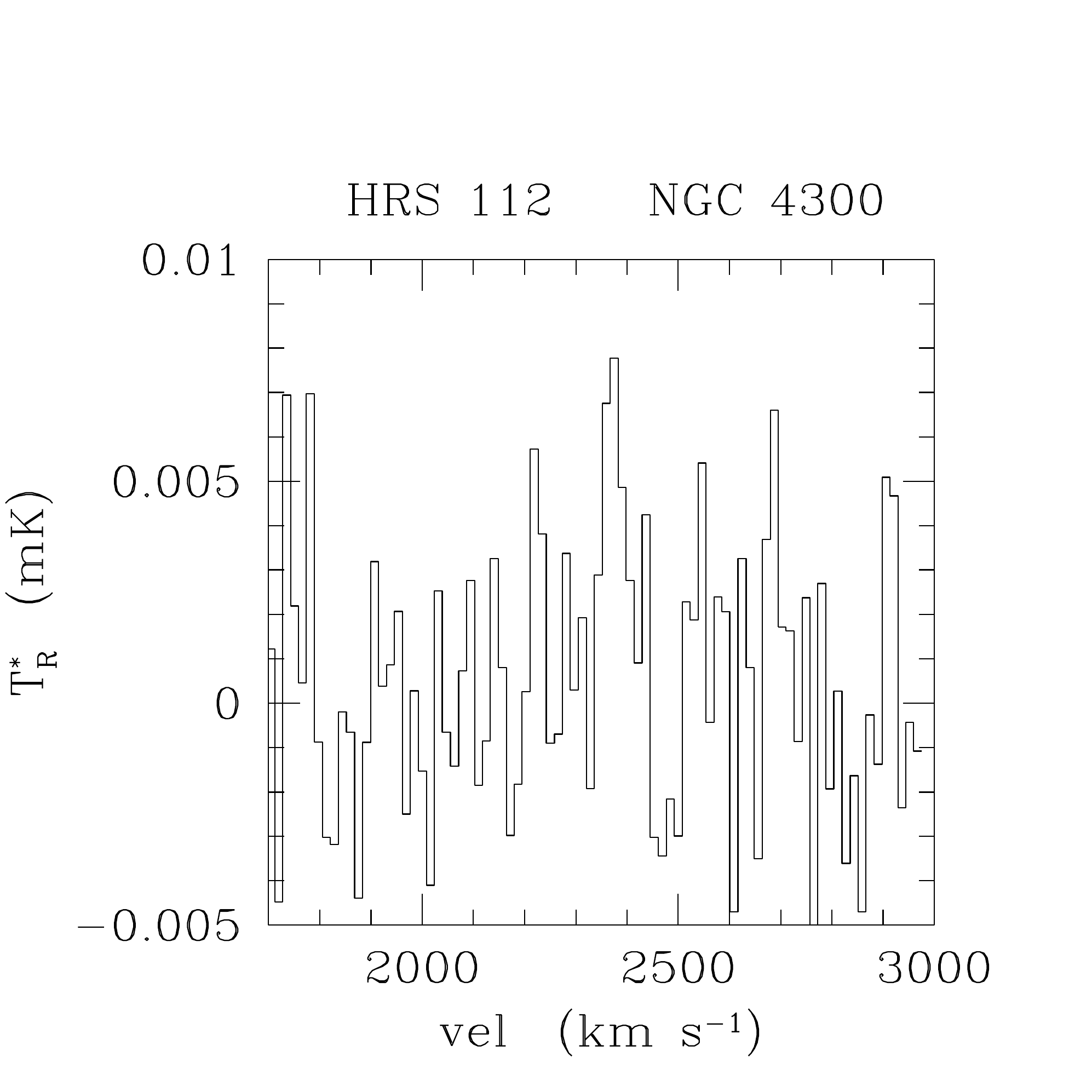}\\
   \includegraphics[width=0.22\textwidth]{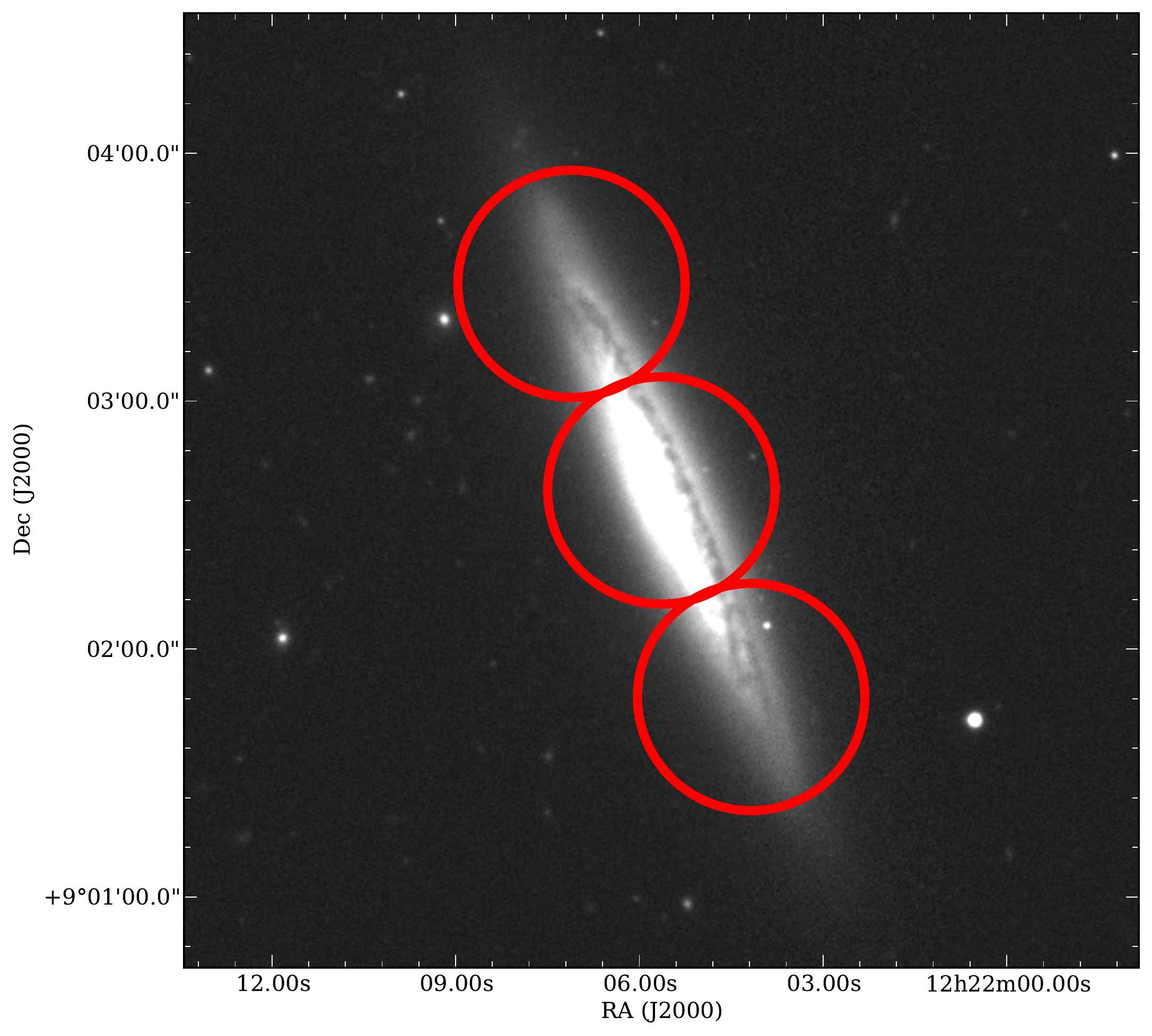}
   \includegraphics[width=0.22\textwidth]{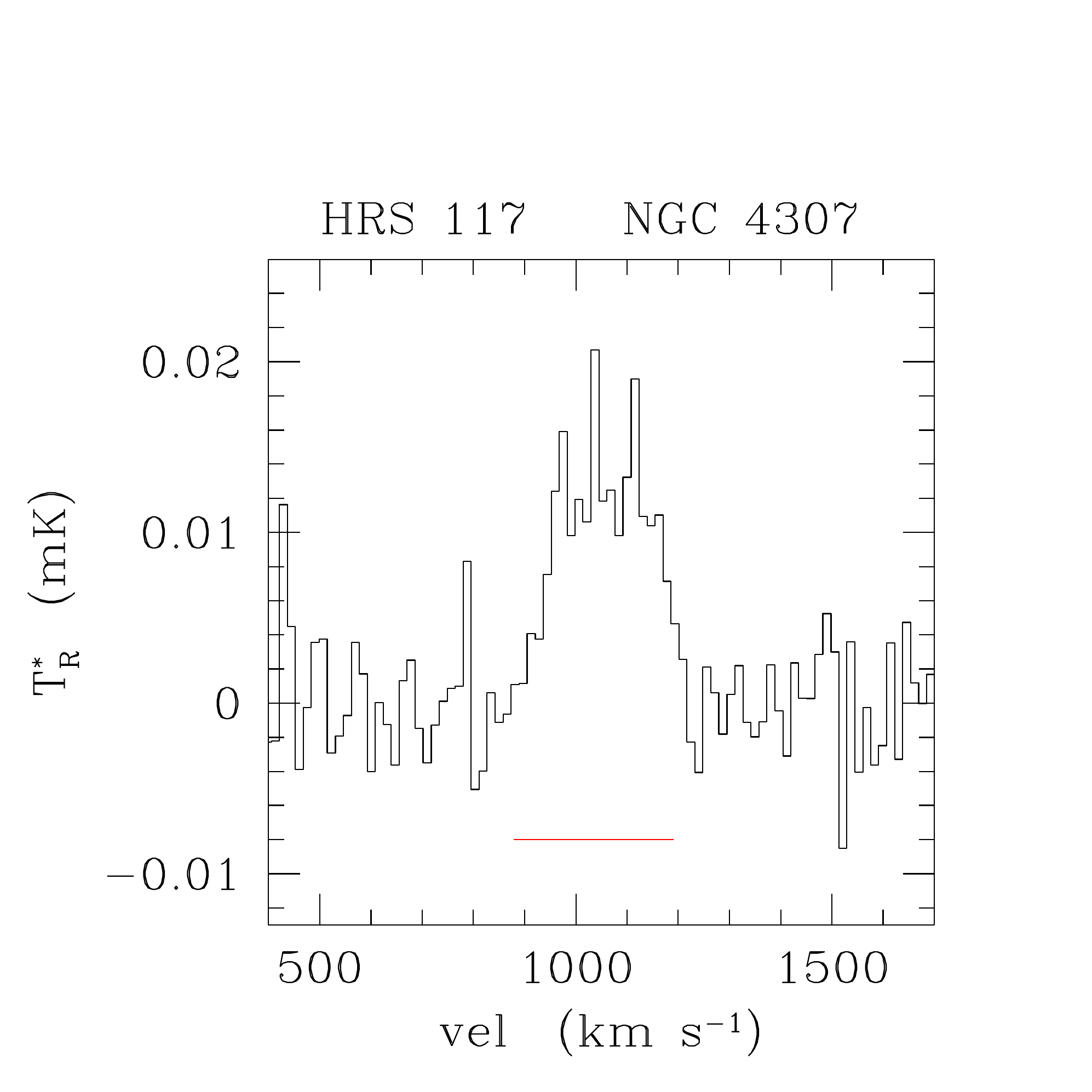}
   \includegraphics[width=0.22\textwidth]{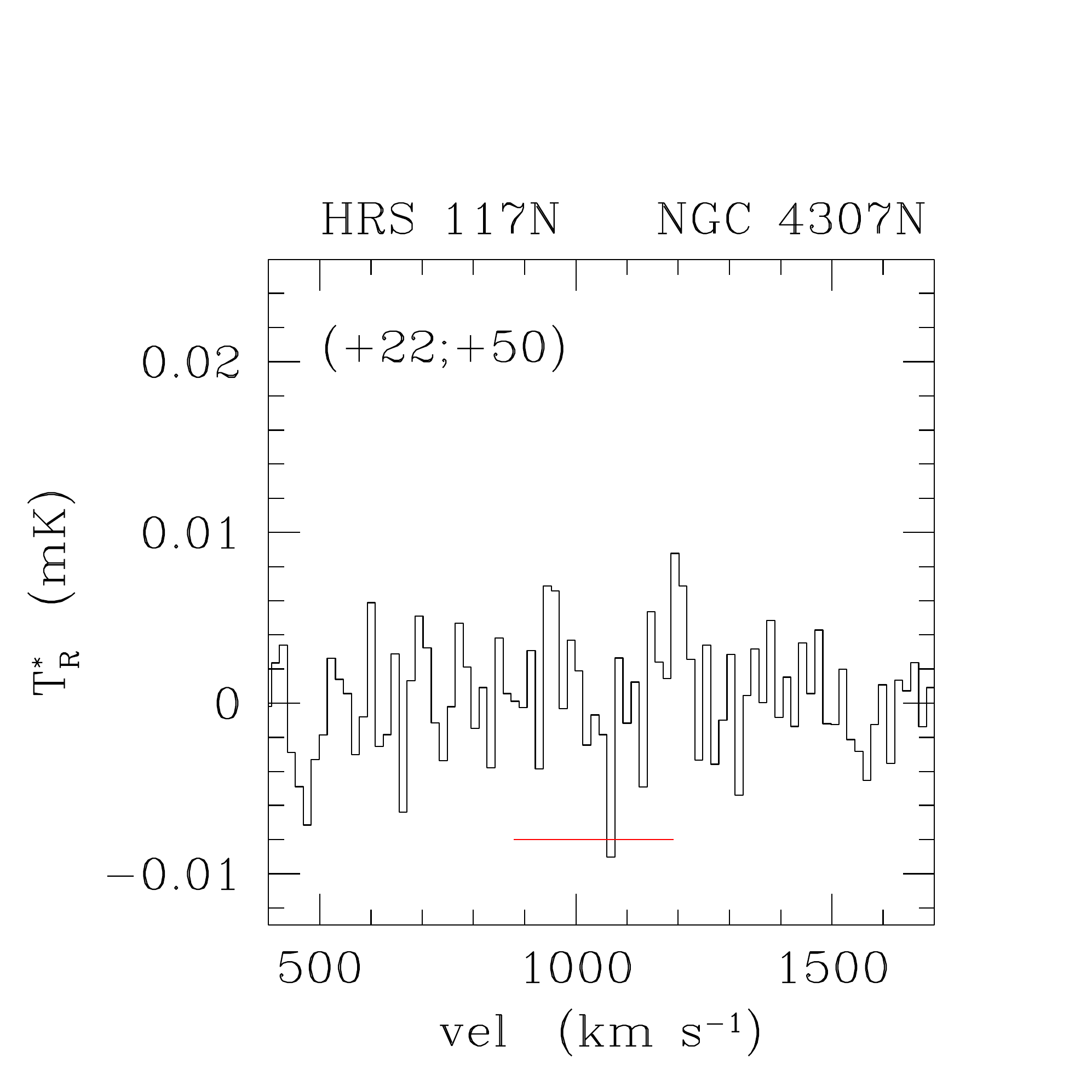}
   \includegraphics[width=0.22\textwidth]{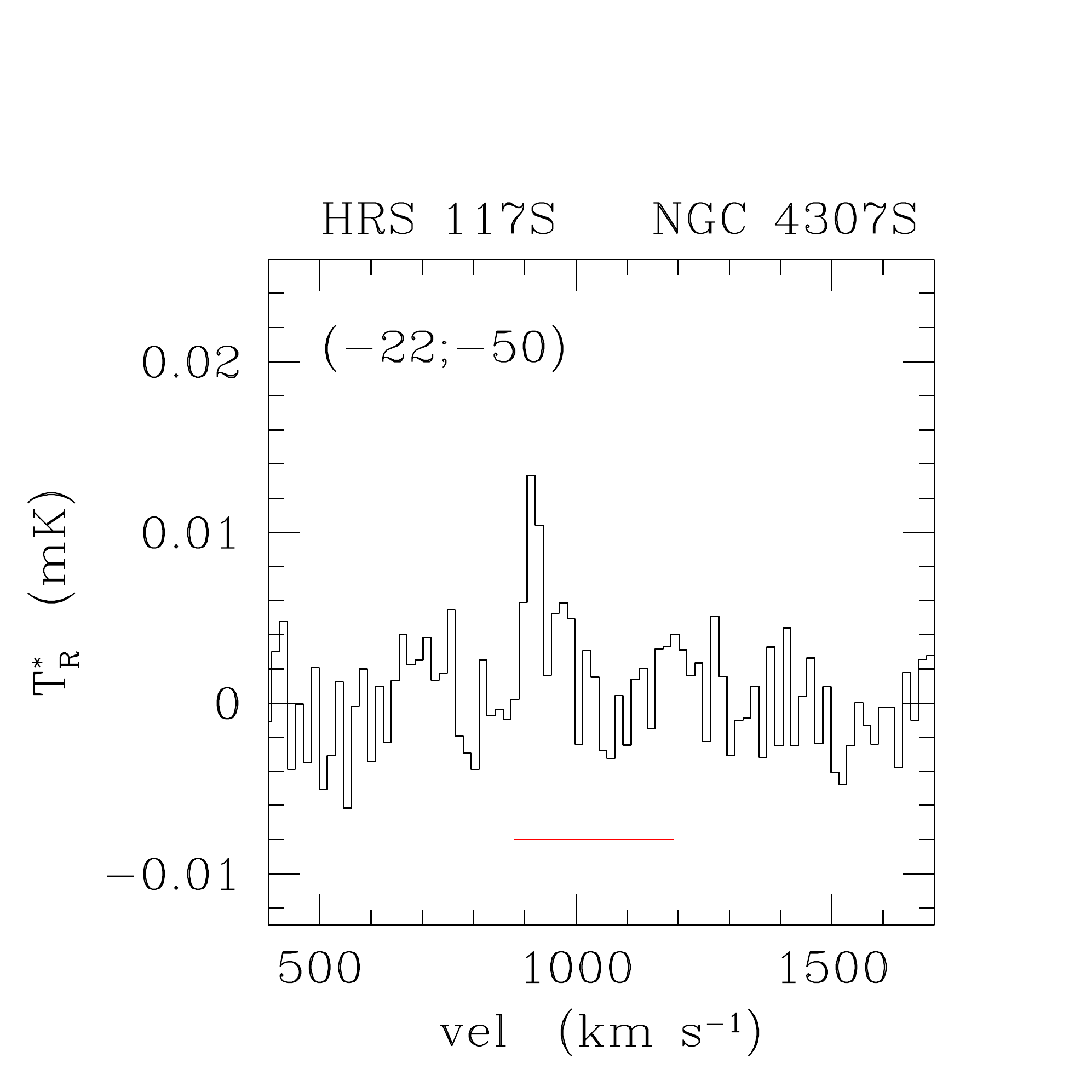}\\
   \includegraphics[width=0.22\textwidth]{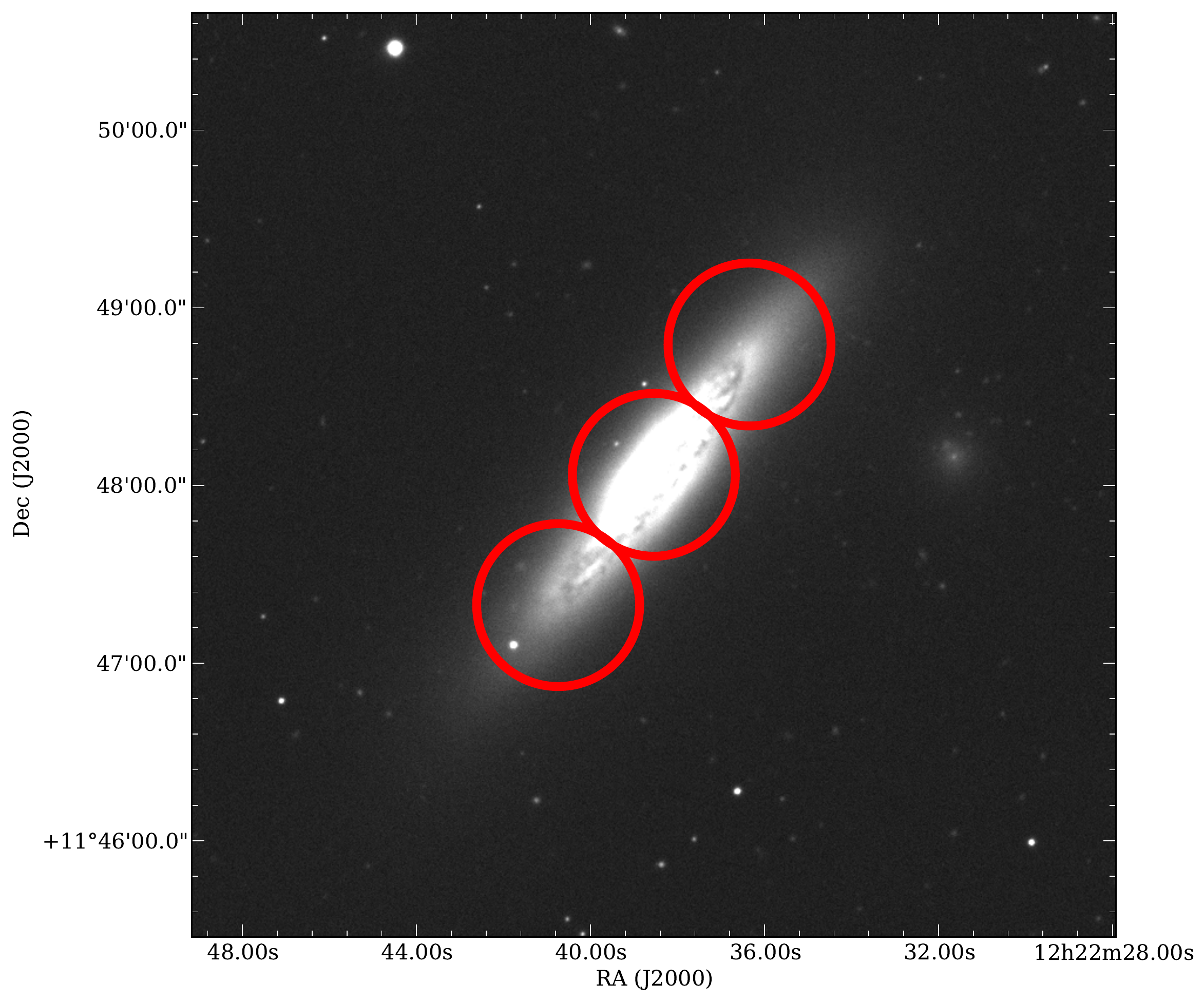}
   \includegraphics[width=0.22\textwidth]{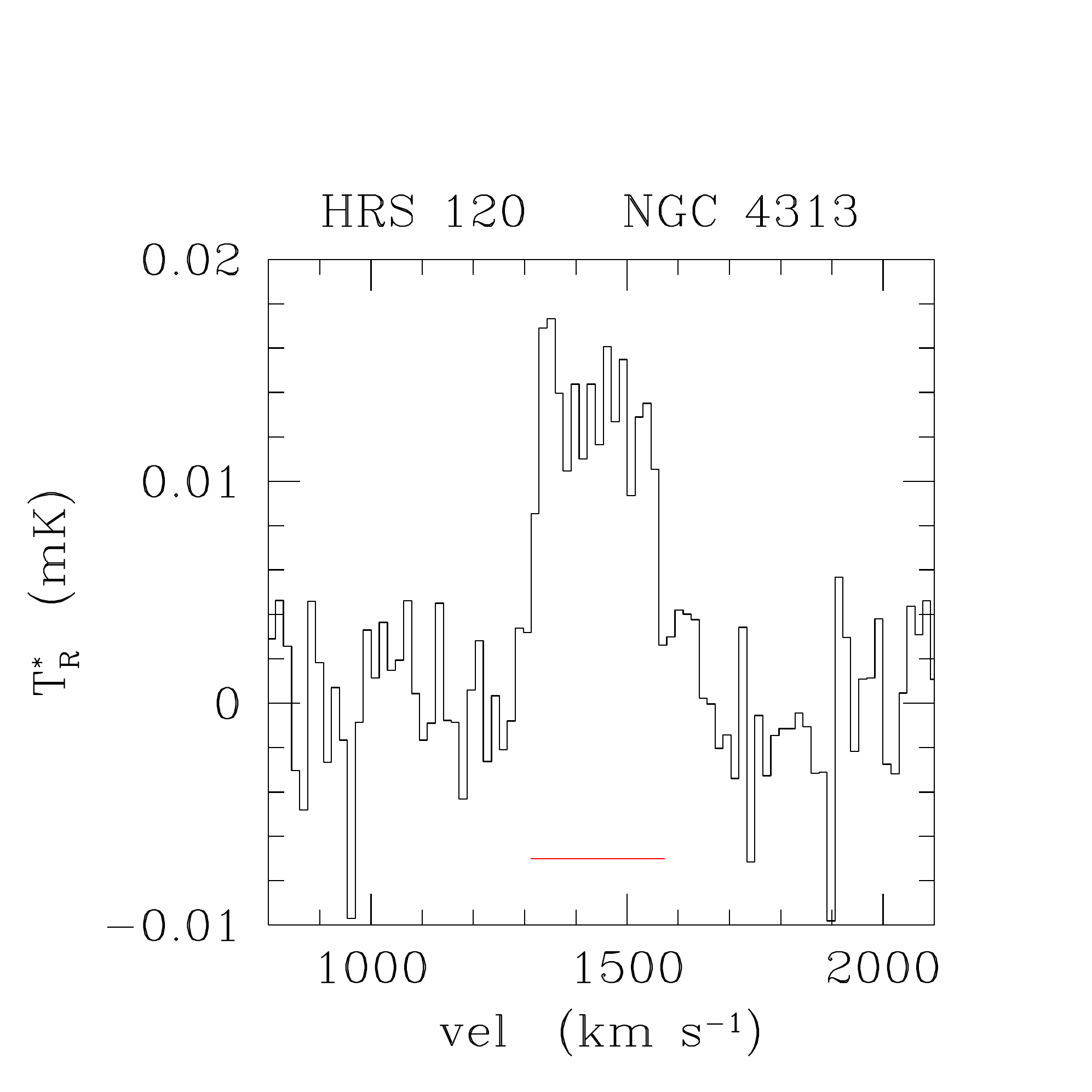}
   \includegraphics[width=0.22\textwidth]{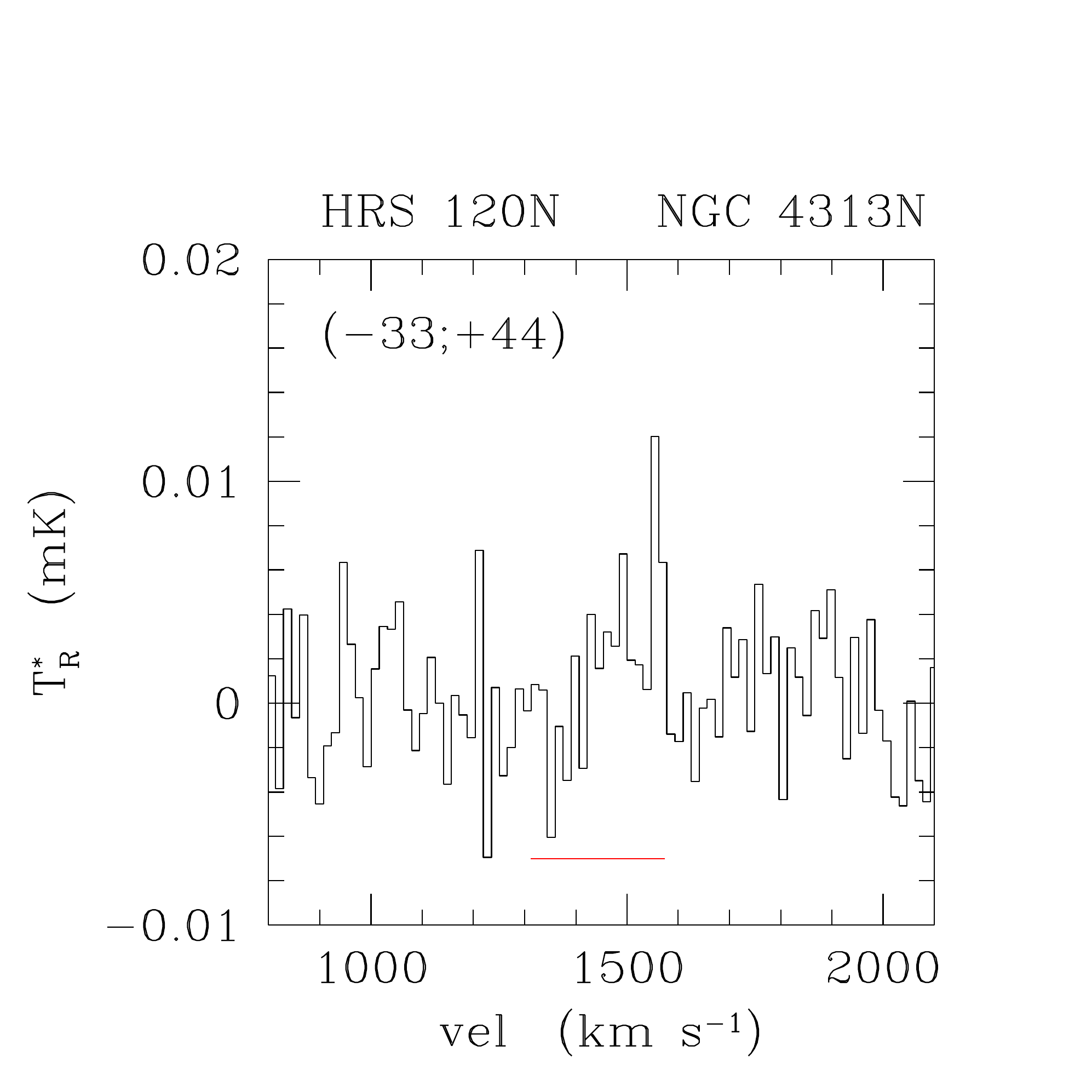}
   \includegraphics[width=0.22\textwidth]{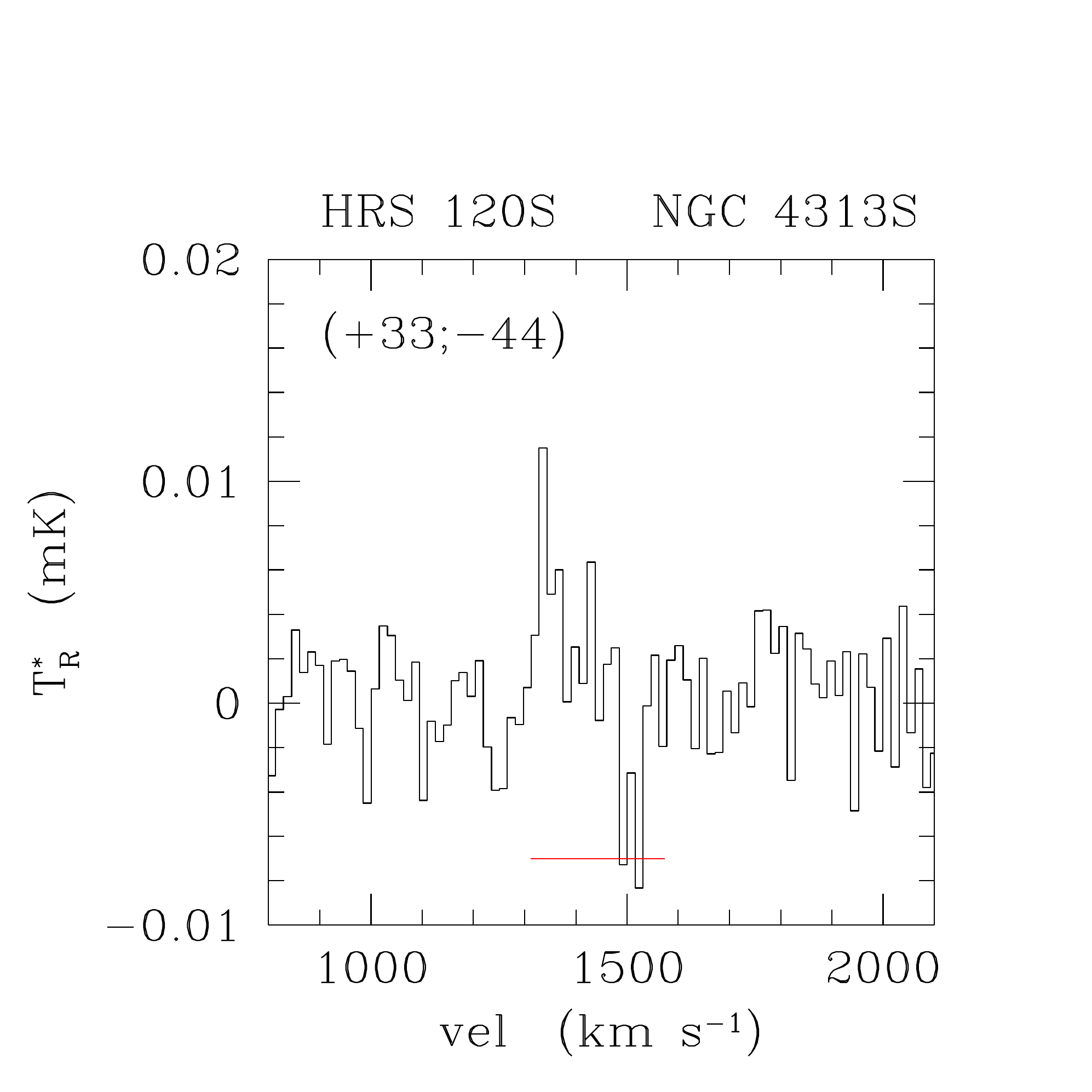}\\
   \includegraphics[width=0.22\textwidth]{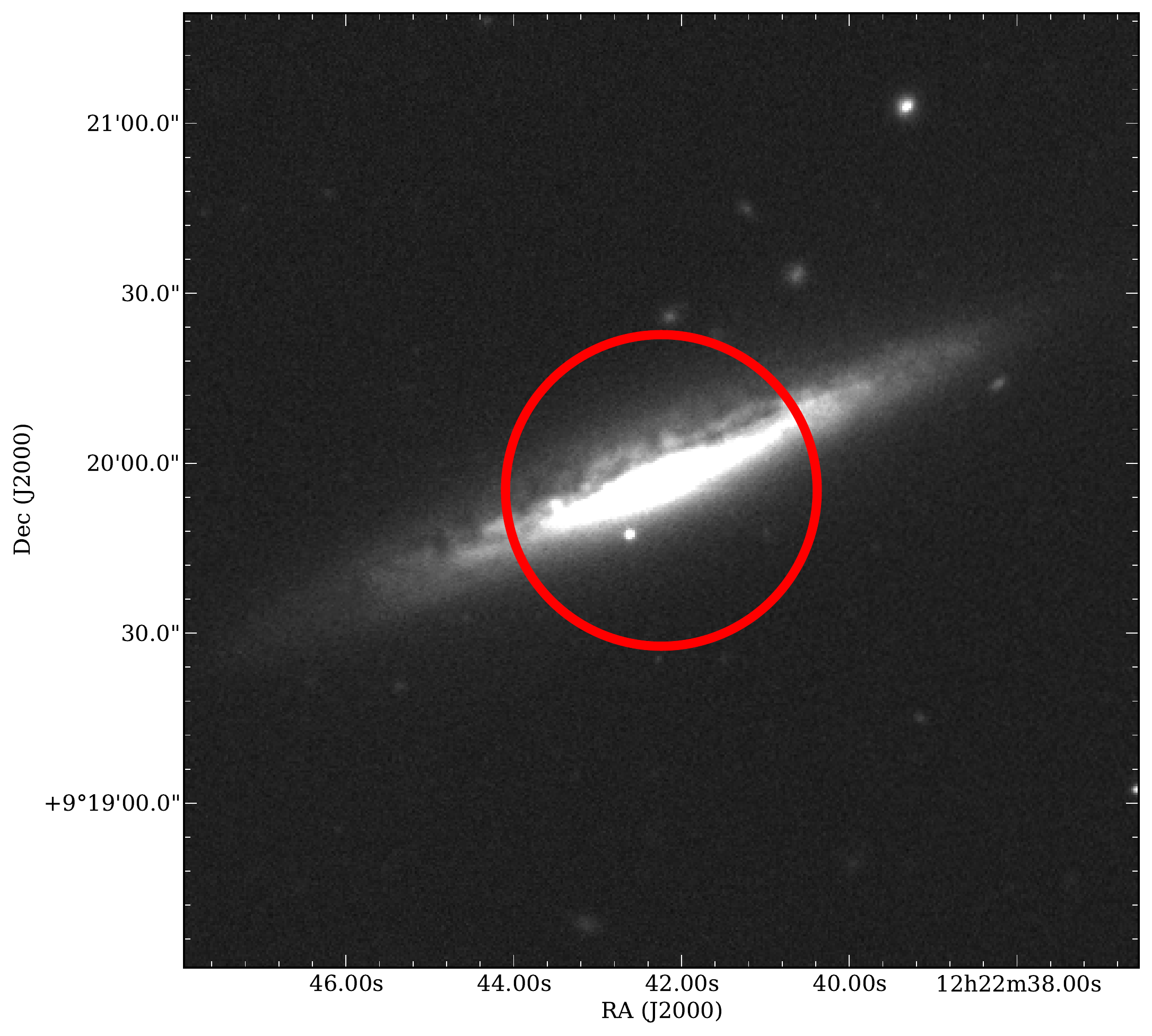}
   \includegraphics[width=0.22\textwidth]{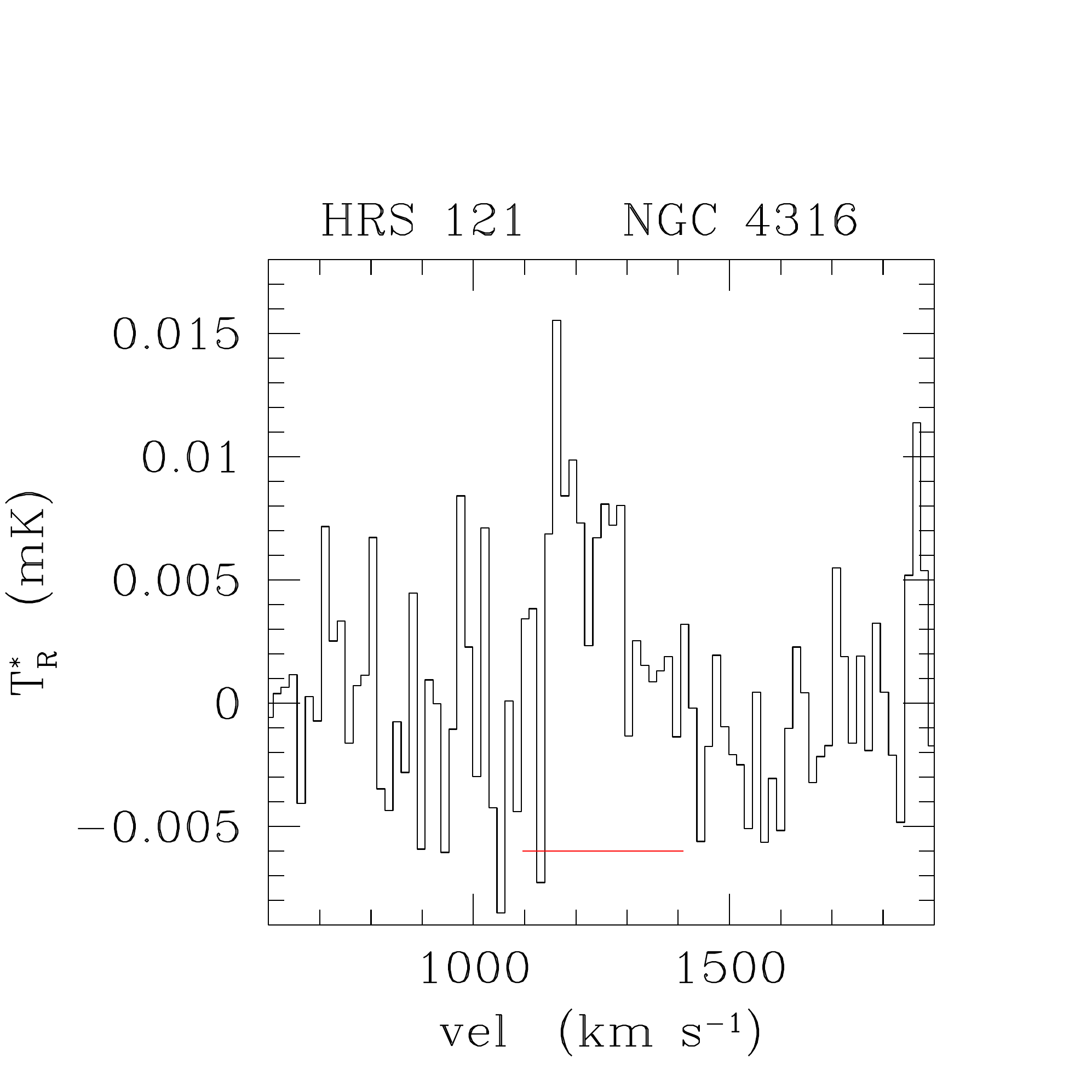}\\
   \caption{Continued.}
   \label{spettri}%
   \end{figure*}
   \clearpage

   \addtocounter{figure}{-1}
   \begin{figure*}
   \centering
   \includegraphics[width=0.22\textwidth]{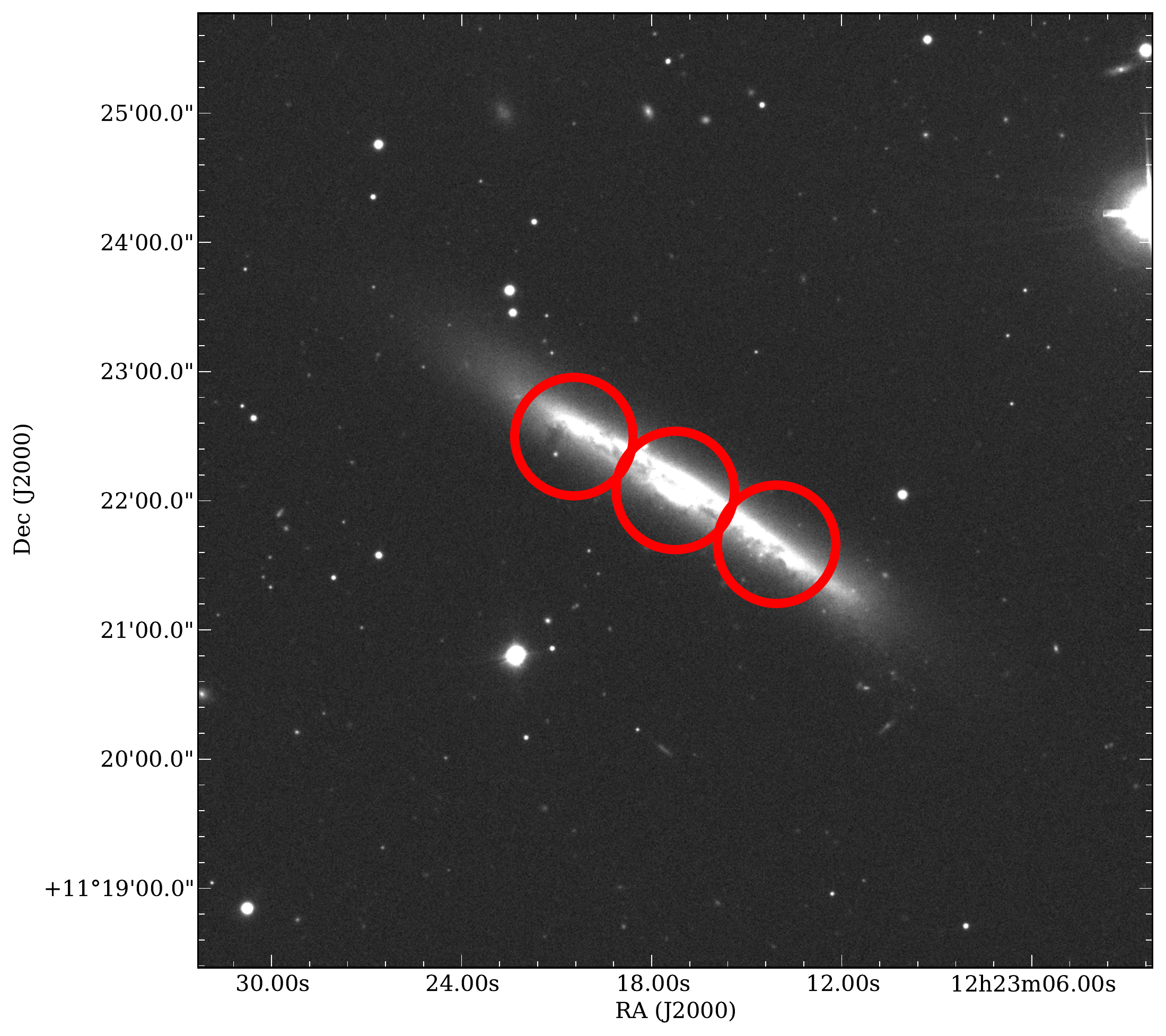}
   \includegraphics[width=0.22\textwidth]{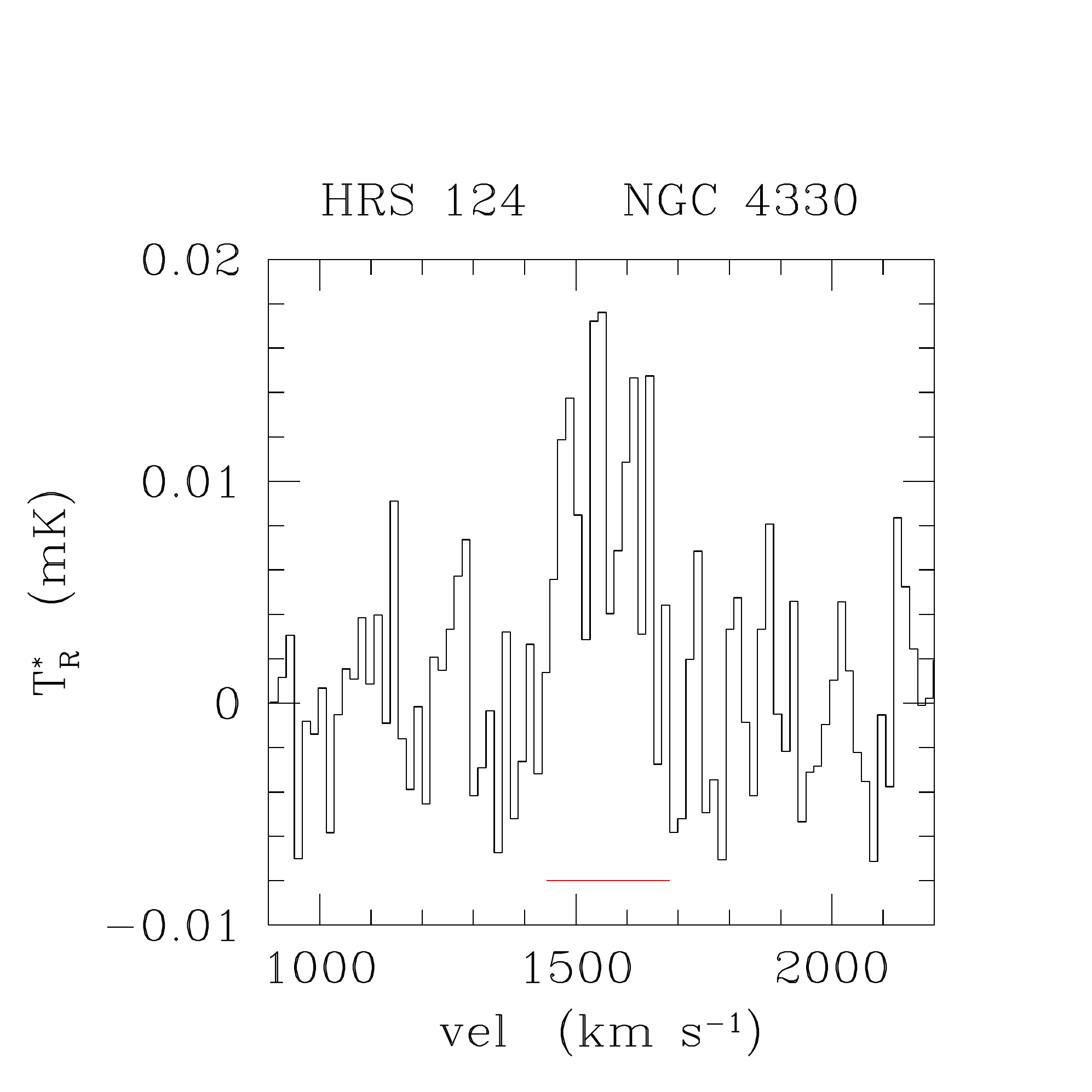}
   \includegraphics[width=0.22\textwidth]{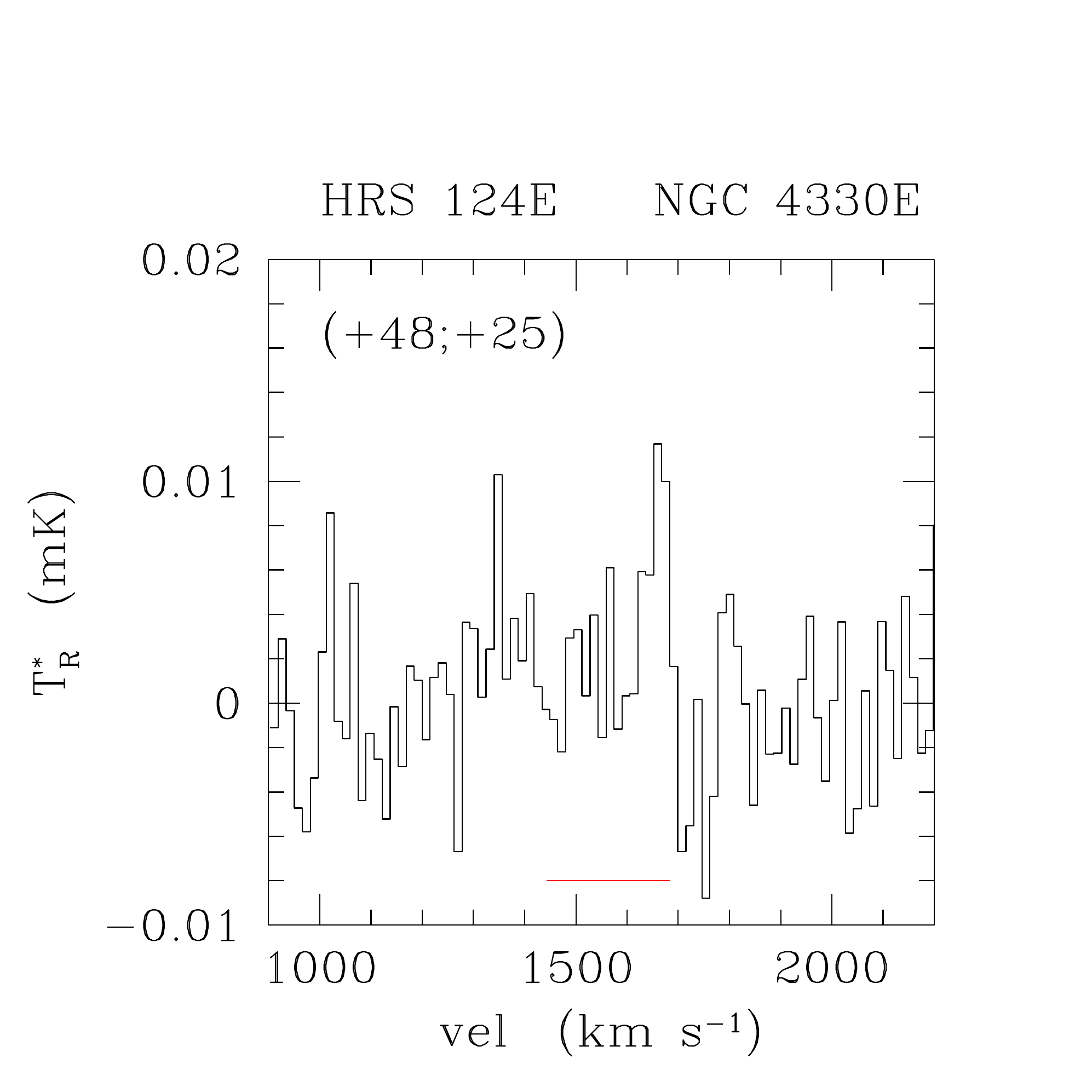}
   \includegraphics[width=0.22\textwidth]{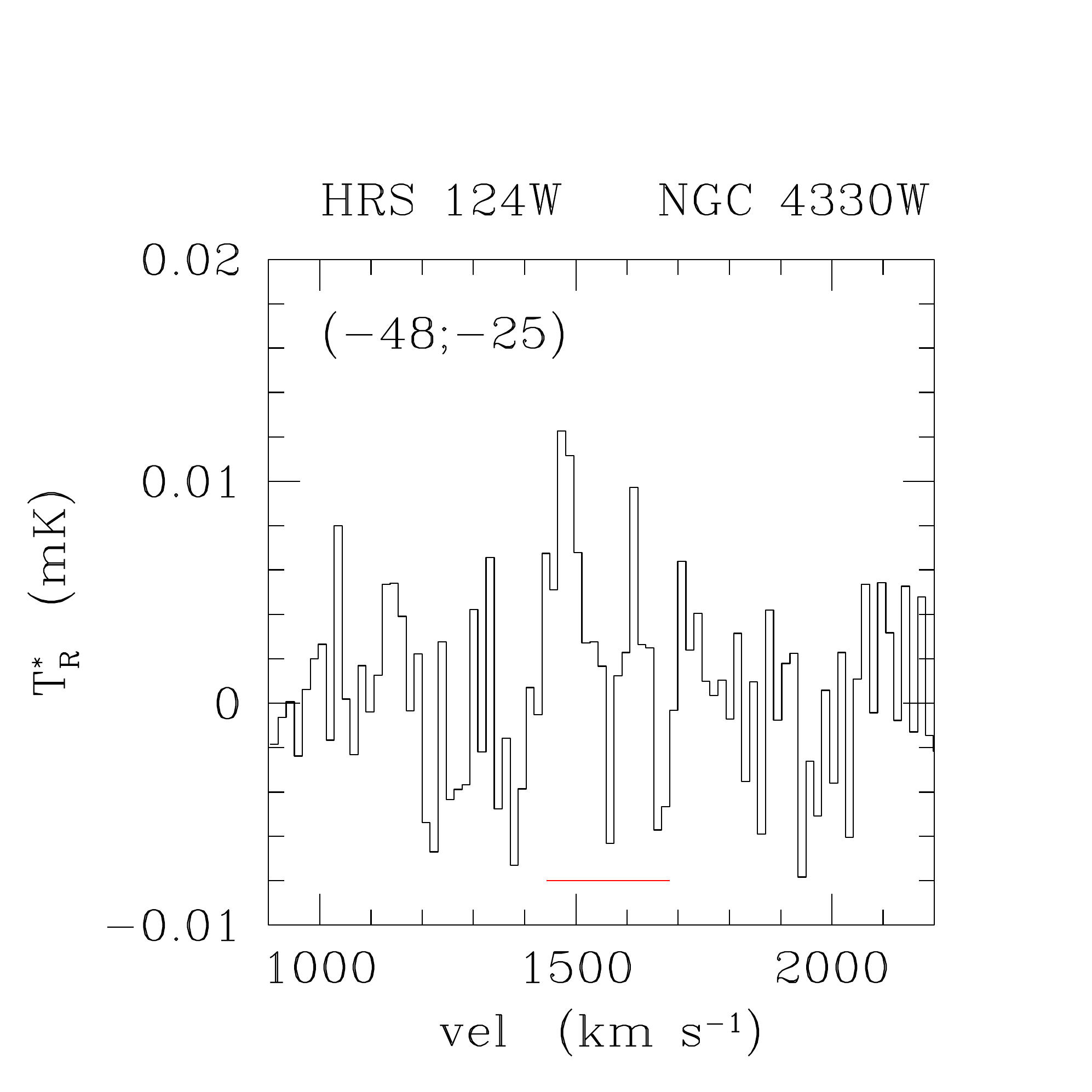}\\
   \includegraphics[width=0.22\textwidth]{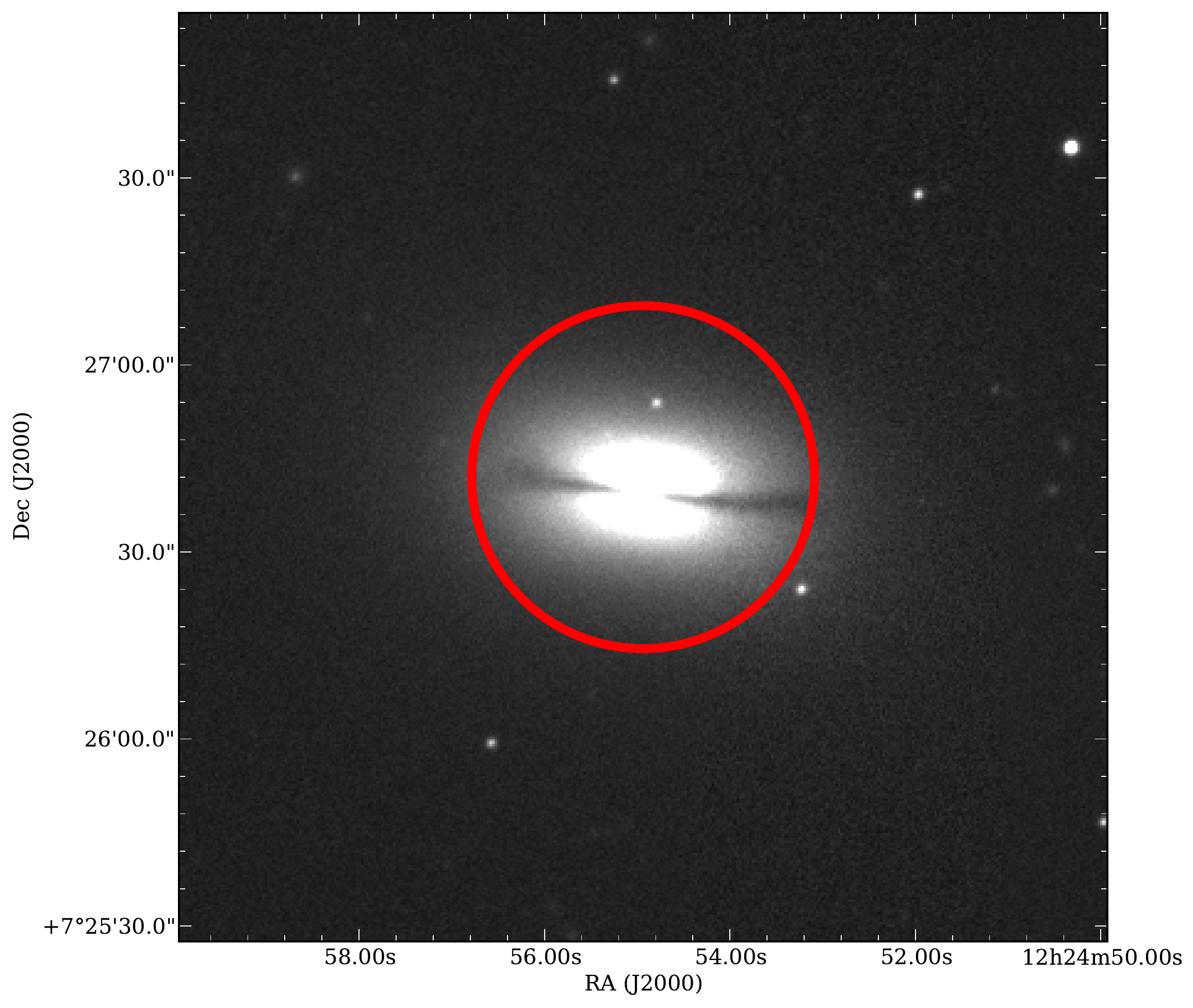}
   \includegraphics[width=0.22\textwidth]{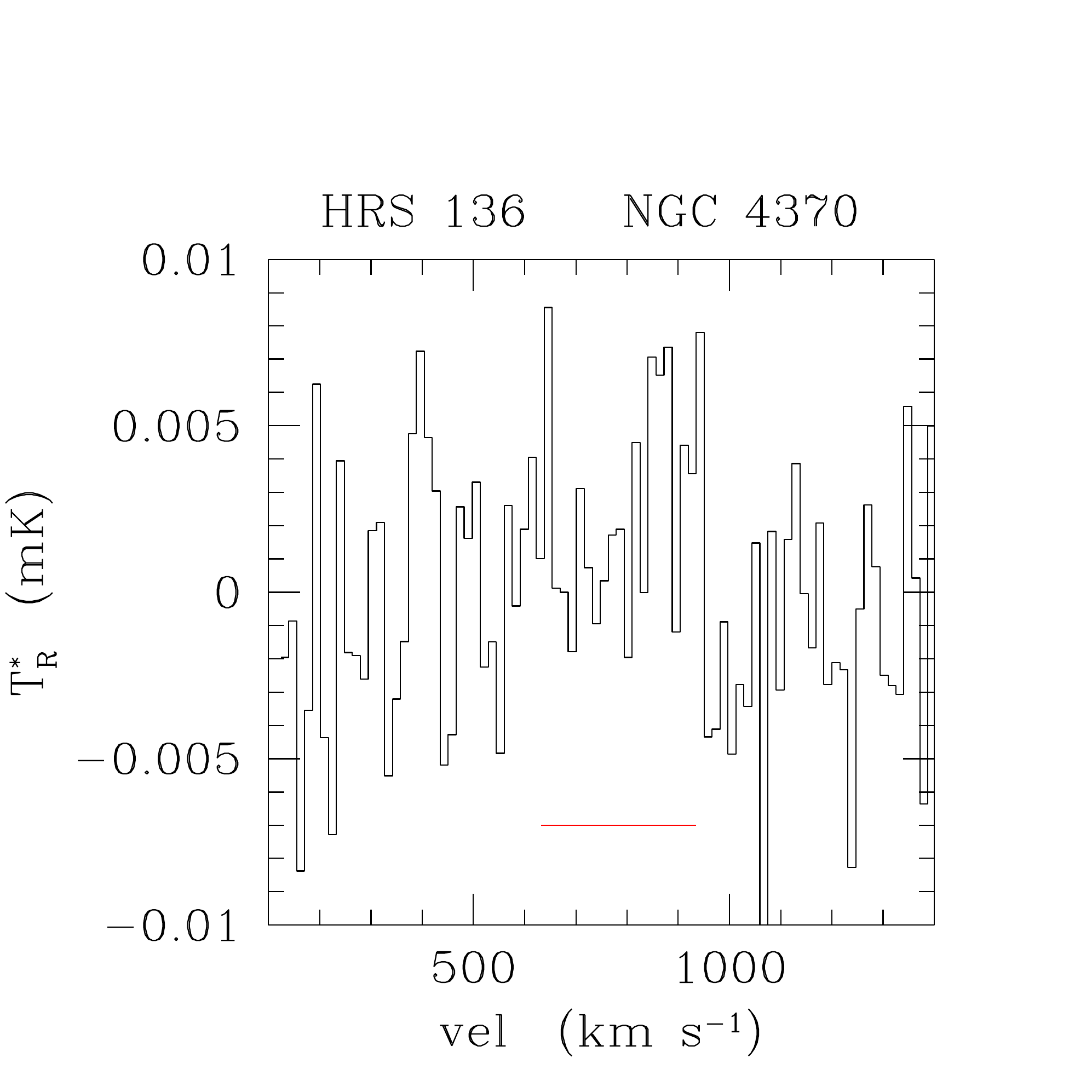}\\
   \includegraphics[width=0.22\textwidth]{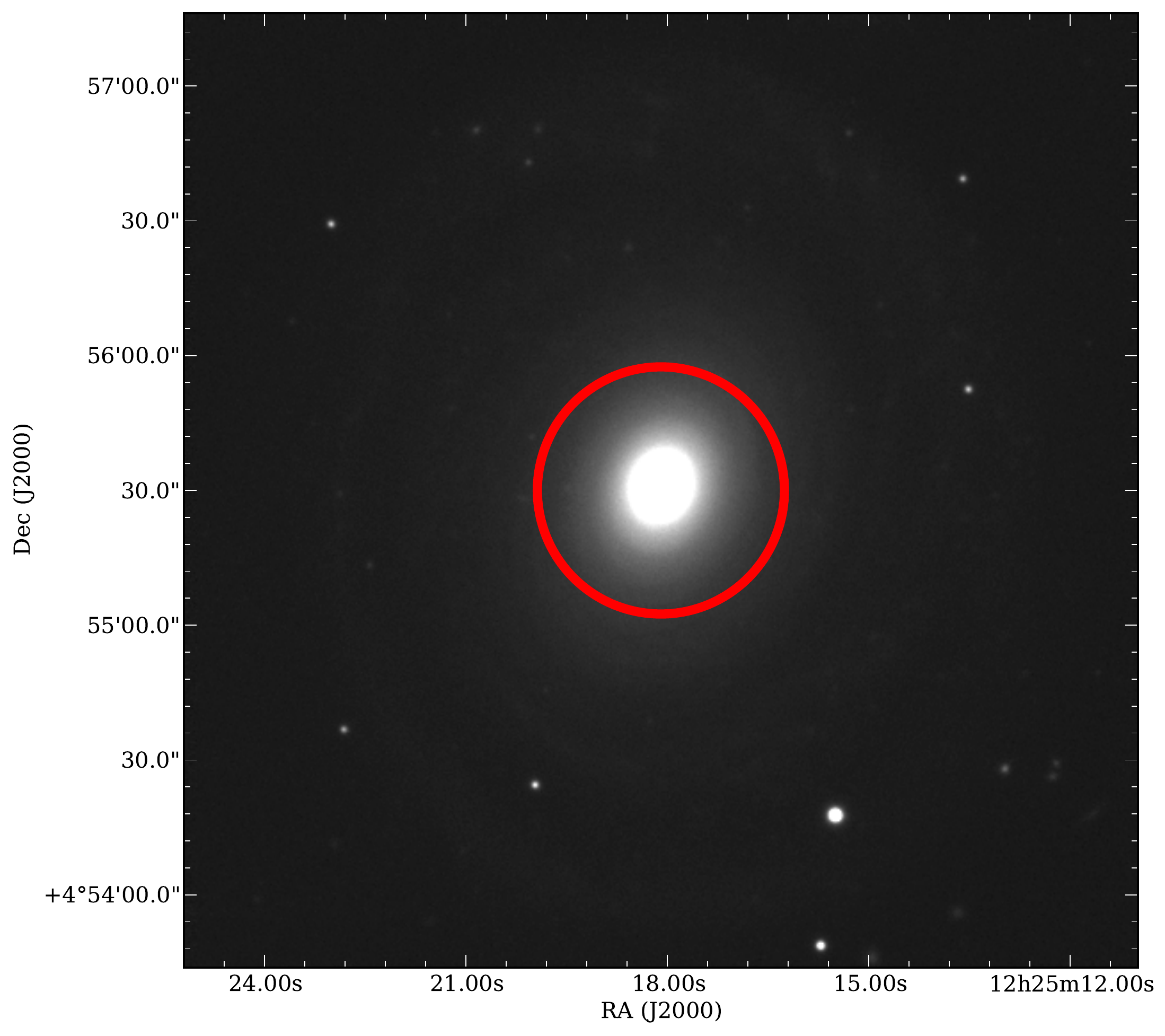}
   \includegraphics[width=0.22\textwidth]{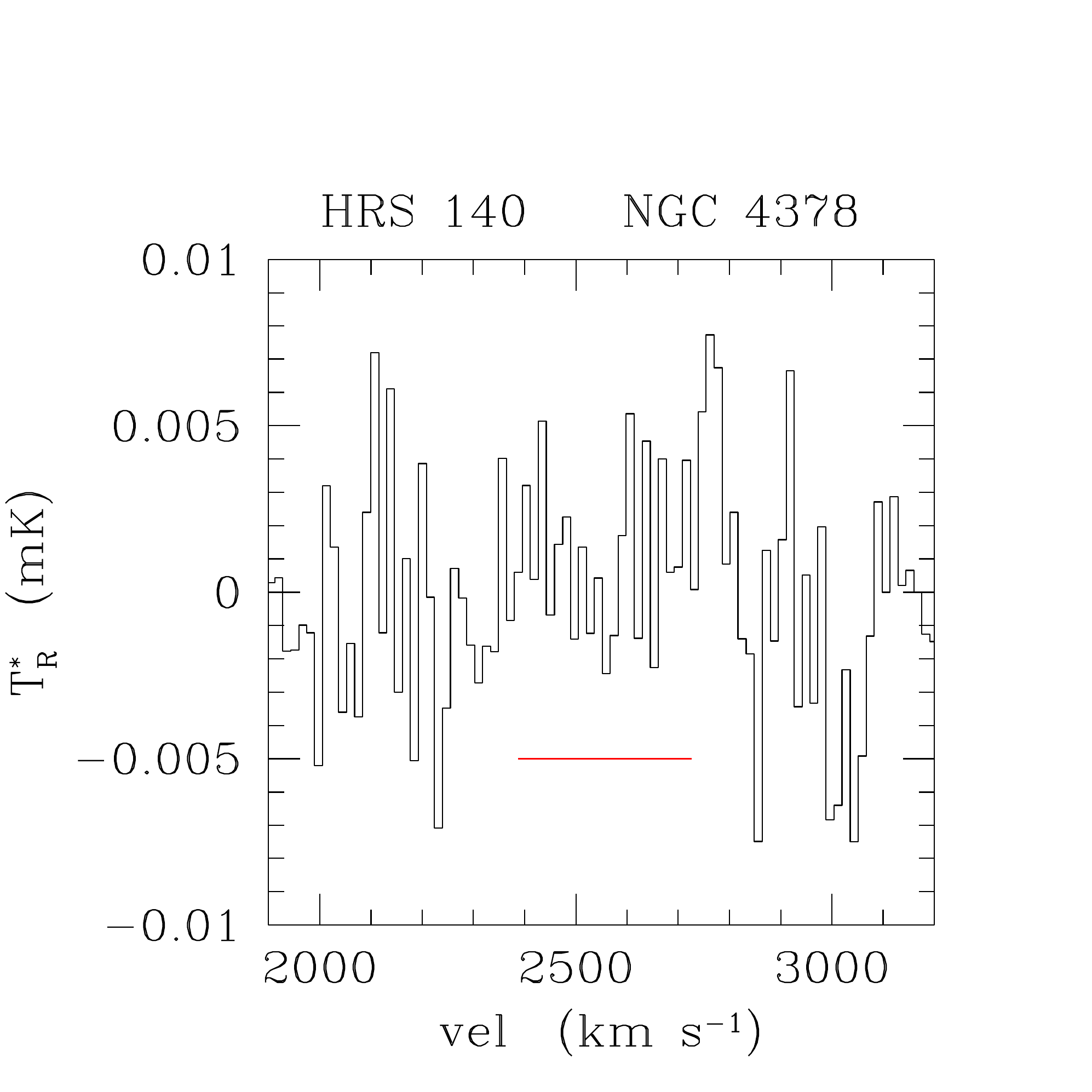}\\
   \includegraphics[width=0.22\textwidth]{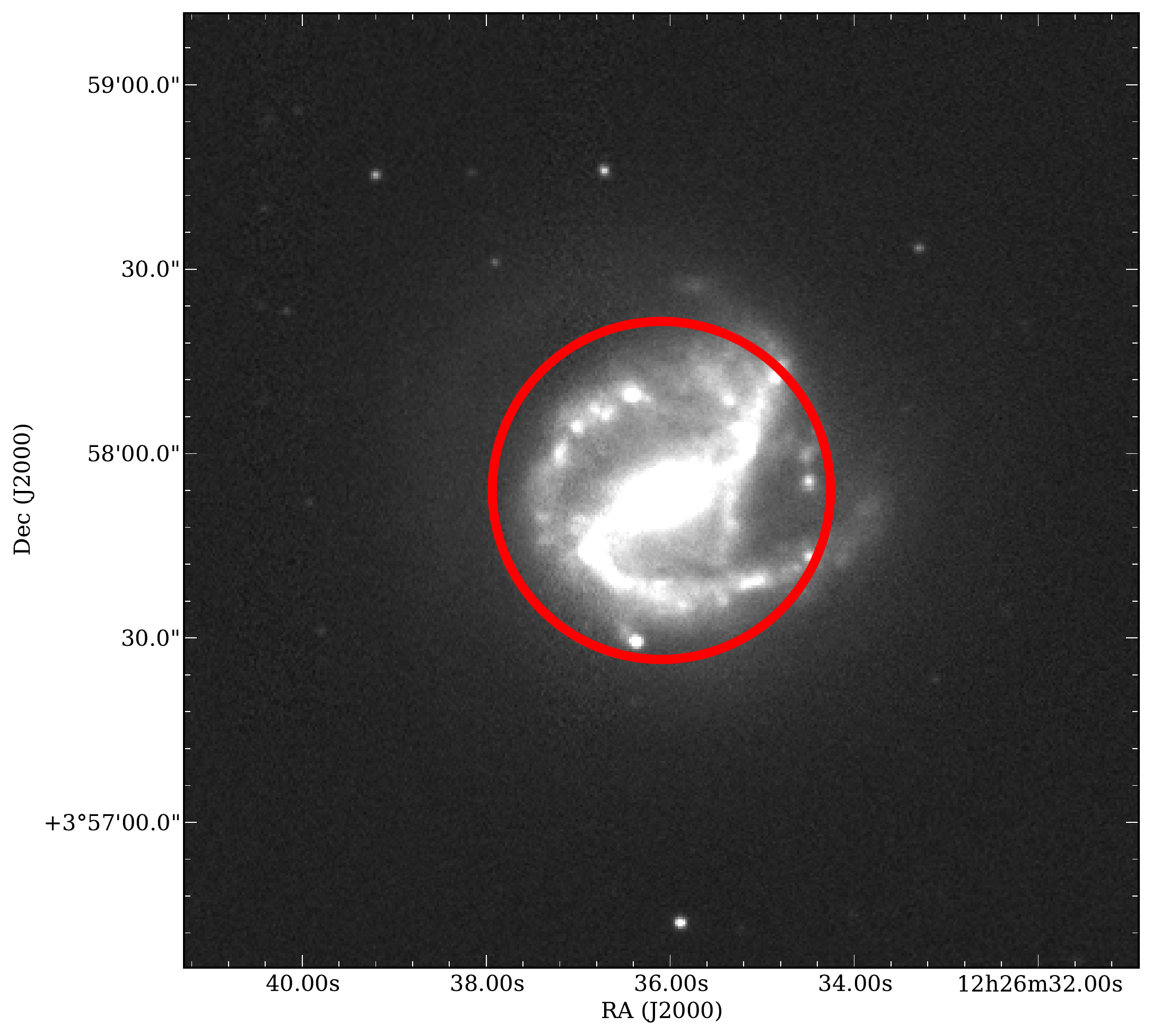}
   \includegraphics[width=0.22\textwidth]{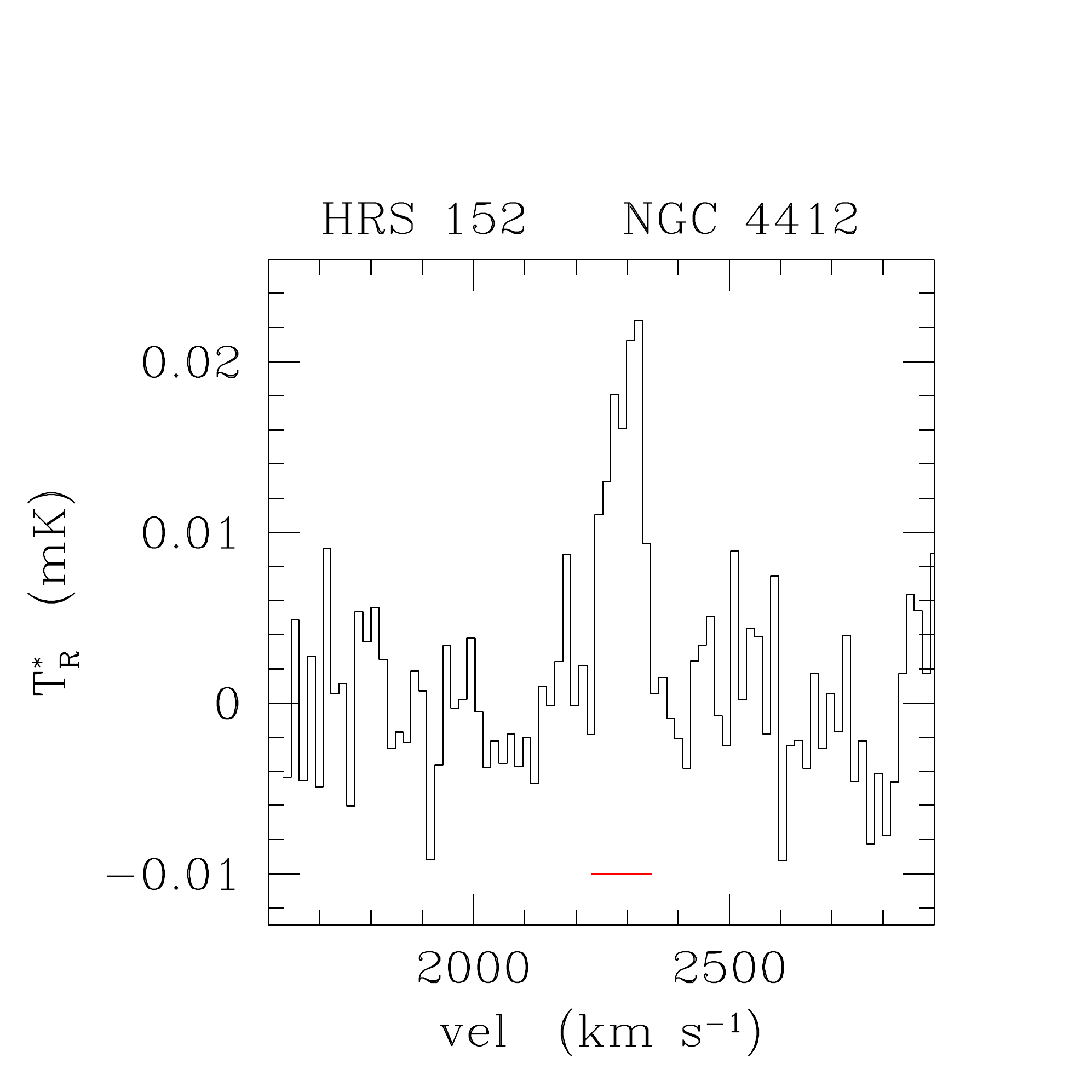}\\
   \includegraphics[width=0.22\textwidth]{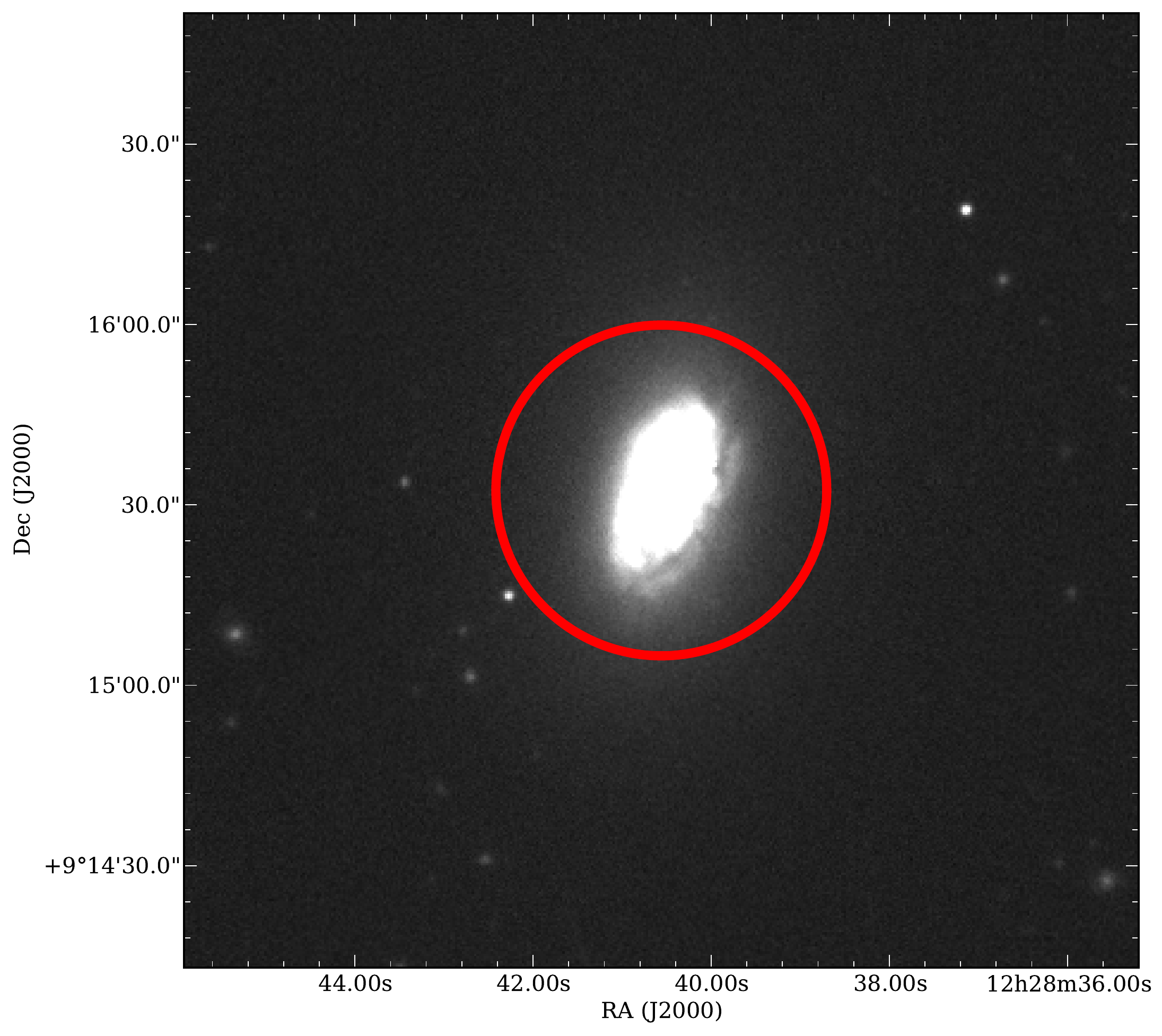}
   \includegraphics[width=0.22\textwidth]{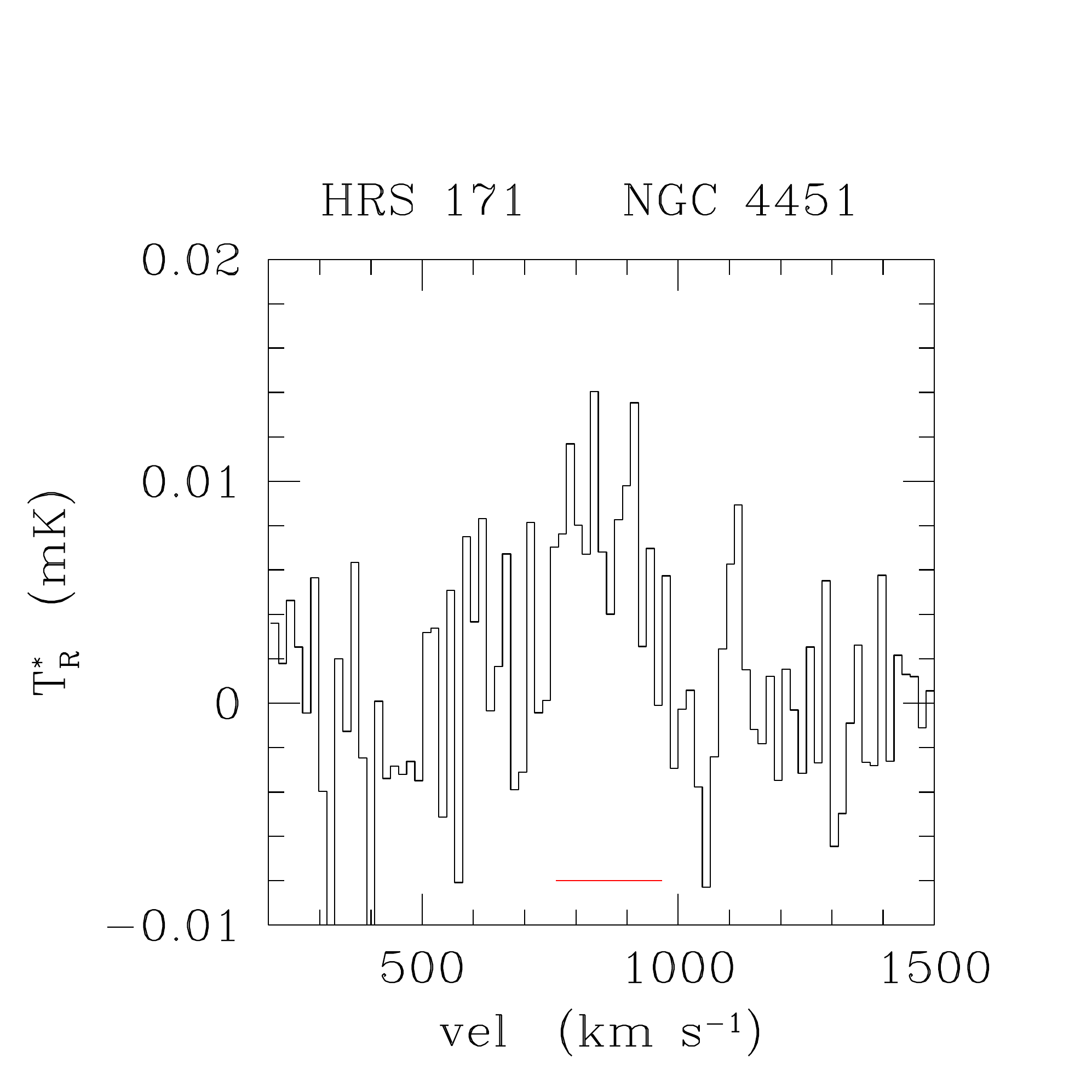}\\
   \caption{Continued.}
   \label{spettri}%
   \end{figure*}
   \clearpage

   \addtocounter{figure}{-1}
   \begin{figure*}
   \centering
   \includegraphics[width=0.22\textwidth]{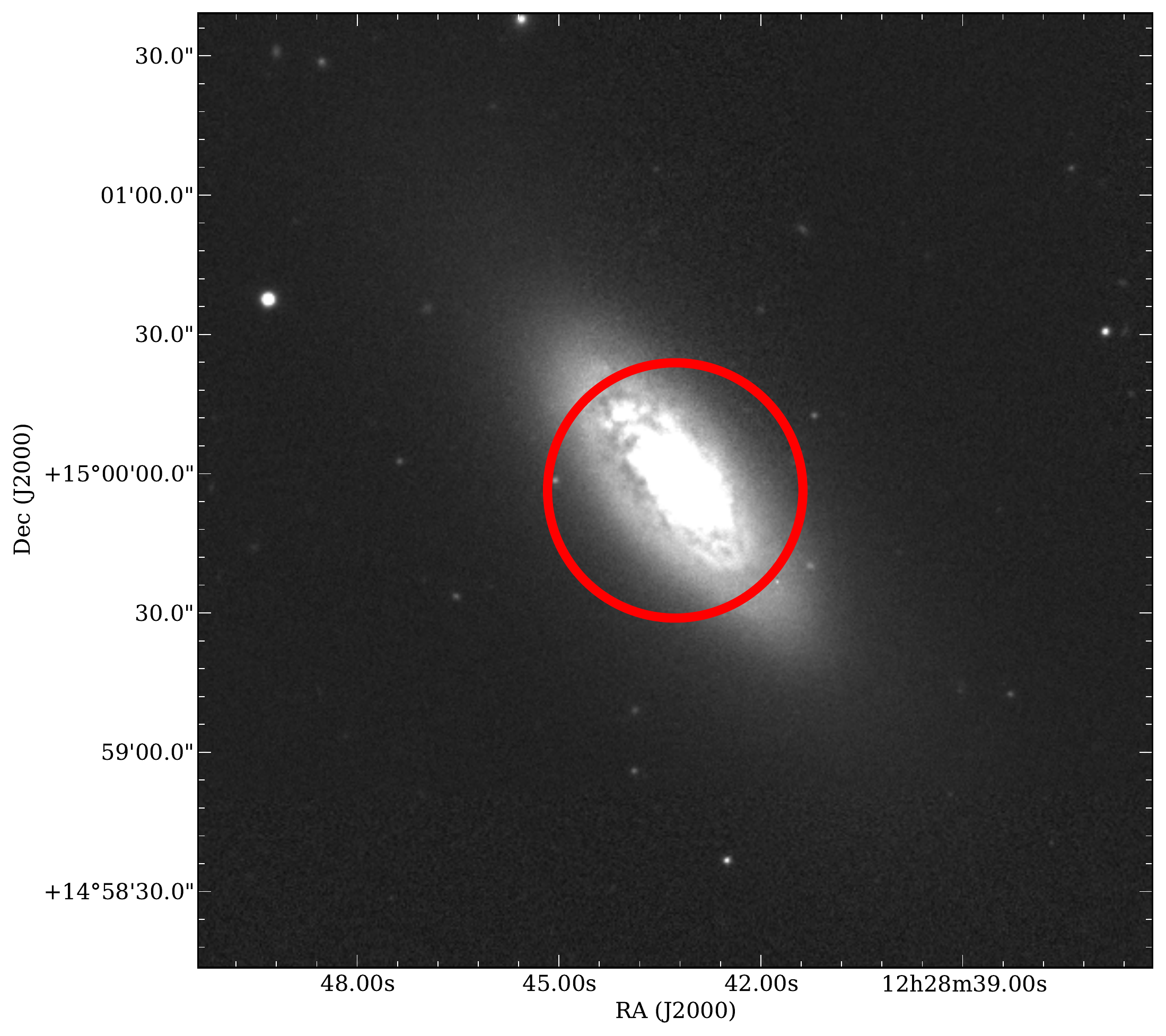}
   \includegraphics[width=0.22\textwidth]{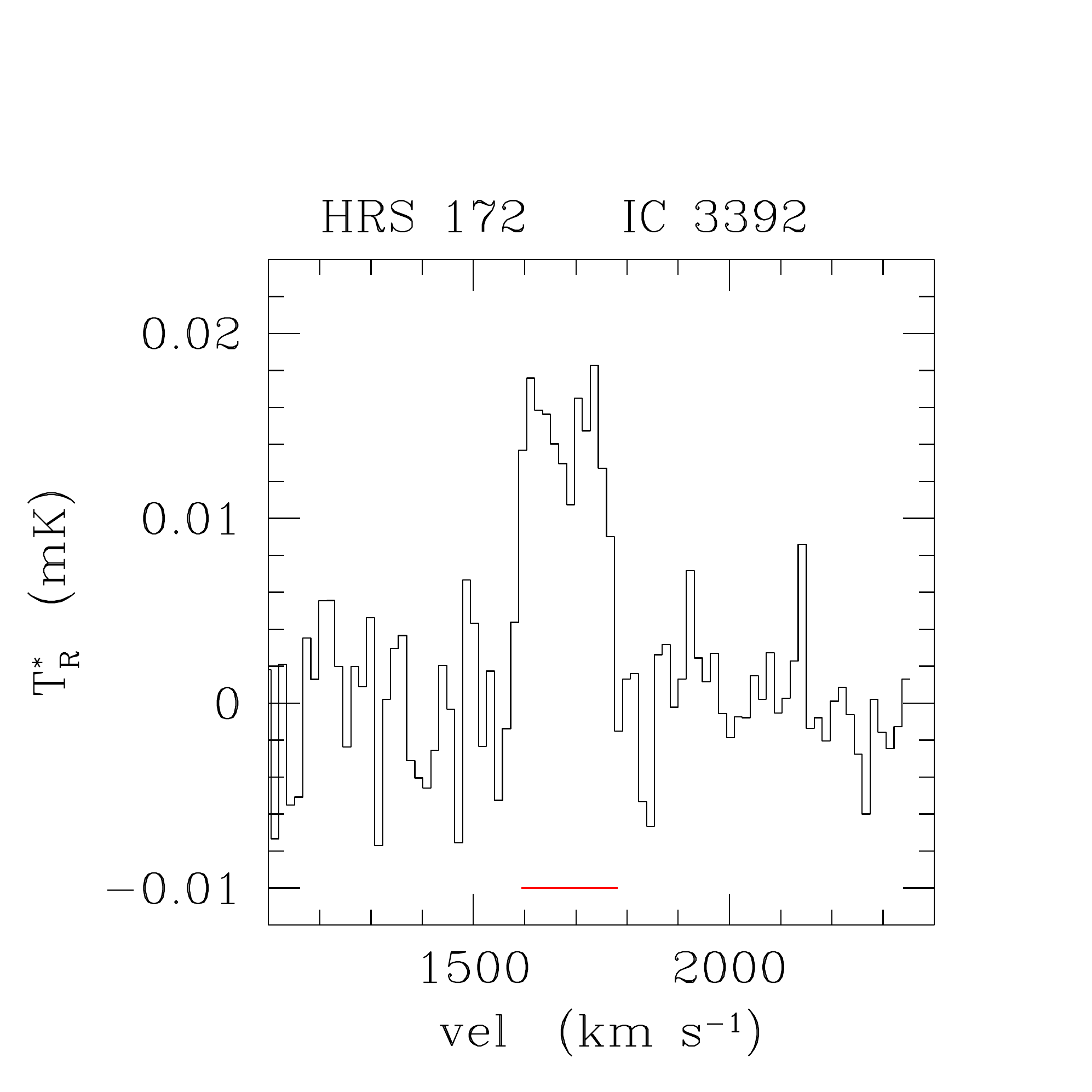}\\
   \includegraphics[width=0.22\textwidth]{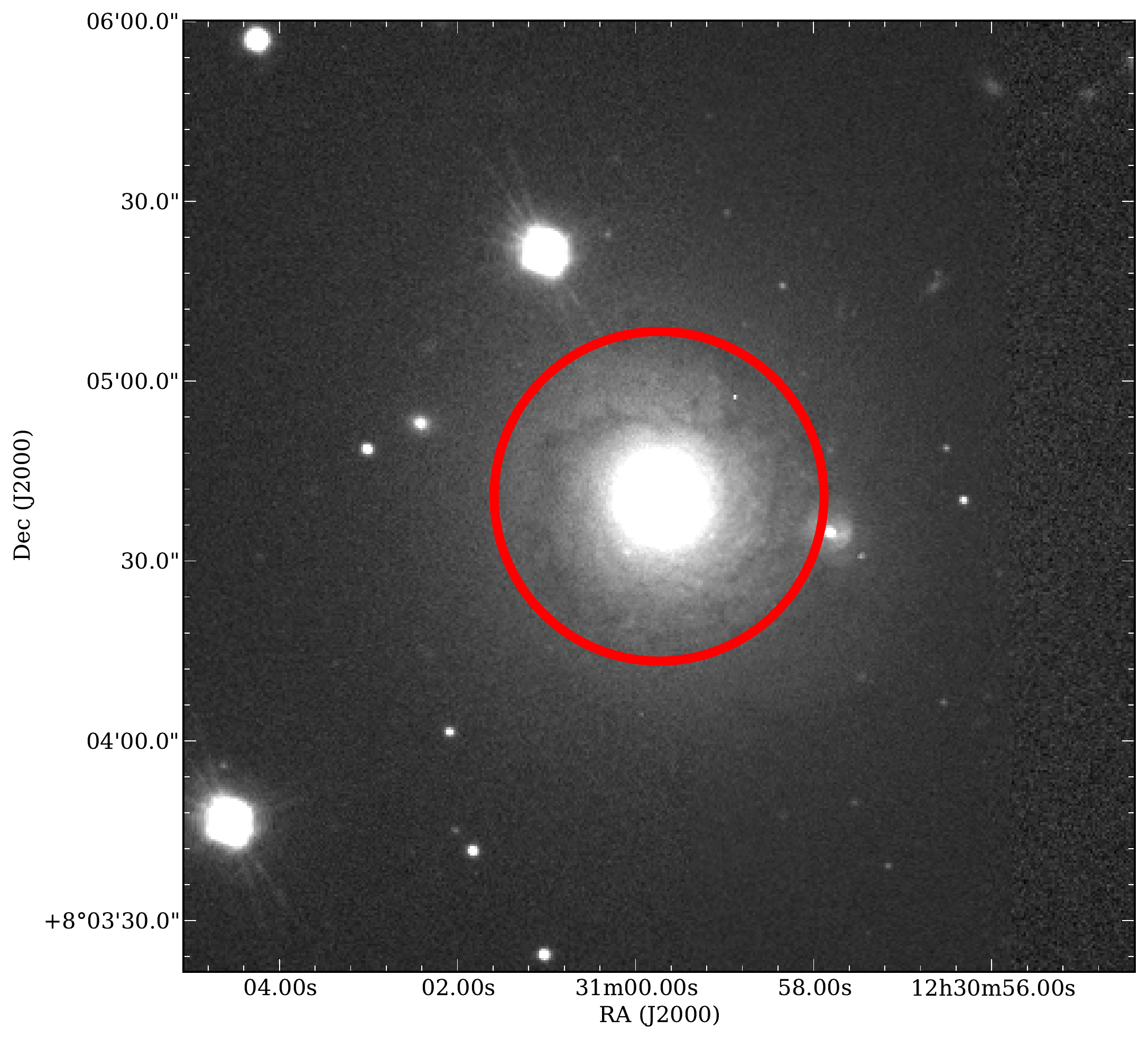}
   \includegraphics[width=0.22\textwidth]{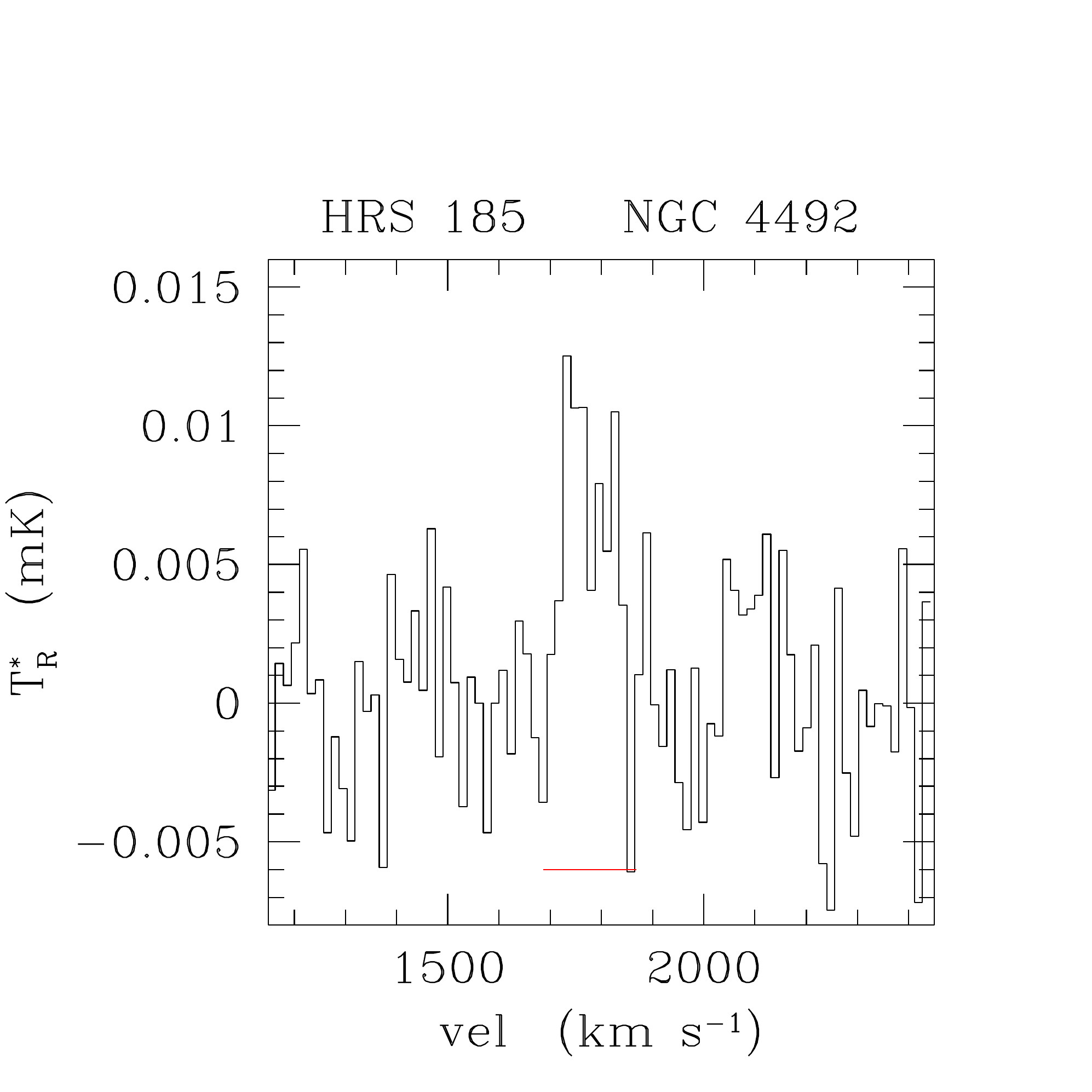}\\
   \includegraphics[width=0.22\textwidth]{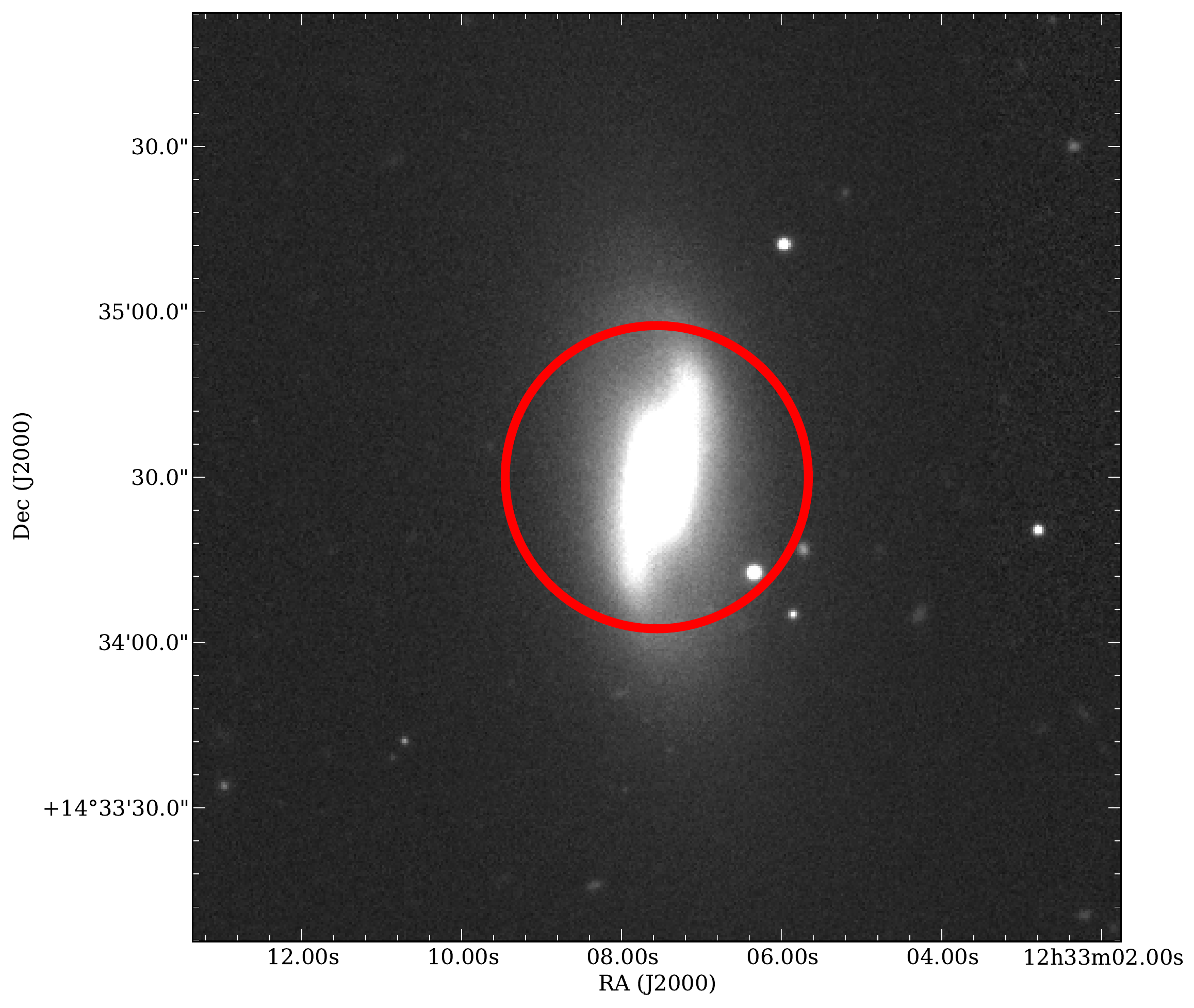}
   \includegraphics[width=0.22\textwidth]{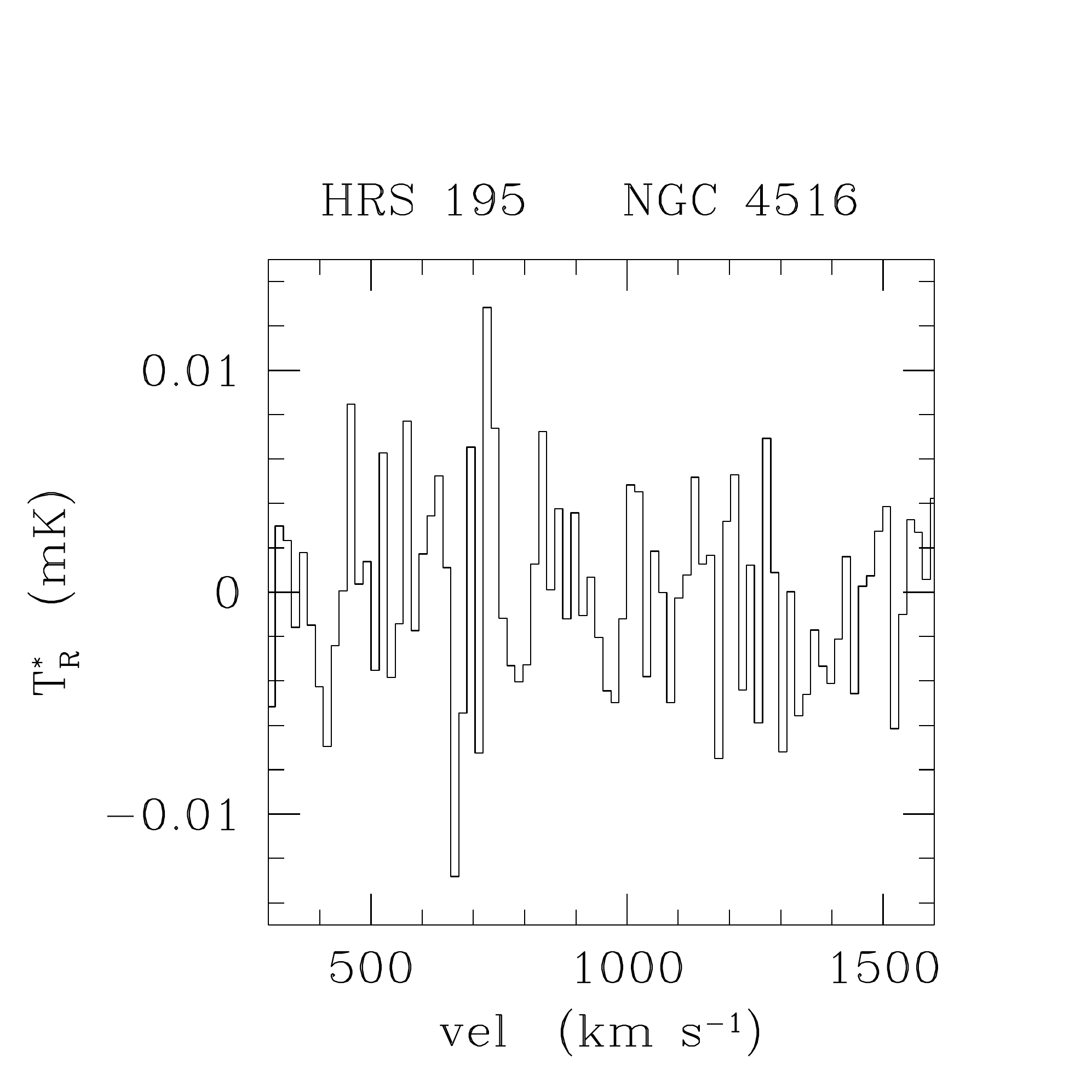}\\
   \includegraphics[width=0.22\textwidth]{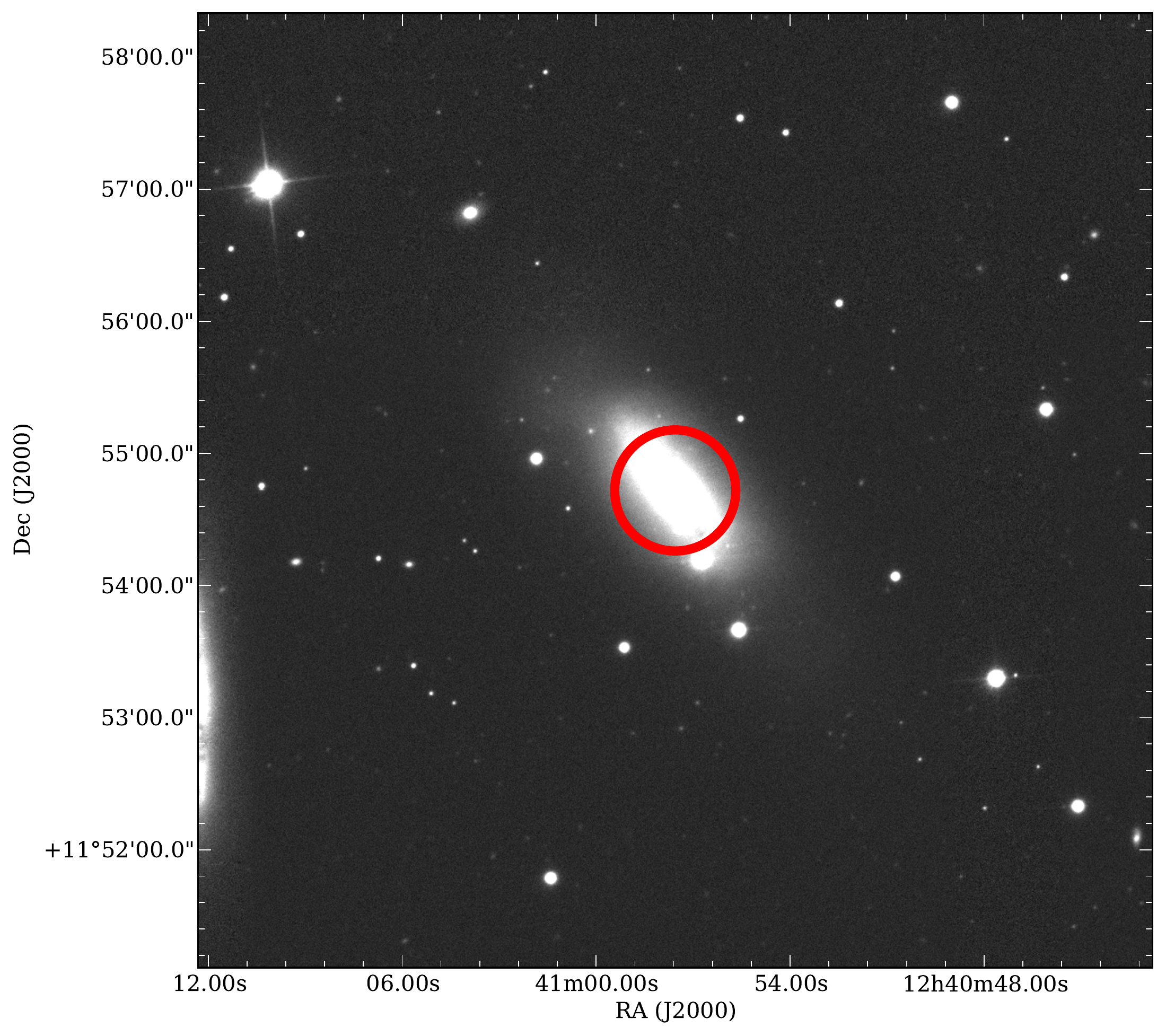}
   \includegraphics[width=0.22\textwidth]{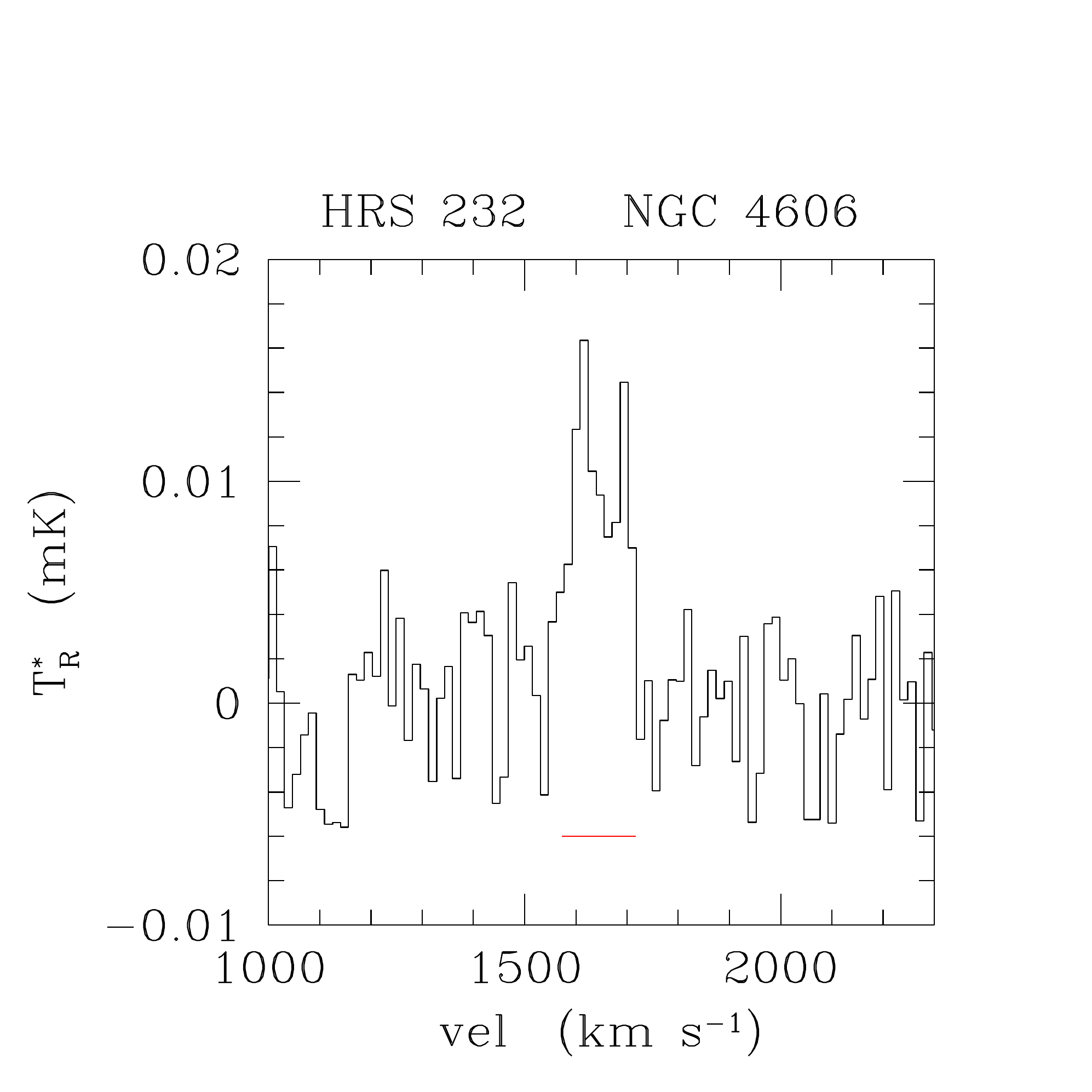}\\
   \includegraphics[width=0.22\textwidth]{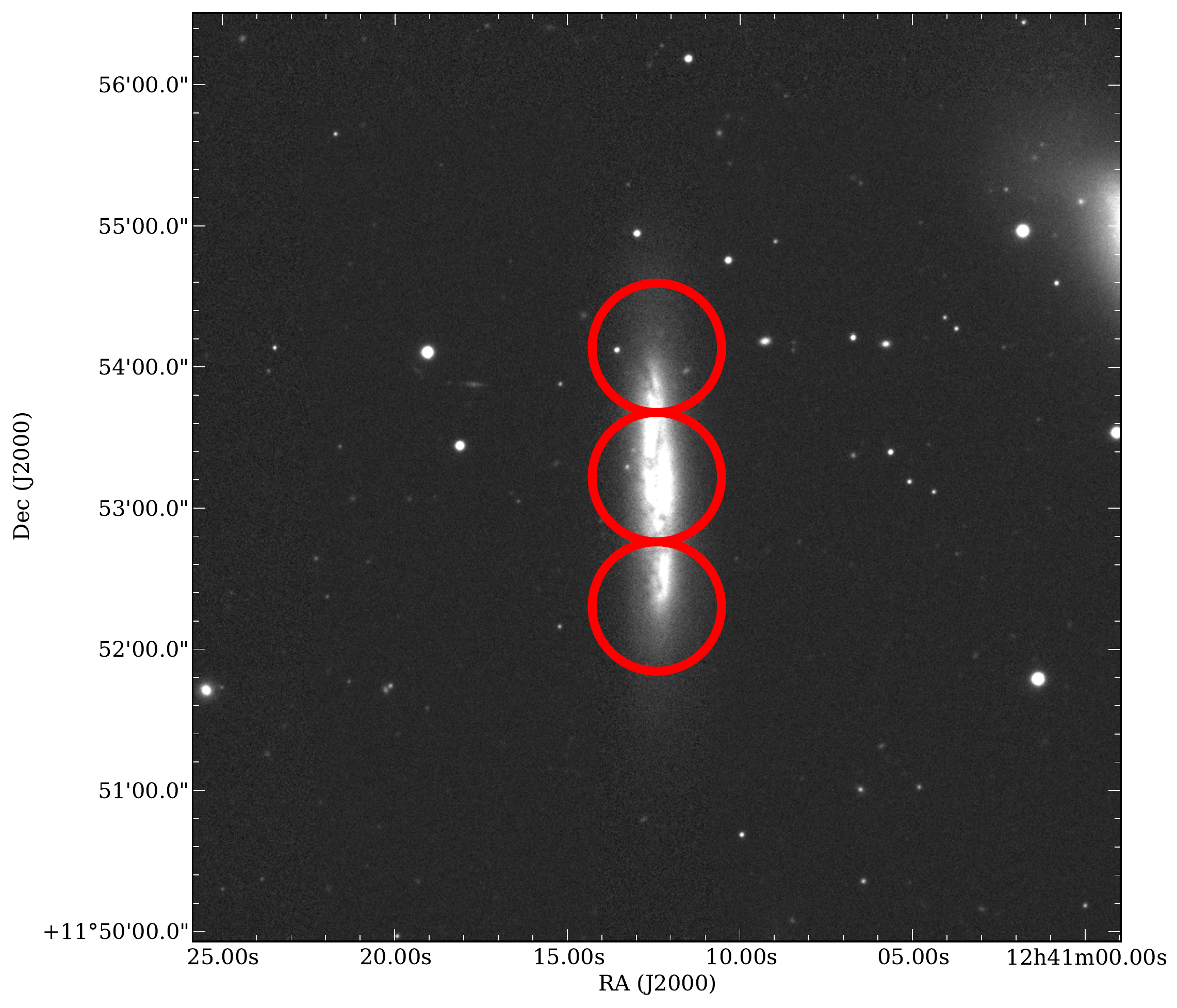}
   \includegraphics[width=0.22\textwidth]{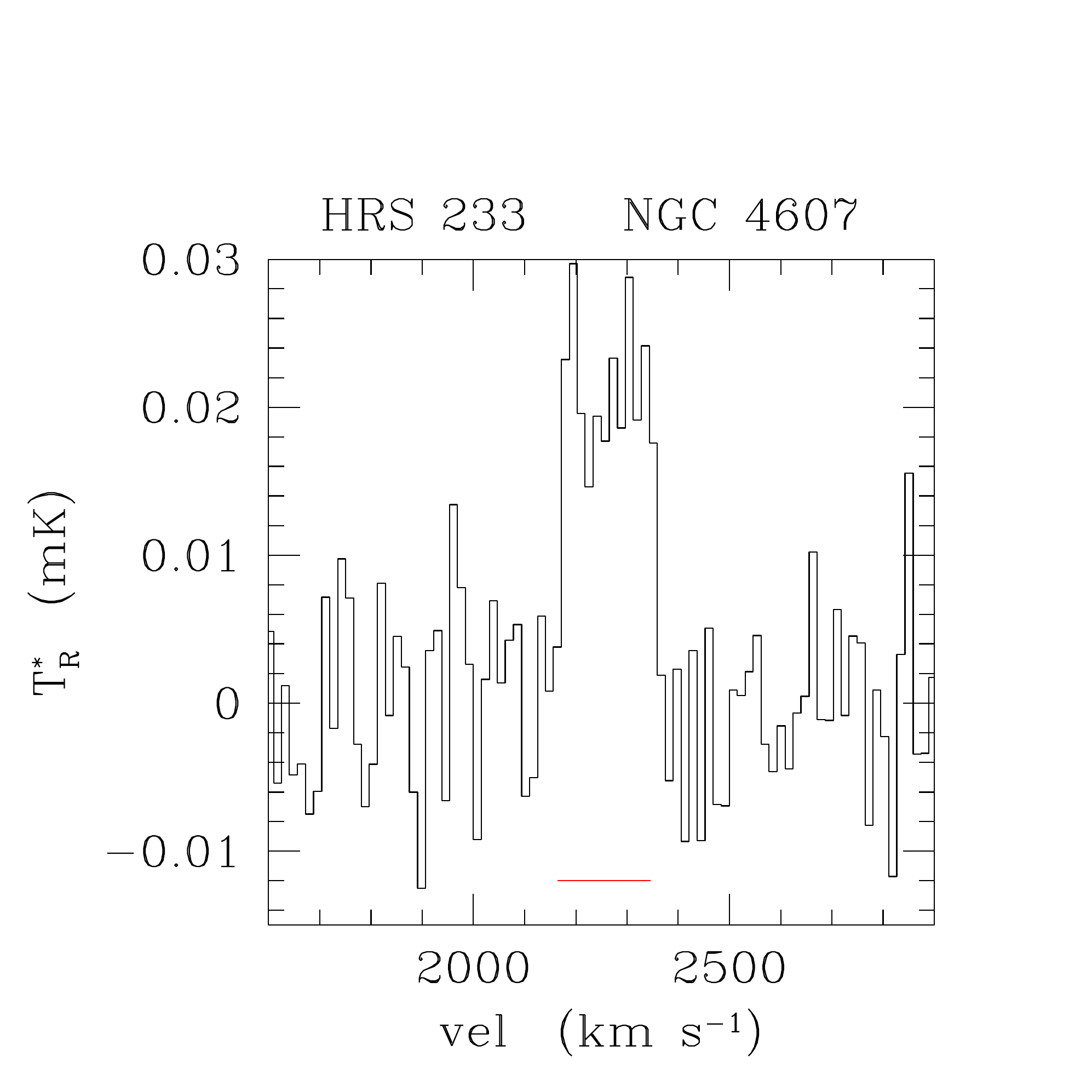}   
   \includegraphics[width=0.22\textwidth]{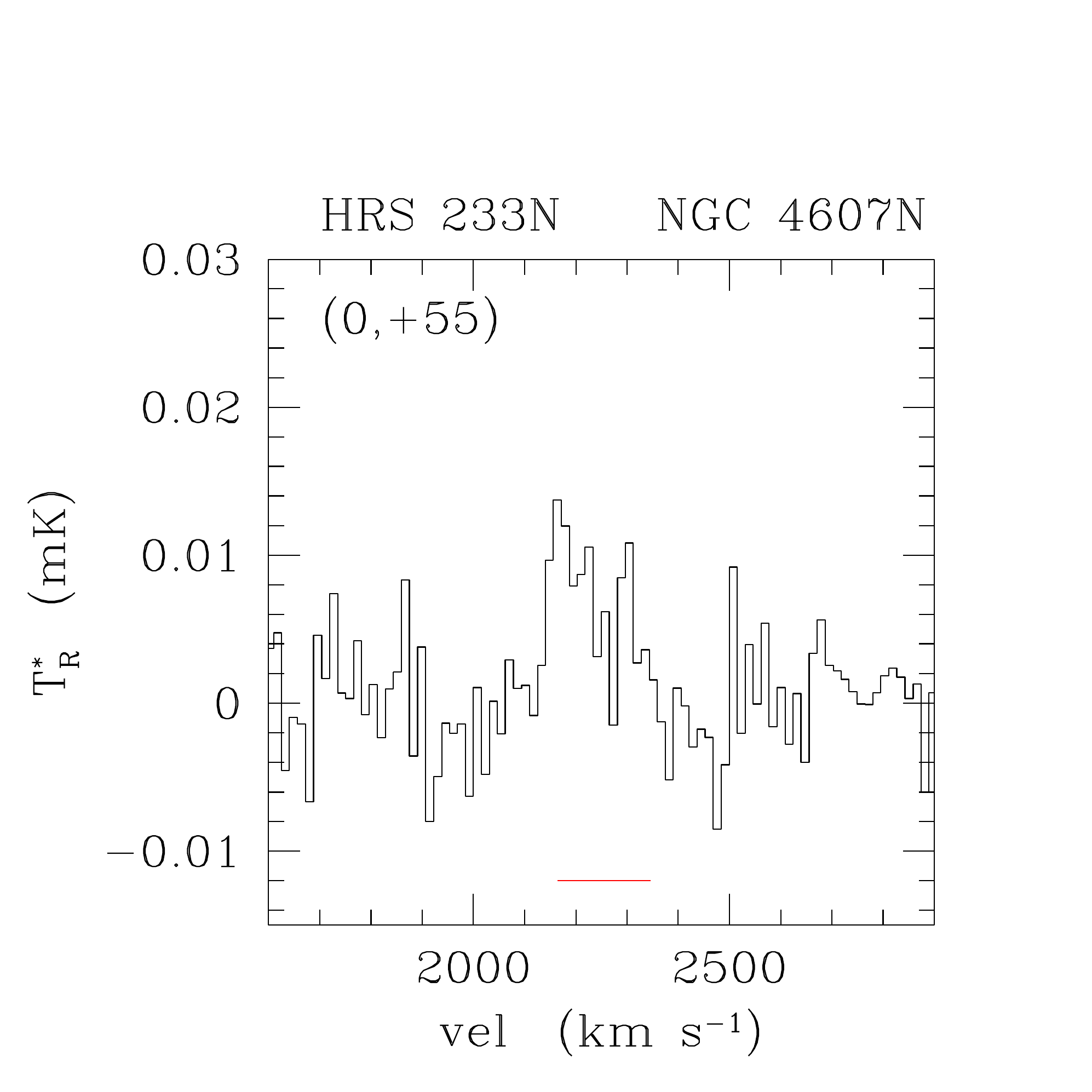}
   \includegraphics[width=0.22\textwidth]{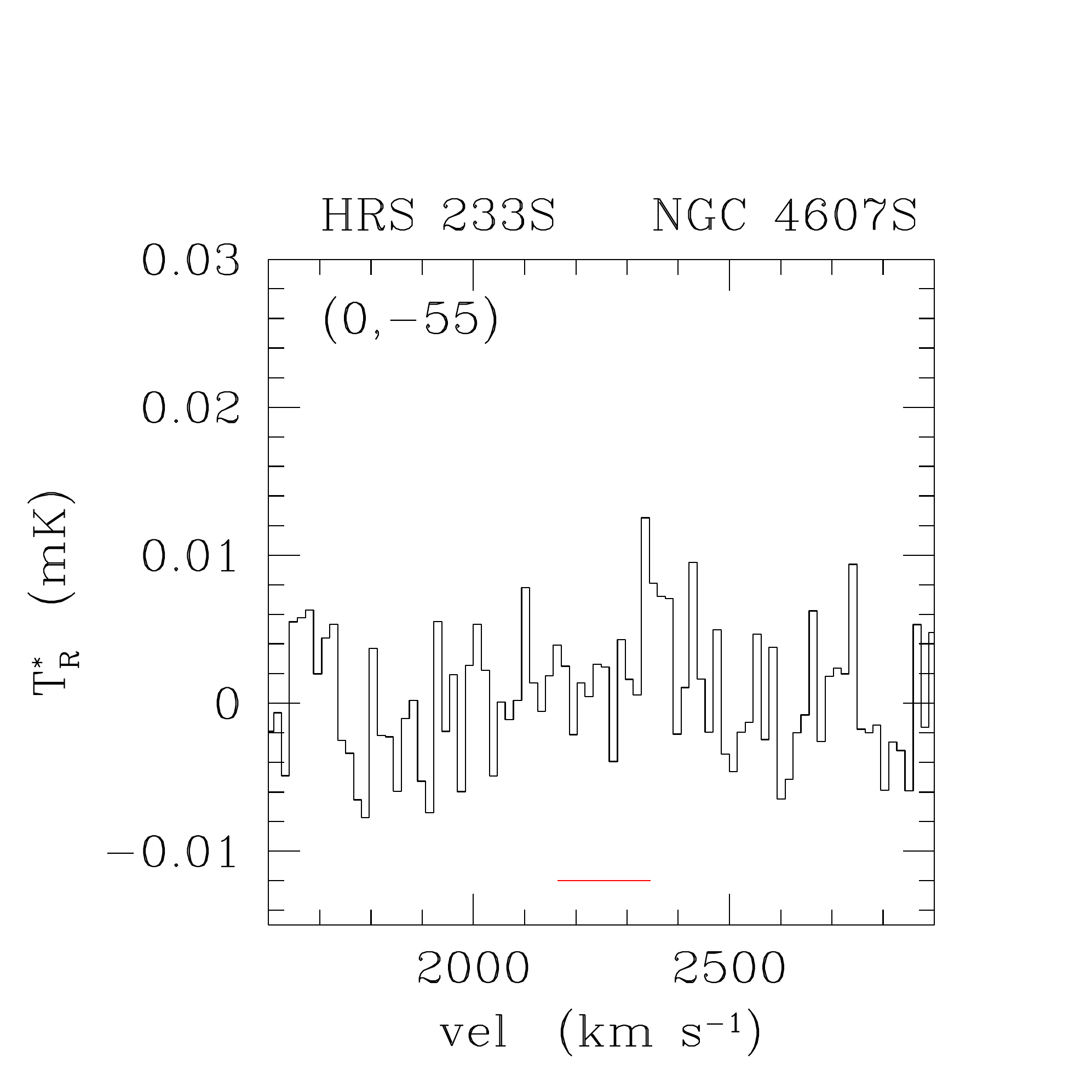}\\
   \caption{Continued.}
   \label{spettri}%
   \end{figure*}
   \clearpage

   \addtocounter{figure}{-1}
   \begin{figure*}
   \centering
   \includegraphics[width=0.22\textwidth]{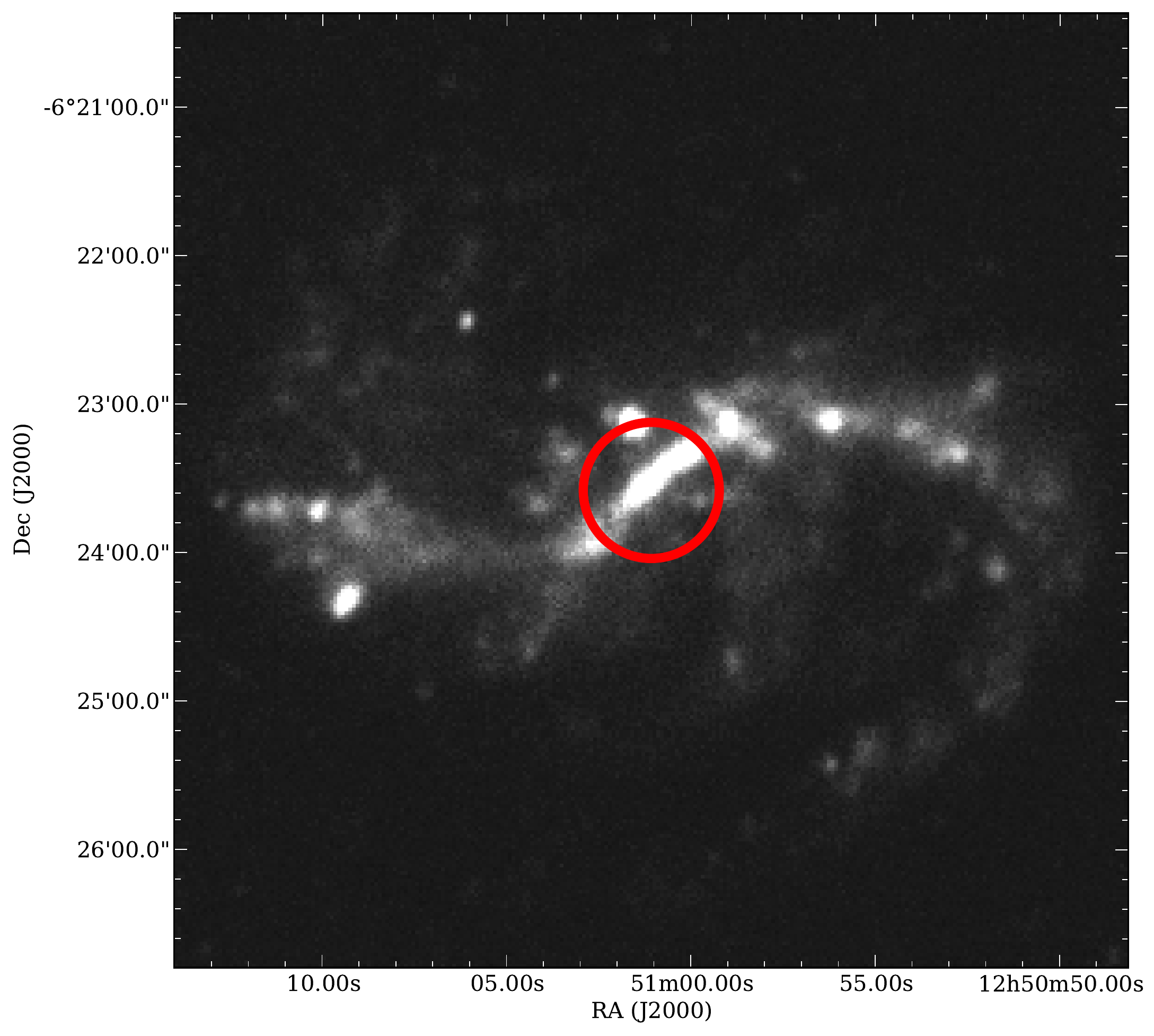}
   \includegraphics[width=0.22\textwidth]{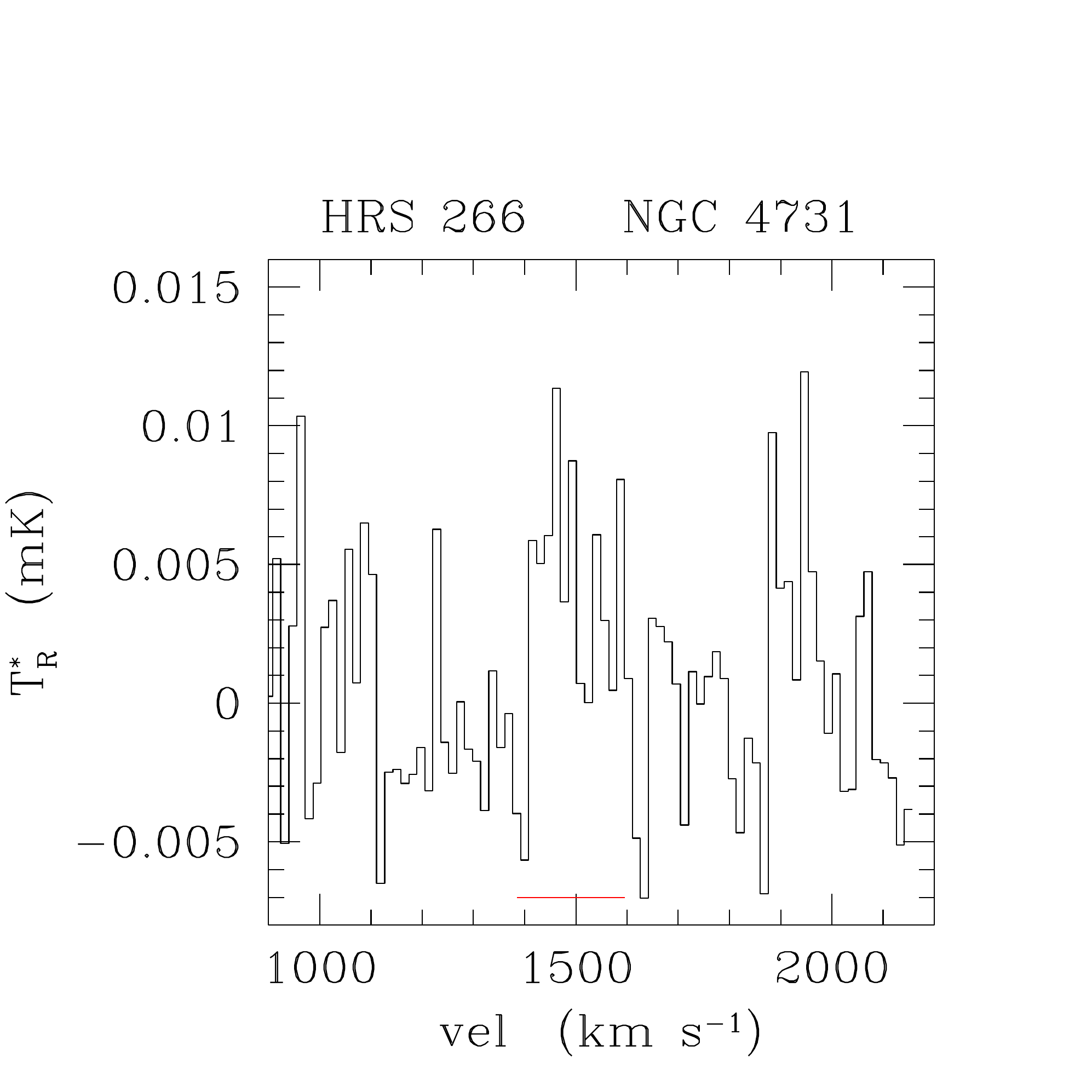}\\
   \includegraphics[width=0.22\textwidth]{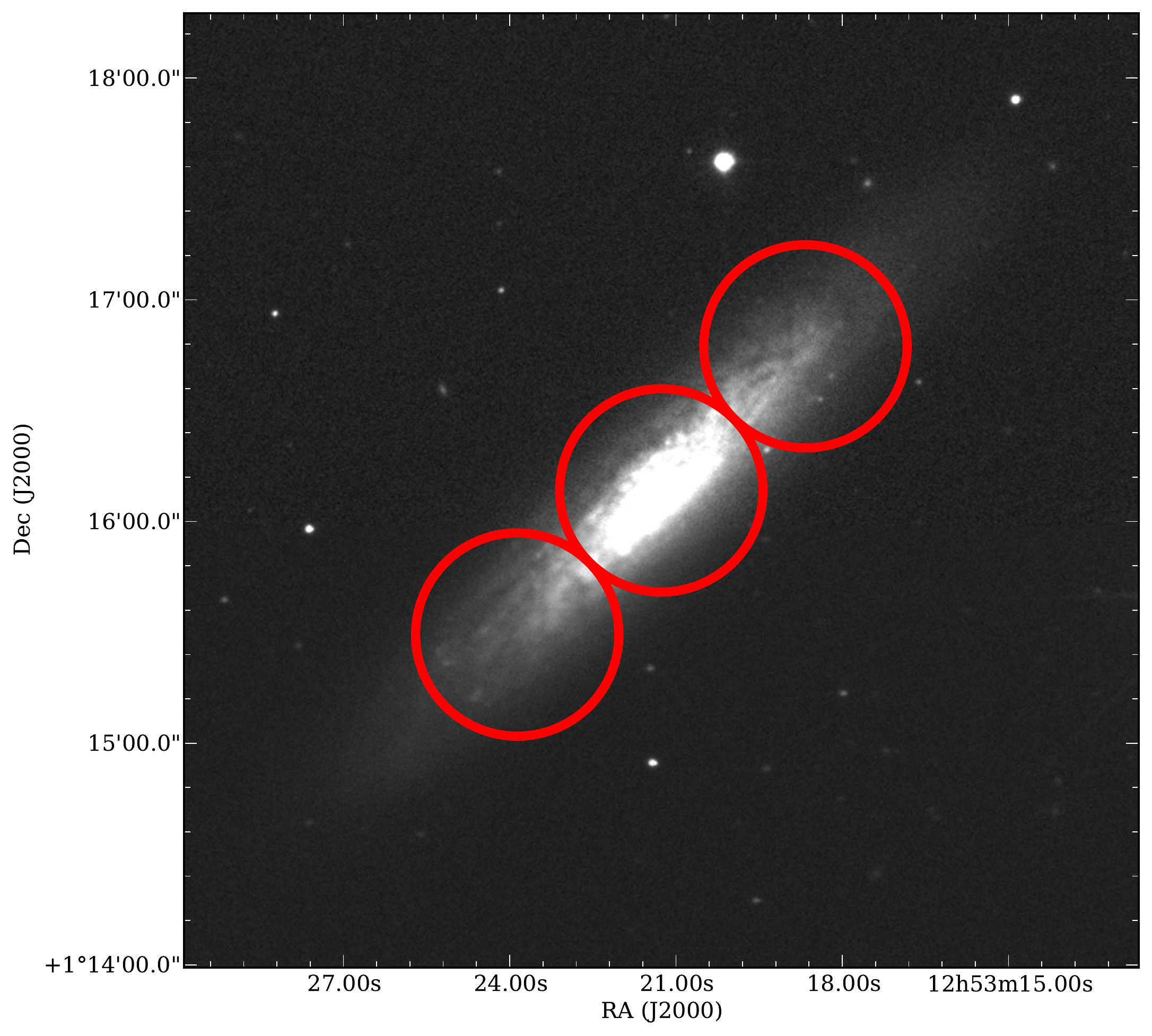}
   \includegraphics[width=0.22\textwidth]{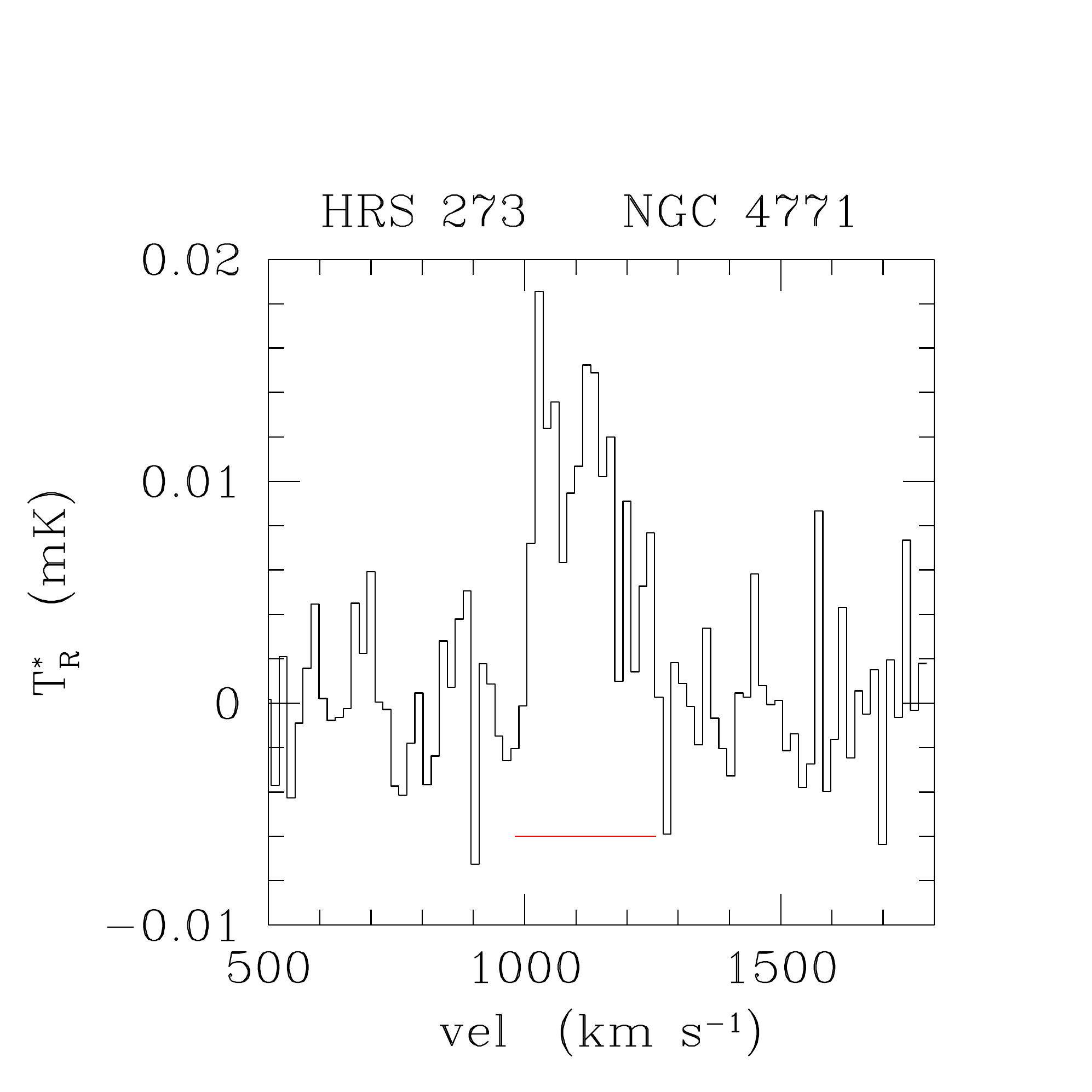}
   \includegraphics[width=0.22\textwidth]{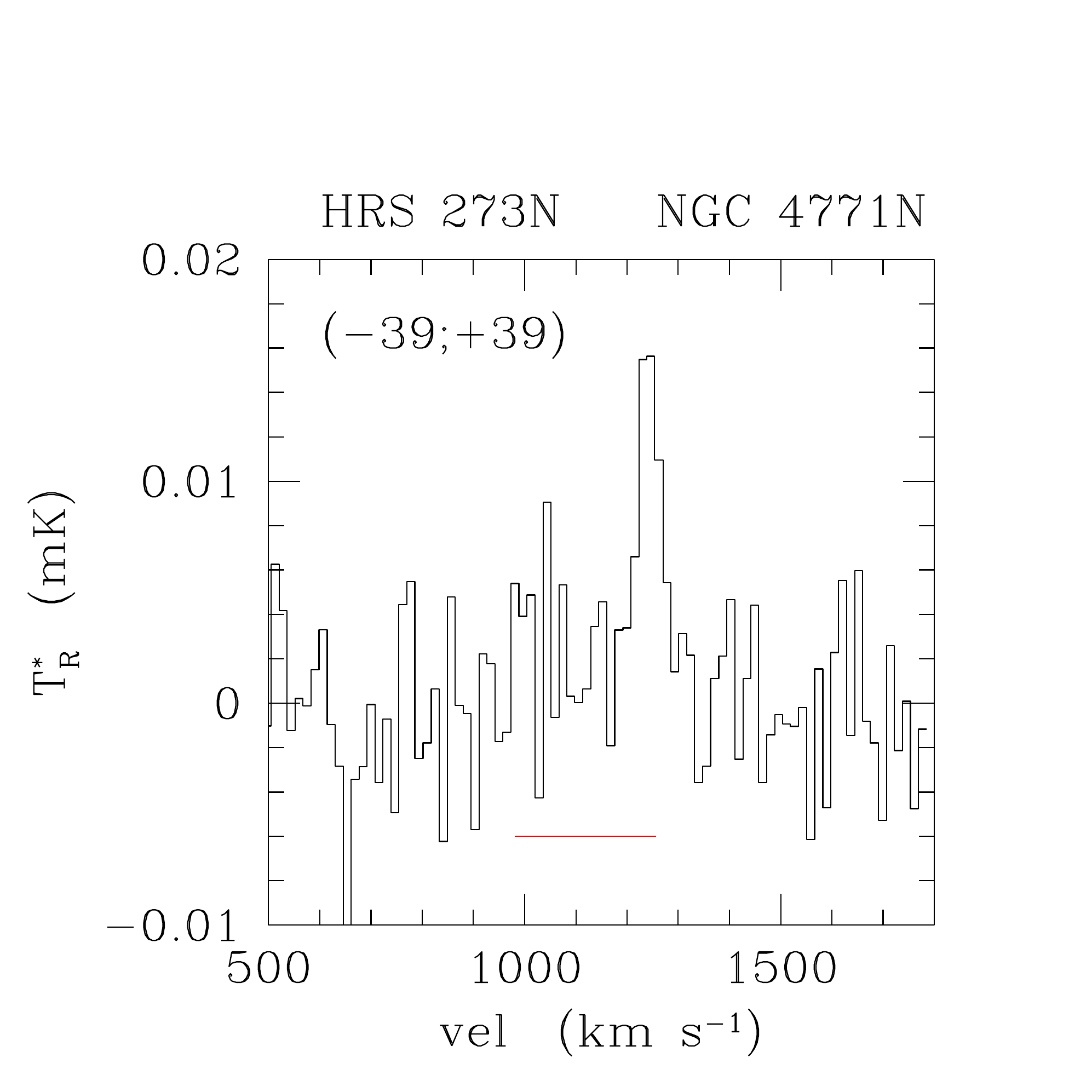}
   \includegraphics[width=0.22\textwidth]{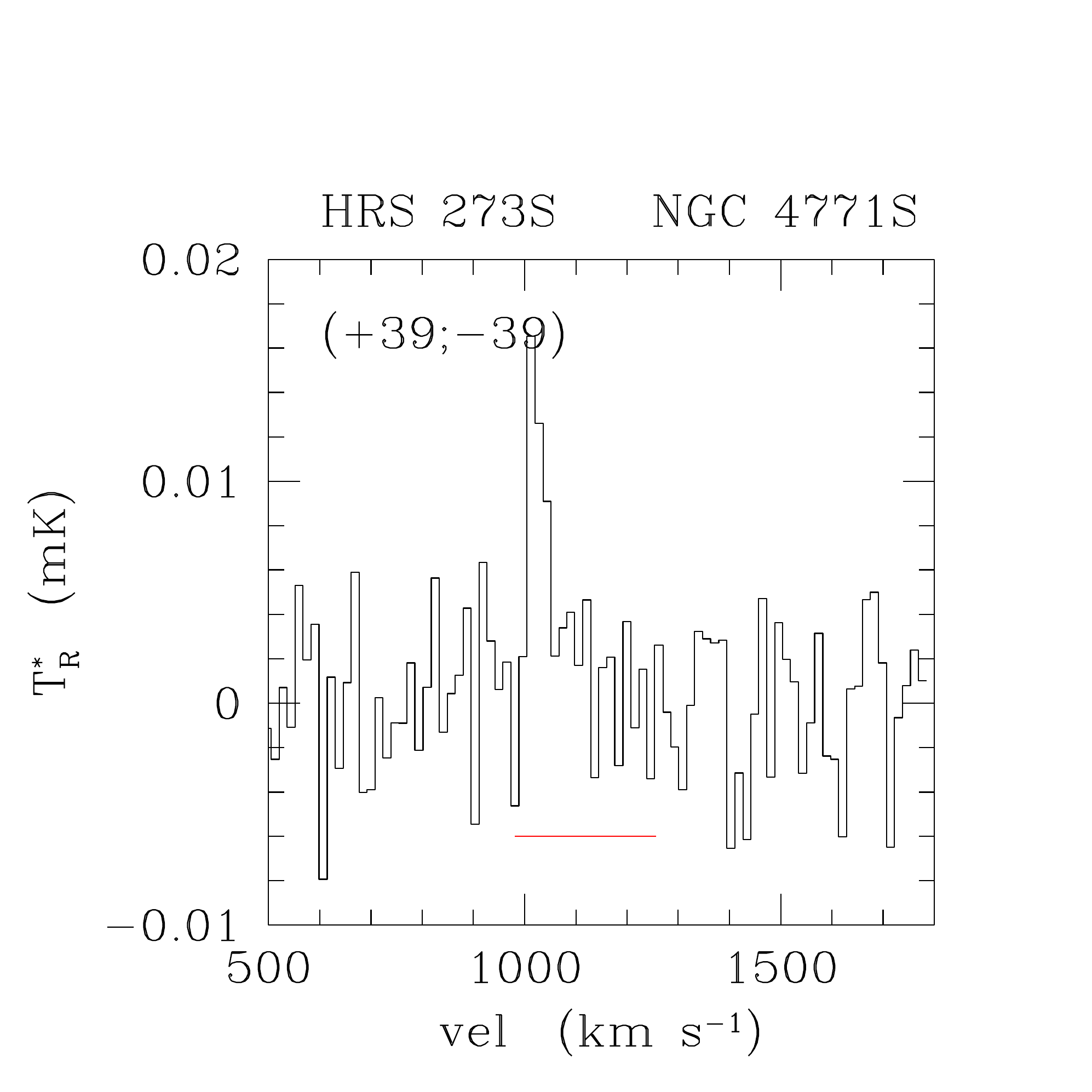}\\
   \includegraphics[width=0.22\textwidth]{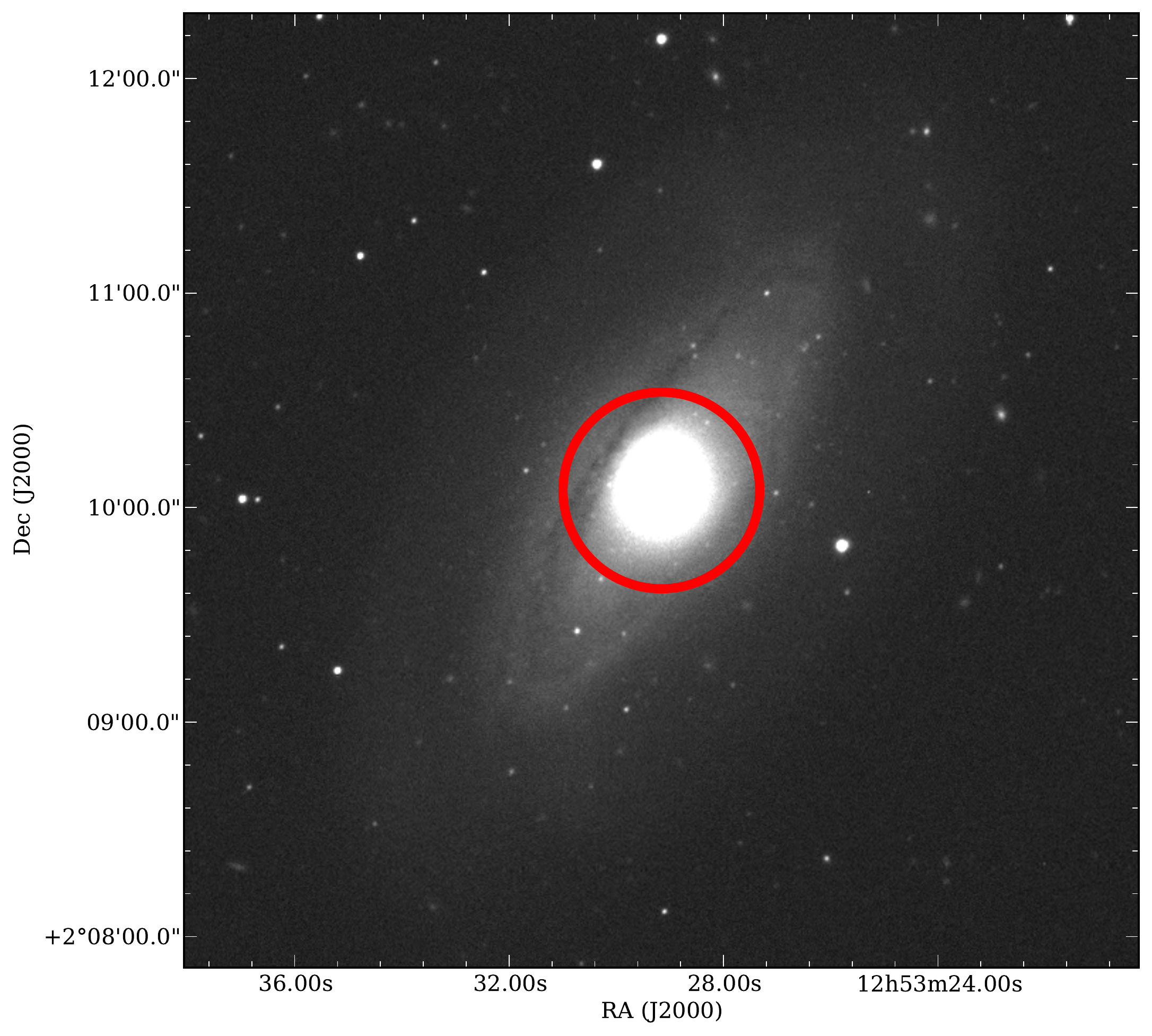}
   \includegraphics[width=0.22\textwidth]{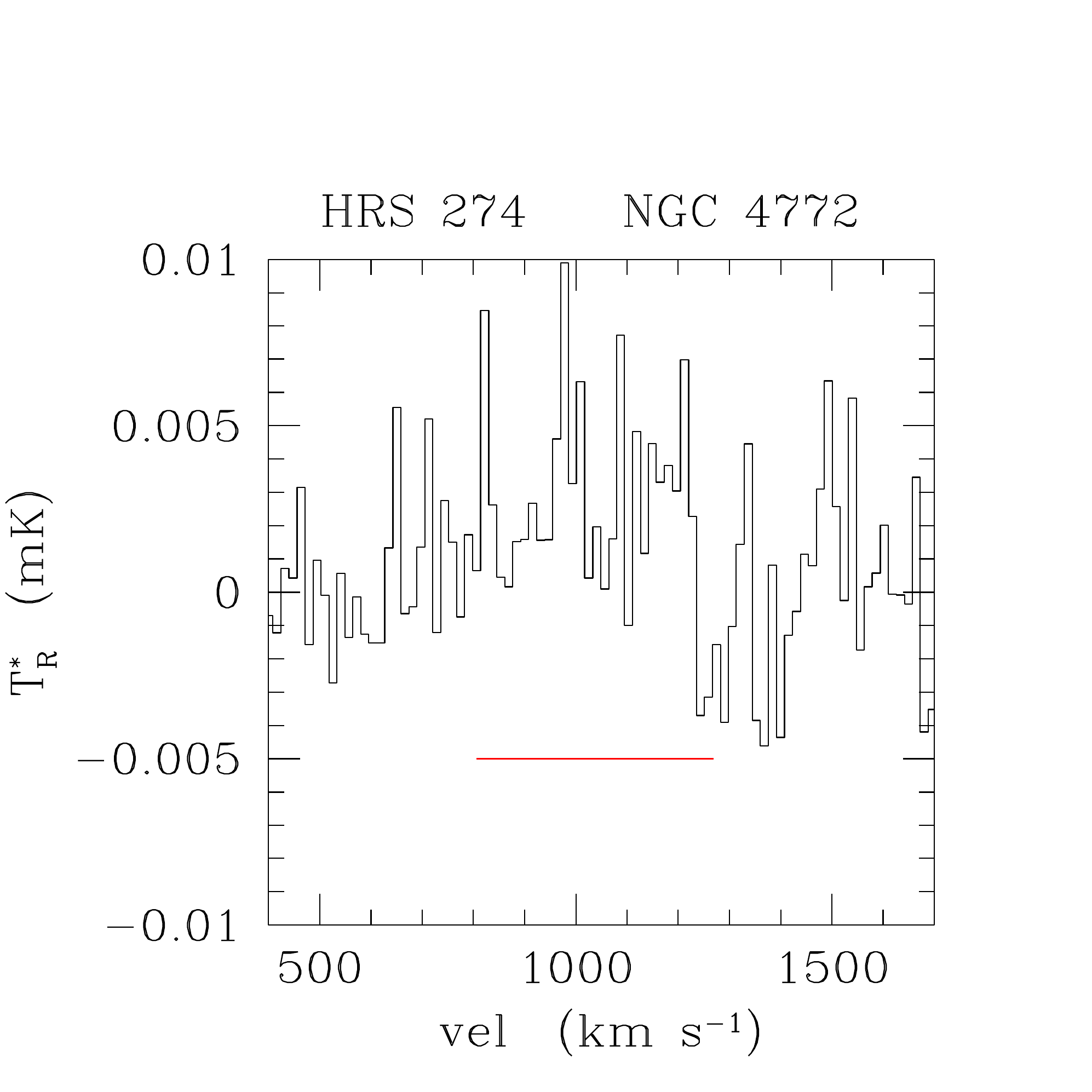}\\
   \includegraphics[width=0.22\textwidth]{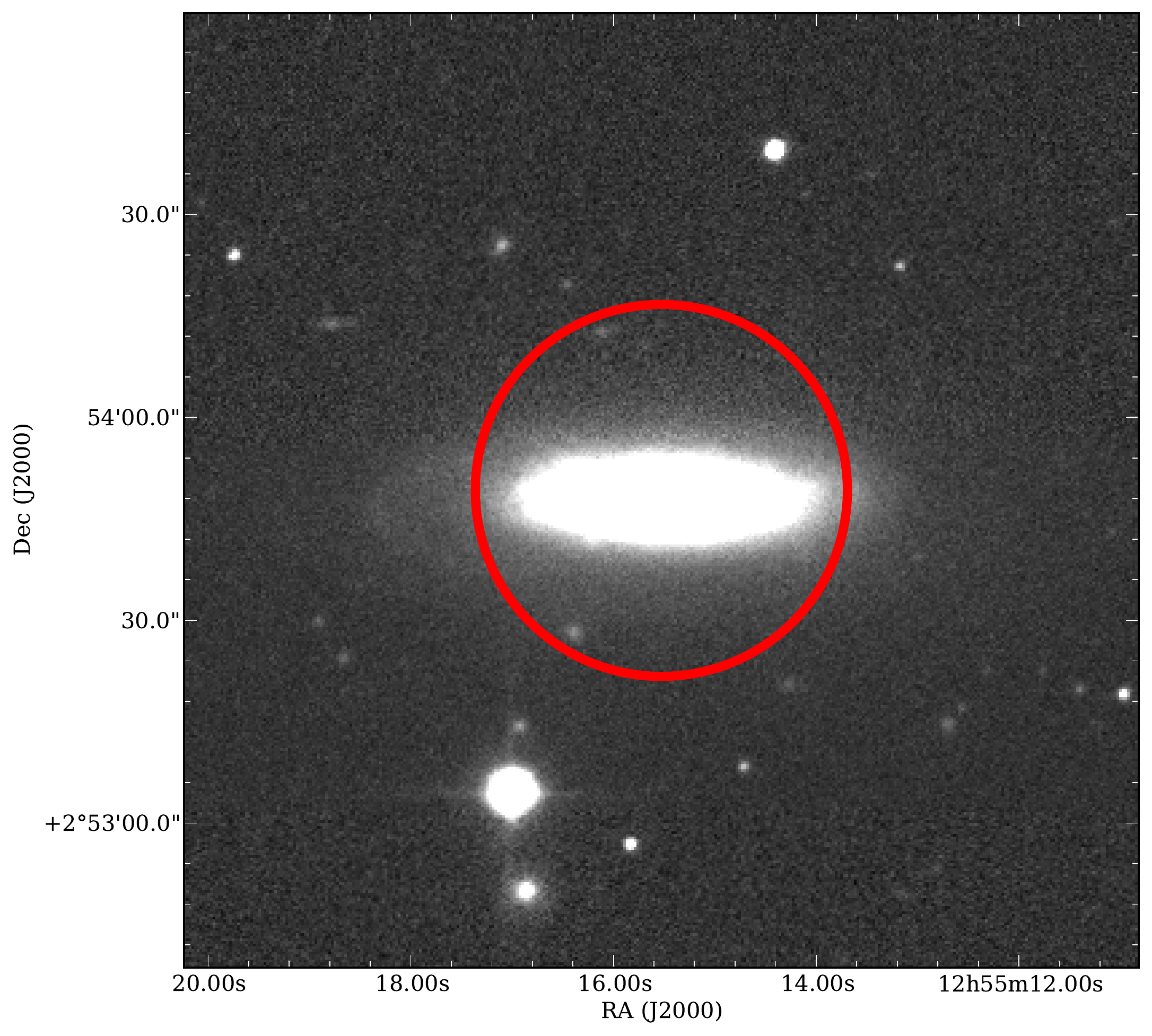}
   \includegraphics[width=0.22\textwidth]{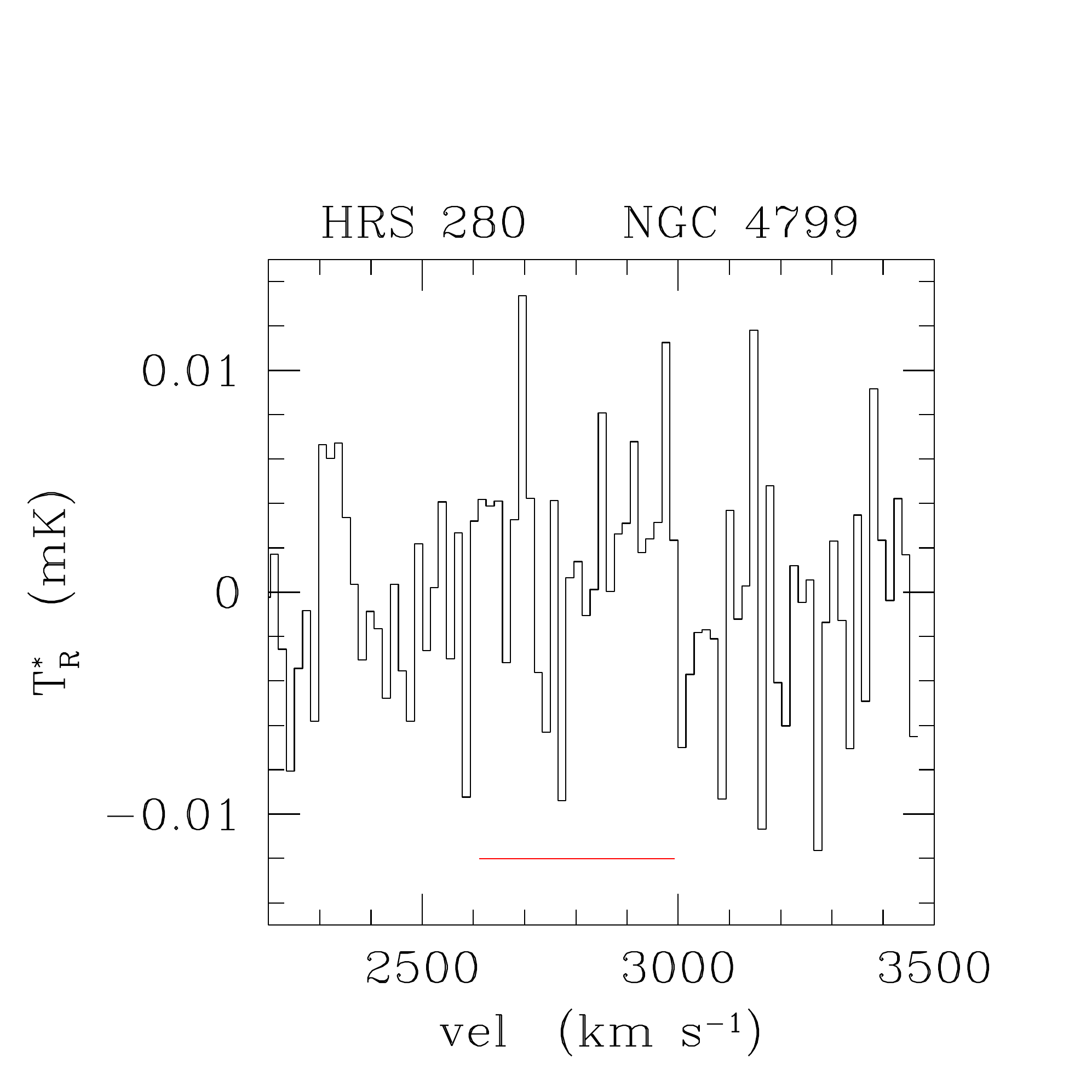}\\
   \includegraphics[width=0.22\textwidth]{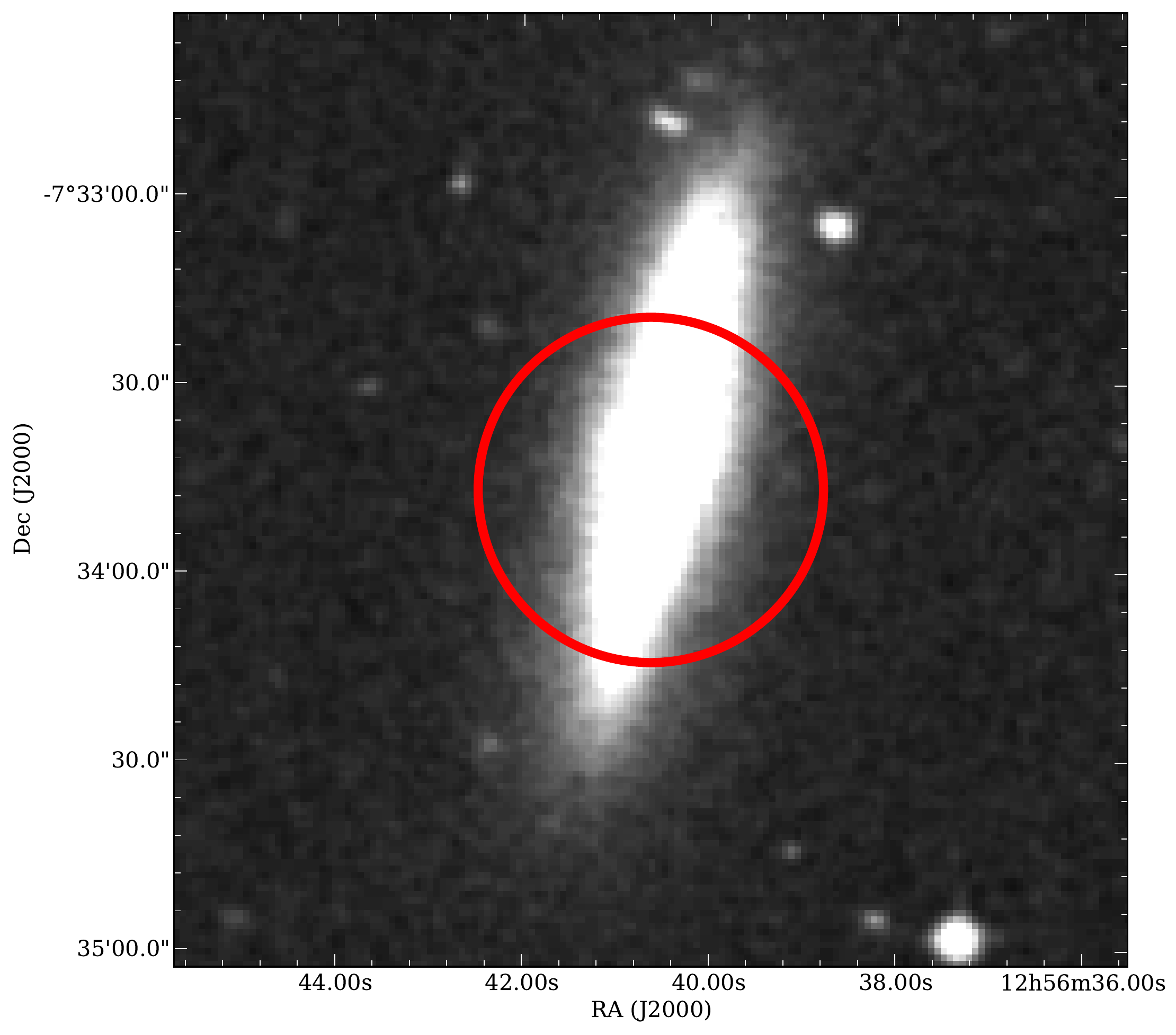}
   \includegraphics[width=0.22\textwidth]{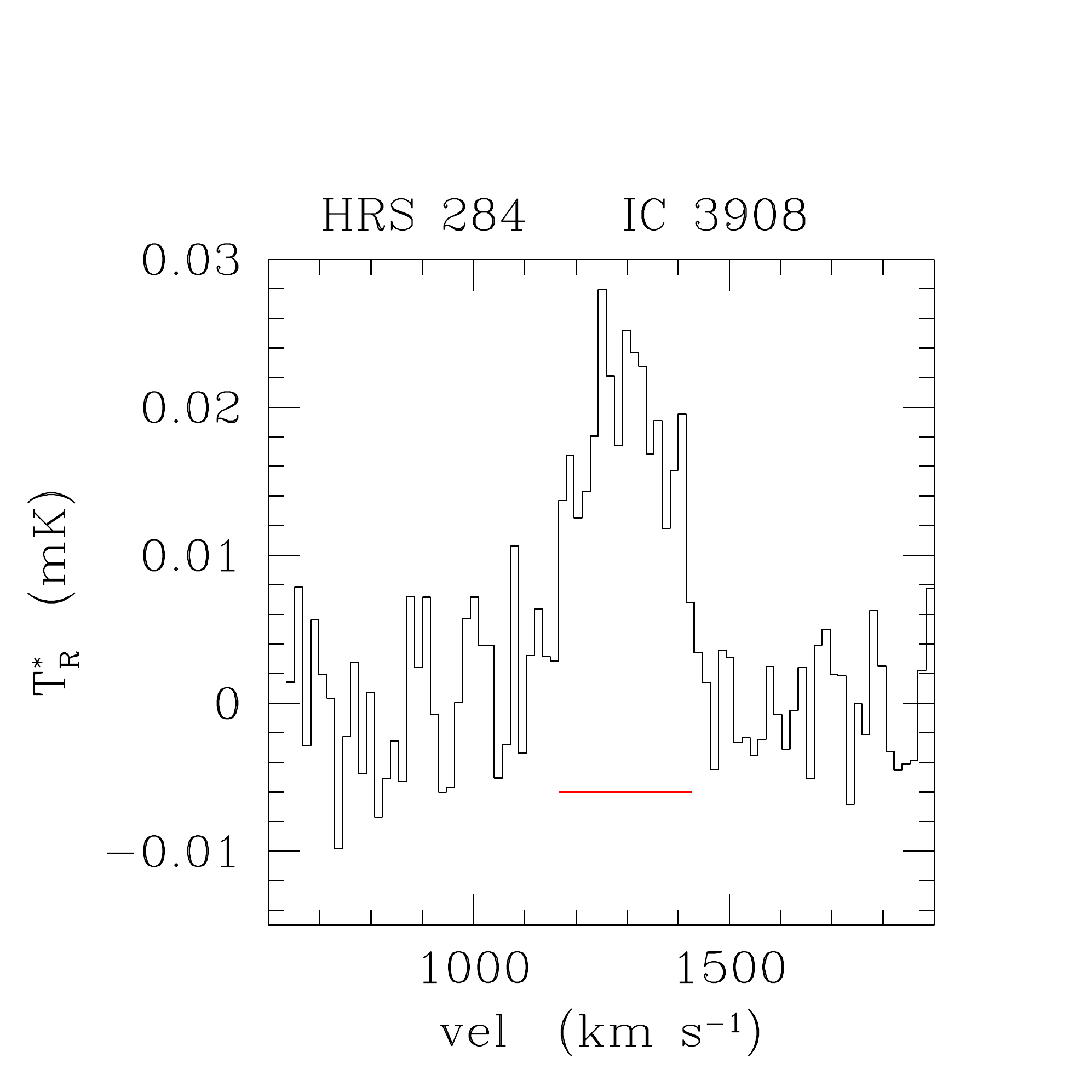}\\
   \caption{Continued.}
   \label{spettri}%
   \end{figure*}
   \clearpage

   \addtocounter{figure}{-1}
   \begin{figure*}
   \centering   
   \includegraphics[width=0.22\textwidth]{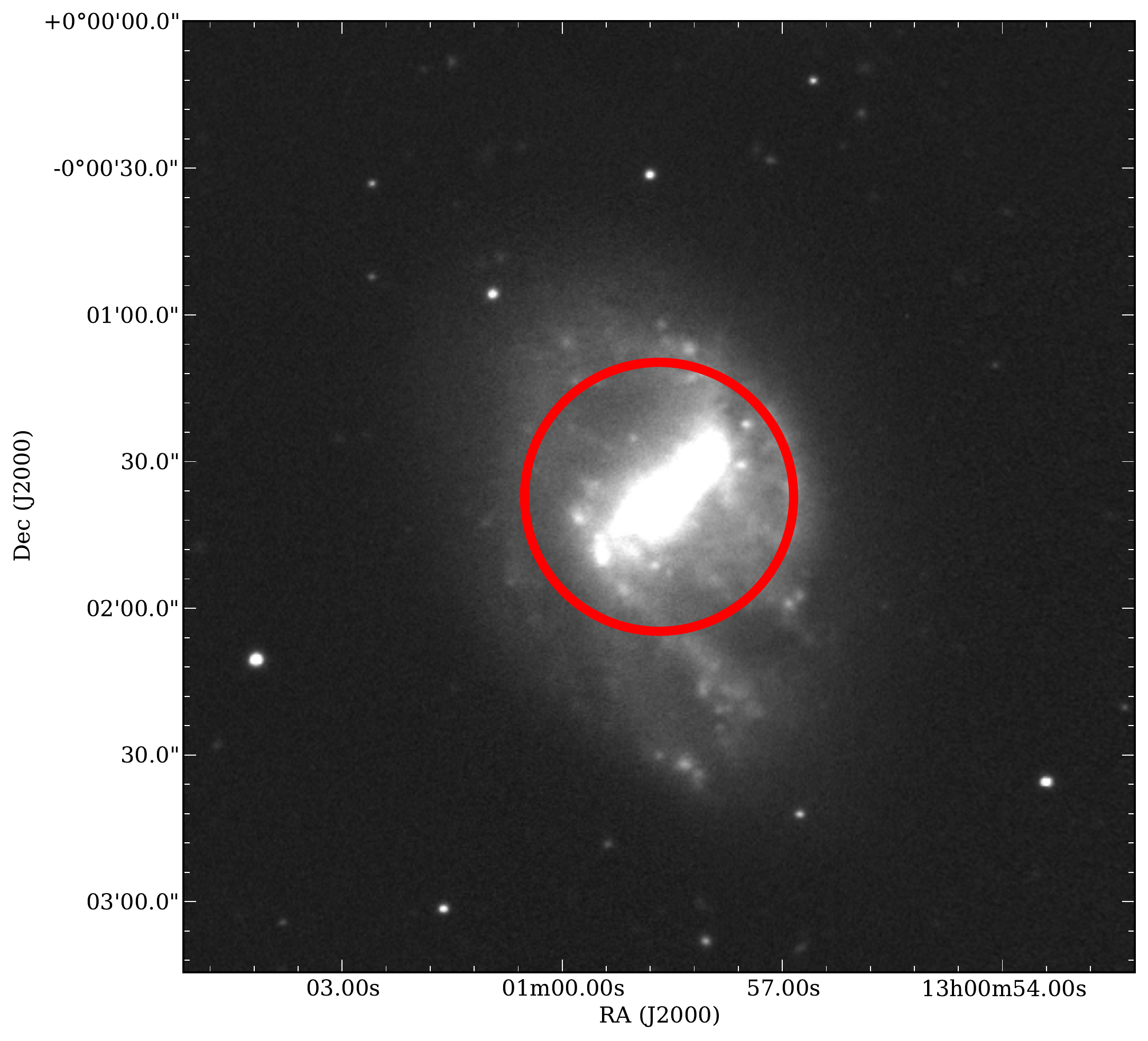}
   \includegraphics[width=0.22\textwidth]{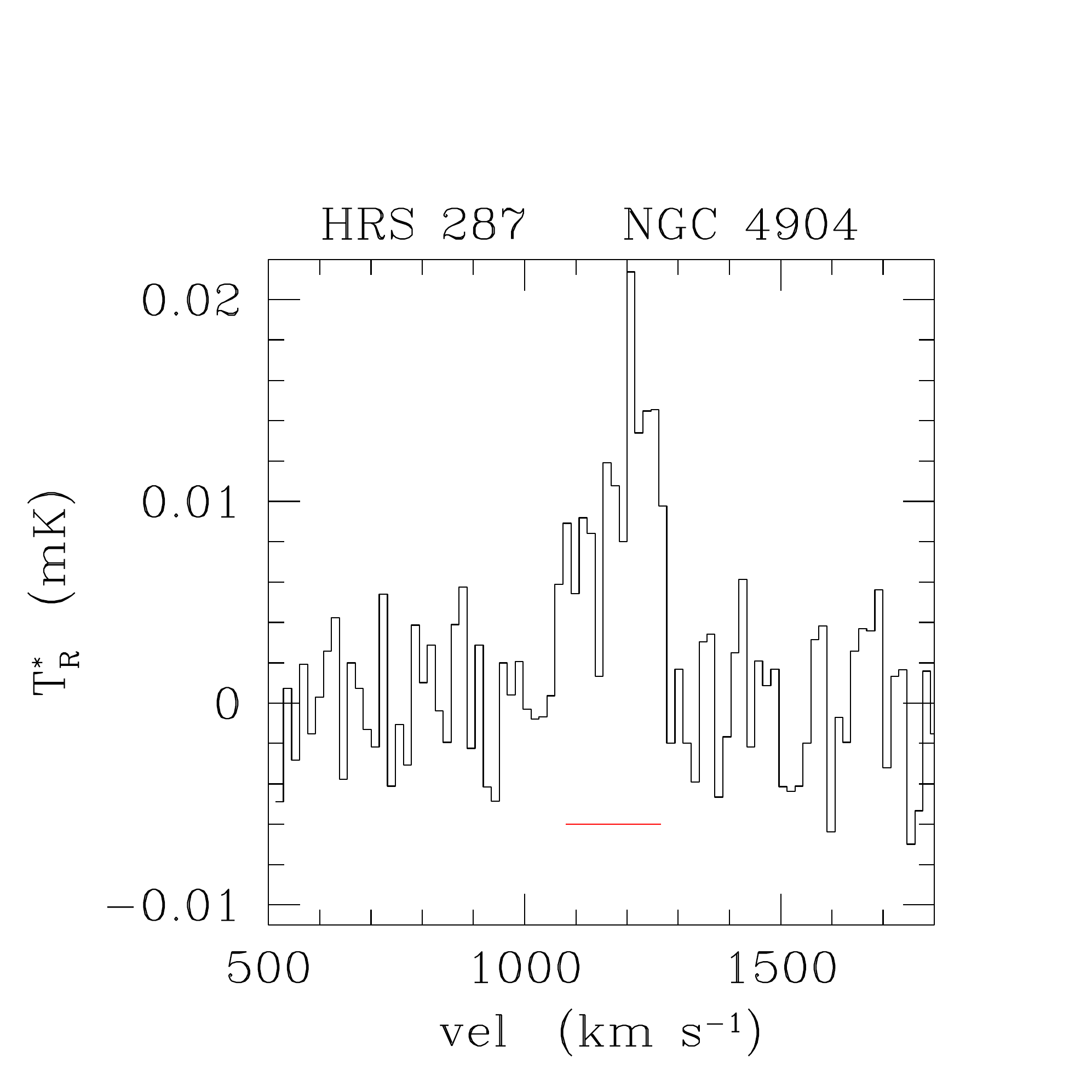}\\
   \includegraphics[width=0.22\textwidth]{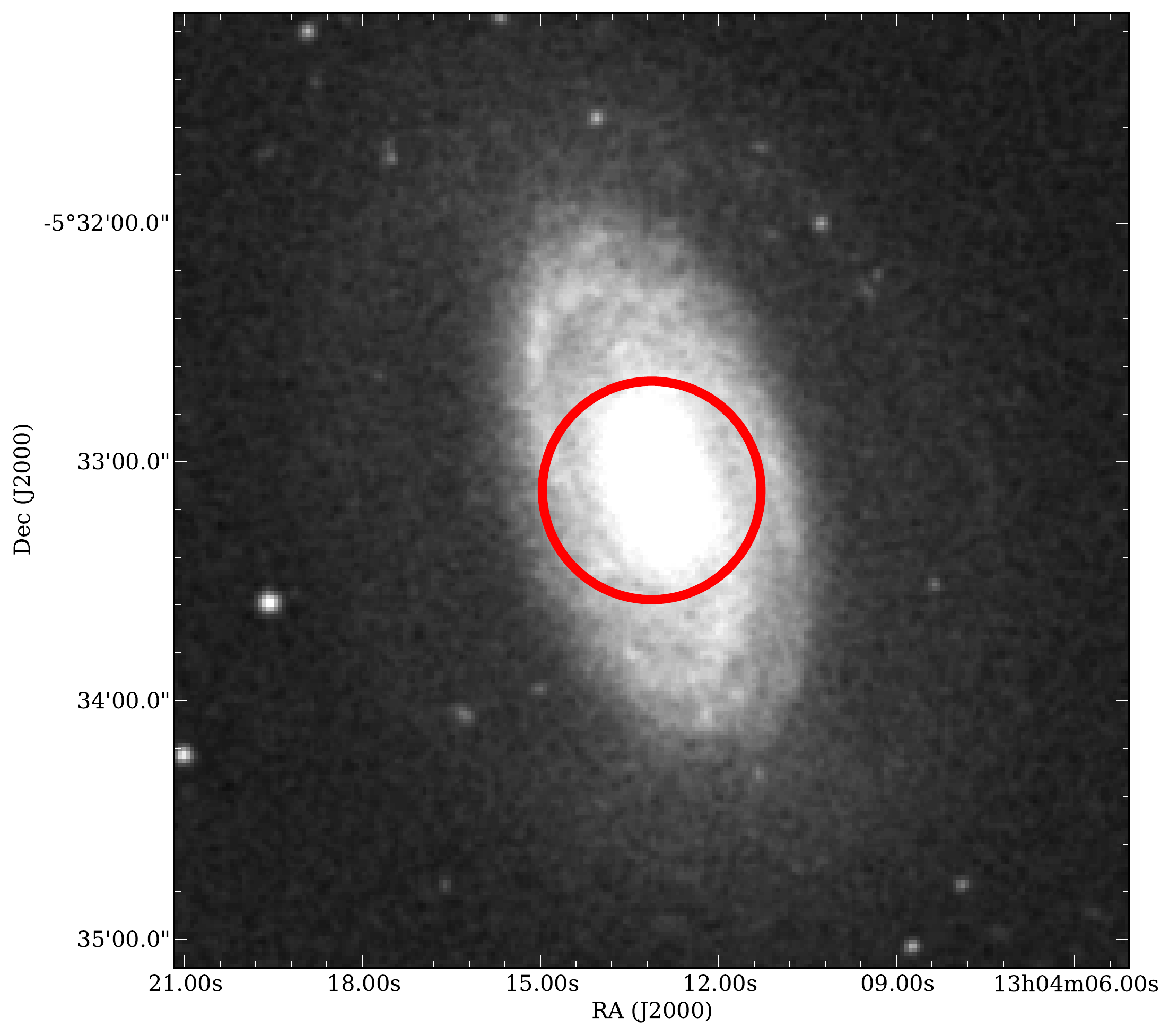}
   \includegraphics[width=0.22\textwidth]{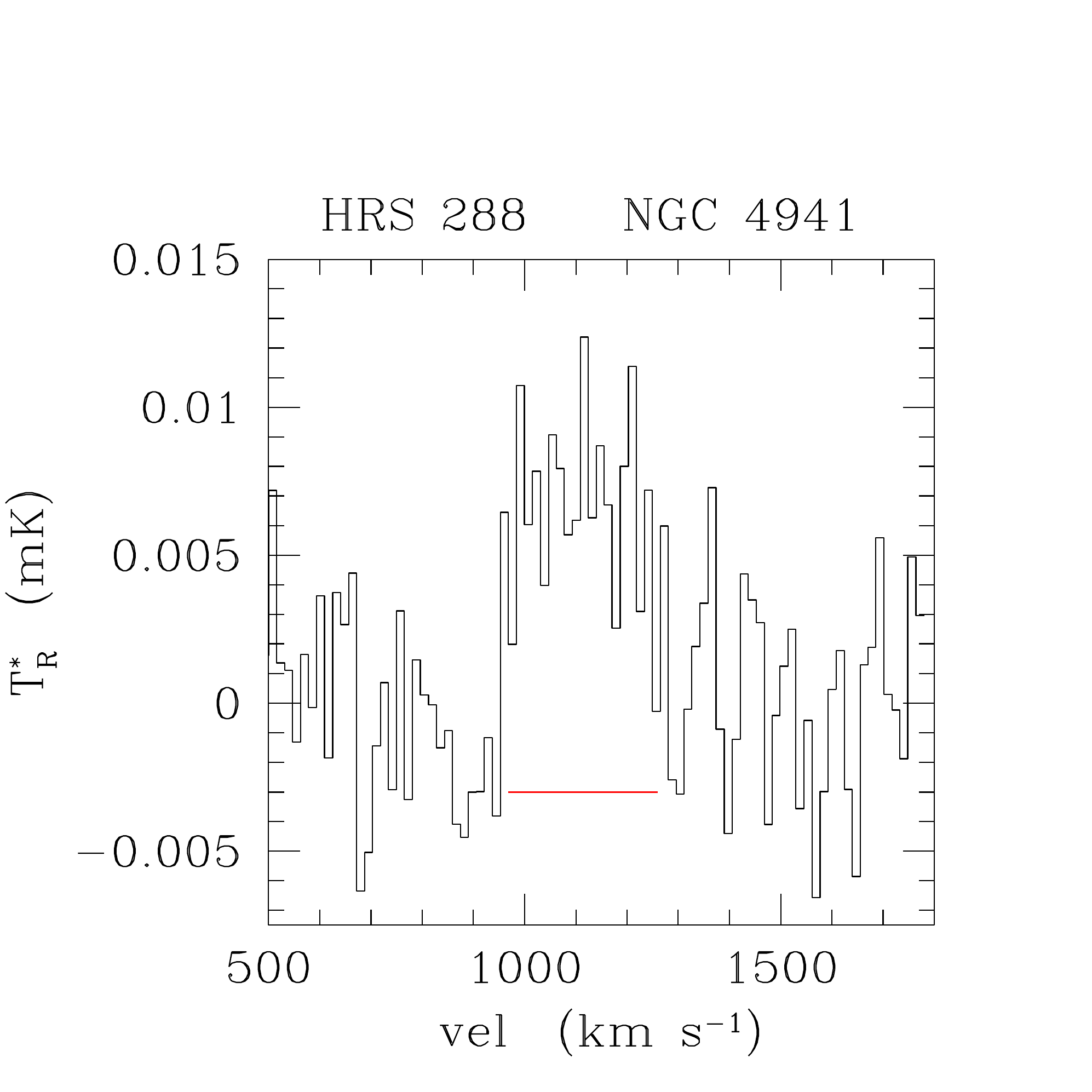}\\
   \includegraphics[width=0.22\textwidth]{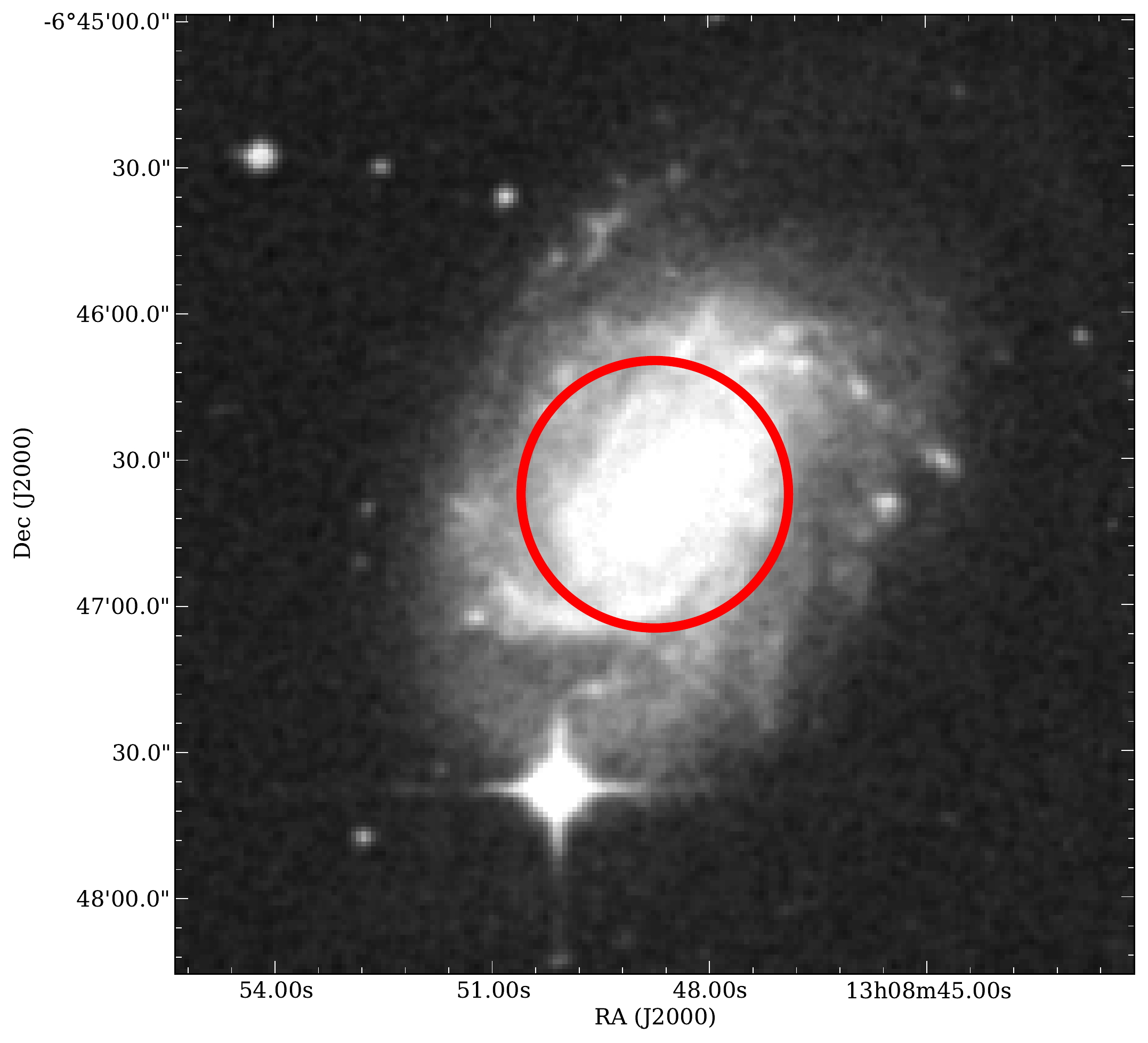}
   \includegraphics[width=0.22\textwidth]{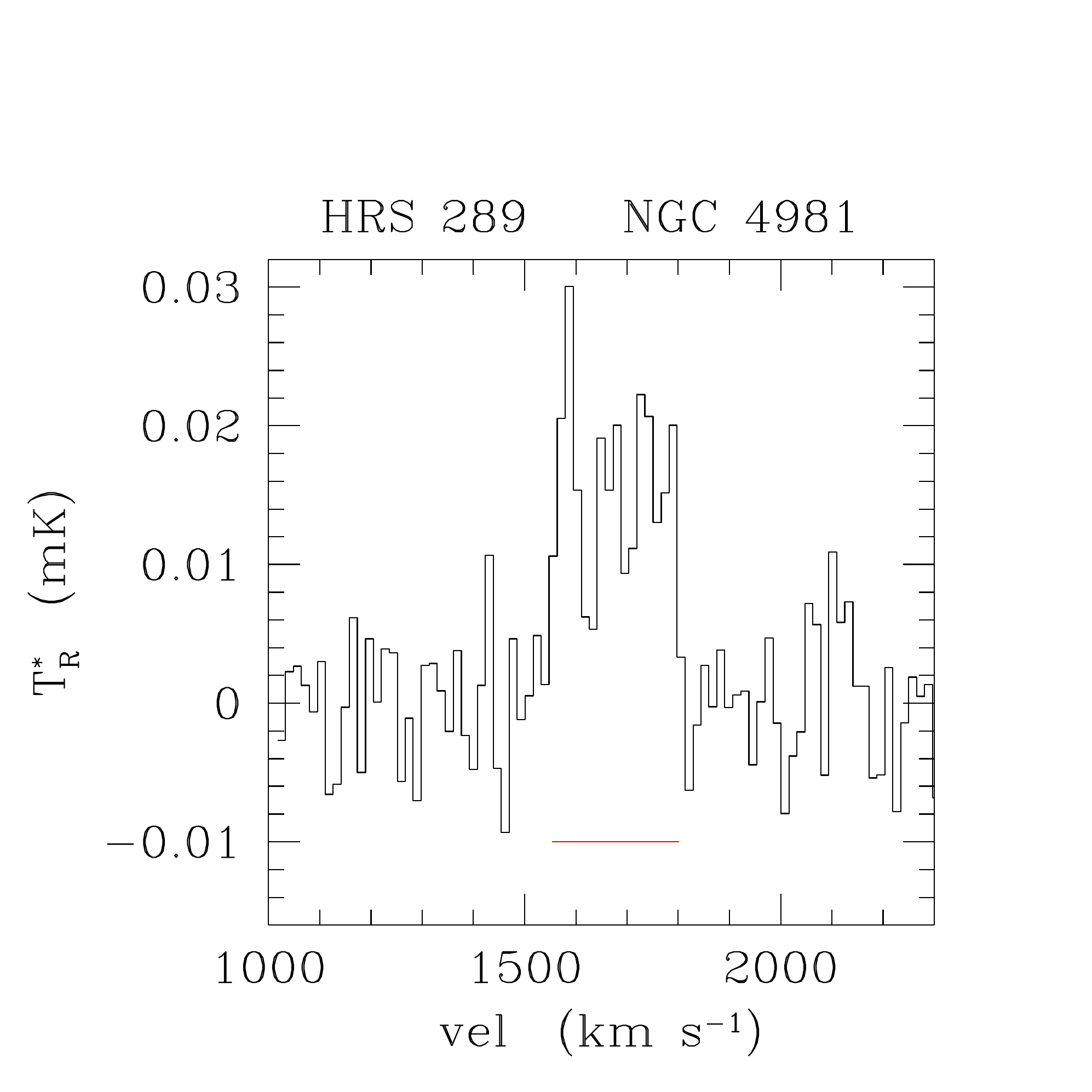}\\
   \includegraphics[width=0.22\textwidth]{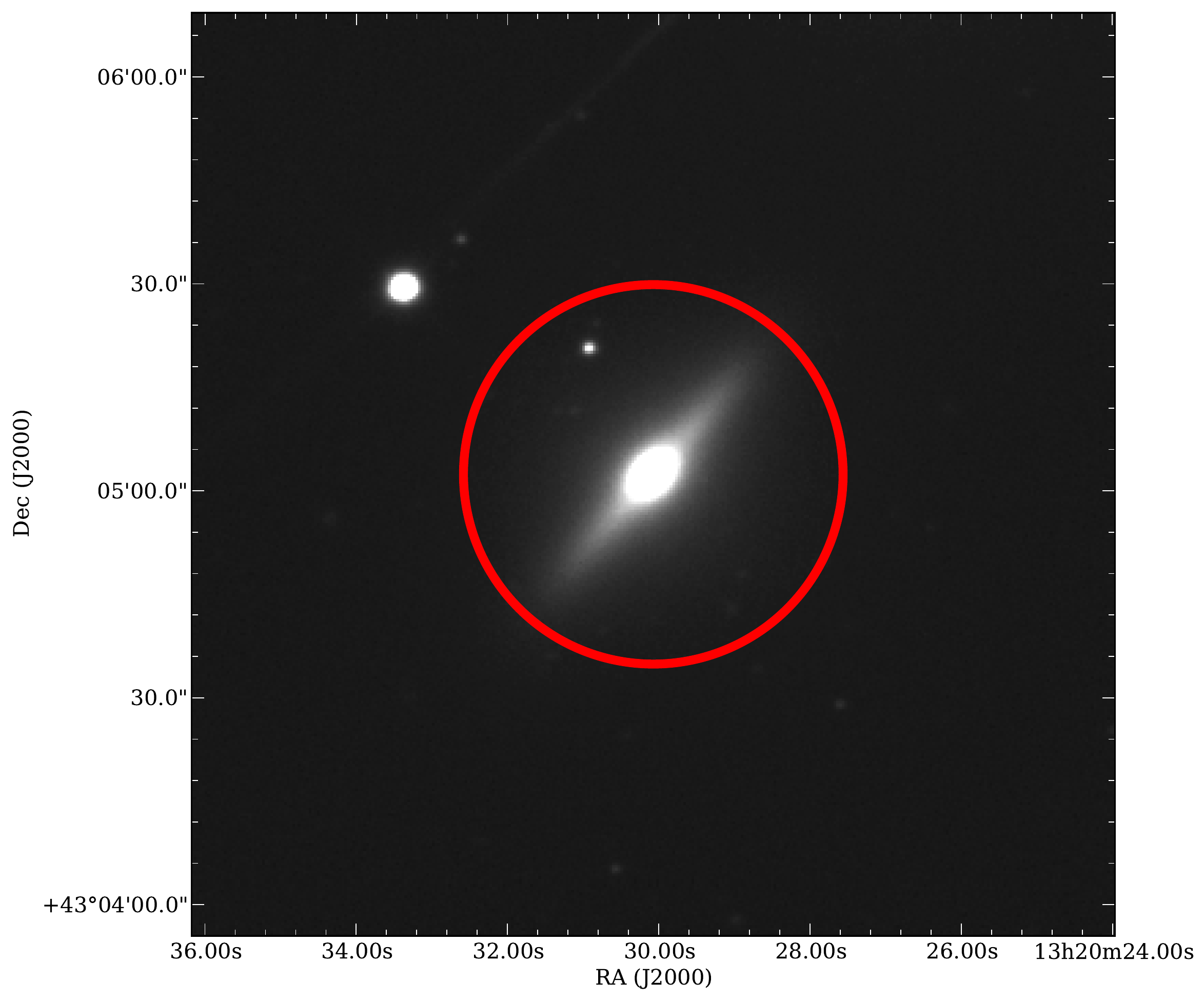}
   \includegraphics[width=0.22\textwidth]{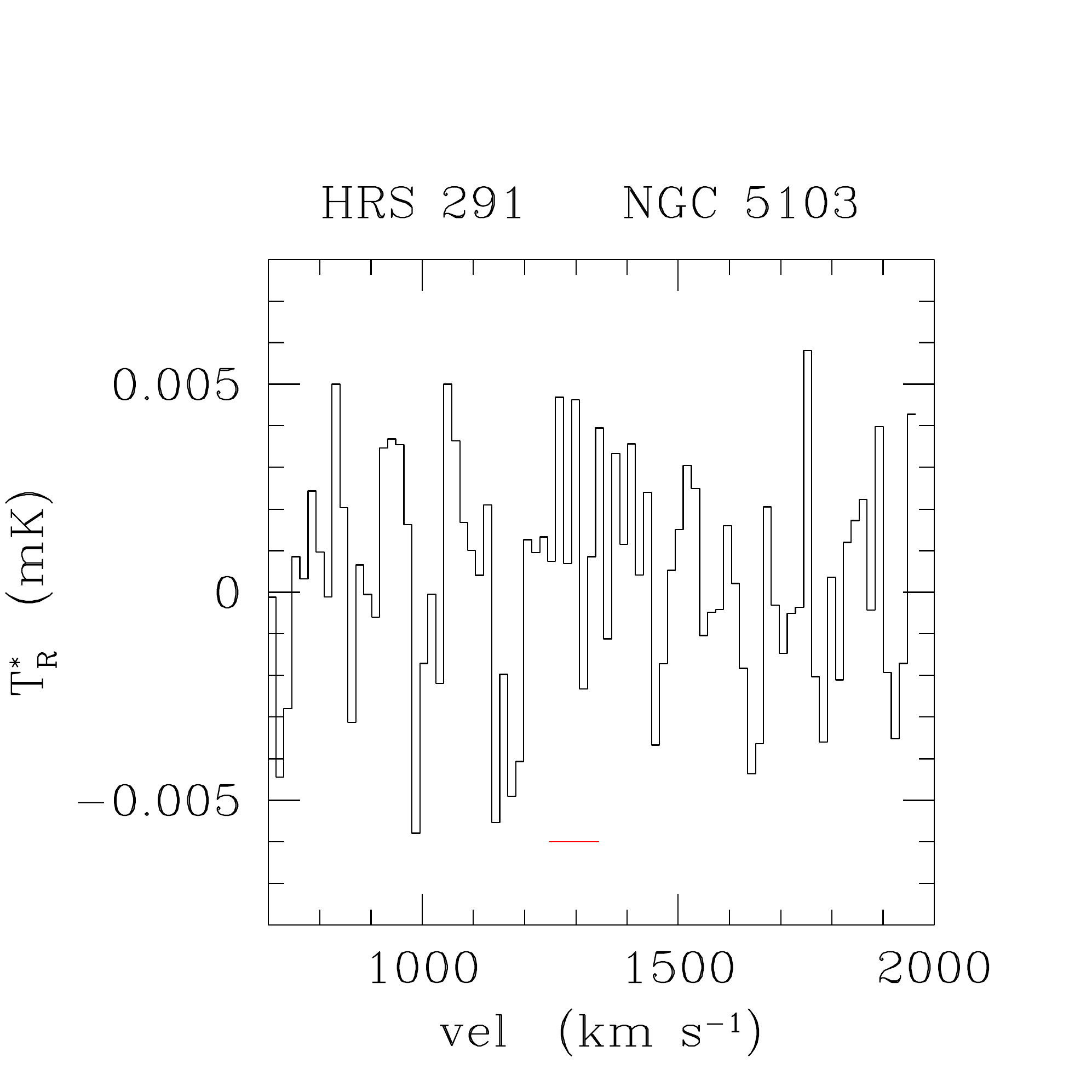}\\
   \includegraphics[width=0.22\textwidth]{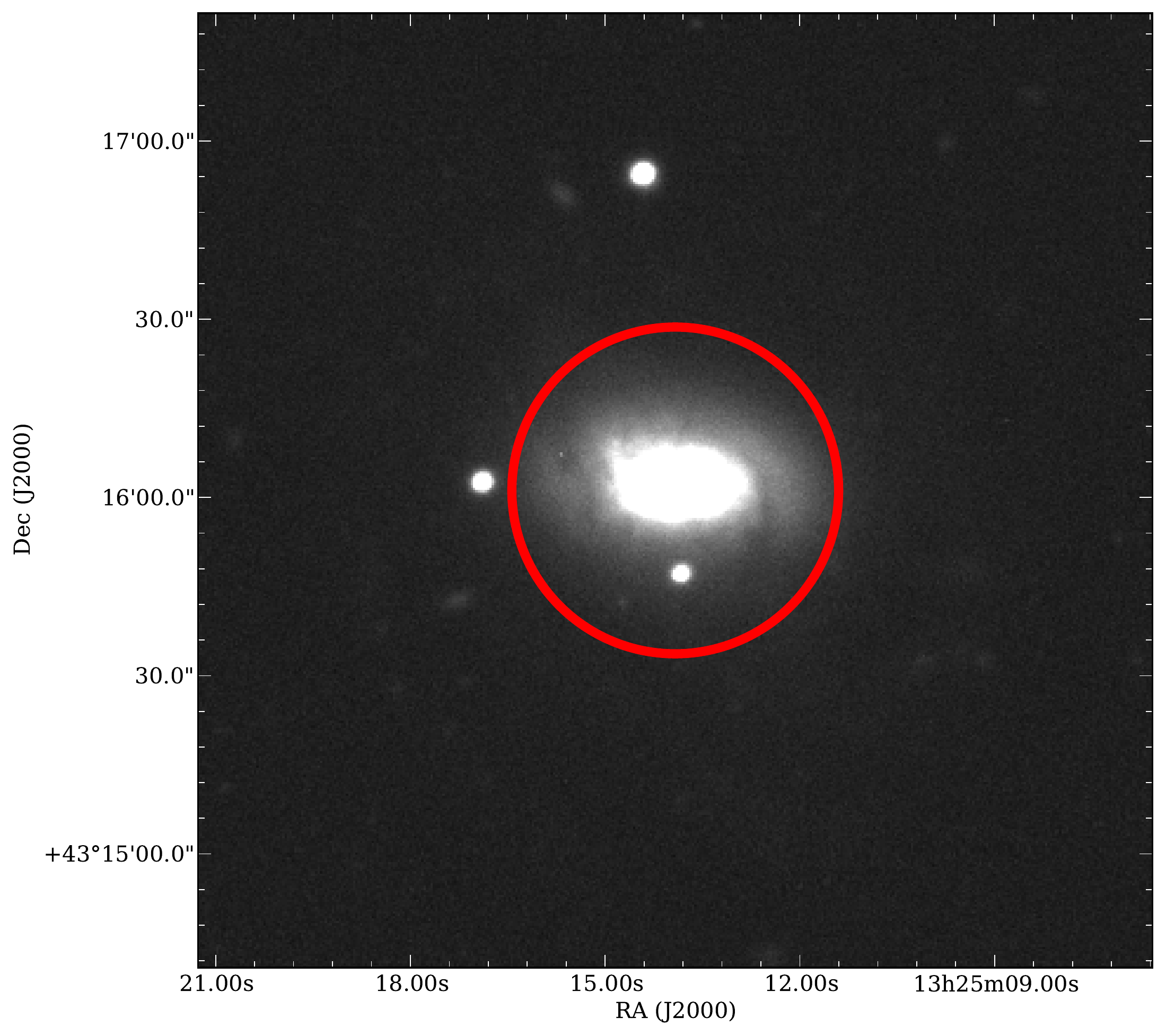}
   \includegraphics[width=0.22\textwidth]{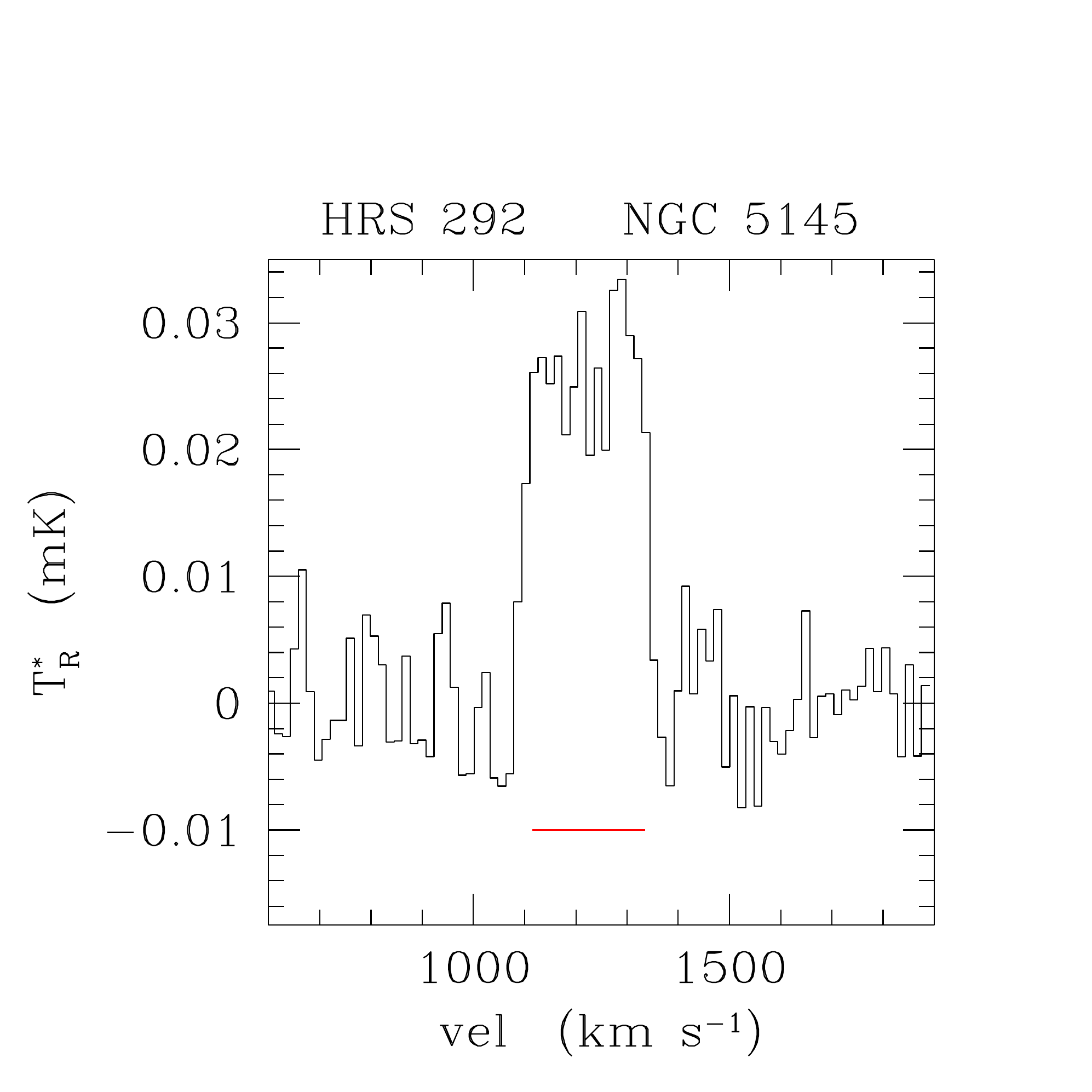}\\
   \caption{Continued.}
   \label{spettri}%
   \end{figure*}
   \clearpage

   \addtocounter{figure}{-1}
   \begin{figure*}
   \centering
   \includegraphics[width=0.22\textwidth]{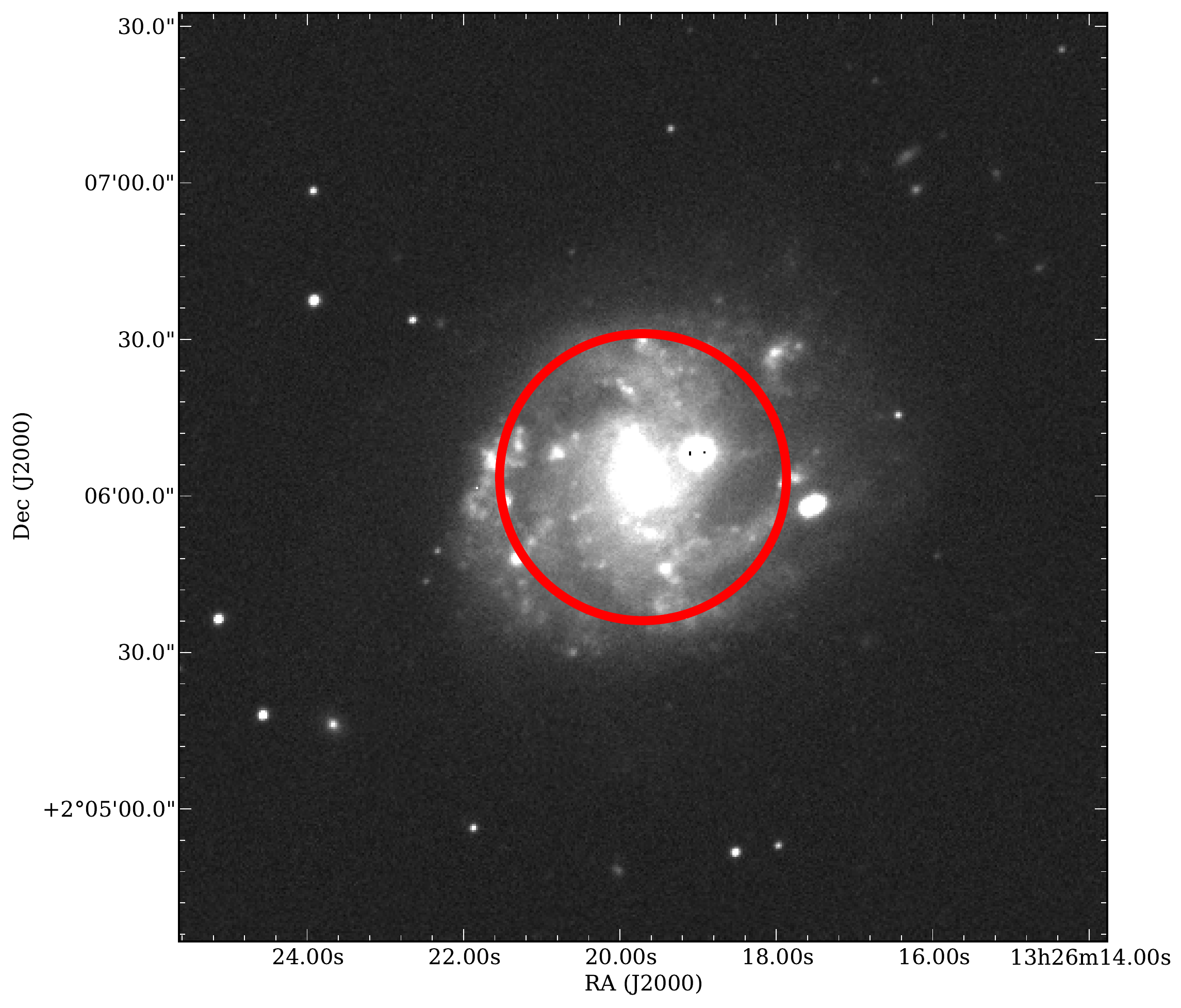}
   \includegraphics[width=0.22\textwidth]{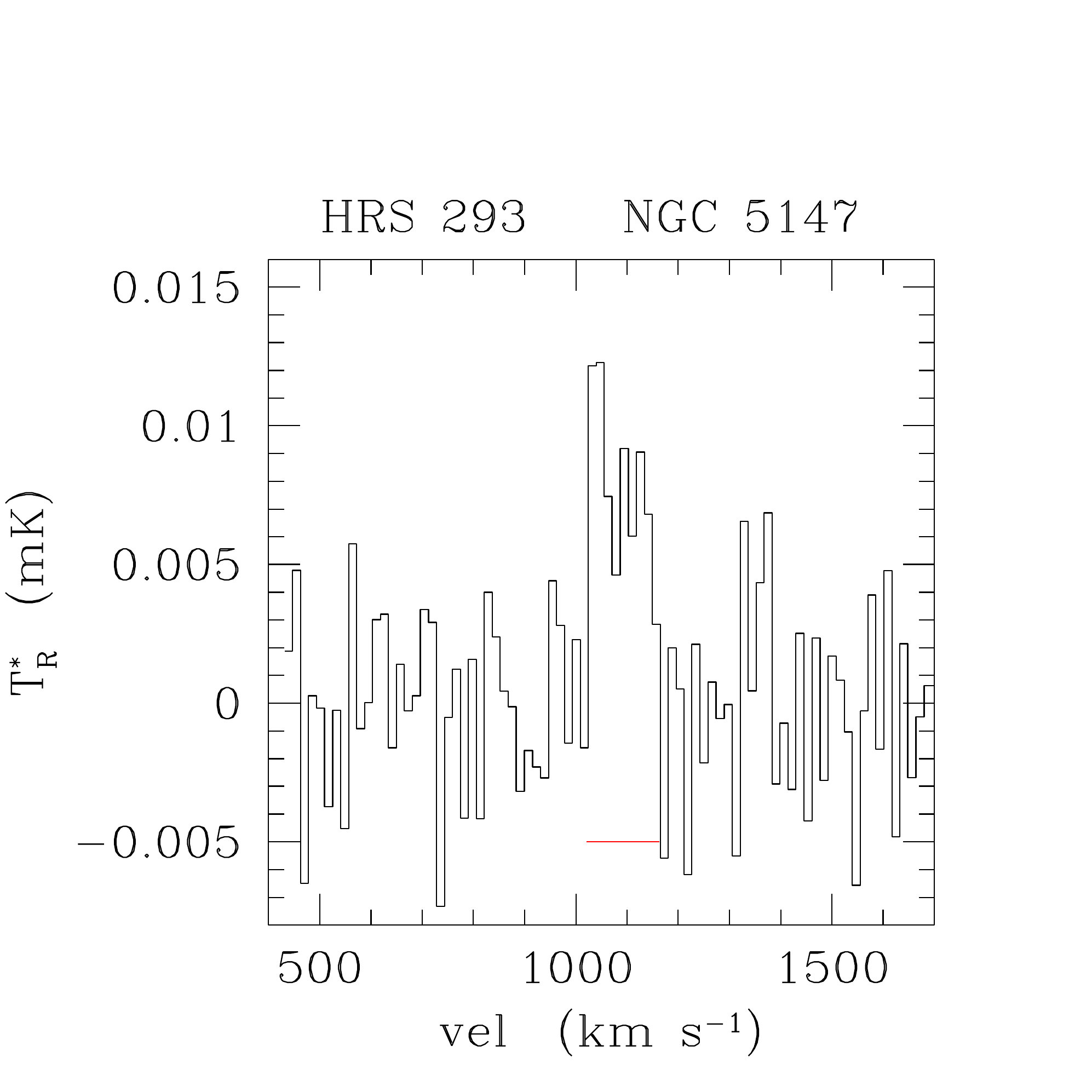}\\
   \includegraphics[width=0.22\textwidth]{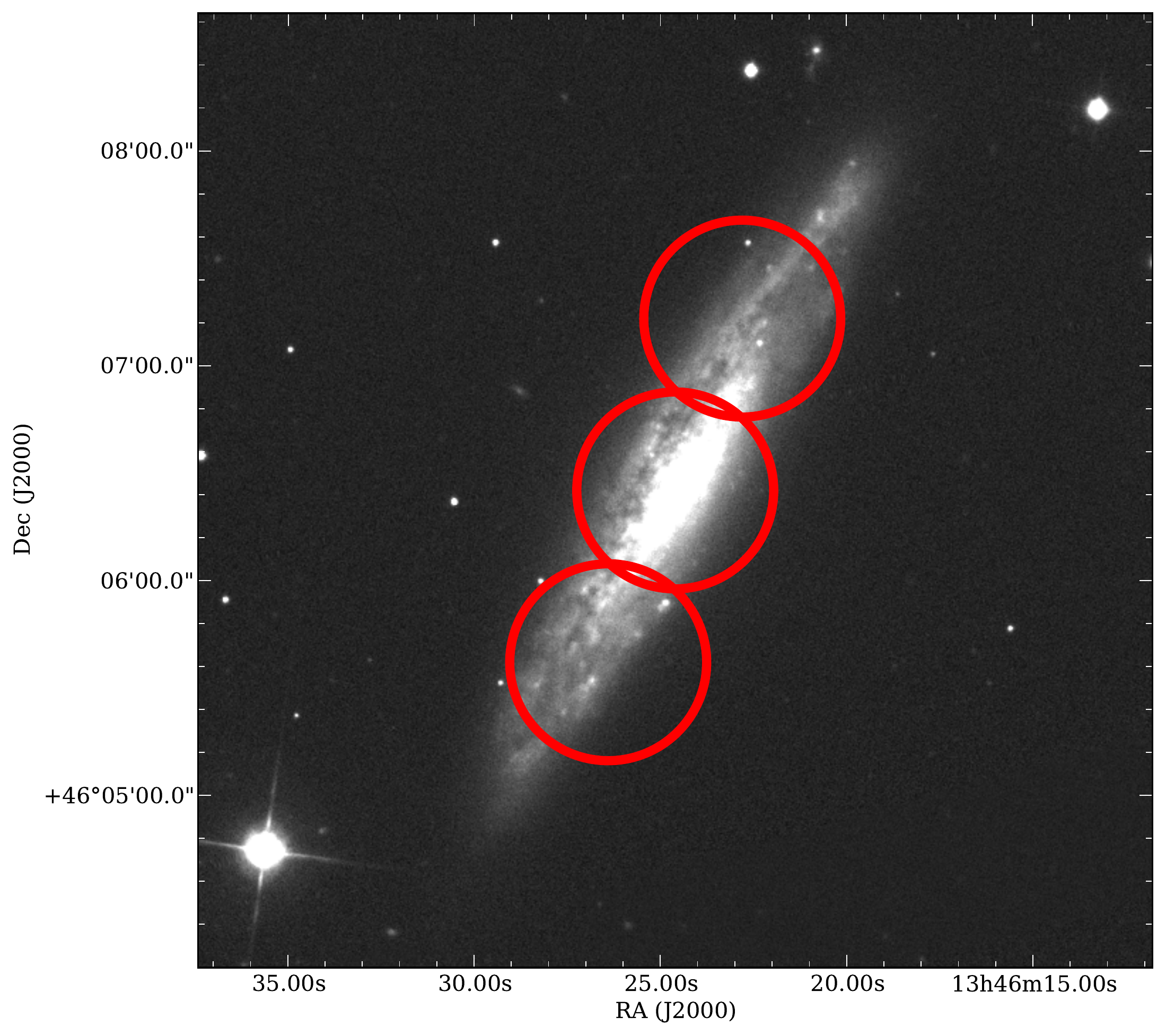}
   \includegraphics[width=0.22\textwidth]{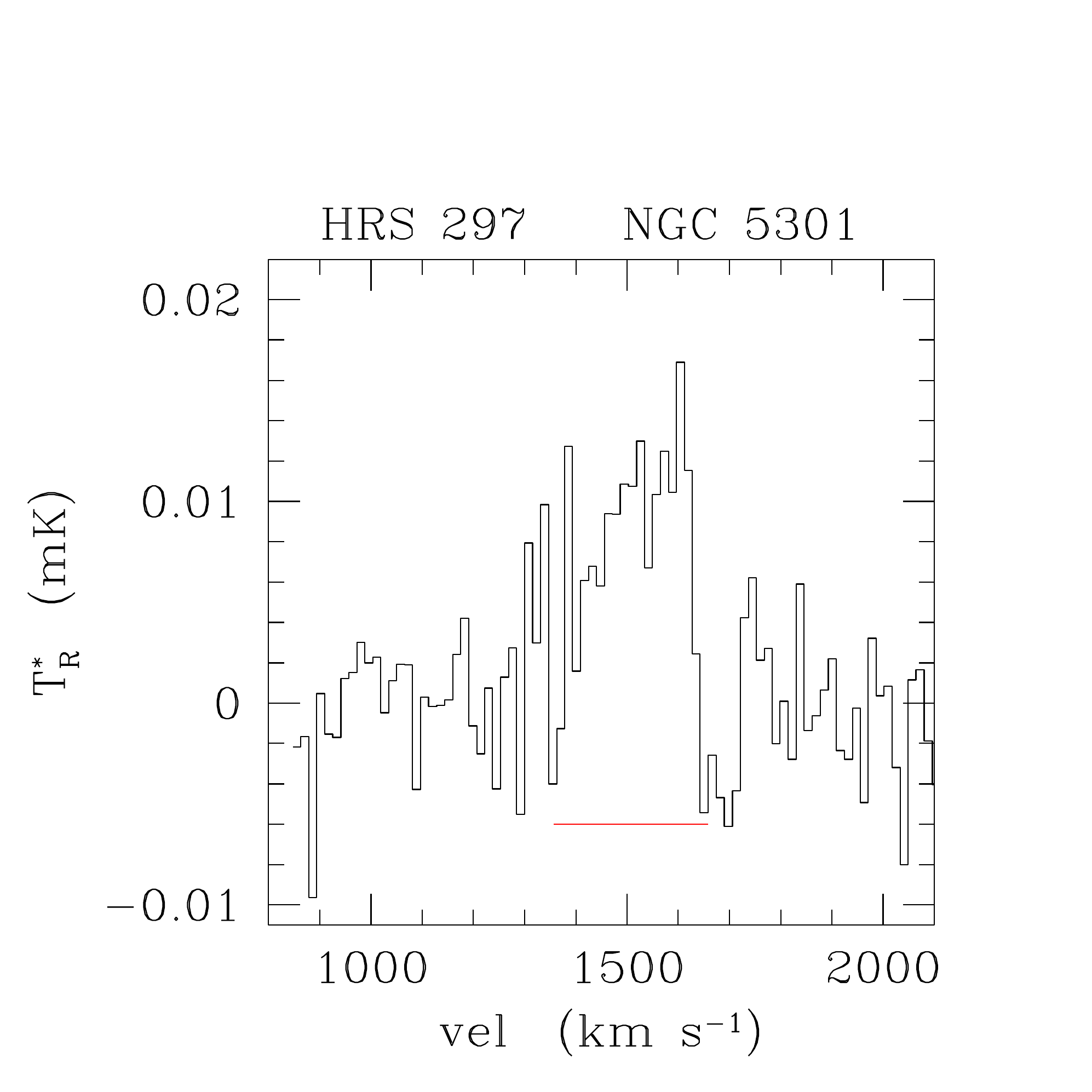}
   \includegraphics[width=0.22\textwidth]{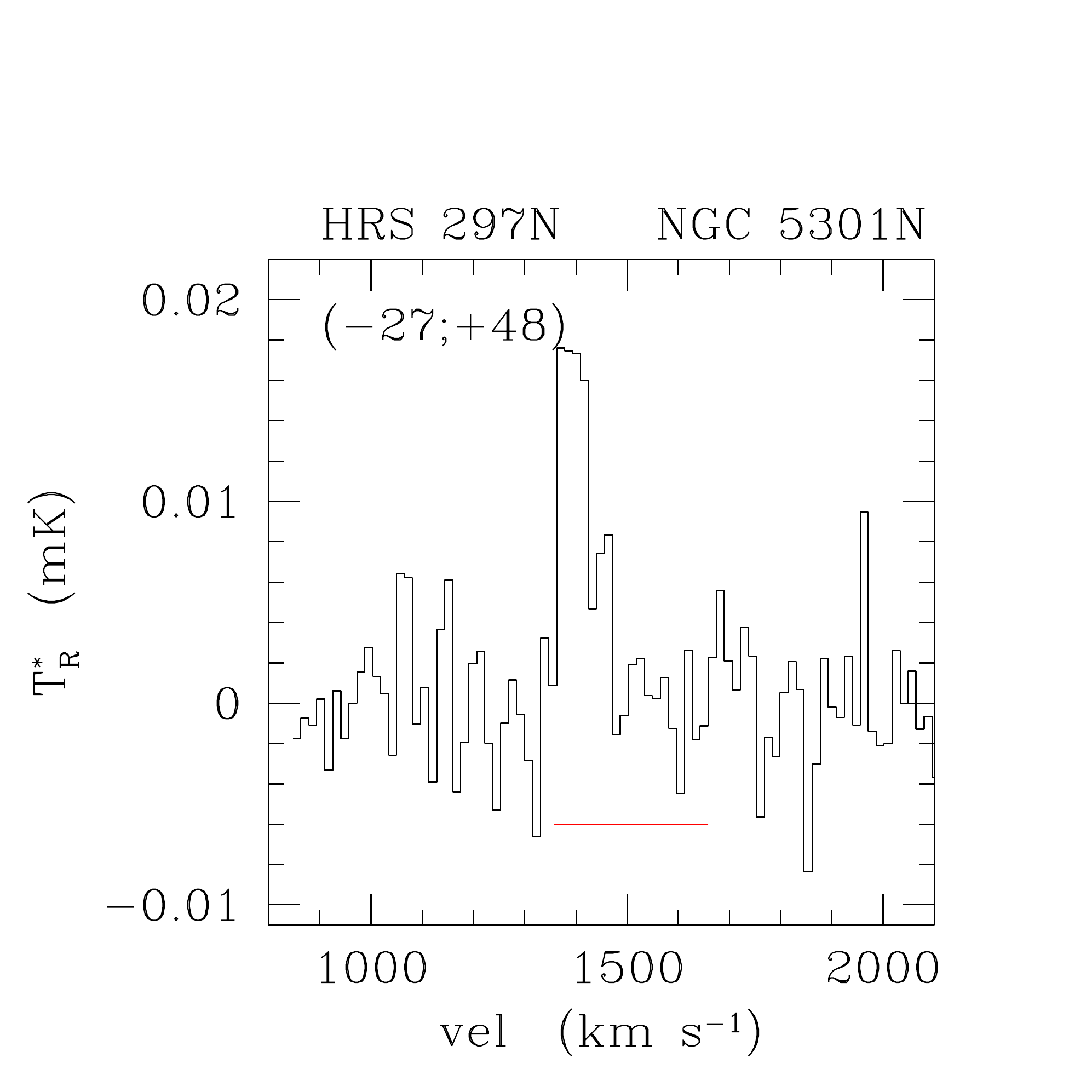}
   \includegraphics[width=0.22\textwidth]{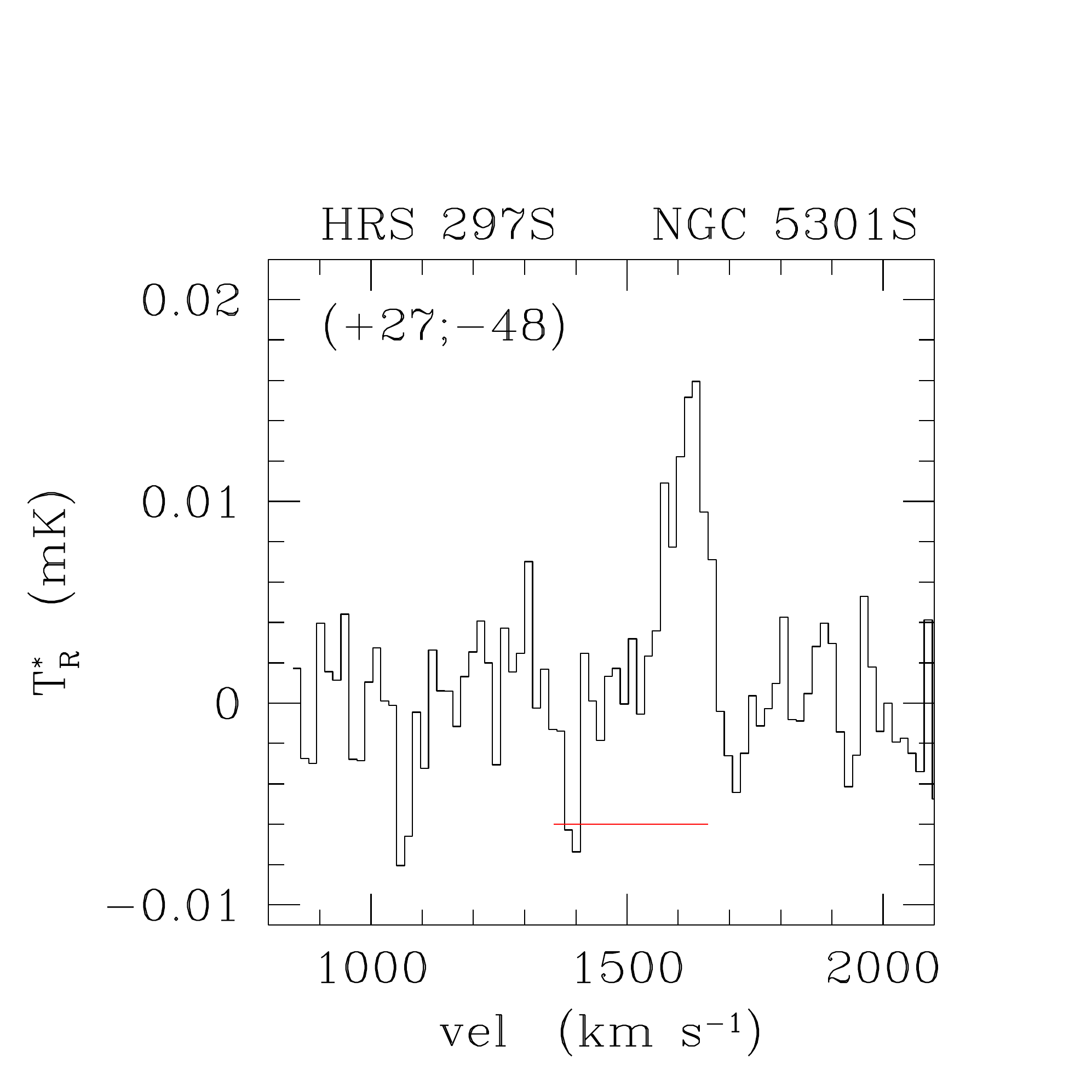}\\
   \includegraphics[width=0.22\textwidth]{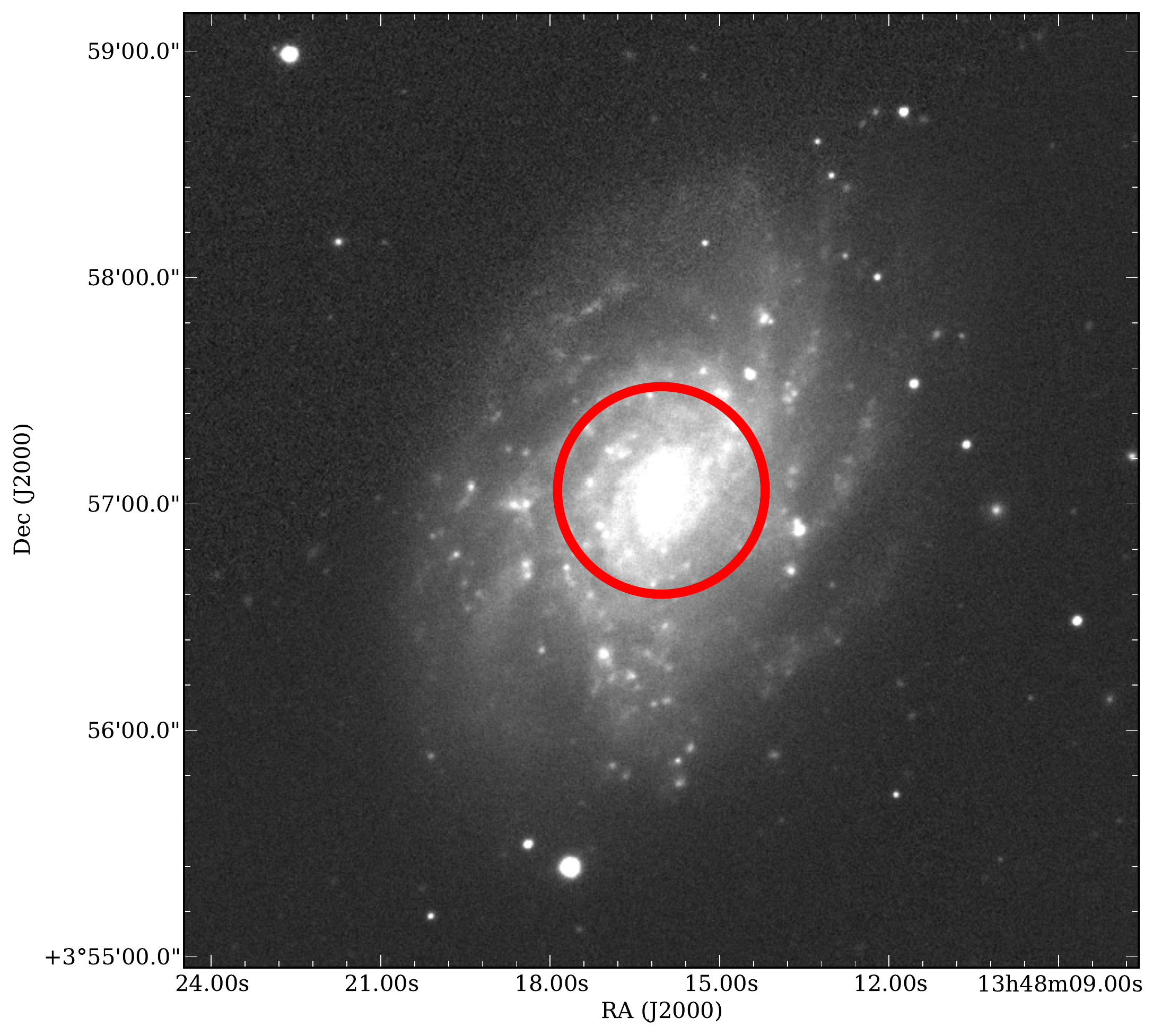}
   \includegraphics[width=0.22\textwidth]{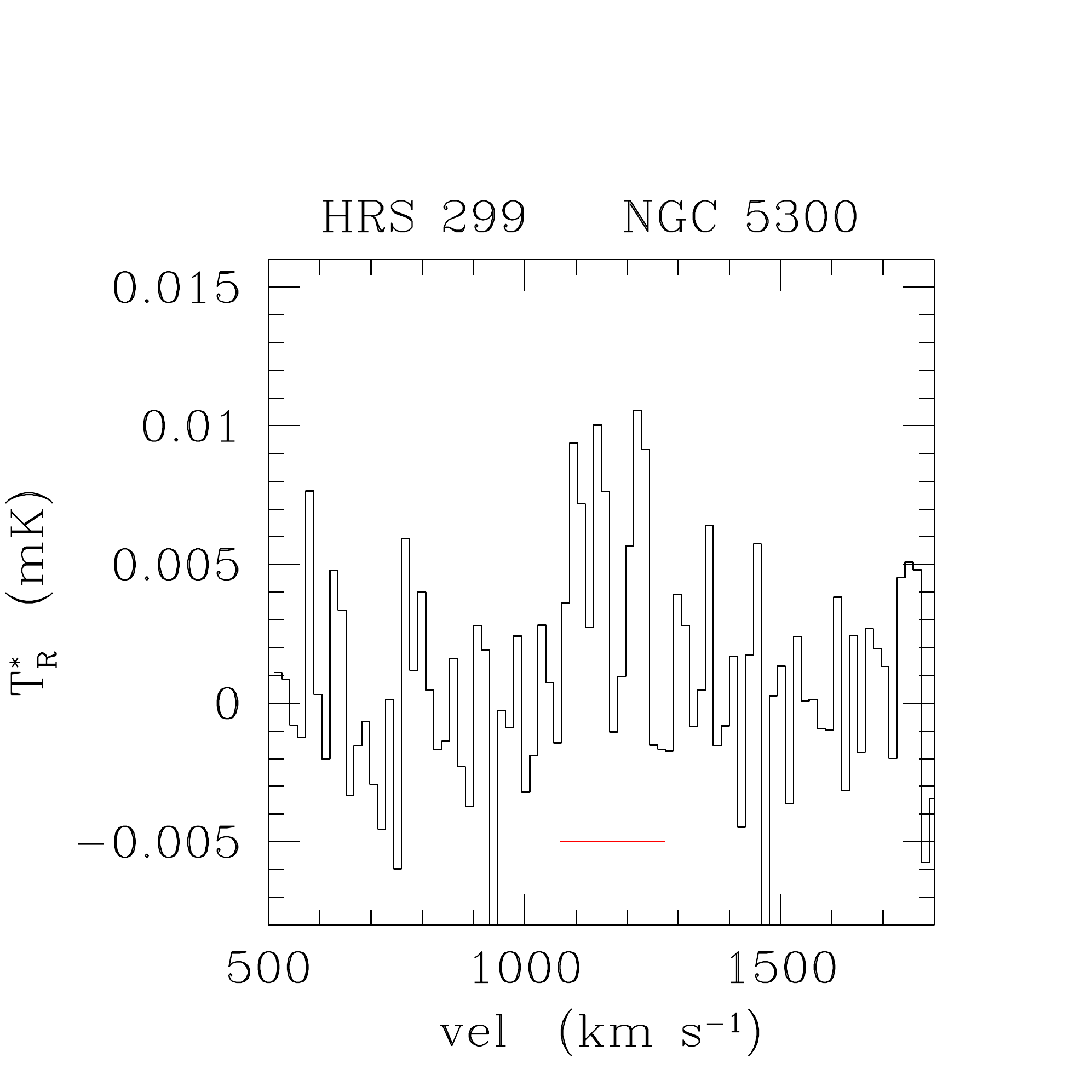}\\
   \includegraphics[width=0.22\textwidth]{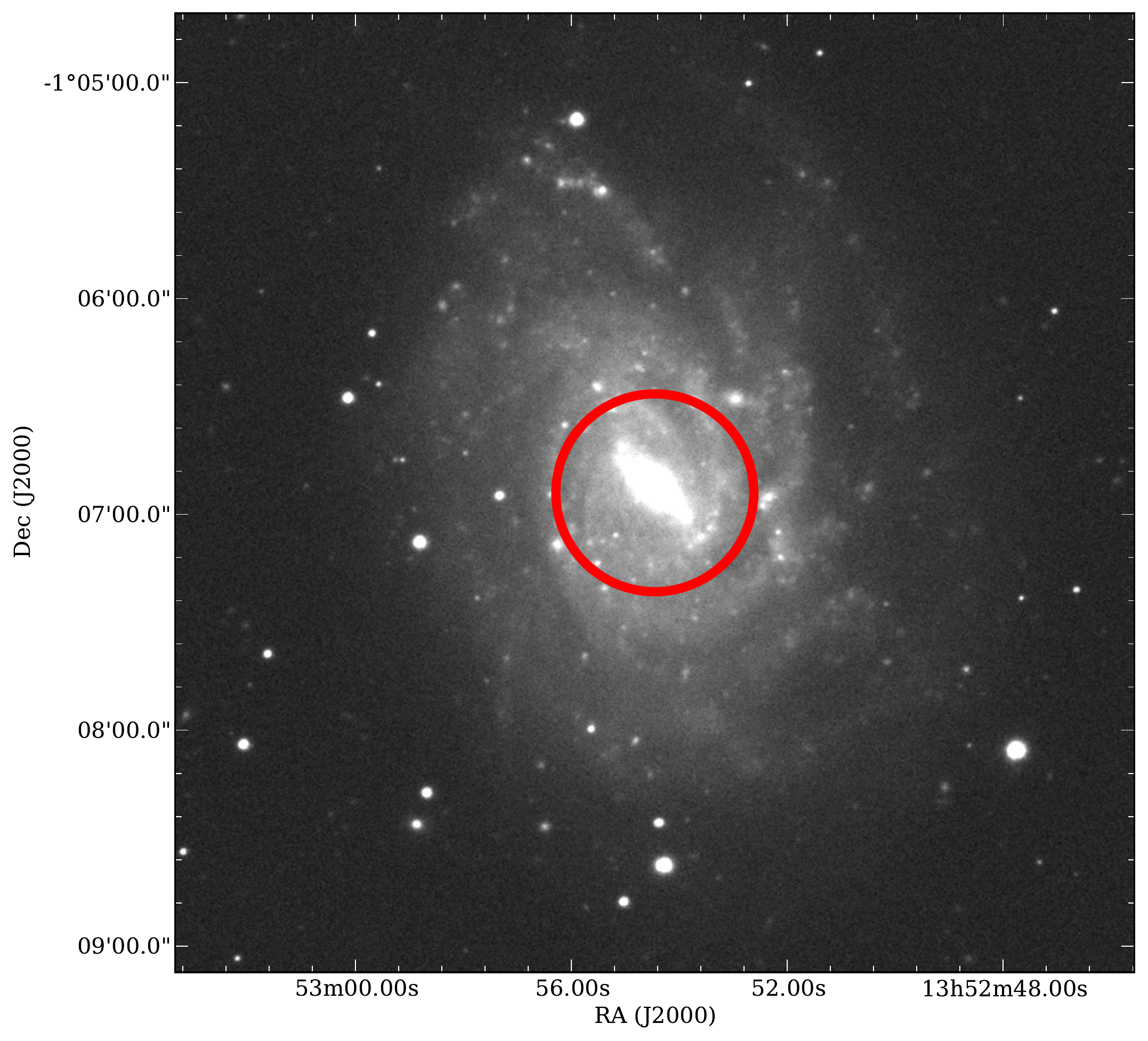}
   \includegraphics[width=0.22\textwidth]{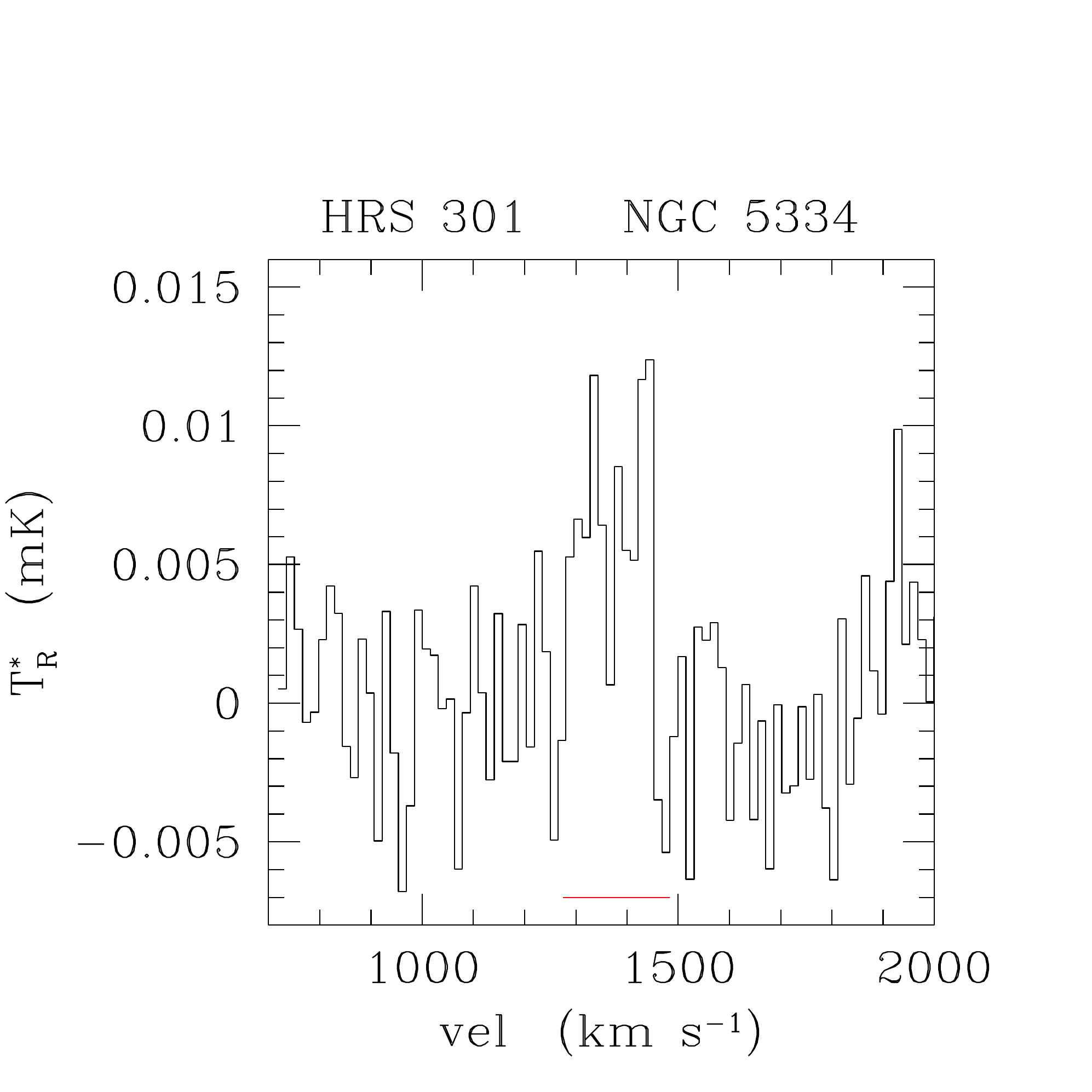}\\
      \includegraphics[width=0.22\textwidth]{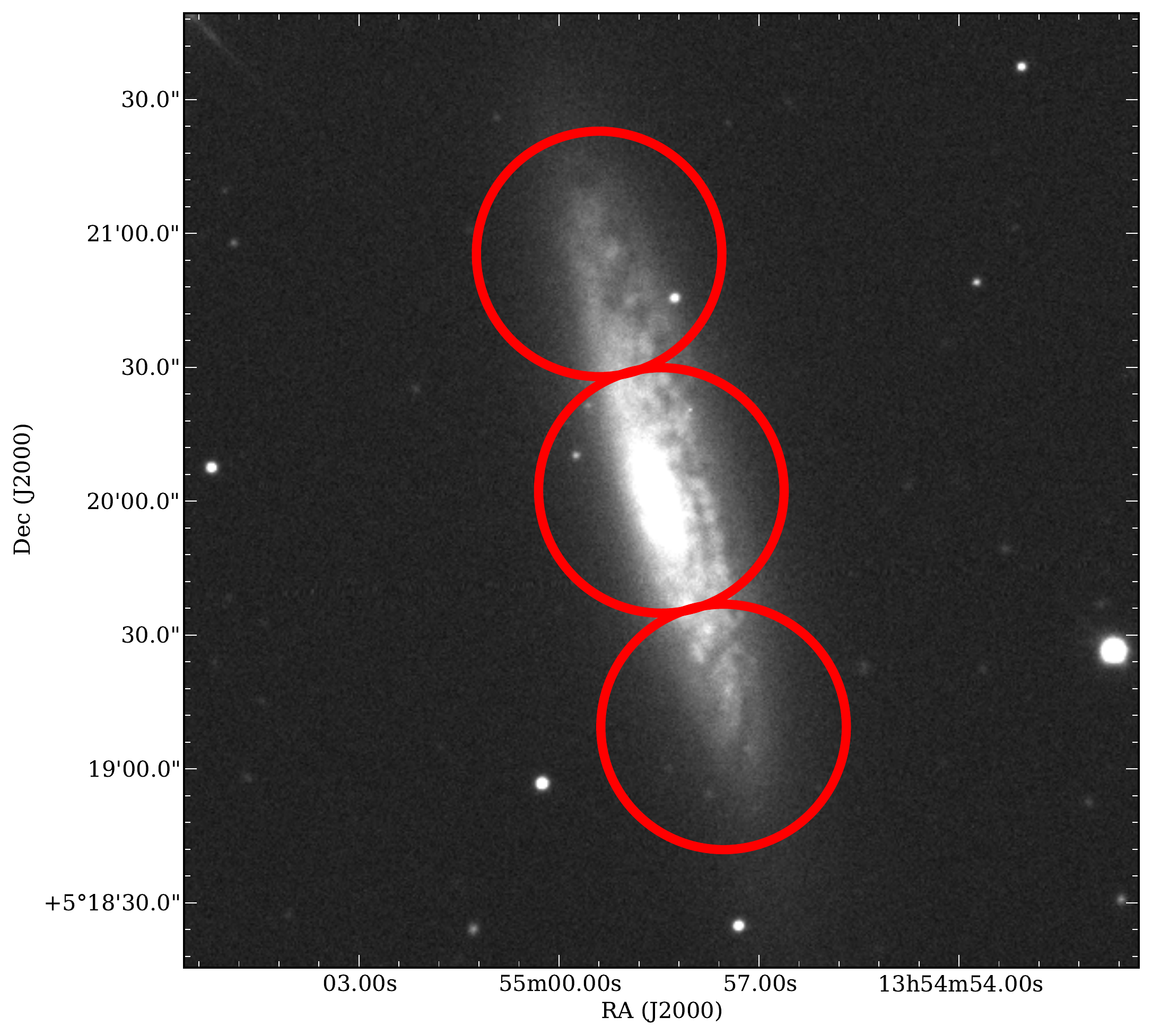}
   \includegraphics[width=0.22\textwidth]{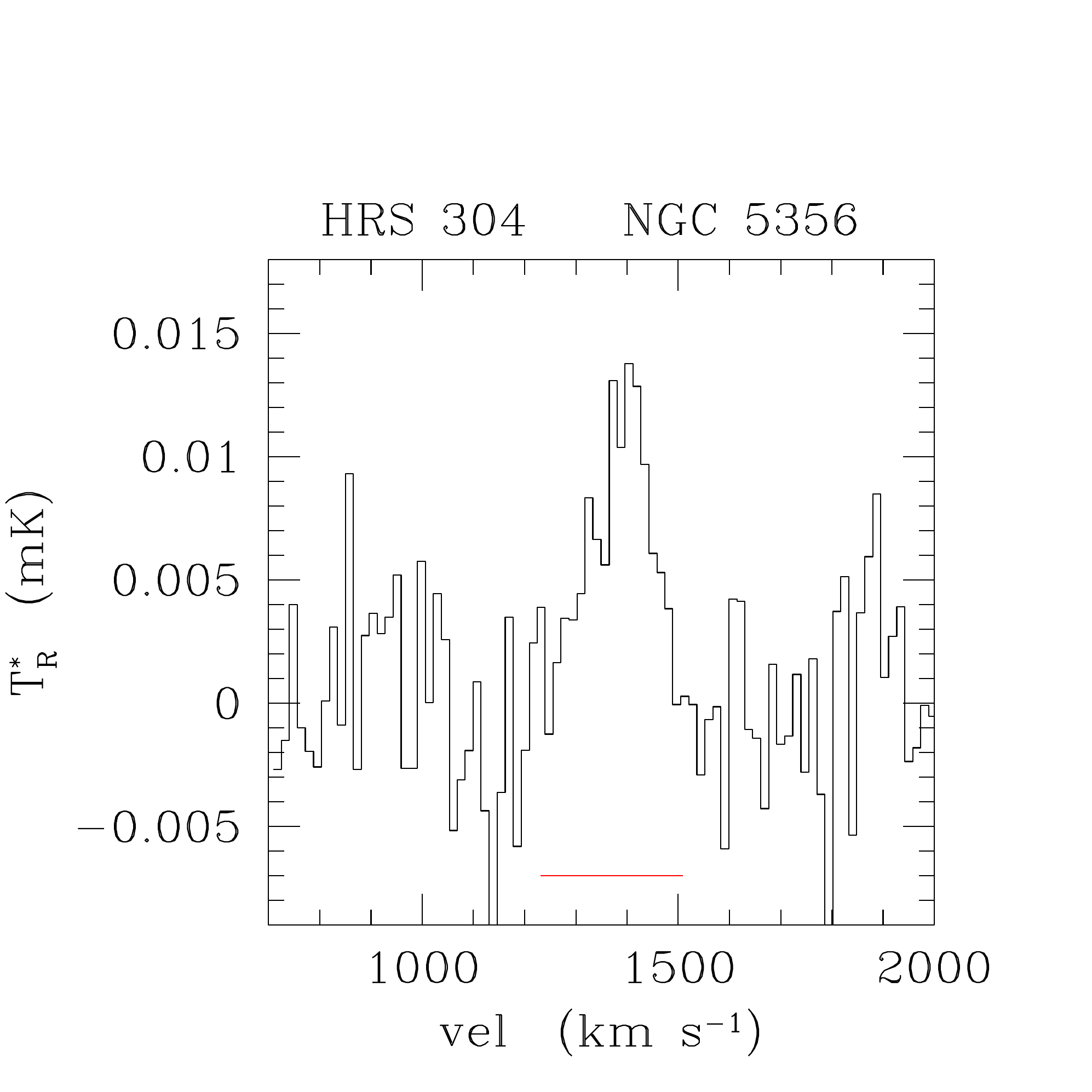}
   \includegraphics[width=0.22\textwidth]{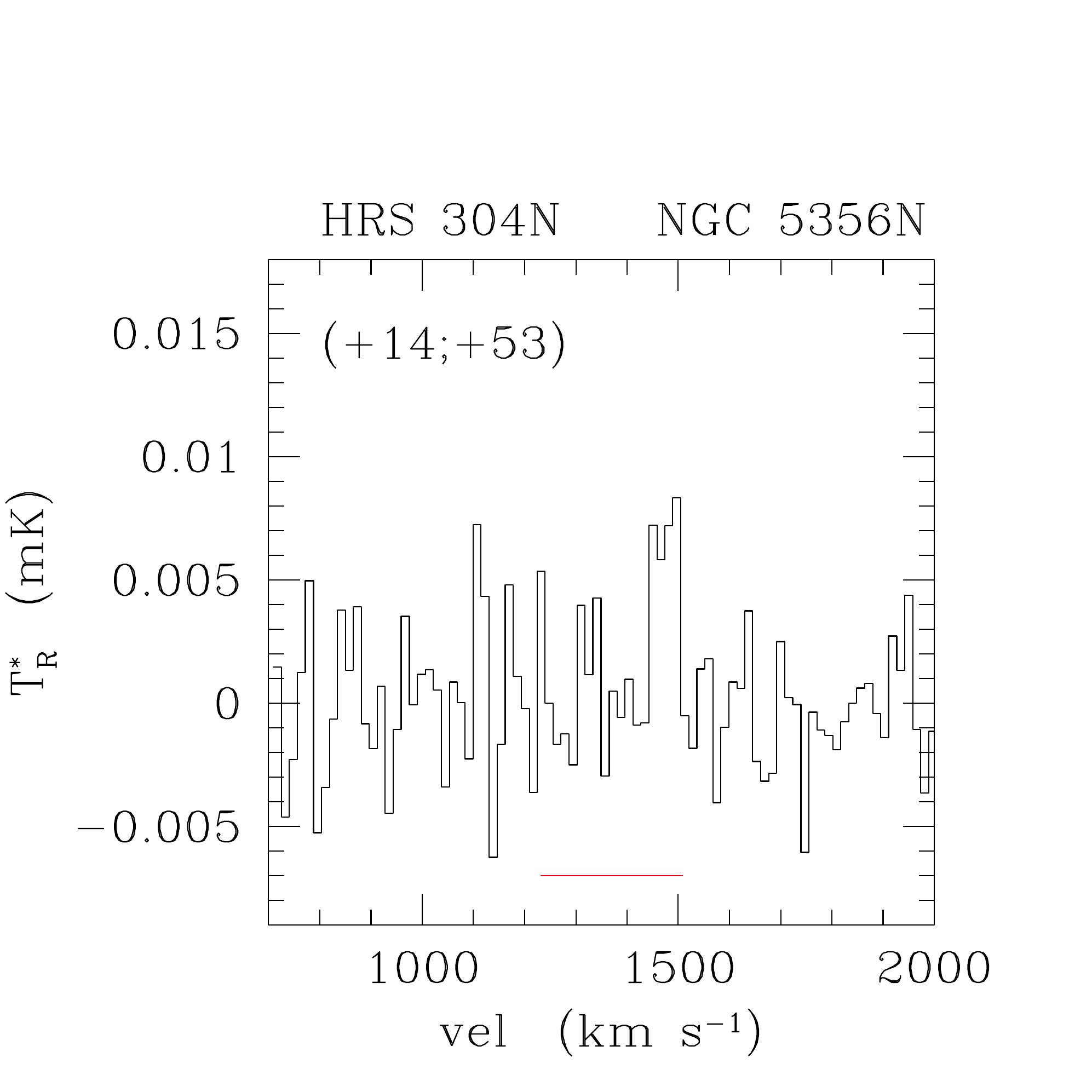}
   \includegraphics[width=0.22\textwidth]{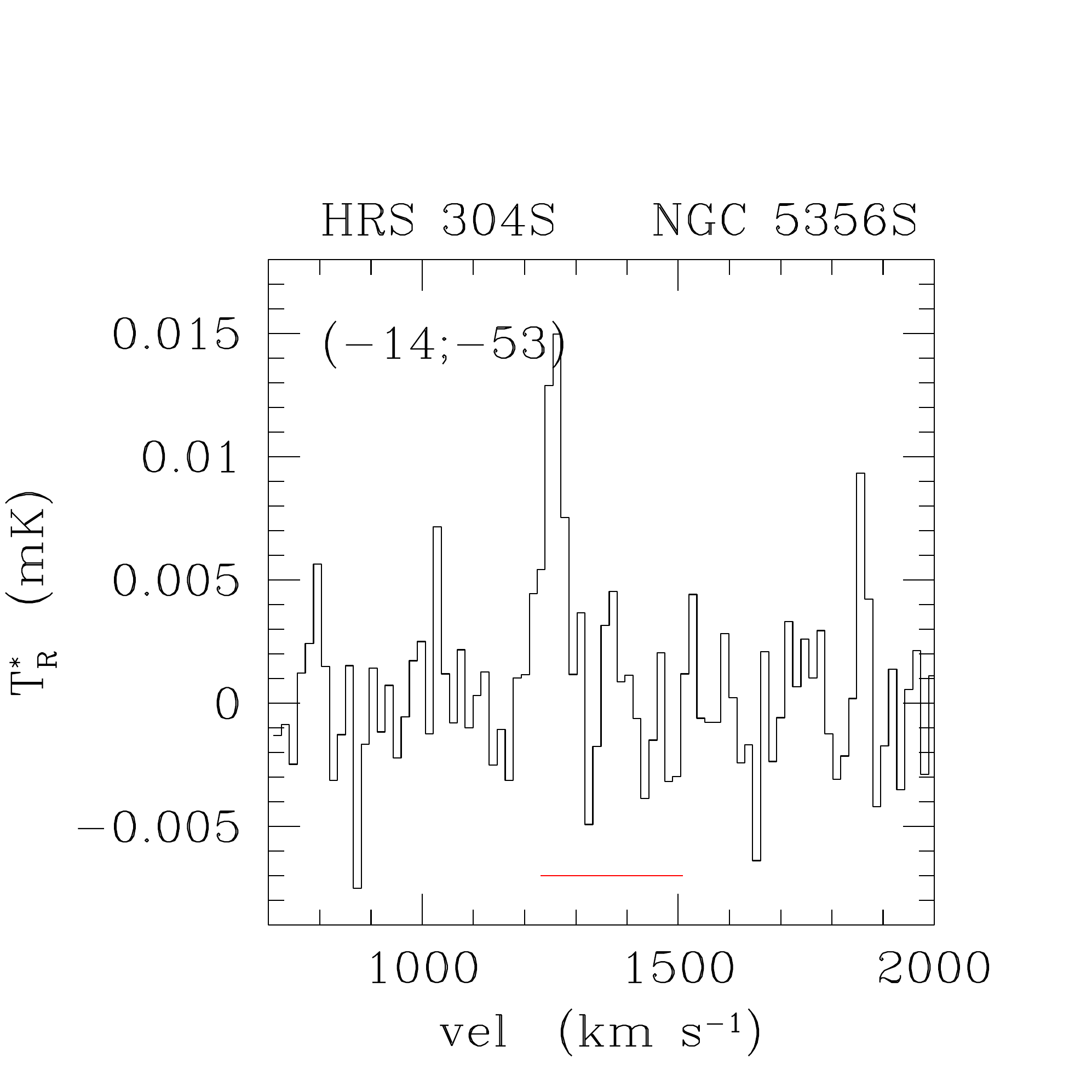}\\

   \caption{Continued.}
   \label{spettri}%
   \end{figure*}
   \clearpage

   \addtocounter{figure}{-1}
   \begin{figure*}
   \centering   
   \includegraphics[width=0.22\textwidth]{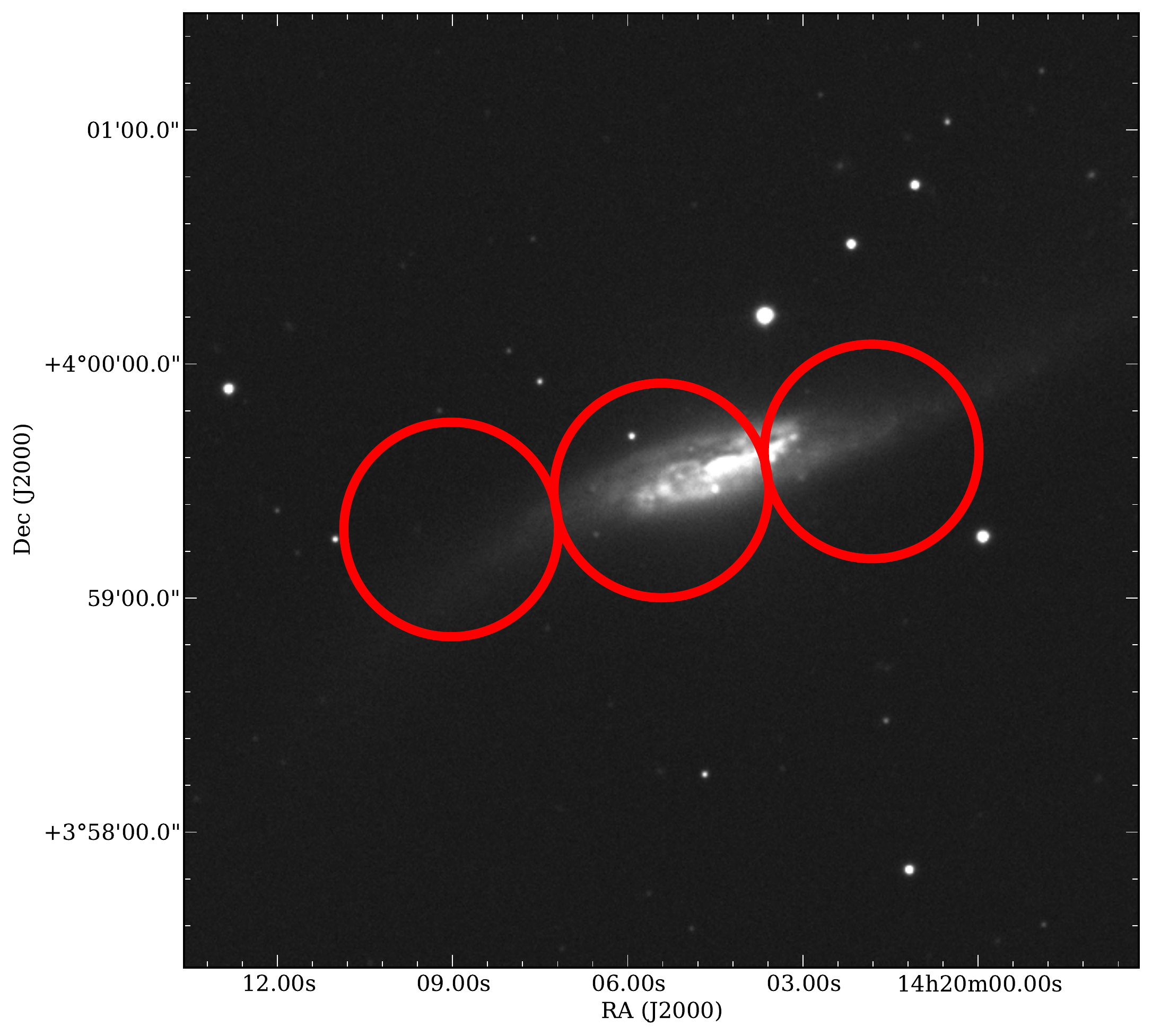}
   \includegraphics[width=0.22\textwidth]{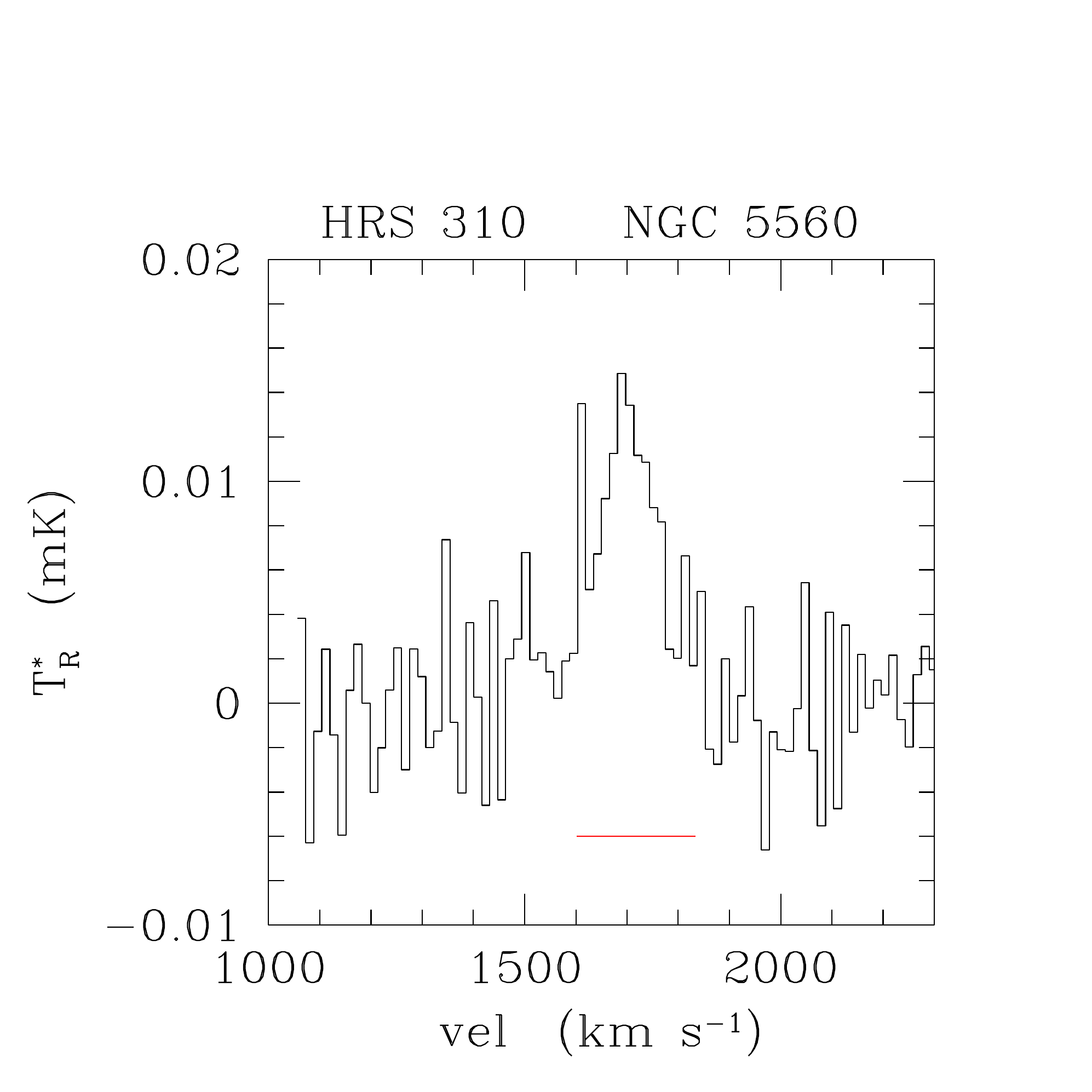}
   \includegraphics[width=0.22\textwidth]{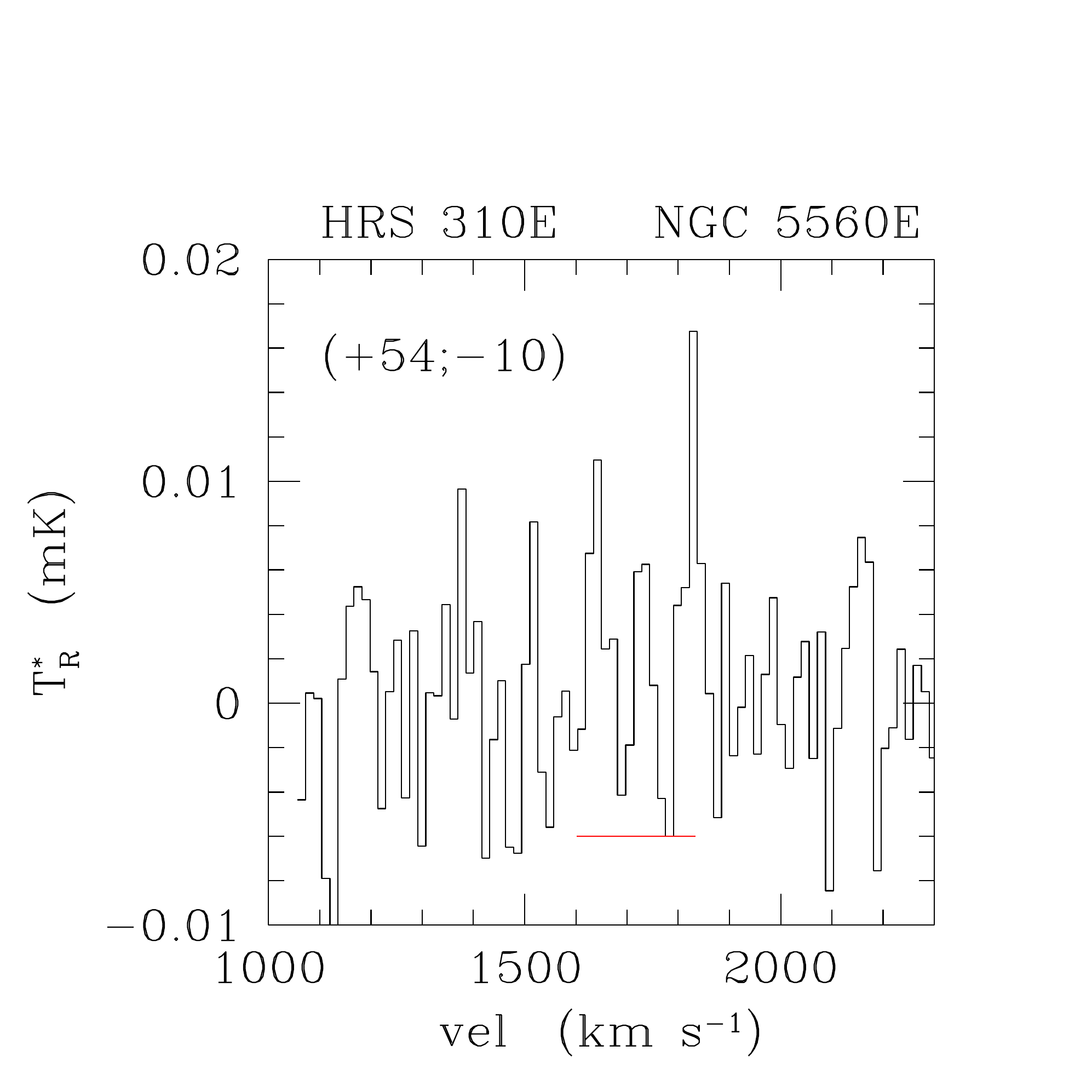}
   \includegraphics[width=0.22\textwidth]{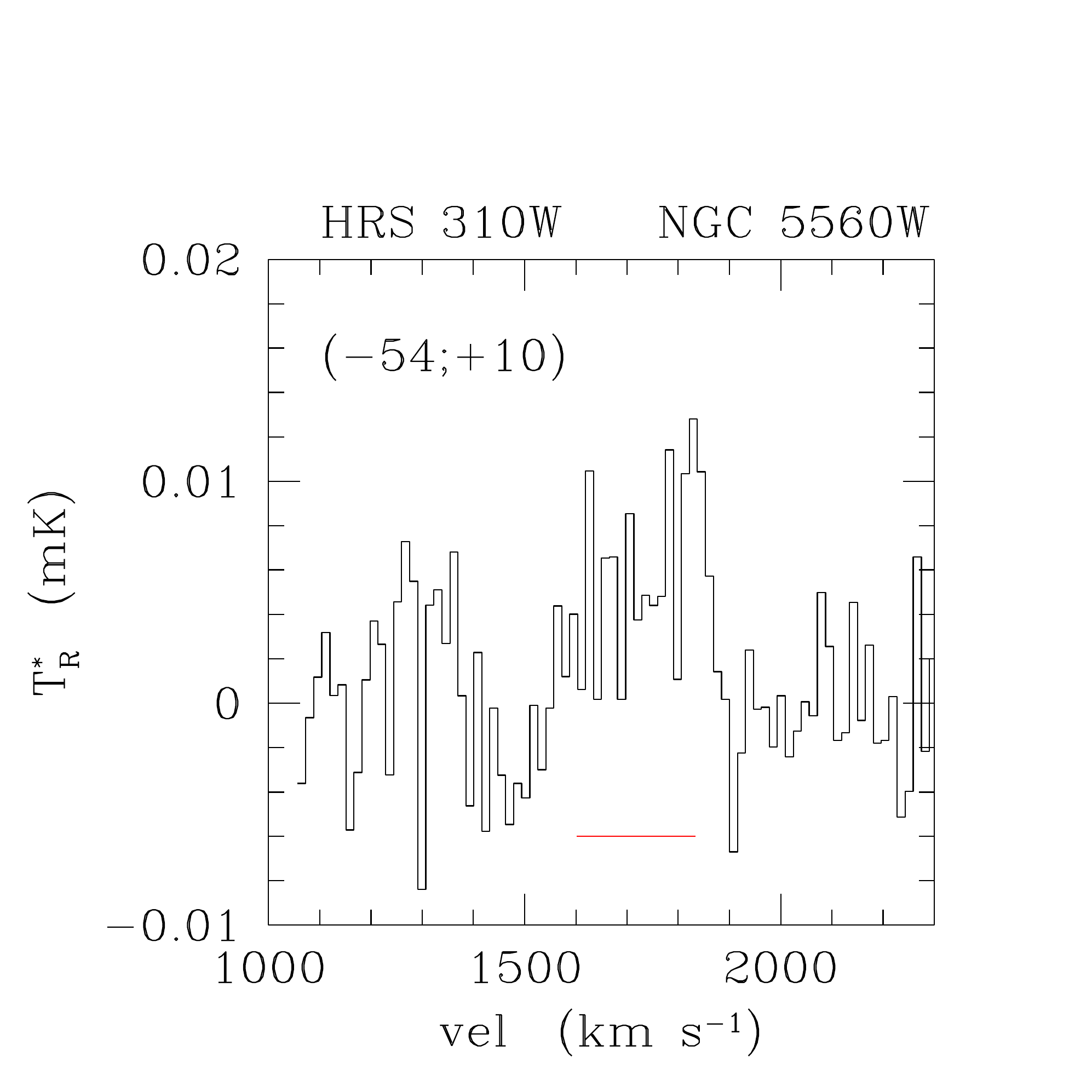}\\
   \includegraphics[width=0.22\textwidth]{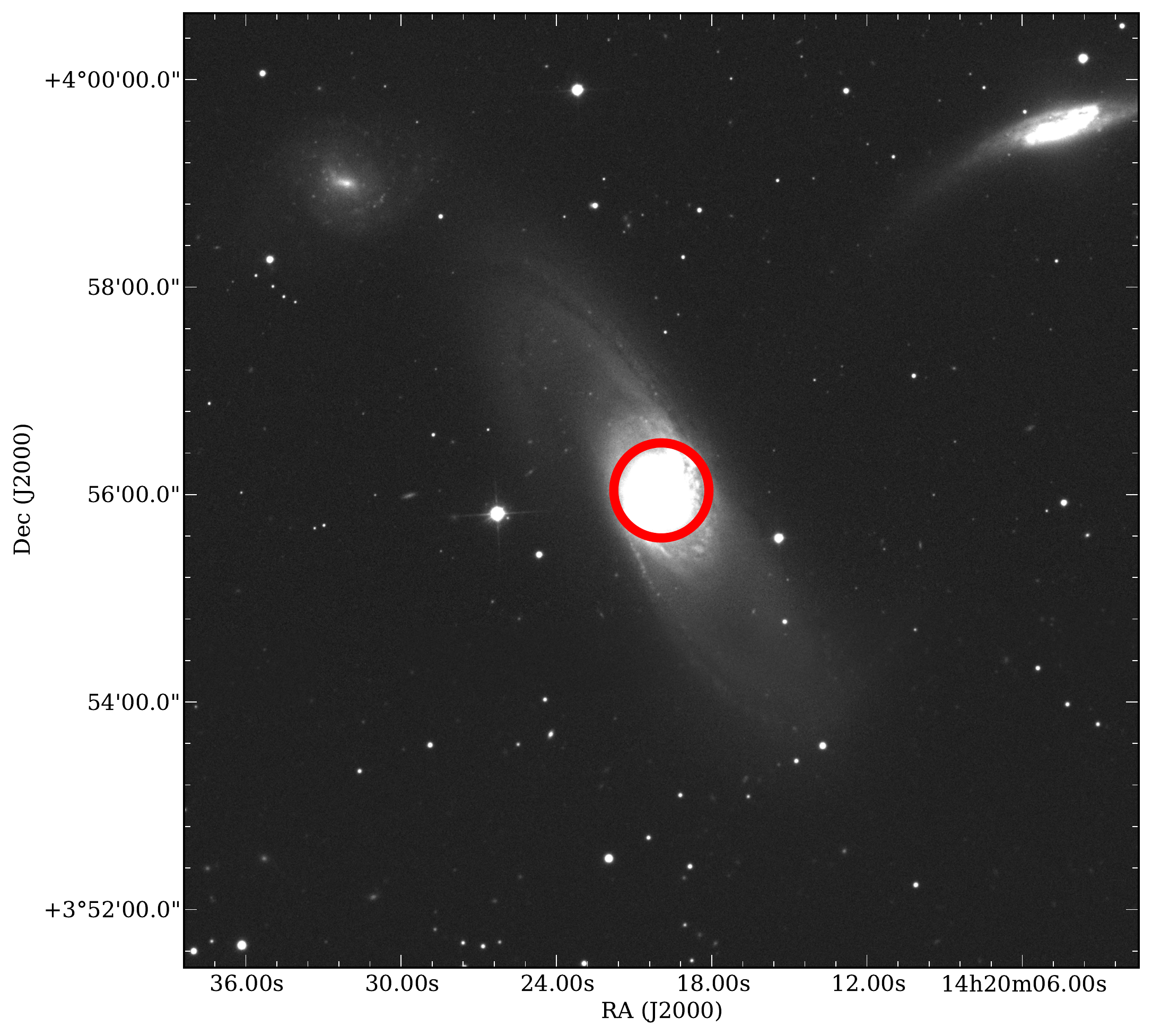}
   \includegraphics[width=0.22\textwidth]{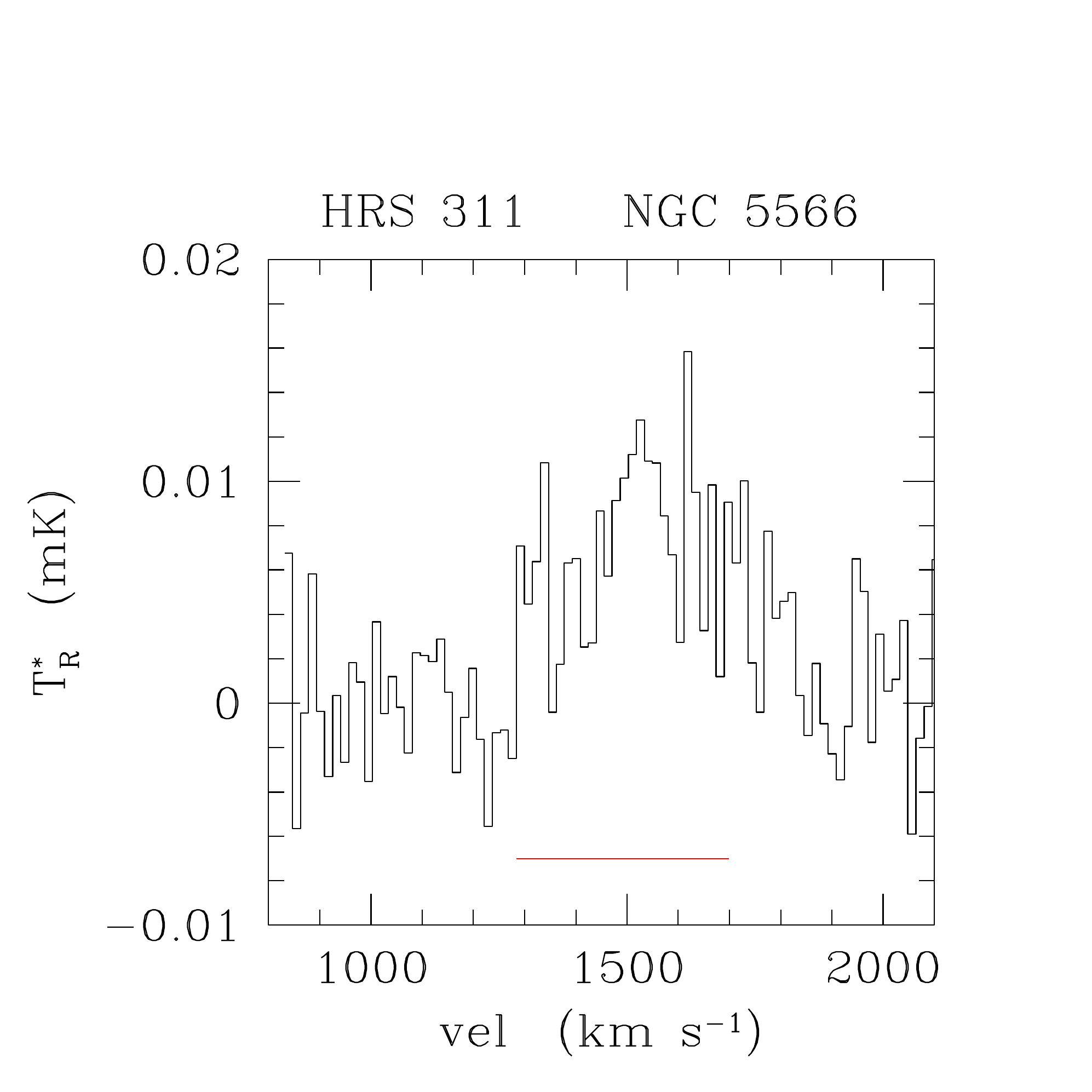}\\
   \includegraphics[width=0.22\textwidth]{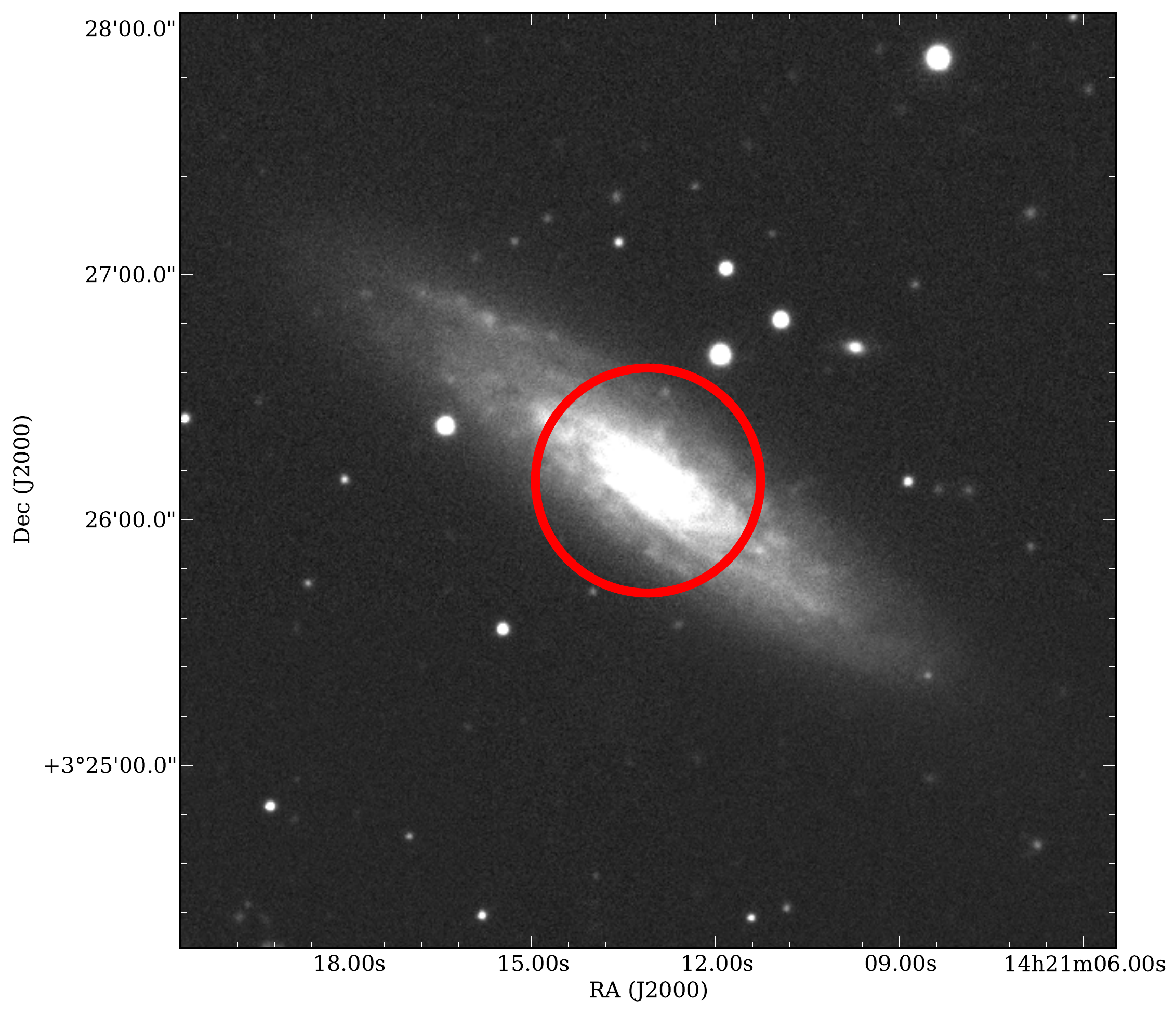}
   \includegraphics[width=0.22\textwidth]{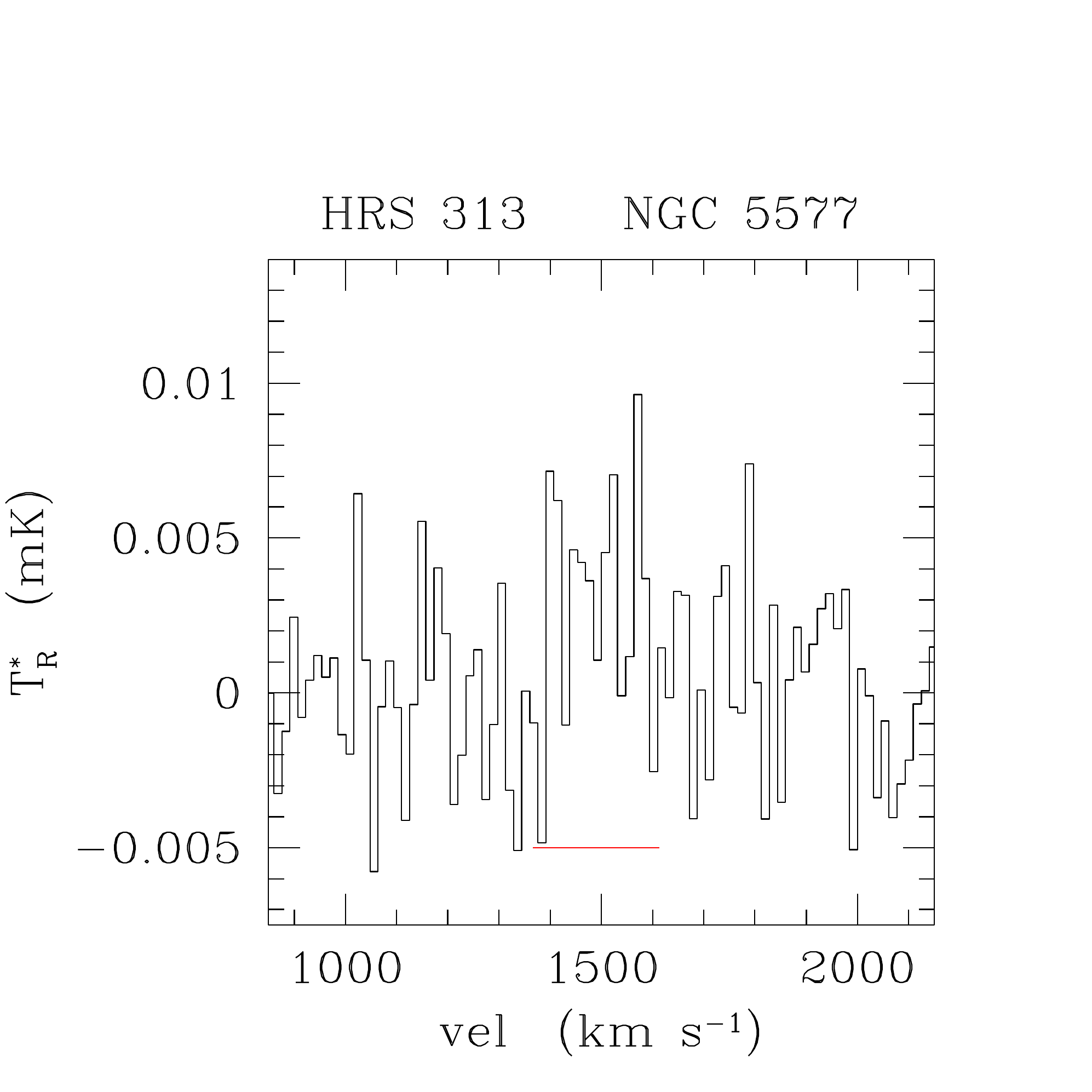}\\
   \includegraphics[width=0.22\textwidth]{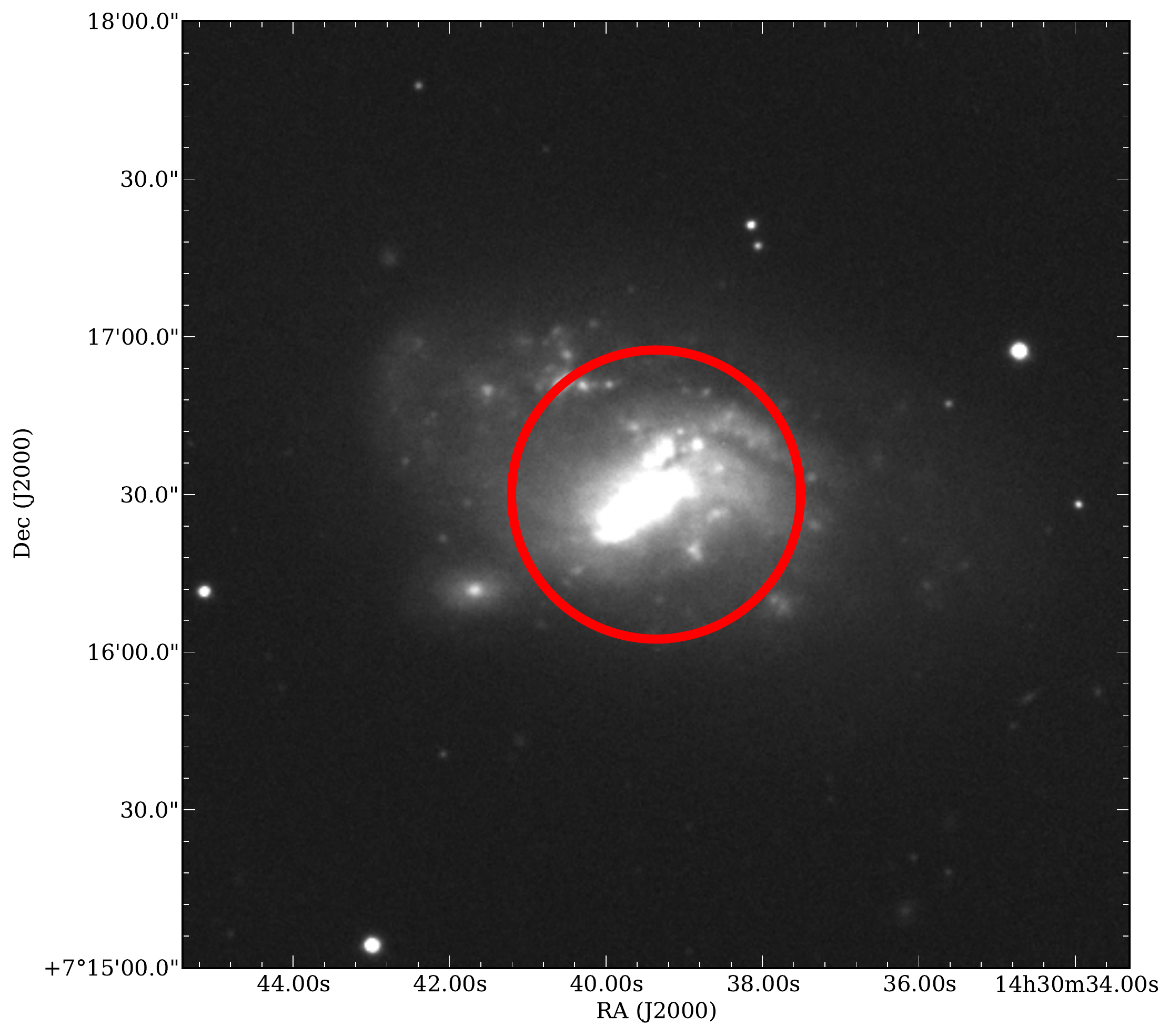}
   \includegraphics[width=0.22\textwidth]{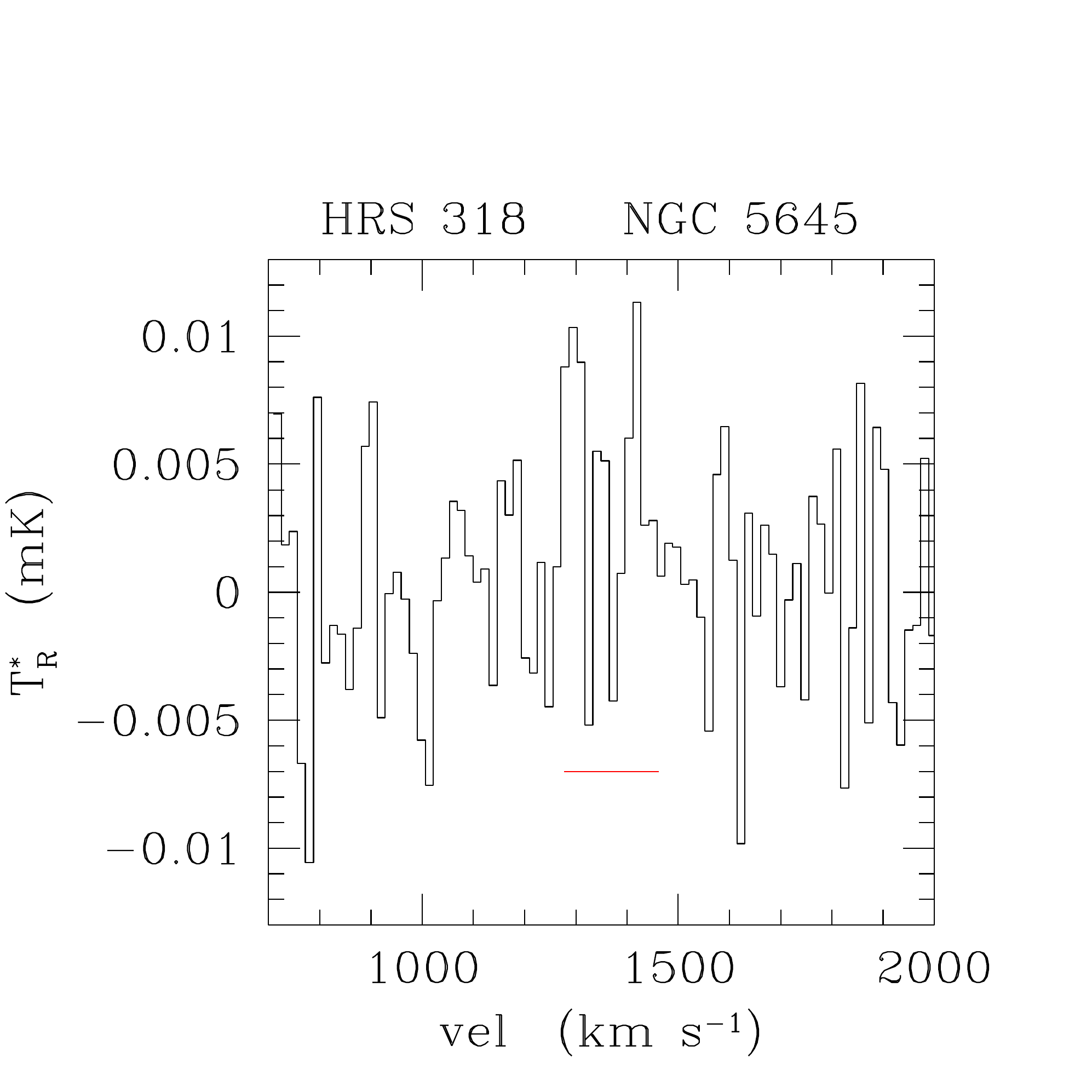}\\
   \includegraphics[width=0.22\textwidth]{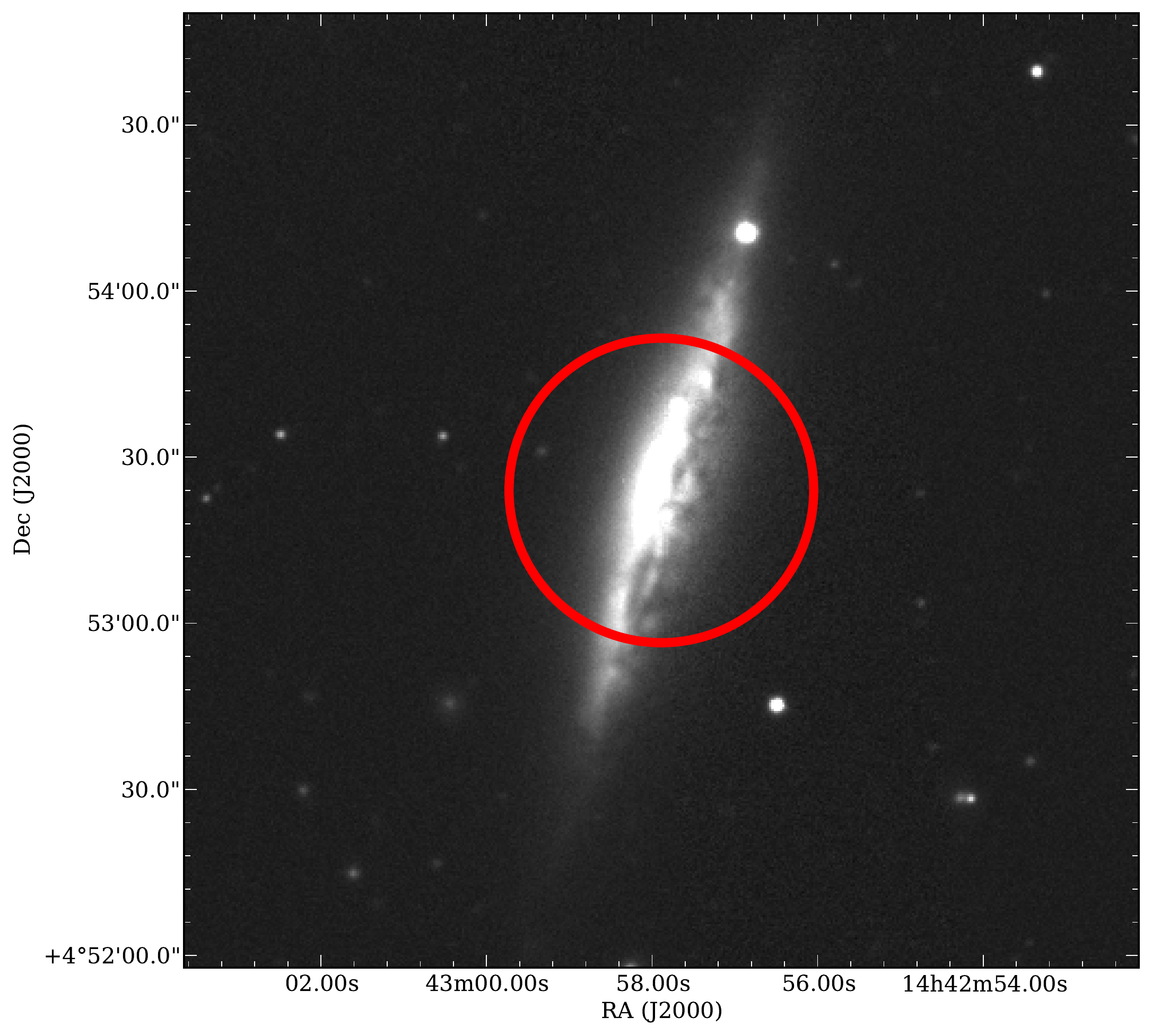}
   \includegraphics[width=0.22\textwidth]{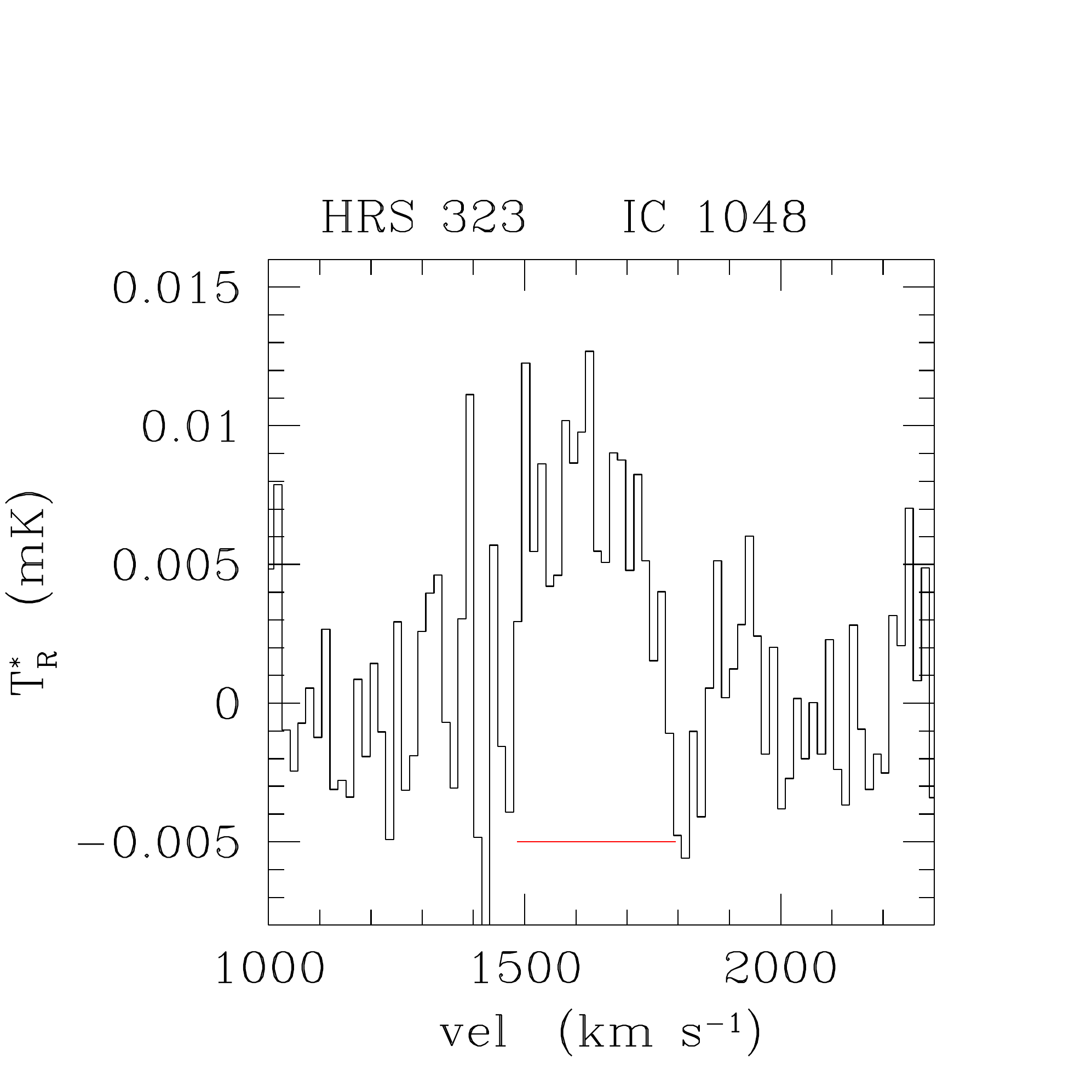}\\
   \caption{Continued.}
   \label{spettri}%
   \end{figure*}
   \clearpage

\end{document}